\documentclass[a4paper,11pt]{book}
\usepackage[utf8]{inputenc}
\usepackage[T1]{fontenc}
\usepackage{imakeidx}
\usepackage{lmodern}
\usepackage{xspace}			   

\usepackage[english]{babel}

\usepackage[margin=28mm,includeheadfoot,bindingoffset=0mm]{geometry}
\usepackage{graphicx,color} 
\usepackage{subfigure} 
\usepackage{adjustbox} 

\usepackage[svgnames]{xcolor}

\addto\captionsfrench{}
\addto\captionsfrench{}
\addto\captionsfrench{}

\usepackage{pdfpages} 
\usepackage{amsmath, mathtools}
\usepackage{float}
\usepackage{amssymb}
\mathtoolsset{showonlyrefs}
\usepackage{amsfonts,bm,bbm}     
\usepackage{upgreek}             
\usepackage{etoolbox}          
\usepackage{calc}              
\usepackage{datetime}                
\usepackage{slantsc} 
\makeatletter
\patchcmd{\chaptermark}{\MakeUppercase}{\scshape\slshape}{}{}%
\patchcmd{\sectionmark}{\MakeUppercase}{\scshape\slshape}{}{}%
\patchcmd{\sectionmark}{\thesection.}{\thesection}{}{}       
\makeatother
\numberwithin{equation}{chapter}                     
\numberwithin{figure}{chapter}                       
\numberwithin{table}{chapter}                        
\mathtoolsset{showonlyrefs}                          
\renewcommand{\thechapter}{\Roman{chapter}}          
\usepackage{titlesec}  
\usepackage{etoolbox}
\makeatletter
\patchcmd{\ttlh@hang}{\parindent\z@}{\parindent\z@\leavevmode}{}{}
\patchcmd{\ttlh@hang}{\noindent}{}{}{}
\makeatother

\titleformat{\section}[hang]{\Large\sffamily\bfseries}	{\rlap{\thesection}}{2em}{}
\titleformat{\subsection}[hang]{\large\sffamily\bfseries}{\rlap{\thesubsection}}{3em}{}
\titleformat{\chapter}[block]{\Huge\sffamily\bfseries\filcenter\MakeUppercase}{\thechapter\ --}{1ex}{}

\usepackage[square, numbers, comma, sort&compress]{natbib}
\usepackage{hyperref}
\hypersetup{colorlinks,linkcolor=DarkBlue,,anchorcolor=red,pdfdisplaydoctitle=true,pdfpagemode=UseOutlines, bookmarksnumbered=true,bookmarksopen=true}

\usepackage{lipsum} 
\usepackage{blindtext}
\blindmathtrue

\newcommand{\gap}{\textnormal{\scriptsize{gap}}}
\newcommand{\num}{\textnormal{\scriptsize{num}}}
\newcommand{\new}{\textnormal{\scriptsize{new}}}

\DeclarePairedDelimiter\bra{\langle}{\rvert}
\DeclarePairedDelimiter\ket{\lvert}{\rangle}
\DeclarePairedDelimiterX\braket[2]{\langle}{\rangle}{#1 \delimsize\vert #2}

\usepackage{thcover}
\thesisname{Th\`{e}se de Doctorat}
\gradename{docteur}
\univ{de l’Universit\'{e} Sorbonne Paris Cit\'{e}} 
\logos{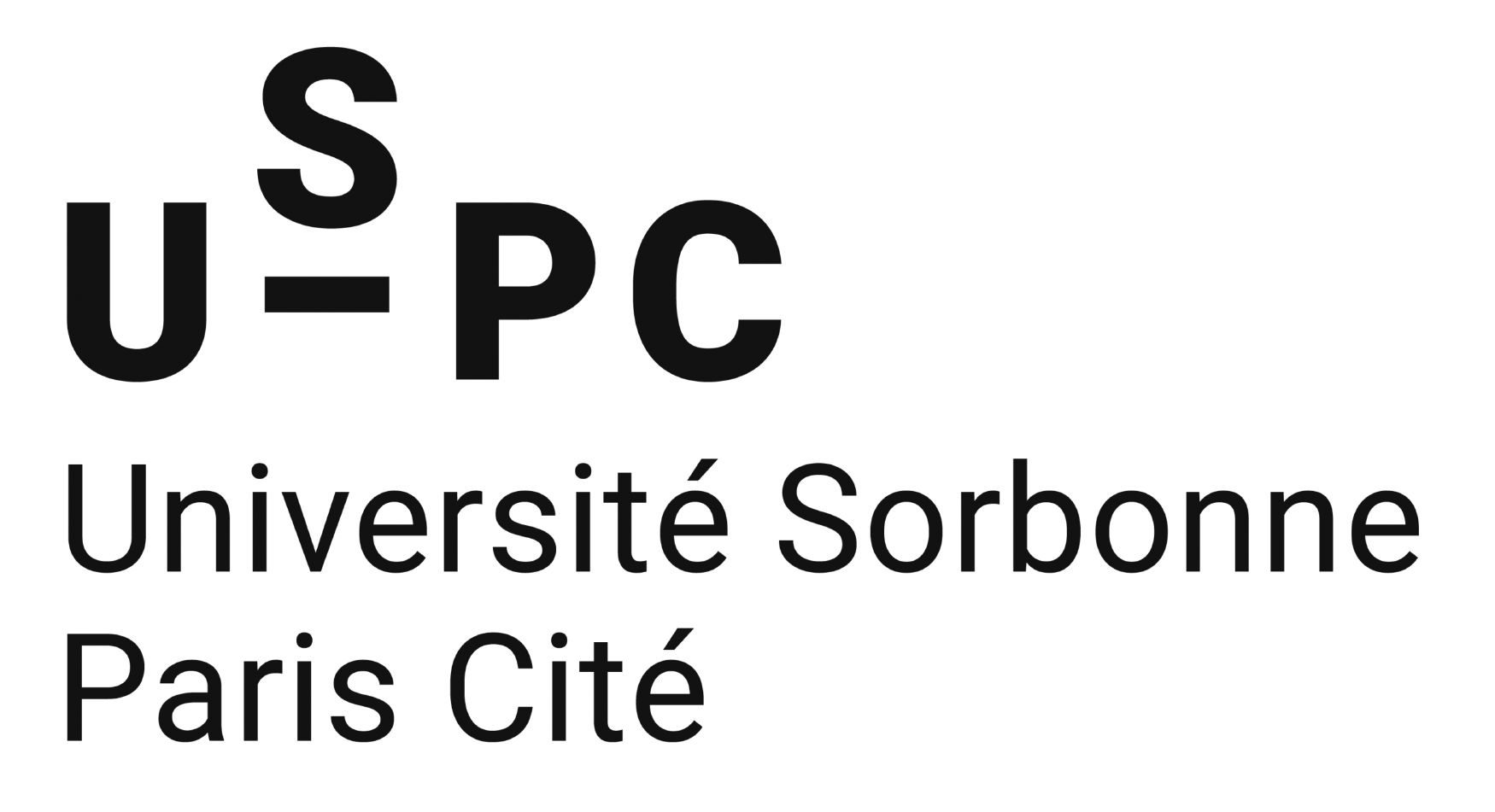}{}{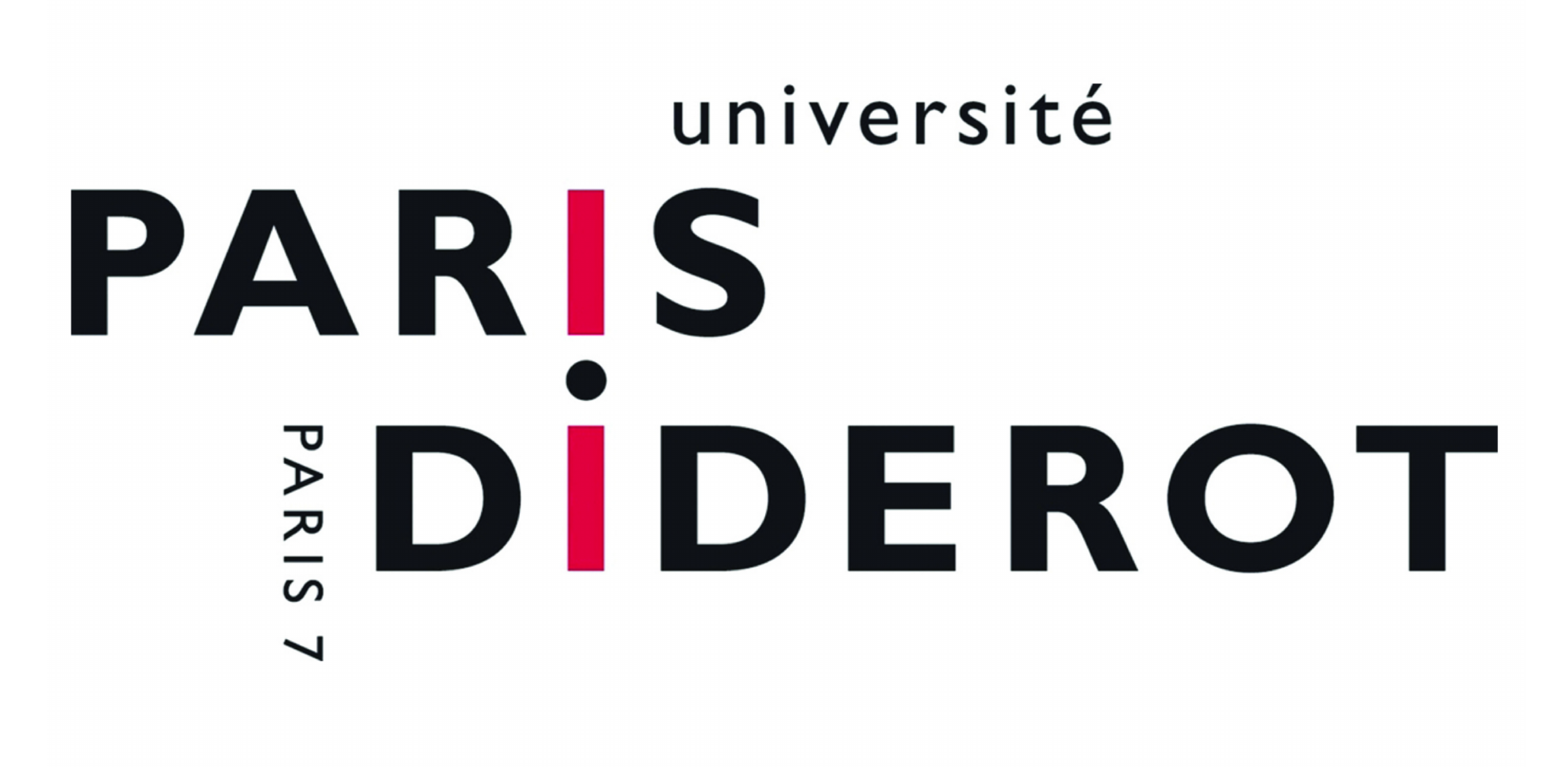}                   
\license{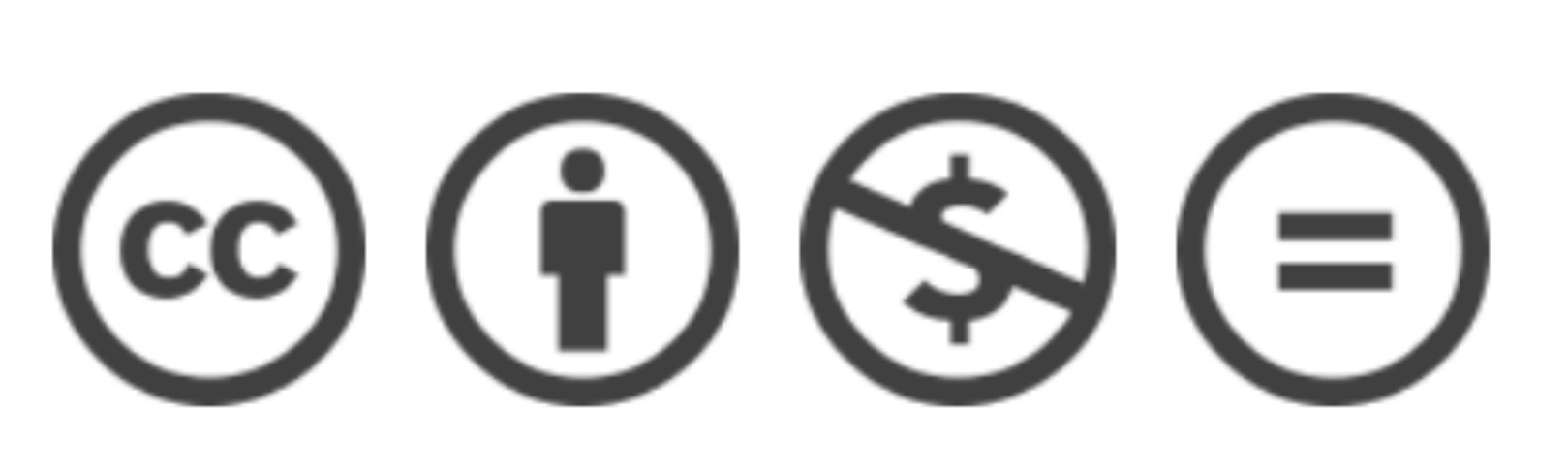}
\ecoledoctnum{393}
\ecoledoct{Pierre Louis de Sant\'{e} Publique \`{a} Paris : \\ Epid\'{e}miologie et Sciences de l'Information Biom\'{e}dicale}
\title{Cloning Algorithms: \\ from Large Deviations to Population Dynamics }

\titlefr{Algorithmes de Clonage: \\ des Grandes Déviations \`{a} la Dynamique des Populations}

\date{1 Juin 2018}
\author{Esteban GUEVARA HIDALGO}
\advisor{Khashayar PAKDAMAN }
\atlab{Equipe de Biologie Computationnelle et Biomath\'{e}matiques \`{a} l'IJM}
\jury{ 
 \small Pr\'{e}sident du jury: & \small Mme. & \small Leticia Cugliandolo & \small Université Pierre et Marie Curie\\
 \small Rapporteur:  & \small M. & \small Juan P. Garrahan & \small University of Nottingham \\
 \small Rapporteur: & \small M. & \small Hugo Touchette & \small Stellenbosch University \\
 \small Examinateur: & \small M. & \small Julien Tailleur & \small Université Paris Diderot \\
 \small Examinateur: & \small M. & \small Vivien Lecomte & \small Université Grenoble-Alpes \\
 \small Directeur de th\`{e}se: &  \small M. &  \small Khashayar Pakdaman &  \small Université Paris Diderot
}

\resume{La dynamique des populations fournit un outil numérique qui permet l'étude des événements rares grâce à la simulation d'un grand nombre de copies du système.
Le processus est muni d'une règle qui favorise les trajectoires rares d'intérêt.
La méthode de l'algorithme de clonage  permet l'estimation de la fonction de grandes déviations (en anglais, LDF) pour les observables additives pour les processus de Markov.
Cependant, cette méthode doit être soigneusement utilisée car il existe des effets de temps de simulation $t$ finie et de taille de population $N_c$ finie.
Premièrement, nous analysons les effets de petit $N_c$ dans un régime transitoire initial en utilisant une approche de population non constante. Ces effets jouent un rôle important dans la détermination numérique de la LDF.
Pour surmonter ces effets, nous avons introduit un délai dans l'évolution des populations, en plus de l'exclusion du régime initial de la croissance de la population où ces effets sont forts.
Ensuite, l'étude des lois d'échelle de $t$ et $N_c$ finie dans l’évaluation de LDF est faite en utilisant deux versions différentes de l’algorithme, en temps discret et en temps continu.  
Nous montrons que ces échelles se comportent comme $1/N_c$ et $1/t$ dans le régimes asymptotiques de grand $N_c$ et de grand-$t$ respectivement.
En outre, nous montrons qu'il est possible d'utiliser cette vitesse de convergence pour extraire le comportement asymptotique des limites de $t$ et $N_c$ infinis, fournissant ainsi une meilleure estimation de la LDF. Enfin, ces lois d'échelles sont généralisées et les indications de leurs limites dans les systèmes de grandes dimensions sont présentées.}
\motscles{Evénements Rares, Grandes Deviations, Dynamique des Populations}

\abstract{Population dynamics provides a numerical tool
allowing for the study of rare events by means of simulating a large number of copies of the system, supplemented with a
selection rule that favours the rare trajectories of interest. 
The cloning algorithm allows the estimation of a large deviation function (LDF) of additive observables in Markov processes.
However, such algorithms are plagued by finite simulation time $t$ and finite population size $N_c$ effects that can render their use delicate.
First, using a 
non-constant population approach, we analyze the small-$N_c$ effects in the initial transient regime. These effects play an important role in the numerical
determination of LDF. 
We show how to overcome these effects by introducing a  time delay in the evolution of populations, additional to the discarding of the initial regime of the population growth where these discreteness effects are strong. 
Then, the study of the finite-$t$ and finite-$N_c$ scalings in the LDF evaluation is done using two different versions of the algorithm, in discrete and continuous-time. We 
show that these scalings behave as $1/N_c$ and $1/t$ in the large-$N_c$ and large-$t$ asymptotics respectively. Moreover, we show that one can make use of this convergence speed in order to 
 extract the asymptotic behavior in the infinite-$t$ and infinite-$N_c$  limits resulting in a better LDF estimation. These scalings are later generalized and evidence of a breakdown for large-size systems is presented.

}

\usepackage{versionswitch}
\begin{document}
\frontcover
\backcover
\clearpage
\thispagestyle{empty}
\phantom{a}
\frontmatter



\clearpage
\thispagestyle{plain}
\par\vspace*{.3\textheight}
{
\centering \huge{\textbf{Acknowledgements}}
\vspace{20px}

\normalsize Many thanks to Khashayar Pakdaman and Vivien Lecomte for their support and discussions. 
%
Thanks to the team of Biologie Computationnelle et Biomath\'{e}matiques at Institut Jacques Monod where this thesis was developed. 
Special thanks to the Ecuadorian State and the Secretar\'ia Nacional de Educaci\'on Superior, Ciencia, Tecnolog\'ia e Innovaci\'on, SENESCYT, which supported financially this PhD program and the French Republic for welcoming me.
Also to my friends and family in Ecuador who were always present.
Finally, I thank my friends in France, especially, my girlfriend who was my companion during these years making the stay in Paris not only easier but also wonderful. 
}


\clearpage
\thispagestyle{empty}
\phantom{a}

\clearpage
\thispagestyle{plain}
\par\vspace*{.08\textheight}
\begin{flushright}
\large \textit{A mi mam\'{a},\\ mis hermanos\\ y mis perros}
\end{flushright}

\clearpage
\thispagestyle{empty}
\phantom{a}


\chapter*{Résumé}
\addcontentsline{toc}{chapter}{ \qquad Résumé}

L'occurrence \textbf{d'événements rares} peut grandement contribuer à l'évolution des systèmes physiques en raison de leur effets dramatiques.
La théorie des grandes déviations fournit un ensemble d'outils qui permettent leur traitment~\cite{touchette_large_2009,TouchetteRev2,TouchetteRev3}. Ces probabilités et fluctuations ont la propriété de décroître exponentiellement en fonction d'un paramètre (comme le temps ou la température). Cela signifie que lorsque le paramètre se devient plus grand, l'événement dévient moins probable~\cite{Varadhan1351}. 
%
%
D'un point de vue pratique, la théorie des grandes déviations peut être vue comme une collection de méthodes qui permettent de déterminer si un principe de grandes déviations existe pour une variable aléatoire donnée et pour déterminer sa fonction de taux ou ``fonction des grandes déviations'' (en anglais, LDF).

\medskip

Seulement dans quelques cas simples, il est possible d'obtenir des expressions exactes et expressions explicites pour la fonction de taux~\cite{BodineauDerrida2, Mehl}.
Pour la plupart des processus stochastiques, l'évaluation de ces fonctions est faite en utilisant des approches analytiques et des méthodes numériques~\cite{touchette_large_2009, TouchetteRev2,TouchetteRev3,giardina_simulating_2011, bucklew_introduction_2013}. 
Ils vont de ``importance sampling method''~\cite{kahn1951estimation},
``adaptive multilevel splitting''~\cite{cerou_adaptive_2007} à ``transition path sampling''~\cite{cochran1977sampling,Hedges1309,Speck,bolhuis_transition_2002}
et des algorithmes ``go with the winner''~\cite{aldous1994go,GRASSBERGER200264}
aux méthodes de dynamique des populations~\cite{tailleur_simulation_2009, giardina_simulating_2011} en temps discret~\cite{giardina_direct_2006} ou temps continu~\cite{lecomte_numerical_2007}. 
Ces méthodes ont été généralisées à de nombreux contextes~\cite{delmoral,HGprl09,hurtado_current_2009,lelievre,Vanden}. 
En physique, ceux-ci sont de plus en plus utilisés dans l'étude des systèmes complexes, par exemple dans l'étude des fluctuations réelles des modèles de transport~\cite{DerridaLebowitz,Derrida,MFT}, glasses~\cite{Hedges1309}, du repliement de protéines~\cite{Weber1} et des réseaux de signalisation~\cite{Vaikuntanathan,Weber2}. 
Mathématiquement, la procédure revient à determiner la fonction des grandes déviations associée à la distribution d'une observable dépendant de la trajectoire, qui peut être reformulée à son tour en la détermination de l'état fondamental d'un opérateur linéaire~\cite{hugoraphael}, une question commune à la physique statistique et à la physique quantique~\cite{DMC}.
%

\medskip 

Dans cette thèse, nous accordons une attention particulière à algorithmes basés sur  \textbf{dynamique des populations}~\cite{giardina_direct_2006,tailleur_probing_2007,
lecomte_numerical_2007,tailleur_simulation_2009,giardina_simulating_2011} afin d'étudier les trajectoires rares en biaisant exponentiellement leur probabilité. 
Dans ce contexte, la procédure numérique introduite par Giardin\`{a}, Kurchan et Peliti~\cite{giardina_direct_2006} surmonte  la difficulté d'observer les fluctuations d'une observable (dont la probabilité diminue exponentiellement dans le temps) pour les chaînes de Markov à temps discret. La fonction des grandes déviations peut être obtenue comme la plus grande valeur propre d'une matrice d'évolution d'une dynamique modifiée~\cite{giardina_direct_2006, tailleur_simulation_2009} qui peut être calculé numériquement~\cite{BodineauDerrida2, Mehl, Baiesi} seulement pour les petits systèmes car la matrice d'évolution est exponentiellement grande dans la taille du système. 
Ensuite, une modification de cette procédure a été proposée~\cite{lecomte_numerical_2007,Lecomte2007} pour lequel les problèmes de discrétisation lié à l'approche originale~\cite{giardina_direct_2006} sont contournés avec une approache directe en temps continu.

\newpage

L'évolution du système a été représentée par une dynamique de population du type de diffusion Monte Carlo~\cite{DMC}.
Cet algorithme a été appliqué pour calculer avec succès les grandes déviations du courant total dans le processus d'exclusion symetrique et asymétrique~\cite{Spohn,Spohn2}, et de l'activité dans le processus de contact~\cite{CP}.
Aussi, pour analyser la dynamique~\cite{garrahanjacklecomtepitardvanduijvendijkvanwijland, garrahan_first-order_2009} des modèles cinètiquement contraints (KCM)~\cite{RitortSollich, FAmodel, Jackle1, Jackle2, Kob, Kronig, KurchanKCM, Sollich, Einax, GarrahanChandler, AldousDiaconis, Jung, Toninelli, Pan, Geissler}, des ``glassy systems''~\cite{Ediger, Angell, Binder} à travers les statistiques de trajectoires de la dynamique montrant que ces modèles présentent une transition dynamique de premier ordre entre les phases dynamiques active et inactive.
Il a également été utilisé pour étudier les symétries dans les fluctuations loin de l'équilibre~\cite{HurtadoPerez} et dans les modèles de transport
~\cite{HGprl09, hurtado_current_2009, hurtado_large_2010}. Ces études permettent non seulement de tester les prédictions de l'hydrodynamique fluctuante~\cite{HGprl09, HurtadoSpon}, mais aussi les limites de la méthode elle-même~\cite{hurtado_current_2009}. 
Il a été également suggéré~\cite{giardina_simulating_2011} que la méthode pourrait être appliquée pour étudier en détail de possible future et l'évolution passée des systèmes planétaires, et aussi l'auto-organisation de la stabilité de notre système solaire. 

\medskip

L'idée de la dynamique de populations est de traduire l'étude d'une classe de trajectoires rares (par rapport à une contrainte globale déterminée) dans l'évolution de plusieurs copies de la dynamique originale, avec un processus de sélection local-dans-temps rendant l'occurrence des trajectoires rares typiques dans la population évoluée.
La distribution de la classe des trajectoires rares dans la dynamique originale est liée à la croissance (ou décroissance) exponentielle de la population des clones du système et la LDF  peut être estimée à partir de leur taux de croissance. 
Les procédures numériques visant à simuler efficacement des événements rares, en utilisant un schéma de dynamique de population sont communément appelées~\textbf{algorithmes de clonage}. Dans de tels algorithmes, les copies du système sont évoluées en parallèle et celles qui montrant le comportement rare d'intérêt sont multipliées itérativement~\cite{DMC,Glasserman_1996,2001IbaYukito,GRASSBERGER200264,
4117599,CappeGuillinetal,Forwardinterfacesampling,PhysRevLett.94.018104,
delmoral2005,Dean2009562,lelievre,giardina_direct_2006,lecomte_numerical_2007,
tailleur_probing_2007,tailleur_simulation_2009,giardina_simulating_2011,
1751-8121-46-25-254002_2013}.

\medskip

Les différentes versions de l'algorithme de clonage utilisées dans cette thèse sont détaillées dans le~\hyperref[chap:I]{Introduction}. Là, nous partons de la construction de l'équation maîtresse, de sa solution et de son interprétation. 
Nous définissons le principe des grandes déviations pour certains observables $\mathcal O$ ce qui peut être interprété comme la probabilité d'observer une valeur atypique de celle observables après une longue échelle de temps.
La fonction de taux de ce principe des grandes déviations correspond à la LDF et c'est un équivalent dynamique de l'entropie intensive dans l'ensemble microcanonique~\cite{touchette_large_2009}. Il code non seulement les fluctuations gaussiennes mais aussi les fluctuations non-gaussiennes (ou les grandes déviations) de l'observable $\mathcal O / t$ qui peut être obtenu par son expansion au-delà de l'ordre quadratique.
Dans la limite du temps infini, le LDF peut ne pas être analytique, ce qui peut être interprété comme une signature d'hétérogénéités dynamiques (transition de phase dynamique)~\cite{Merolle10837, jackST}.

\medskip

Le problème de la détermination de la fonction de taux est en général une tâche difficile dans l'ensemble microcanonique, on préfère donc aller à l'ensemble dynamique canonique (ou espace de Laplace). Au lieu de fixer la valeur de l'observable $\mathcal O $ afin de déterminer la LDF on introduit un paramètre $s$ (intensif dans le temps) qui biaise le poids statistique des histoires et fixe la valeur moyenne de $\mathcal O $, de sorte que $s \neq 0$ favorise ses valeurs non-typiques.
Le paramètre $s$ involves une modification (exponentielle) du poids statistique des histoires du système. Valeurs pour $s=0$ correspondent aux moyennes de l'état stable de $\mathcal O$. Pendant ce temps, les valeurs des $s \neq 0$ favorisent les histoires avec des valeurs non-typiques de l'observable $\mathcal O$.

\medskip

Pour des raisons pratiques, il est convenient de calculer la fonction génératrice des cumulants (par ses sigles en anglais CGF) au lieu de LDF (qui sont reliées par une transformée de Legendre), ce que nous calculons en pratique tout au long de la thèse. 
Nous montrons comment estimer le CGF à partir de l'interprétation de la dynamique des populations modifiée 
ou à partir de la plus grande valeur propre de l'opérateur d'évolution modifiée.
Dans le premier cas, l'équation d'évolution temporelle qui décrit la dynamique modifiée peut être interprétée pas comme l'évolution d'un système unique, mais comme une dynamique des populations sur un grand nombre $N_c$ des copies du système qui évolue de manière couplée~\cite{giardina_direct_2006,tailleur_probing_2007}. C'est-à-dire, comme un processus stochastique avec des taux de transition fourni par un mécanisme de sélection
où un clone du système est copié s'il est rare ou tué sinon.

\medskip

Nous détaillons également les approches avec population totale \textbf{non-constante} et \textbf{constante} de l'algorithme de clonage et les estimateurs CGF obtenus à partir de ceux-ci. Nous expliquons comment pour l'approche de la population totale constante, un uniforme élagage/clonage est appliqué au-dessus de la dynamique de clonage afin d'éviter l'explosion ou la disparition exponentielle de la population.
Alors que la dernière version est évidemment plus ``computer-friendly'', l'ancienne version présente des caractéristiques intéressantes: Premièrement, il est directement lié à l'évolution des systèmes biologiques (sauts stochastiques représentant des mutations, les règles de sélection étant interprétées comme une pression darwinienne); Seconde, le uniforme élagage/clonage de la population, bien que non biaisé, induit des corrélations dans la dynamique que l'on pourrait vouloir éviter; Enfin, dans certaines situations où les taux de sélection sont très fluctuants, l'algorithme de population constante ne peut pas être utilisé dans la pratique en raison des effets de population finie (la population étant éliminée par un seul clone), et on doit recourir à la non-constante. Un exemple de la mise en ouvre de cette version peut être trouvé dans Ref.~\cite{ViscoTrizac}.

\medskip

À la fin de l'introduction, nous présentons les exemples de modèles utilisés pour notre analyse: une simple dynamique d'annihilation-création à deux états, et un processus de contact sur un treillis périodique unidimensionnel.
Le premier système (chapitres:~\ref{chap:Discreteness},~\ref{chap:DiscreteTime},~\ref{chap:ContinuousTime} et~\ref{chap:CGF}) a été choisi pour sa simplicité et la possibilité de comparer les prédictions numériques avec les valeurs exactes de CGF. D'autre part, le processus de contact (chapitres:~\ref{chap:ContinuousTime},~\ref{chap:CGF} et~\ref{chap:LargeL}) est utilisé pour étendre l'analyse et vérifier les résultats  sur a (plus complexe) système de ``many body'' où la dépendance avec la taille du système peut également être analysée.
%
Dans les deux cas, nous considérons l'activité dynamique $K$~\cite{garrahanjacklecomtepitardvanduijvendijkvanwijland,garrahan_first-order_2009,Hedges1309,ThermoCP,PhysRevE.81.011111,ChandlerGarrahan,QuantumJump,
Micromaser,Genway,Ates,Hickey,jackEast} comme l'additif observable $\mathcal O$~\eqref{eq:obs} d'intérêt. L'expression analytique du CGF est obtenue (lorsque cela est possible) en résolvant la plus grande eigenvalue de l'opérateur modifié comme discuté dans la Sec.~\ref{sec:largest}. 

\bigskip
Dans le chapitre~\ref{chap:Discreteness}: \textbf{Discreteness Effects in Population Dynamics}~\cite{hidalgo_discreteness_2016}, nous appliquons l'algorithme de population non-constante afin d'analyser numériquement les effets dus à la petite taille de population dans le régime transitoire initial sur un  modèle simple d'annihilation-création (Sec.~\ref{sec:bdp}) où sa mise en ouvre et ses propriétés peuvent être examinées dans les moindres détails.
Au cours du régime transitoire initial de l'évolution des populations, il y a une grande distribution des temps où la première série de sauts se produisent. Cela signifie que les fluctuations au moment initial produisent que certaines populations restent dans leur état initial beaucoup plus longtemps que d'autres, produisant un écart dans leur évolution individuelle. Cela induit un décalage relatif qui dure pour toujours. Ces effets jouent un rôle important en particulier pour la détermination de la fonction de grandes déviations.
L'estimation de CGF provient de la détermination du taux de croissance d'une   log-population moyenne (Sec.~\ref{sec:NCPA}) sur de nombreuses réalisations de la dynamique.

\medskip

Afin de réaliser cette moyenne de manière systématique, nous définissons une procédure que nous avons appelée \textbf{merging}.
Toutefois, cette moyenne 
est fortement dépendante non seulement du nombre de réalisations,
et sur la taille de la population initiale mais aussi sur le temps (ou population) de coupure considéré pour arrêter leur évolution.
C'est-à-dire, en limitant l'évolution de nos populations jusqu'à un maximum de temps $T_{\max}$ (ou population $N_{\max}$) ce qui n'est pas ``assez grand'', la population moyenne (et la détermination du CGF) peut être influencée par \textbf{effets de discrétion aux temps initiaux}, causés par une petite taille de la population.
Nous avons proposé comme alternative afin de surmonter l'influence des effets de discrétisation se débarrasser des régions des populations où ces effets sont présents. Autrement dit, couper le régime transitoire initial des populations. Dans ce cas, nous voyons que la moyenne des populations est limitée à un intervalle qui peut être très petit et cela peut induire une mauvaise estimation de CGF.

\medskip

En complément, nous avons trouvé un moyen de souligner les effets du régime de croissance exponentielle dans la détermination du CGF en utilisant le fait que les log-populations après un temps assez long deviennent parallèles (Fig.~\ref{fig:distance1}(a)) et qu'une fois que les populations ont surmonté le régime des effets de discrétion, la distance entre elles devient constante (Fig.~\ref{fig:distance1}(b)) et ces effets ne sont plus forts  (Sec.~\ref{Parallel Behaviour in Log-Populations}). 
D'autre part, nous soutenons en Sec.~\ref{Time Delay Correction} que ces effets de discrétion initiale ou le décalage initiale entre les populations pourrait être compensée en effectuant sur les populations un déplacement du temps (Eq.~\eqref{eq:13}). Cette procédure est choisie de façon à chevaucher les évolutions de la population dans leur régime de grande durée large-time régime (Fig.~\ref{fig:delayedPOP}(b)). 
Ceci avec un rejet des régimes initiaux dans l'évolution de la population surpasse l'influence des effets de discrétion améliorant l'estimation de CGF.

\medskip

Nous montrons que c'est vrai, indépendamment de la méthode utilisée pour calculer le taux de croissance de la population moyenne, comme le Fig.~\ref{fig:psi2}. En outre, il est montré que si en plus, nous effectuons la transformation temporaire, l'estimation de CGF est encore améliorée et plus proche de la valeur théorique (Sec.~\ref{``Bulk'' and ``Fit'' Slopes}). Ce résultat est confirmé plus tard en Sec.~\ref{Relative Distance and Estimator Error} en calculant la distance relative des estimateurs numériques à la valeur théorique et leurs erreurs. Comme peut être observé dans Fig.~\ref{fig:relativeD}, l'écart de la valeur théorique est plus élevé pour les valeurs de $s$ proches de $0$, mais est plus petit après la ``correction du temps'' pour presque chaque valeur de $s$. De même pour l'estimateur d'erreur (Fig.~\ref{fig:error}).
De plus, nous étudions les propriétés de ces retards. Nos résultats numériques supportent également un comportement ``loi de puissance'' dans le temps de la variance de retard. Par ailleurs, la distribution des délais prend une forme universelle, après avoir rééchelonné la variance à 1. 

\bigskip

Le chapitre~\ref{chap:Discreteness}~\cite{hidalgo_discreteness_2016} est structuré comme suit: en Sec.~\ref{sec:avepopldf} nous décrivons les problèmes liés à la moyenne des réalisation distincts,
que nous quantifions en Sec.~\ref{Parallel Behaviour in Log-Populations}.
En Sec.~\ref{sec:time_correct} nous proposons la méthode pour augmenter l'efficacité de l'algorithme de dynamique de la population en appliquant un temps de retarde dépendant de la réalisation, et
nous présentons les résultats de son application en Sec.~\ref{sec:psi_timedelay}.
Nous caractérisons numériquement la distribution de ces temps de retarde en Sec.~\ref{sec:time-delay-prop}.
Nos conclusions et perspectives sont réunis en Sec.~\ref{sec:discussionDE}.

\bigskip

En dehors des approches de population constante, les mécanismes de sélection dans l'algorithme de clonage peuvent être mis en ouvre de différentes manières. L'un d'eux, avec chaque évolution des copies des copies \textbf{(Continuous-Time)}~\cite{tailleur_simulation_2009, lecomte_numerical_2007, giardina_simulating_2011} ou bien, pour chaque intervalle de temps pré-fixé \textbf{(Discrete-Time)}~\cite{giardina_direct_2006}. Les différences importantes entre les deux techniques sont discutées dans Secs.~\ref{subsubsec:differenceContinuousTime} et~\ref{Discrete-time_algorithm}. 

\medskip

L'algorithme proposé par Giardin\`a 
et al.~\cite{giardina_direct_2006,lecomte_numerical_2007,
tailleur_probing_2007,tailleur_simulation_2009,giardina_simulating_2011,
1751-8121-46-25-254002_2013} (un technique à temps discrete ) est utilisé pour évaluer numériquement le CGF d'additif (ou extensif dans le temps) observables dans les processus de Markov~\cite{opac-b1093895,touchette_large_2009}. 
Le CGF est obtenu comme le taux de croissance exponentiel que la population présenterait si elle n'était pas maintenue constante. 
Il a été appliqué à de nombreux systèmes physiques, y compris les systèmes chaotiques, la dynamique vitreuse et les modèles de gaz en treillis sans équilibre, a permis l'étude de nouvelles propriétés, telles que le comportement des respirateurs dans la chaîne de Fermi-Pasta-Ulam-Tsingou ~\cite{tailleur_probing_2007}, transitions de phase dynamiques dans des modèles cinétiquement contraints~\cite{garrahanjacklecomtepitardvanduijvendijkvanwijland}, et un principe d'additivité pour les processus d'exclusion simples~\cite{PhysRevLett.92.180601,hurtado_large_2010}. 
Sous cette approche, le correspondant estimateur CGF n'est valide que dans les limites du temps de simulation infinie~$t$ et taille de la population infinie~$N_c$. La stratégie habituelle suivie pour obtenir ces limites est d'augmenter le temps de simulation et la taille de la population jusqu'à ce que la moyenne de l'estimateur sur plusieurs réalisations ne dépende pas de ces deux paramètres, jusqu'à des incertitudes numériques. 

\medskip

Bien que la méthode a été largement utilisée, il y a eu moins d'études axées sur la justification analytique de l'algorithme. De plus, il introduit deux paramètres supplémentaires en considération:
la taille de la population $ N_c $ et le temps de simulation $ t $, tous le deux affectent considérablement la précision de l'estimation de CGF. Même si l'on croit heuristiquement que l'estimateur LDF converge vers le résultat correct à mesure que le nombre de copies $ N_c $ augmente, il n'y a pas de preuve de sa convergence. Relatif à ce manque de preuve, bien que nous utilisions l'algorithme en supposant sa validité, nous n'avons aucune idée de la vitesse à laquelle l'estimateur converge en $ N_c \rightarrow \infty$.  
Nous discutons de cette convergence en effectuant une étude analytique en temps discret dans le chapitre~\ref{chap:DiscreteTime} et en utilisant une approche numérique en temps continu dans le chapitre~\ref{chap:ContinuousTime}. Il est important de remarquer que les deux versions de l'algorithme (temps discret et continu) diffèrent sur un point crucial qui fait qu'une extension de l'analyse développée au chapitre~\ref{chap:DiscreteTime} ne peut pas être faite directement pour comprendre le cas de temps continu dans chapitre~\ref{chap:ContinuousTime}. 

\medskip

Dans le chapitre~\ref{chap:DiscreteTime}: \textbf{Finite-Time and -Size Scalings in the Evaluation of Large Deviation Functions: I. Analytical Study using a Birth-Death Process}~\cite{partI}, 
%
nous discutons de cette convergence en définissant deux types d'erreurs numériques.
Premièrement, pour un nombre fini et fixe de clones $ N_c $, en faisant la moyenne sur un grand nombre de réalisations, l'estimateur CGF converge vers une valeur incorrecte, qui est différente du résultat de grande déviation souhaité. Nous appelons cette déviation de la valeur correcte, \textbf{erreurs systématiques}. Par rapport à ces erreurs, nous considérons également les fluctuations de la valeur estimée. Plus précisément, pour une valeur fixe de $ N_c $, les résultats obtenus dans différentes réalisations sont répartis autour de cette valeur incorrecte. Nous appelons les erreurs associées à ces fluctuations \textbf{erreurs stochastiques}. Bien que les deux erreurs soient importantes dans les simulations numériques, le derniere peut conduire cet algorithme à produire de mauvais résultats. Par exemple, l'erreur systématique croît exponentiellement à mesure que la température diminue~\cite{nemoto_population-dynamics_2016}. 

\medskip

Pour étudier ces erreurs, nous avons utilisé une description de processus de naissance-mort~\cite{tagkey2007iii,opac-b1079113} por l'algorithme de dynamique de population comme il est expliqué ci-dessous: Nous nous concentrons sur les systèmes physiques décrits par une dynamique de Markov~\cite{giardina_direct_2006,giardina_simulating_2011,lecomte_numerical_2007} avec un nombre fini d'états $M$. Nous dénotons par $i$ ($i=0,1,\cdots M-1$) les états du système. Ce processus de Markov a sa propre dynamique stochastique, décrite par les taux de transition $w(i\rightarrow j)$.
Dans les algorithmes de dynamique de population, afin d'étudier ses trajectoires rares, on prépare $ N_c $ copies du système, et simule ces copies en fonction de (\textit{i}) la dynamique de $w(i\rightarrow j)$ (suivi indépendamment pour toutes les copies) et (\textit{ii}) la étape de ``clonage'' dans laquelle l'ensemble des copies est directement manipulé, i.e., certaines copies sont éliminées pendant que d'autres sont multipliées (See Table~\ref{tablecorrespondence_}). 
Formellement, la dynamique des populations représente pour une \emph{unique}  copie du système, un processus qui ne préserve pas la probabilité, comme mentionné dans Sec.~\ref{mut}. Ce fait a motivé l'étude des processus auxiliaires~\cite{jack_large_2010}, des processus efficaces~\cite{1742-5468-2010-10-P10007} et  
driven processes~\cite{PhysRevLett.111.120601} pour construire une dynamique modifiée (et leurs approximations~\cite{PhysRevLett.112.090602}) qui préserve la probabilité.  

\medskip

Au chapitre~\ref{chap:DiscreteTime}, nous formulons explicitement la~\textbf{méta-dynamique} des copies elles-mêmes en utilisant un processus stochastique de naissance-mort qui préserve la probabilité, et il nous permet d'étudier les erreurs numériques de l'algorithme lors de l'évaluation CGF.  
Nous considérons la dynamique des copies comme un processus stochastique  de naissance-mort dont l'état est noté par $n=(n_0,n_1,n_2,\ldots, n_{M-1})$, où $0\leq n_i\leq N_c$ représente le nombre de copies qui sont dans l'état $ i $ dans l'ensemble des copies. Nous introduisons explicitement les taux de transition décrivant les dynamiques de $ n $, que nous désignons par $\sigma (n\rightarrow \tilde n)$. Nous montrons que la dynamique décrite par ces taux de transition conduit en général à l'estimation CGF correcte du système original $w(i\rightarrow j)$ dans la limit  $N_c \rightarrow \infty$.
Nous montrons aussi que les erreurs systématiques sont de l'ordre $\mathcal O(1/N_c)$, alors que les erreurs numériques sont de l'ordre $\mathcal O(1/(\tau N_c))$ (où $\tau$ is an averaging duration). Ce résultat contraste nettement avec les méthodes Monte-Carlo standard, où les erreurs systématiques sont toujours 0. 
La formulation développée au chapitre~\ref{chap:DiscreteTime}~\cite{partI} nous donne la possibilité de calculer exactement les expressions des coefficients de convergence, comme nous le faisons en Sec.~\ref{Section:Demonstrations} sur un exemple simple. 

\medskip

Chapitre~\ref{chap:DiscreteTime}~\cite{partI} est structuré comme suit. Nous définissons d'abord le problème CGF au début de Sec.~\ref{sec:Birth-deathprocess}, ensuite, nous formulons le processus de naissance-mort utilisé pour décrire l'algorithme en Sec.~\ref{subsec:TransitionmatrixAlgorithm}. En utilisant ce processus de naissance-mort, nous démontrons que l'estimateur de l'algorithme converge vers la correct fonction des grandes desviations en Sec.~\ref{Subsec:DerivationLargeDeviationEstimator}. À la fin de cette section, en Sec.~\ref{subsec:systemsizeexpansion}, nous discutons de la vitesse de convergence de cet estimateur (les erreurs systématiques) et nous dérivons son échelle $\sim 1/N_c$. In Sec.~\ref{Sec:largedeviation_largedeviation}, nous passons aux erreurs stochastiques. Pour discuter de cela, nous introduisons la LDF de l'estimateur, à partir de laquelle nous dérivons que la vitesse de convergence des erreurs stochastiques est proportionnelle à $1/(\tau N_c)$. Dans la section suivante, Sec.~\ref{Section:Demonstrations}, nous introduisons un modèle simple à deux états, auquel nous appliquons les formulations développées dans les sections précédentes. Nous dérivons les expressions exactes des erreurs systématiques en Sec.~\ref{subsection:systematicerrors_twostate} et des erreurs stochastiques en Sec.~\ref{subsection:largedeviationsInPopTwoState}. 
Ensuite, en Sec.~\ref{subsec:Different large deviation estimator}, nous proposons un autre grand estimateur de déviation et 
enfin, en Sec.~\ref{sec:discussion}, nous résumons les résultats obtenus.

\newpage

À partir de cette formulation, nous avons déduit les scalings finies en $ N_c $ et $t$ des erreurs systématiques de l'estimateur CGF, montrant que ceux-ci se comportent comme $1/N_c$ et $1/t$ dans le grand-$N_c$ et grand-$t$ asymptotiques respectivement. 
En principe, connaissant l'échelle {\it a priori} signifie que la limite asymptotique de l'estimateur dans le $t \to \infty$ et $N_c \to \infty$ limites peuvent être interpolées à partir des données à fini $t$ et $N_c$. 
Toutefois, si cette idée est réellement utile ou non est une question non triviale, comme il y a toujours une possibilité que les valeurs de début des  $\mathbf{N_{c}^{-1}}$- et $\mathbf{t^{-1}}$\textbf{-scalings} sont trop volumineux pour les utiliser. 

\bigskip

Dans le chapitre~\ref{chap:ContinuousTime}: \textbf{Finite-Time and -Size Scalings in the Evaluation of Large Deviation Functions: II. Numerical Approach in Continuous Time}~\cite{partII}, nous considérons une version continue dans le temps des algorithmes de dynamique des populations~\cite{lecomte_numerical_2007,tailleur_simulation_2009}. 
Nous montrons numériquement que on peut en effet faire usage de ces propriétés afin de concevoir une méthode originale et simple qui prenne en compte les échelles exactes du corrections de $t$ et $N_c$ finies afin de fournir des estimateurs CGF significativement meilleurs (\textbf{scaling method}) dans  l'application à un système avec des interactions à plusieurs corps (un processus de contact). 
%
Nous soulignons que les deux versions de l'algorithme diffèrent sur un point crucial qui fait qu'une extension de l'analyse développée au chapitre~\ref{chap:DiscreteTime}~\cite{partI} ne peut pas être fait directement afin de comprendre le cas en temps continu (Sec.~\ref{Discrete-time_algorithm}). Nous soulignons donc que l'observation de ces échelles eux-mêmes est également non triviale.

\medskip

Nous notons que les scalings qui régissent la convergence aux limites des temps temps et de taille infinies (avec corrections en $ 1 / N_c $ et en $ 1 / t $) doivent être pris en compte correctement: en effet, en tant que lois de puissance, elles ne présentent pas de taille et de temps caractéristiques au-delà desquelles les corrections seraient négligeables.
La situation est très similaire à l'étude de la force de dépinçage critique dans les driven random manifolds: la force critique présente une correction de 1 sur la taille du système~\cite{kolton_uniqueness_2013} qui doit être considéré correctement afin d'extraire sa valeur réelle.
Génériquement, de telles échelles fournissent également un critère de convergence aux régimes asymptotiques de l'algorithme: il faut confirmer que l'estimateur CGF présente des corrections (premièrement) $1/t$ et (en second lieu) dans $1/N_c$ par rapport à une valeur asymptotique afin de s'assurer que cette valeur représente une évaluation correcte de la CGF.

\medskip

Le chapitre~\ref{chap:ContinuousTime}~\cite{partII} est organisé comme suit. 
%
En Sec.~\ref{sec:LDFT} nous étudions le comportement de l'estimateur CGF en fonction du temps d'observation (pour une population fixe~$N_{c}$) et nous voyons comment sa limite de temps infinie peut être extraite à partir des données numériques.  En Sec.~\ref{sec:LDFN} nous analysons le comportement de l'estimateur en augmentant le nombre de clones (pour un temps de simulation final donné) et la limite de taille infinie de l'estimateur CGF. Sur la base de ces résultats, nous présentons en Sec.~\ref{sec:LDFinfTinfN} a une méthode qui nous permet d'extraire les limit infinies de temps et taille de l'estimator de la fonction des grandes déviations à partir d'une analyse d'échelle à taille finie et à temps fini.
%
%
En Sec.~\ref{Discrete-time_algorithm}, nous discutons la difficulté d'une approche analytique de l'algorithme en temps continu. Enfin, nos conclusions sont faites en Sec.~\ref{sec:conclusionsP3}. 

\medskip

Afin de compléter la discussion principale effectuée au chapitre~\ref{chap:ContinuousTime}~\cite{partII}, en chapitre~\ref{chap:CGF}~\cite{partII} nous étudions les fluctuations de l'estimateur CGF (défini dans la version continue dans le temps). Ceci est fait en étudiant sa distribution et sa dépendance avec le temps de simulation et le nombre de clones. Compatible avec le théorème de la limite centrale, nous montrons comment un redimensionnement approprié de l'estimateur CGF produit un effondrement des distributions dans une distribution standard normale pour différentes valeurs de $N_c$ et des temps de simulation. 
De plus, nous discutons dans Sec.~\ref{sec:two-estimators} une autre façon de le définir, qui a déjà été introduit dans Sec.~\ref{subsec:Different large deviation estimator} pour la version à temps discret.

\medskip

L'analyse d'échelle de $t$ et $N_c$ finies dans l'évaluation de CGF
a été réalisée suivant deux approches différentes: un analytique, au chapitre~\ref{chap:DiscreteTime}~\cite{partI}, en utilisant une version à temps discret de l'algorithme de dynamique des populations~\cite{giardina_direct_2006}, et numérique, dans le chapitre~\ref{chap:ContinuousTime}~\cite{partII}, en utilisant une version à temps continu~\cite{lecomte_numerical_2007,tailleur_simulation_2009}. 
Dans les deux cas, les erreurs systématiques de ces échelles ont été trouvés à se comporter comme $1/t$ et $1/N_c$ dans les asymptotiques de $ t $ et $ N_c $ grandes respectivement. 
De plus, il a été montré comment ces propriétés d'échelle peuvent être utilisées pour améliorer l'estimation CGF par la mise en ouvre d'une scaling method (Sec.~\ref{ssec:SM}). 
Ceci a été fait en considérant que le comportement asymptotique de l'estimateur dans $t \to \infty$ et $N_c \to \infty$ limits peut être interpolé à partir des données obtenues à partir de simulations à temps de simulation et nombre de clones \textbf{finies et relativement petites}. 
%
%
Cependant, la validité de ces échelles et l'efficacité de la méthode n'ont été prouvées 
que dans les cas où le nombre de sites $L$ (où la dynamique se produit) 
était petit: une simple dynamique d'annihilation-création à deux états~(Sec.~\ref{sec:bdp}) (dans un site) et un processus de contact unidimensionnel~(Sec.~\ref{sec:CP}) (avec $L=6$ sites).

\bigskip

En chapitre~\ref{chap:LargeL}: \textbf{Breakdown of the Finite-Time and Finite-$N_c$ Scalings in the Large-$L$ Limit}~\cite{largeLCP}, nous complétons les résultats présentés dans les chapitres~\ref{chap:DiscreteTime}~\cite{partI} et~\ref{chap:ContinuousTime}~\cite{partII} en étendant l'analyse à un processus de contact de grand-$ L $
Afin de le faire, 
nous redéfinissons ces échelles de manière plus générale.
Nous supposons le comportement de l'estimateur CGF décrit par un 
$t^{-\gamma_{t}}$-scaling (Eq.~\eqref{eq:tScal2}) et un $N_{c}^{-\gamma_{N_{c}}}$-scaling (Eq.~\eqref{eq:nScal2}). Cette redéfinition nous permet de vérifier dans les systèmes de grand-$L$ si effectivement $\gamma_{t} \approx 1$, $\gamma_{N_{c}} \approx 1$ et
si les termes $\chi_{\infty}^{(N_{c})}$ et $\chi_{\infty}^{\infty}$ représentent les limites en $t \to \infty$ et $N_{c} \to \infty $ de l'estimateur CGF.

\medskip

Ceci est fait d'abord en Sec.~\ref{sec: CGFL100} où nous avons considéré un processus de contact avec $L=100$ sites et deux valeurs représentatives du paramètre $s$ ($s=-0.1$ et $s=0.2$).
Bien que le $t^{-1}$-scaling et le $N_{c}^{-1}$-scaling ont été prouvés à tenir pour $s = 0.1$, ce n'était pas le cas pour $s=0.2$. 
Comment ce changement d'échelle est-il produit en fonction du paramètre $s$ est présenté en détail dans Sec.~\ref{sec: gamma_tn} où les exposants $\gamma_{t}(s)$ et $\gamma_{N_{c}}(s)$ sont caractérisés.
En particulier, pour $\gamma_{N_{c}}(s)$, nous avons été en mesure de distinguer trois étapes dans son comportement, où, le $N_{c}^{-1}$-scaling était valide jusqu'à $s=s^{*}$, puis $\gamma_{N_{c}}$ diminue à $0$ at $s=s^{**}$ et enfin, il devient négatif pour $s>s^{**}$.
En Sec.~\ref{sec: SML100} nous montrons comment ces échelles affectent la détermination de  la limite infini en $t$ et $N_c$ de l'estimateur CGF. Cela se produit parce que le scaling method reposait sur la validité du $t^{-1}$- et $N_{c}^{-1}$-scalings. Comme pour $L=100$ ce n'est pas le cas, il est possible de voir comment les différents estimateurs correspondaient les uns aux autres jusqu'à $s=s^{*}$ à partir de laquelle ils divergent jusqu'à $s = s^{**}$  où il y a une discontinuité.
Cette analyse est étendue au plan $s-L$ en Sec.~\ref{sec: planeSL}  où les exposants $\gamma_{t}$ et $\gamma_{N_c}$ ont été calculé pour une grille de valeurs des paramètres $(s,L)$.
Leur caractérisation est faite en introduisant une dépendance du $s'$, $s^{*}$ et $s^{**}$ avec le nombre de sites précédemment défini en Sec.~\ref{sec: CPL100} ainsi que l'utilisation du nombre de zéros de l'exposant $\gamma_{N_{c}}^{(L)}(s)$ afin de caractériser les différents groupes de $L$.

\newpage

Le chapitre~\ref{chap:LargeL}~\cite{largeLCP} est organisé comme suit:
%
%
La généralisation du scalings du $t$ et $N_c$ fini du CGF pour systèmes avec grand $L$ 
%
est fait en Sec.~\ref{subsec: tnScalingL}. 
Nous utilisons ces résultats dans Sec.~\ref{sec: CPL100} où nous vérifions la validité du $t^{-1}$- et $N_c^{-1}$-scalings (Sec.~\ref{sec: CGFL100}), leur comportement dans la dynamique modifié par $s$ (Sec.~\ref{sec: gamma_tn}) ainsi que l'applicabilité du scaling method (Sec.~\ref{sec: SML100}) pour un processus de contact avec $L=100$ sites.
Cette analyse est généralisée en Sec.~\ref{sec: planeSL} où nous caractérisons les échelles de $t$ et $N_c$ fini de la CGF dans le plan $s-L$. Avant de présenter nos conclusions en Sec.~\ref{sec:conclusion}, nous discutons des effets de la transition de phase dynamique 
sur les scalings en Sec.~\ref{sec:DPTcp}. 

\bigskip

Alternativement aux méthodes mentionnées au début de cette thèse, on peut faire usage d'une approche complètement différente afin d'étudier des événements rares.
Ceci est l'étude empirique des modèles qui se cachent derrière les \textbf{ données} correspondant à certains phénomènes naturels ou sociaux
 (e.g., tremblements de terre, marchés boursiers, météo, épidémies, etc). 
Dans un contexte de séries temporelles financières, ces modèles sont connus comme \textbf{stylized facts} ou \textbf{seasonalities}~\cite{1, 2, 3, 4, 5, 6, 7} et les rares événements d'intérêt pourraient correspondre, par exemple, à des chutes de marché ou à des bulles financières~\cite{11,12}. 
Ces stylized facts ont la caractéristique d'être communs et persistants sur différents marchés, périodes de temps et actifs, éventuellement~\cite{7} parce que les marchés fonctionnent en synchronisation avec les activités humaines qui laissent une trace dans les séries temporelles financières.

\medskip

Cependant, l'utilisation de la ``bonne horloge'' pourrait être d'une importance primordiale lorsqu'il s'agit de propriétés statistiques et les modèles pourraient varier en fonction si nous utilisons des données quotidiennes ou  ``intra-day data'' et ``event time'', temps commercial ou des intervalles de temps arbitraires (e.g., $T = 1$, $5$, $15$ minutes, etc). 
Par exemple, il est un fait bien connu que les distributions empiriques des rendements financiers et  log-returns sont ``fat tailed''~\cite{8, 9}. Cependant, comme on augmente l'échelle de temps du fat-tail la propriété devient moins prononcée et la distribution approche la forme gaussienne~\cite{10}. Commel'a indiqué dans Ref.~\cite{4}, le fait que la forme de la distribution change avec le temps indique clairement que le processus aléatoire sous-jacent aux prix doit avoir une structure temporelle non triviale.

\medskip

Dans un travail précédent, Allez et al.~\cite{7} a établi plusieurs nouveaux stylized facts concernant les intra-day seasonalities de la dynamique des stocks individuels et transversaux. Cette dynamique est caractérisée par l'évolution des moments de ses retours au cours d'une journée type. 
%
%
Basé sur les travaux de Allez et al.~\cite{7} et Kaisoji~\cite{11}, au chapitre~\ref{chap:Intraday}, 
nous effectuons une analyse statistique sur les rendements et les prix relatifs des CAC~$40$ et S\&P~$500$. 
Nous analysons les \textbf{intra-day seasonalities} de la dynamique du individuels et transversal stocks en le caractérisant 
%
par l'évolution des moments des retours (et des prix relatifs) au cours d'une journée typique. 
Pour ``single stock intra-day seasonalities'' nous nous référons au comportement moyen des moments des retours (et prix relatifs) d'un stock moyen dans une journée moyenne. De même, la cross-sectional intra-day seasonality n'est pas plus que le comportement moyen d'un moment d'index. 
Nous présentons ces saisonnalités pour les retours (Figs.~\ref{fig:Fig2E} et~\ref{fig:Fig3E}) et prix relatifs (Figs.~\ref{fig:Fig7E} et~\ref{fig:Fig8E}) et comparé la moyenne des stocks de la volatilité des stocks individuels $\left[\sigma _{\alpha }(k)\right]$, la moyenne temporelle de la cross-sectional volatility $\langle\sigma _{d}(k,t)\rangle$ et la valeur absolue moyenne du equi-weighted index $\langle|\mu _{d}|\rangle$ (Figs.~\ref{fig:Fig4E} et~\ref{fig:Fig9E}). 

\medskip

Notamment, dans le cas des retours, ces modèles dépendent réellement de la taille de la boîte. This fact is well illustrated with $5$ différentes valeurs de la taille de la boîte à travers de Fig.~\ref{fig:Fig11E} pour les volatilités et Fig.~\ref{fig:Fig12E} pour le kurtosis dans lequel son inversé U-pattern est évident au moment nous considérons petites tailles de boîte.
Dans le cas des prix relatifs, les volatilités présentent également le même type de intra-day pattern (Fig.~\ref{fig:Fig9E}), mais contrairement aux retours, il est indépendant de la taille de la boîte, et l'indice que nous considérons, mais caractéristique pour chaque indice. Nous suggérons dans Sec.~\ref{AP} comment cette indépendance de taille de boîte le intra-day patterns en prix relatif pourrait être utilisé pour caractériser \textbf{atypical days} pour les index et \textbf{anomalous behaviors} en stocks. Ceci est présenté dans Figs.~\ref{fig:Fig13E} et~\ref{fig:Fig14E} où nous avons présenté nos intra-day seasonalites pour le (a) moyenne et (b) la volatilité en bleu et les respectifs les cross-sectional moments pour 3 jours (et les moments de stock unique pour 3 stocks) pris au hasard en bleu clair et nous avons vu comment le comportement moyen de leurs moments se déplacent avec nos intra-day patterns ce qui n'était pas le cas pour la journée $11$ et le stock $228$.
Comme cette thèse est axée sur l'algorithme de clonage, nous avons préféré laisser cette étude dans le dernier chapitre~\ref{chap:Intraday}:~\textbf{Intra-day Seasonalities in High Frequency Financial Time Series}~\cite{binsize}.

\bigskip

Comme déjà suggéré par le placement de citations à côté des chapitres, séparé de~\hyperref[chap:I]{Introduction}, où nous établissons nos définitions, le reste de cette thèse est basée sur les résultats qui sont apparus dans~\hyperref[pubs]{Publications} produit pendant ce programme de doctorat. 
La recherche actuelle et prospective, ainsi que quelques questions ouvertes, sont présentées dans le~\hyperref[chap:Perspectives]{Perspectives} après le~\hyperref[chap:Conclusion]{Conclusion}.



\tableofcontents	

	\clearpage
	\thispagestyle{empty}
	\phantom{a}

\listoffigures


\chapter[ \qquad Preface]{Preface}
\label{chap:Presentation}
The occurrence of \textbf{rare events} can vastly contribute to the evolution of physical systems because of their potential dramatic effects. Their understanding has gathered a strong interest and, focusing on stochastic dynamics, a large variety of numerical methods have been developed to study their properties~\cite{touchette_large_2009,giardina_simulating_2011, bucklew_introduction_2013}.
They range from 
importance sampling~\cite{kahn1951estimation},
adaptive multilevel splitting~\cite{cerou_adaptive_2007} to transition path sampling~\cite{cochran1977sampling,Hedges1309,Speck,bolhuis_transition_2002}
and from ``go with the winner'' algorithms~\cite{aldous1994go,GRASSBERGER200264}
to discrete-time~\cite{giardina_direct_2006} or continuous-time~\cite{lecomte_numerical_2007} population dynamics~\cite{tailleur_simulation_2009, giardina_simulating_2011}. 
These methods have been generalized to many contexts~\cite{delmoral,HGprl09,hurtado_current_2009,lelievre,Vanden}. 
In Physics, those are being increasingly used in the study of complex systems, for instance in the study of current fluctuations in models of transport~\cite{DerridaLebowitz,Derrida,MFT}, glasses~\cite{Hedges1309}, protein folding~\cite{Weber1} and signalling networks~\cite{Vaikuntanathan,Weber2}. 
Mathematically, the procedure amounts to determining a large deviation function (LDF) associated to the distribution of a given trajectory-dependent observable, which in turns can be reformulated in finding the ground state of a linear operator~\cite{hugoraphael}, a question common to both statistical and quantum physics~\cite{DMC}.
%
\medskip

In this thesis, we will give particular attention to \textbf{population dynamics algorithms}~\cite{giardina_direct_2006,tailleur_probing_2007,
lecomte_numerical_2007,tailleur_simulation_2009,giardina_simulating_2011} which aim at studying rare trajectories by exponentially biasing their probability. 
%
%
The idea of population dynamics is to translate the study of a class of rare trajectories (with respect to a determined global constraint) into the evolution of several copies of the original dynamics, with a local-in-time selection process rendering the occurrence of the rare trajectories typical in the evolved population.
The distribution of the class of rare trajectories in the original dynamics is related with the exponential growth (or decay) of the population of clones of the system and LDF can be estimated from its growth rate. 
%
\medskip

The numerical procedures aimed at simulating rare events efficiently, using a  population dynamics scheme are commonly refereed as~\textbf{cloning algorithms}. In such algorithms, copies of the system are evolved in parallel and the ones showing the rare behavior of interest are multiplied iteratively \cite{DMC,Glasserman_1996,2001IbaYukito,GRASSBERGER200264,
4117599,CappeGuillinetal,Forwardinterfacesampling,PhysRevLett.94.018104,
delmoral2005,Dean2009562,lelievre,giardina_direct_2006,lecomte_numerical_2007,
tailleur_probing_2007,tailleur_simulation_2009,giardina_simulating_2011,
1751-8121-46-25-254002_2013}. Some of the limitations and associated improvements of the population dynamics algorithms have been studied in~\cite{hurtado_current_2009,tchernookov_list-based_2010,kundu_application_2011,nemoto_population-dynamics_2016}.
\medskip

Two versions of such algorithms exist: the \textbf{non-constant} and the \textbf{constant total population} approaches. For the last one, a uniform pruning/cloning is applied on top of the cloning dynamics so as to avoid the exponential explosion or disappearance of the population.
While the later version is obviously more computer-friendly, the former version presents interesting features: First, it is directly related to the evolution of biological systems (stochastic jumps representing mutations, selection rules being interpreted as Darwinian pressure); second, the uniform pruning/cloning of the population, although unbiased, induces correlations in the dynamics that one might want to avoid; last, in some situations where the selection rates are very fluctuating, the constant-population algorithm cannot be used in practice because of finite-population effects (population being wiped out by a single clone), and one has to resort to the non-constant one. An example of the implementation of this version can be found in Ref.~\cite{ViscoTrizac}.
\medskip

In chapter~\ref{chap:Discreteness}: \textbf{Discreteness Effects in Population Dynamics}~\cite{hidalgo_discreteness_2016}, we apply the non-constant population algorithm in order to analyze numerically the small population-size effects in the initial transient regime. 
These effects play an important role for the numerical determination of large deviation functions of additive observables for stochastic processes. 
The LDF estimation, in this case, reduces to the determination of the growth rate of a population, averaged over many realizations of the dynamics. 
However, this averaging 
is highly dependent not only on the number of realizations,
and on the initial population size but also on the cut-off time (or population) considered to stop their numerical evolution. This may result in an over-influence of \textbf{discreteness effects at initial times}, caused by small population size. We show how to overcome these effects by introducing a (realization-dependent) time delay in the evolution of populations, additional to the discarding of the initial transient regime of the population growth where these discreteness effects are strong. We show that the improvement in the estimation of the large deviation function comes precisely from these two main contributions.
\medskip

Apart from the population-constraint approaches we just mentioned, the selection mechanisms within the cloning algorithm can be implemented in different ways. One of them, along with each evolution of the copies \textbf{(Continuous-Time)}~\cite{tailleur_simulation_2009, lecomte_numerical_2007, giardina_simulating_2011} or alternatively, for each pre-fixed time-interval \textbf{(Discrete-Time)}~\cite{giardina_direct_2006}. The important differences between both techniques are discussed in Secs.~\ref{subsubsec:differenceContinuousTime} and~\ref{Discrete-time_algorithm}. 
\medskip

The algorithm  proposed by Giardin\`a 
et al.~\cite{giardina_direct_2006,lecomte_numerical_2007,
tailleur_probing_2007,tailleur_simulation_2009,giardina_simulating_2011,
1751-8121-46-25-254002_2013} (a discrete-time approach) is used to evaluate numerically the LDF of additive (or ``time-extensive'') observables in Markov processes~\cite{opac-b1093895,touchette_large_2009}. 
The LDF is obtained as the exponential growth rate that the population would present if it was not kept constant. 
It has been applied to many physical systems, including chaotic systems, glassy dynamics and non-equilibrium lattice gas models, and it has allowed the study of novel properties, such as the behavior of breathers in the Fermi-Pasta-Ulam-Tsingou chain~\cite{tailleur_probing_2007}, dynamical phase transitions in kinetically constrained models~\cite{garrahanjacklecomtepitardvanduijvendijkvanwijland}, and an additivity principle for simple exclusion processes~\cite{PhysRevLett.92.180601,hurtado_large_2010}. 
Under this approach, the corresponding LDF estimator is in fact valid only in the limits of infinite simulation time~$t$ and infinite population size~$N_c$. The usual strategy that is followed in order to obtain those limits is to increase the simulation time and the population size until the average of the estimator over several realizations does not depend on those two parameters, up to numerical uncertainties. 
\medskip

While the method has been used widely, there have been less studies focusing on the analytical justification of the algorithm. Moreover, it introduces two additional parameters into consideration:
the population size $N_c$ and the simulation time $t$, both of which affect considerably the accuracy of the LDF estimation. Even though it is heuristically believed that the LDF estimator converges to the correct result as the number of copies $N_c$ increases, there is no proof of this convergence. Related to this lack of proof, although we use the algorithm by assuming its validity, we do not have any clue how fast the estimator converges as $N_c \rightarrow \infty$.  
We discuss this convergence performing an analytical study in discrete time in chapter~\ref{chap:DiscreteTime} and using a numerical approach in continuous time in chapter~\ref{chap:ContinuousTime}. It is important to remark that the two versions of the algorithm (discrete- and continuous-time) differ on a crucial point which implies that an extension of the analysis developed in chapter~\ref{chap:DiscreteTime} cannot be done straightforwardly in order to comprehend the continuous-time case in chapter~\ref{chap:ContinuousTime}. 
\medskip

In chapter~\ref{chap:DiscreteTime}: \textbf{Finite-Time and -Size Scalings in the Evaluation of Large Deviation Functions: I. Analytical Study using a Birth-Death Process}~\cite{partI}, in order to 
to study the numerical errors of this algorithm, we explicitly devise a stochastic birth-death process that describes the time evolution of the population probability. 
From this formulation, we derived the finite-$N_c$ and finite-$t$ scalings of the systematic errors of the LDF estimator, showing that these behave as $1/N_c$ and $1/t$ in the large-$N_c$ and large-$t$ asymptotics respectively. 
In principle, knowing the scaling {\it a priori} means that the asymptotic limit of the estimator in the $t \to \infty$ and $N_c \to \infty$ limits may be interpolated from the data at finite $t$ and $N_c$. 
However, whether this idea is actually useful or not is a non-trivial question, as there is always a possibility that onset values of the $\mathbf{N_{c}^{-1}}$- and $\mathbf{t^{-1}}$\textbf{-scalings} are too large to use these scalings. 
\medskip
%

In chapter~\ref{chap:ContinuousTime}: \textbf{Finite-Time and -Size Scalings in the Evaluation of Large Deviation Functions: II. Numerical Approach in Continuous Time}~\cite{partII}, we consider a {continuous-time} version of the population dynamics algorithms~\cite{lecomte_numerical_2007,tailleur_simulation_2009}. 
We show numerically that one can indeed make use of these properties in order to devise an original and simple method that takes into account the exact scalings of the finite-$t$ and finite-$N_c$ corrections in order to provide significantly better LDF estimators (\textbf{scaling method}). We study the fluctuations of the standard estimator in chapter~\ref{chap:CGF}~\cite{partII} and additionally, we discuss an alternative way of defining the LDF estimator. 
%
%
However, the validity of these scalings and the method efficiency is proved
in chapter~\ref{chap:ContinuousTime} 
only in cases for which the number of sites $L$ (where the dynamics occurs) 
was small: a simple two-states annihilation-creation dynamics (in one site) and a one-dimensional contact process~\cite{CP, GRASSBERGER1979373, liggett2012interacting} (with $L=6$ sites).
%
\medskip

In chapter~\ref{chap:LargeL}: \textbf{Breakdown of the Finite-Time and Finite-$N_c$ Scalings in the Large-$L$ Limit}~\cite{largeLCP}, we complement the results presented in chapter~\ref{chap:ContinuousTime} 
by extending the analysis of the finite-scalings of the LDF to a large-$L$ contact process. 
The dependence of these scalings with the number of sites is analyzed by introducing the exponents $\gamma_{t}$ and $\gamma_{N_c}$. The generalized $\mathbf{t^{-\gamma_{t}}}$\textbf{-} and $\mathbf{N_{c}^{-\gamma_{N_c}}}$\textbf{-scalings} allow to characterize the behavior in the large-$L$ limit where we verify that $t^{-1}$ and $N_c^{-1}$-scalings are no longer valid.
\medskip

Alternatively to the methods mentioned at the beginning of this introduction, one can make use of a completely different approach in order to study rare events.
This is the empirical
study of the patterns that hide behind the \textbf{data} corresponding to some natural or social phenomena (e.g., earthquakes, stock markets, weather, epidemics, etc). 
In a financial time series context, these patterns are known as \textbf{stylized facts} or \textbf{seasonalities}~\cite{1, 2, 3, 4, 5, 6, 7} and the rare events of interest could correspond, for example, to market crashes or financial bubbles~\cite{11,12}. 
These properties have the characteristic of being common and persistent across different markets, time periods and assets possibly~\cite{7} because markets operate in synchronization with human activities which leave a trace in the financial time series.
\newpage
Following specially the works by Allez et al.~\cite{7} and Kaisoji~\cite{11}, in chapter~\ref{chap:Intraday} 
we perform a statistical analysis over the returns and relative prices of the CAC~$40$ and the S\&P~$500$. 
We analyze the \textbf{intra-day seasonalities} of single and cross-sectional stock dynamics by characterizing it 
%
by the evolution of the moments of the stock returns (and relative prices) during a typical day. 
We show the bin-size and index independence 
for the case of the relative prices but not for the returns. 
However, we suggest how this fact could be used in order to characterize \textbf{atypical days} for indexes and \textbf{anomalous behaviours} of stocks. As this thesis is focused on the cloning algorithm, we have preferred to leave this study in the last chapter~\ref{chap:Intraday}:~\textbf{Intra-day Seasonalities in High Frequency Financial Time Series}~\cite{binsize}.
\medskip
\medskip

As already suggested by the placement of citations next to the chapters, apart from the~\hyperref[chap:I]{Introduction}, where we establish our definitions, the rest of this thesis is based on results that have appeared in~\hyperref[pubs]{Publications} produced during this PhD program. 
The current and prospective research, as well as some open questions, are presented in the~\hyperref[chap:Perspectives]{Perspectives} after the~\hyperref[chap:Conclusion]{Conclusion}.

\clearpage
\thispagestyle{plain}
\par\vspace*{.35\textheight}{\centering 
\Large \guillemotleft  Le secret de la libert\'{e} est d'\'{e}clairer les hommes,\\
comme celui de la tyrannie est de les retenir dans l'ignorance\guillemotright \\
\medskip 
\large Maximilien Robespierre \par}

\clearpage
\thispagestyle{empty}
\phantom{a}

\mainmatter
\chapter[\quad Introduction]{Introduction }
\label{chap:I}
\section[LDT: From Boltzmann to Cloning Algorithms]{\large{Large Deviation Theory: From Boltzmann to Cloning Algorithms}} 
The theory of large deviations deals with probabilities 
of rare
events~\cite{touchette_large_2009,TouchetteRev2,TouchetteRev3}. 
These probabilities or fluctuations have the characteristic of decaying exponentially as a function of
some parameter (like the time or the temperature) meaning that, as the parameter becomes larger, the event becomes less probable~\cite{Varadhan1351}. 
They are of important interest in many fields like statistics, queuing theory, finance, engineering and in equilibrium and non-equilibrium statistical physics.
From a practical point of view, large deviation theory can be seen as a collection of methods 
which allow to determine if a large deviation principle exists for a given random variable and to determine its respective rate (or large deviation) function (LDF). 

The first large deviation result is due to Boltzmann in 1877~\cite{Boltzmann1877, EllisRev1}. 
He showed how the relative entropy expresses the asymptotic behavior of multinomial probabilities presenting the entropy as a bridge between the microscopic level, of physical interactions, and a macroscopic one, where the physics laws are formulated. 
This constituted a probabilistic interpretation of the Second Law of Thermodynamics~\cite{Boltzmann1877}
and the basis which led to the development of the classical equilibrium statistical mechanics~\cite{Gibbs}. Ellis~\cite{EllisRev1} describes this 
interpretation 
as ``a revolutionary moment in human culture during which both statistical mechanics and the theory of large deviations were born''. 

Some large deviation results like Cram\'{e}r's theorem~\cite{Cramer} (who initiated a mathematical theory of large deviations in the 30's), Chebyshev’s inequality~\cite{opac-b1093895} and the Sanov’s theorem~\cite{Sanov}, 
were also anticipated by Boltzmann~\cite{Boltzmann1877, EllisRev1}. However, there was not a unified or general framework that dealt with them until the 60's and 70's when 
this theory was developed 
by Donsker and
Varadhan~\cite{DonskerVar1, DonskerVar2, DonskerVar3, DonskerVar4, Varadhan} and by Freidlin and Wentzell~\cite{Freidlin}.

In some cases, the large deviation principle can be determined directly from the probability distribution of a random variable. This is done by deriving a large deviation
approximation using Stirling’s or other asymptotic formulae. 
However, a more general result was provided 
by the G\"{a}rtner-Ellis theorem~\cite{Gartner, EllisGartner} which is the product of a result proved by G\"{a}rtner~\cite{Gartner} and later generalized by Ellis~\cite{EllisGartner,EllisRev1,EllisRev2,ellis2006theory,Ellis} which explicitly refers to the construction of the currently adopted large deviation principle.
This was inspired from the work of Varadhan~\cite{Varadhan}. However, meanwhile the G\"{a}rtner-Ellis theorem is used to prove the existence of a large deviation principle and the determination of the corresponding rate function from the knowledge of the scaled cumulant generating function (CGF), the Varadhan’s theorem is used to calculate the CGF knowing the rate function. Moreover, the contraction principle~\cite{DonskerVar4} introduced by Donsker and Varadhan allows to compute a rate function from the knowledge of another rate function. 
A direct application of the G\"{a}rtner-Ellis theorem or of the contraction principle
allows to formulate a large deviation principle for many problems like 
 sums of (binary, symmetric Levy, totally skewed Levy, etc) i.i.d. random variables (Cram\'{e}r theorem~\cite{Cramer}), random vectors (Sanov's theorem~\cite{Sanov}), Markov processes~\cite{DonskerVar1, DonskerVar2, DonskerVar3, DonskerVar4, Varadhan}, among others. 
 
Donsker and Varadhan defined three levels of large deviation results~\cite{Ellis}. Level-1 is the level of sample means, Level-2 is the level of empirical distributions and Level-3 is the level of the empirical processes. 
The latter distributions can be derived from the former using the contraction principle~\cite{DonskerVar4}. For example, the Level-2 rate function of Markov chains can be derived by
contracting the large deviations of the pair empirical matrix~\cite{touchette_large_2009}.

Large deviation theory has been suggested to be a generalization of the Central Limit Theorem~\cite{Martin, Bryc} because it provides information not only about the small but also its large fluctuations of a random variable far away from its typical values. 
It is also considered that it extends the Law of Large Numbers~\cite{Lanford} providing information of how fast a random variable converges in probability to its mean.
In fact, the existence of a Law of Large Numbers for a random variable is a good sign that there also holds a large deviation principle and also it can be used as a departure point~\cite{O'Connell1,O'Connell2}.

Some physicists consider large deviation theory as a natural generalization of the entropy-probability relation fully exploited by Einstein in his theory of thermodynamic fluctuations~\cite{Einstein1, Einstein2}.
According to this theory, the probabilities can be expressed in
terms of entropy functions. This fact allows to use large deviation theory to understand the foundations of statistical mechanics. In this way, large deviation theory explains for example why the entropy and free energy are related through a Legendre transform and why equilibrium states can be calculated via extremum principles (maximum entropy for the microcanonical ensemble and minimum free energy for the canonical ensemble) 
generalizing them to arbitrary macrostates and
arbitrary many-particle systems~\cite{touchette_large_2009}.
On the other hand, the well-known maximum entropy principle of Jaynes~\cite{Jaynes1,Jaynes2,Jaynesbook} can be obtained by considering the Level-2 large deviations of systems of independent particles. 
Einstein fluctuation formula was used by Varadhan and Donsker~\cite{DonskerVar1, DonskerVar2, DonskerVar3, DonskerVar4, Varadhan} as the basis of the standard theory for static equilibrium fluctuations. Additionally, Onsager and Machlup~\cite{OnsagerM1,OnsagerM2,Falkoff1,Falkoff2} used it in order to propose a reformulation of linear fluctuation theory about equilibrium.

The implementation of large deviation techniques for studying the equilibrium properties of many-particle systems described at a probabilistic level by statistical mechanical ensembles 
has its roots in the work of Ruelle~\cite{Ruelle1}, Lanford~\cite{Lanford}, and
especially Ellis~\cite{Ellis, EllisRev1, EllisRev2}. Ellis is considered to provide (in Ref.~\cite{Ellis}) the most complete framework in which the large deviation theory is introduced to physics stressing in the connections between probability, large deviations and equilibrium statistical mechanics.
The first work on large deviations and statistical mechanics is attributed to Lanford~\cite{Lanford} which uses concepts from large deviation theory to explain the fact that while matter is extremely complicated at microscopic level, it can be described at the macroscopic level by a small number of parameters. Moreover, 
using a large deviation approach on the ensembles in statistical mechanics,
the study of equilibrium states and
their fluctuations can be reduced to the study of properly defined rate functions (entropy functions)~\cite{touchette_large_2009,Ellis,Lewis1, Lewis2, Lewis3, Lewis4, Lewis5, Lewis6}.
Many other links  between statistical mechanics and large deviations also has been discussed by Lewis, Pfister, and Sullivan~\cite{Lewis1, Lewis2, Lewis3, Lewis4, Lewis5, Lewis6}, as well as Oono~\cite{Oono}, Amann~\cite{Amann} 
and in several reviews~\cite{touchette_large_2009, TouchetteRev2,TouchetteRev3, EllisRev1, EllisRev2}.

\newpage
Behind the application of large deviation theory to equilibrium statistical physics lies the idea that outcomes of a macrostate involving $n$ particles should concentrate in probability around
certain stable or equilibrium values
even though the state of the particles is described by
a random variable. 
In many cases the outcomes satisfy a large deviation principle due to the probability of observing a departure from these equilibrium
values is exponentially small with $n$. Thus, 
in order to describe the  macrostate of a large
many-particle system it is only necessary to know its equilibrium values~\cite{touchette_large_2009,Ellis, TouchetteRev2, Lewis1, Lewis2, Lewis3, Lewis4, Lewis5, Lewis6, Oono, Amann}.

Additionally to the equivalences already mentioned, we have that the thermodynamic limit is a large deviation limit, and the free energy is the equivalent of a scaled cumulant generating function~\cite{touchette_large_2009,TouchetteRev2,TouchetteRev3}.
Moreover, the Legendre transform which connects the entropy and free energy in thermodynamics is nothing but the Legendre-Fenchel transform connecting the rate function and the scaled cumulant generating function in the G\"{a}rtner-Ellis theorem~\cite{Gartner, EllisGartner} and in Varadhan’s theorem~\cite{Ellis}.
The equilibrium properties of mean field models can be studied as Level-2 or directly at the Level-1 of large deviations. Maximum entropy principles
have been applied successfully to these models which consider all-to-all coupling between particles like 
the Curie-Weiss model~\cite{EllisRev2, Ellis,Eisele} and its parent model, the Potts model~\cite{EllisRev2, Orey, EllisWang, Costeniuc}, the Blume-Emery-Griffiths model~\cite{Barre, EllisTouchette, EllisOtto}, the mean-field Hamiltonian model~\cite{BarreBouchet}, as well as mean-field versions of the spherical model~\cite{Kastner, Casetti}, and the $\phi^{4}$ model~\cite{Hahn1, Hahn2, Campa}.

The microcanonical and canonical ensembles differ from each other in the way their respective microstates are weighted. In the microcanonical ensemble, the control parameter is the energy (or the mean energy), and the microstates are distributed with the same probabilistic weight if they have the same value of control parameter. 
On the other hand, 
in the canonical ensemble, the control parameter is the inverse temperature, the  probability measure is the Gibbs measure and the rate functions are the macrostate free energies (which are the basis of the Ginzburg–Landau theory of phase transitions~\cite{Landau}). Some examples of results derived in the canonical ensemble can be found in Refs.~\cite{EllisRev1,EllisRev2,Ellis,Barre}.
%
%
The thermodynamic equivalence (or non-equivalence) between the microcanonical and canonical ensembles is related to the concavity of the entropy. This comes from the fact that the free energy can always be obtained as the Legendre–Fenchel transform of the entropy, but the entropy
can be obtained as the Legendre–Fenchel transform of the free energy only when the entropy is concave.
%
%
Moreover, the G\"{a}rtner-Ellis theorem can be reformulated (in a physical way) as : ``If there is no first-order phase transition in the canonical ensemble, then the microcanonical entropy is the Legendre transform of the canonical free energy''~\cite{touchette_large_2009}. 
%
 Examples of models with non-concave entropies are the mean-field Blume-Emery-Griffiths model~\cite{Barre, EllisTouchette, EllisOtto}, the
mean-field Potts model~\cite{Costeniuc, Ispolatov}, some models of plasmas~\cite{Kiessling1} and 2D turbulence~\cite{EllisHaven1,EllisHaven2,Kiessling2}, as well as models of gravitational systems~\cite{Chavanis, Lynden}. 
This thermodynamic equivalence is translated in terms of Gibbs’s entropy and Boltzmann’s entropy in the thermodynamic limit, where the Gibbs entropy is equal (up to a constant) to the Boltzmann entropy evaluated at the equilibrium mean energy value~\cite{touchette_large_2009, Lebowitz}.

Large deviation theory is becoming the standard formalism to study non-equilibrium systems~\cite{touchette_large_2009, DerridaNon}, modelled in general by stochastic differential equations or Markov processes~\cite{Gardiner}. They have the characteristic of evolving dynamically in time or to be maintained in out-of-equilibrium steady states under the application of an external force.
It has been suggested that large deviation theory provides the proper basis for building a theory of non-equilibrium systems~\cite{Oono, Eyink}.
This requires, of course the inclusion of the time in the large deviation analysis and the consideration that we do not know the underlying probability distribution states as the concept of ensemble is not defined for non-equilibrium systems.
In spite of this, many large deviation principles have been derived for example for Markovian models of interacting particles~\cite{liggett2012interacting,DerridaNon,Bertini, Spohn}
such as the exclusion process, the zero-range process and their different variants~\cite{DerridaLebowitz,DerridaASEP,Benois,Landim, PhysRevLett.92.180601,BodineauDerrida2,BodineauDerrida3,Kipnis1} in some cases at the level of density field~\cite{Landim, PhysRevLett.87.150601,Derrida2002,PhysRevLett.89.030601,Derrida2003}
or at the level of current~\cite{PhysRevLett.92.180601,BodineauDerrida2,BodineauDerrida3}.

The so called fluctuation theorem~\cite{Evans3} and other non-equilibrium work relations~\cite{KurchanW} concern the large deviations
of work~\cite{Kurchan6}. It states that the probability of observing an entropy production opposite to that dictated by the second law of thermodynamics decreases exponentially.
It was first proposed and tested numerically in 1993~\cite{Evans1}, the first mathematical proof was in 1994~\cite{Evans2} and it was verified experimentally in 2002~\cite{Wang}. 
Gallavotti and Cohen~\cite{GallavottiCohen1,GallavottiCohen2} used these results as a basis to proof a fluctuation
theorem for the entropy rate of chaotic deterministic systems. This was extended later to general Markov processes by
Kurchan~\cite{KurchanFT}, Lebowitz and Spohn~\cite{Lebowitz1999}, and Maes~\cite{Maes}.
%
These results have inspired several experimental studies of
fluctuation relations that appear for example, particles immersed in fluids~\cite{Wang,Andrieux}, electrical circuit~\cite{Zon,Ciliberto}, granular
media~\cite{Aumatre2001,Feitosa,PhysRevLett.95.110202,0295-5075-72-1-055,Visco2006}, turbulent fluids~\cite{Ciliberto1998215,CILIBERTO2004240}, and the effusion of ideal gases~\cite{Cleuren}, among other systems.


Other applications of large deviation theory are related to multifractals~\cite{Beck,Paladin,McCauley, Falconer}, chaotic systems~\cite{GaspardChaos, Alekseev,EckmannRuelle,Lasota}, disordered systems, and quantum systems.
Multifractal analysis can be seen as a large deviation theory of self-similar measures
~\cite{Zohar, Veneziano, Harte}.
Dynamical systems 
often give rise to large deviation principles without a perturbing noise. 
Their study 
in the context of chaotic systems and ergodic theory is the subject of the so-called thermodynamic formalism~\cite{Keller, Beck} developed by Ruelle~\cite{Ruelle2, Ruelle3} and
Sinai~\cite{Sinai1, Sinai2}. 
This formalism introduces the topological pressure and the structure function which play the role of the CGF implying a direct connection between dynamical systems and large deviation theory~\cite{Keller, Young1, Lopes, Waddington, Pollicott}.
Additionally, large deviation
principles can be obtained when studying disordered and quantum systems, for example, in random walks in random environments
~\cite{Gantert, Comets2000, VaradhanRandom, Zeitouni}, 
spin glasses~\cite{Dorlas1, Dorlas2,Talagrand2007}, 
boson gases~\cite{Cegla1988,vandenBerg1988,Dorlas2005},
quantum gases~\cite{doi:10.1063/1.533185,Gallavotti2002}, 
and 
quantum spin systems~\cite{doi:10.1063/1.2812417, Lenci2005, Neto}.

In this point, it is important to remark that only in few simple cases is it possible to obtain exact and explicit expressions for the rate functions~\cite{BodineauDerrida2, Mehl}.
For most stochastic processes, the evaluation of these functions is done by using analytical approximations and numerical methods~\cite{touchette_large_2009, TouchetteRev2,TouchetteRev3}. 
They range from 
importance sampling~\cite{kahn1951estimation},
adaptive multilevel splitting~\cite{cerou_adaptive_2007} 
to transition path sampling~\cite{cochran1977sampling,Hedges1309,Speck,bolhuis_transition_2002}
and ``go with the winner'' algorithms~\cite{aldous1994go,GRASSBERGER200264}.
Kurchan and his collaborators 
generalized a procedure used previously to study rare events in chemical reactions~\cite{Tanase1, Tanase2, TailleurTanase}
in order to compute
large deviation functions in dynamical systems~\cite{tailleur_probing_2007}, discrete-time~\cite{giardina_direct_2006, tailleur_simulation_2009} and continuous-time~\cite{lecomte_numerical_2007, tailleur_simulation_2009} population dynamics~\cite{giardina_simulating_2011}, 
being generalized then to many contexts~\cite{delmoral,HGprl09,hurtado_current_2009,lelievre,Vanden}.
 
The numerical procedure introduced by Giardin\`{a}, Kurchan and Peliti~\cite{giardina_direct_2006} overcomes the difficulty of 
observing the fluctuations of an observable (whose probability decreases exponentially in time) for discrete-time Markov chains. It was known that the large deviation function can be obtained as the largest eigenvalue of a evolution matrix of a modified dynamics~\cite{giardina_direct_2006, tailleur_simulation_2009} which can be computed numerically~\cite{BodineauDerrida2, Mehl, Baiesi} specially for small systems as the evolution matrix is exponentially large in the system size. 
Later, a modification of the procedure was proposed~\cite{lecomte_numerical_2007,Lecomte2007} for which the time discretization issues of the original approach~\cite{giardina_direct_2006} are bypassed with a direct continuous-time approach.
The evolution of the system was represented by a population dynamics of the type of the diffusion Monte Carlo~\cite{DMC}. This \textbf{cloning algorithm} was applied to successfully compute the large deviations of the total current in the symmetric and asymmetric exclusion process~\cite{Spohn,Spohn2}, and of the activity in the contact process~\cite{CP}. 

Among its applications, Garrahan et al. ~\cite{garrahanjacklecomtepitardvanduijvendijkvanwijland, garrahan_first-order_2009} analyzed the dynamics of kinetically constrained models~\cite{RitortSollich, FAmodel, Jackle1, Jackle2, Kob, Kronig, KurchanKCM, Sollich, Einax, GarrahanChandler, AldousDiaconis, Jung, Toninelli, Pan, Geissler} of glassy systems~\cite{Ediger, Angell, Binder} by analyzing the statistics of trajectories of the dynamics. They showed that these models exhibit a first-order dynamical transition between active and inactive dynamical phases.
It also has been used to study symmetries in fluctuations far from equilibrium~\cite{HurtadoPerez} and in transport models 
~\cite{HGprl09, hurtado_current_2009, hurtado_large_2010}. These studies allow not only to test the predictions
of fluctuating hydrodynamics~\cite{HGprl09, HurtadoSpon}, but also the limits of the method
itself~\cite{hurtado_current_2009}. 
It also has been suggested~\cite{giardina_simulating_2011} that the method could be applied to study in detail the possible future and past evolution of planetary systems, and also the self-organization of the stability of our solar system. 

\medskip

In this chapter, we introduce the cloning algorithm. This method will be used through the thesis in order to analyze the issues previously mentioned in the \hyperref[chap:Presentation]{Preface}. We start 
from the construction of the master equation, its solution and interpretation. Then, we introduce the large deviations of additive observables and the $s$-modified dynamics. We show how to estimate these large deviations from the population dynamics interpretation of the modified dynamics 
or from the largest eigenvalue of the modified evolution equation. Finally, we present the example models used for this analysis: a simple two-state annihilation-creation dynamics, and a contact process on a one-dimensional periodic lattice.

\section{Discrete and Continuous Master Equation} 
Consider a system whose dynamics occurs in jumps between configurations. We denote the set of available configurations $\{C\}$ and the transition rates between them $W(C\to C')$, setting $W(C\to C)=0$. We are interested in describing the probability for the system to be in configuration $C$ at time $t$, that we denote $P(C,t)$. 
In order to do that, we start from the following considerations:
During the time interval $\,dt$ the system either stays in the same configuration $C$ or changes configuration to $C'$.
Thus, the transition probability $p$ between configurations can be expressed in terms of 
$dt$ and 
$W$ as
\begin{align}
\label{w1} p(C \to C')  &= \,dt\ W(C\to C')  \qquad  \forall C'\neq C,  \\
\label{w2} p(C \to C)   &= 1 - \,dt\ {\sum_{C'}} W(C\to C').
\end{align}
The tendency of the dynamics to leave from a configuration $C$ to any other is captured in the escape rate $r(C)$, defined as 
\begin{equation}
\label{escaperate}
r(C) = {\sum\limits_{C'}^{ }}  W(C \rightarrow C')
\end{equation}
which appears in the second term of Eq.~\eqref{w2}. 
The probability of being in configuration $C$ at time $t+dt$ is none other that the probability of being at configuration $C$ given that the system was at configuration $C'$ at time $t$, plus the probability of having remained at configuration $C$ between time $t$ and $t+dt$, which can be expressed as 
 \begin{equation}
\label{discretemaster0} 
P(C,t+\,dt) = P(C,t+\,dt\mid C',t) + P(C,t+\,dt\mid C,t),
\end{equation}
where each term in Eq.~\eqref{discretemaster0} is given by
\begin{align}
\label{pC1} P(C,t+\,dt\mid C',t) &= \Big( {\sum\limits_{C'}^{ }} \,dt\ W(C'\to C) \Big) P(C',t), \\
\label{pC2} P(C,t+\,dt\mid C,t) &= \Big( 1 - \,dt\ {\sum\limits_{C'}^{ }} W(C\to C') \Big) P(C,t).
\end{align}
The transition rules~\eqref{w1} and~\eqref{w2} are equivalent to the discrete evolution equation~\eqref{discretemaster0} which after replacing Eqs.~\eqref{pC1} and~\eqref{pC2} reads
\begin{equation}
\label{discretemaster}
P(C,t+\,dt) = {\sum\limits_{C'}^{ }}\ \bigg[ \,dt\ W(C'\to C)P(C',t) + \Big( 1 - \,dt\ W(C\to C') \Big) P(C,t) \bigg].
\end{equation}
Equation~\eqref{discretemaster} is also known as the discrete master equation.
The second right term of Eq.~\eqref{discretemaster} ensures probability conservation as $\sum_{C} P(C,t+\,dt)=\sum_{C} P(C,t)=1$.
Taking the limit $\,dt \to 0$ in Eq.~\eqref{discretemaster} and replacing Eq.~\eqref{escaperate}, we obtain its analogous version in continuous time
%
\begin{equation}\label{eq:1}
\partial_{t}P(C,t)={\sum\limits_{C'}^{ }}\ \Big[ W(C'\rightarrow C)P(C',t) - r(C)P(C,t) \Big].
\end{equation}

\subsection{Conservation of Probability and Equilibrium States}
The probability $P(C,t)$ is conserved at all times, i.e.,
\begin{equation}\label{consP}
\partial_{t} {\sum\limits_{C}} P(C,t) = 0.
\end{equation}
The steady state solution $P_{\text{st}}$ of Eq.~\eqref{eq:1}, which is obtained from $\partial_{t}P(C,t)=0$, verifies the global balance condition 
\begin{equation}\label{eq:gbc}
\sum\limits_{C'} W(C \rightarrow C')P_{\text{st}}(C) = \sum\limits_{C'} W(C'\rightarrow C)P_{\text{st}}(C'),
\end{equation}
for all $C$. If the steady state also satisfies the detailed balance condition 
\begin{equation}\label{eq:dbc}
 W(C \rightarrow C')P_{\text{eq}}(C) = W(C'\rightarrow C)P_{\text{eq}}(C'),
\end{equation}
for all $C$ and $C'$, then the steady steady is an equilibrium state of the system, i.e., $P_{\text{st}} = P_{\text{eq}}$.
The last condition implies that there is no current of probability in the steady state, and that the dynamics starting from  $P_{\text{eq}}$ is reversible. This can be seen (using Eq.~\eqref{eq:dbc}) from 
\begin{equation}
 W(C_0 \rightarrow C_1) \ldots W(C_{K-1} \rightarrow C_K)P_{\text{eq}}(C_0) = W(C_{K} \rightarrow C_{K-1}) \ldots W(C _1 \rightarrow C_0)P_{\text{eq}}(C_K),
\end{equation}
where the probability density of the history $C_0 \rightarrow \cdots \rightarrow C_K$ is the same as its time-reversed history $C_K \rightarrow \cdots \rightarrow C_0$.

\section{Master Equation Matrix Form}
In order to study the properties of the master equation it is convenient to introduce the following vector and operator notations~\cite{dirac_new_1939}:
Consider an orthonormal vector space of basis $\ket{C} $ with scalar product 
$\braket{C'}{C} = \delta_{CC'}$
where $\bra{C} $ is the transpose of $\ket{C} $. A vector is denoted as
$\ket{v} = \sum_{C} v_C \ket{C} $
with $v_C = \braket{C}{v}$. An operator
$\mathbb{A} = \sum_{CC'} \mathbb{A}_{CC'} \ket{C}\bra{C'}$
can be represented in matrix form of elements 
$\mathbb{A}_{CC'} = \bra{C} \mathbb{A} \ket{C'}$.
%
Using this notation, the master equation~\eqref{eq:1} takes the linear form
\begin{equation}
\label{meq}
 \partial_t \ket{P(t)} = \mathbb W \ket{P(t)}
\end{equation}
for the probability vector 
\begin{equation}
 \ket{P(t)} = \sum\limits_{C} P(C,t) \ket{C}.
\end{equation}
The master operator $\mathbb W$ is a matrix of elements
\begin{equation}
  \label{eq:defopW}
  (\mathbb W)_{CC'}= W(C' \to C) -r(C)\ \delta_{CC'},
\end{equation}
i.e., 
\begin{equation}
(\mathbb W)_{CC'}=
    \begin{cases}
W(C\to C')             &\text{if}\  C \neq C' \\
-r( C)                 &\text{if}\  C = C'.    
    \end{cases}
\end{equation}
The diagonal elements of matrix~\eqref{eq:defopW} correspond to waiting times between jumps when the system stays in the same configuration.

\subsection{Conservation of Probability and Equilibrium States Revisited}
Using the vector notation introduced above, the conservation probability~\eqref{consP} reads as
$ \sum_{C} (\mathbb W)_{CC'} = 0$  
which with $\bra{-} = \sum_{C} \bra{C}$ becomes
\begin{equation}
\label{eq:consP3}
\bra{-} \mathbb W = 0,
\end{equation}
meaning that the vector $\bra{-} $ is a left eigenvector of $\mathbb W$ (of eigenvalue $0$).
On the other hand, the global balance condition reads  
\begin{equation}
\label{eq:gbc2}
\mathbb W \ket{P_{\text{st}}} = 0,
\end{equation}
where the vector $\ket{P_{\text{st}}}$ is the right eigenvector of $\mathbb W$ (also of eigenvalue $0$). As $\mathbb W$ and $\mathbb W^{T}$ have the same spectrum, the conservation of probability ensures the existence of a steady state. 
Moreover, all the eigenvalues of $\mathbb W$ are of real part negative.
The detailed balance condition~\eqref{eq:dbc} can be written in terms of a diagonal operator $\hat{P}_{\text{eq}}$ of elements $P_{\text{eq}}(C)$ as 
\begin{equation}
\label{eq:dbc2}
\mathbb W \hat{P}_{\text{eq}}=\hat{P}_{\text{eq}}\mathbb W^{T}.
\end{equation}
By multiplying Eq.~\eqref{eq:dbc2} by the left by $\bra{-}$ and using Eq.~\eqref{eq:consP3} we verify that $\ket{P_{\text{eq}}}$ is indeed a steady state through $\bra{-} \hat{P}_{\text{eq}} = \bra{P_{\text{eq}}}$. Moreover from Eq.~\eqref{eq:dbc2} we also have
\begin{equation}
\hat{P}_{\text{eq}}^{-1/2}\ \mathbb W\ \hat{P}_{\text{eq}}^{1/2} =\hat{P}_{\text{eq}}^{1/2}\ \mathbb W^{T}\ \hat{P}_{\text{eq}}^{-1/2},
\end{equation}
implying that $\mathbb W^{\text{sym}} = \hat{P}_{\text{eq}}^{-1/2} \mathbb W \hat{P}_{\text{eq}}^{1/2} $ is a symmetric operator (self-adjoint in fact) and it can be diagonalized in an orthonormal basis. Given that $\mathbb W $ and $\mathbb W^{\text{sym}} $ have the same spectrum, the last equation also implies that $\mathbb W$ has real eigenvalues.

\section{Solution of the Master Equation }
\label{sec:solME}
Given the initial condition $\ket{P_0} = \ket{P(t=0)} = \sum_{C} P_{0}(C) \ket{C}$, Eq.~\eqref{meq} has as solution
\begin{equation}
\label{eq:sol1}
\ket{P(t)}= e^{t\mathbb W} \ket{P(0)}= \sum\limits_{n \geqslant 0} \frac{t^n}{n!} \mathbb W^{n}\ket{P(0)}.
\end{equation}
However, as the matrix $\mathbb{W}$~\eqref{eq:defopW} has diagonal elements different from zero, Eq.~\eqref{eq:sol1} does not allow a description as a sum over trajectories
of successively different visited configurations. Thus, in order to get rid of the diagonal terms of matrix $\mathbb{W}$~\eqref{eq:defopW}, it is convenient to transform the master equation~\eqref{eq:1} by defining 
\begin{equation}
\label{eq:Q}
Q(C,t)= e^{-tr(C)}P(C,t).
\end{equation}
Equation~\eqref{eq:Q} verifies
\begin{equation}
\label{eq:dQ}
\partial_t \ket{Q(t)} = \mathbb{W}_{Q}(t) \ket{Q(t)},
\end{equation}
where
\begin{equation}
\big(\mathbb{W}_{Q}(t)\big)_{CC'} = W(C'\rightarrow C)\ e^{t (r(C')-r(C))}.
\end{equation}
The solution of Eq.~\eqref{eq:dQ} is given by
\begin{equation}
\ket{Q(t)}= \tau_{exp} \bigg\{ \int_{0}^{t} \,dt'\ \mathbb{W}_{Q}(t') \bigg\} \ket{Q(0)},
\end{equation}
where $\tau_{exp}$ is the time-ordered exponential:
\begin{equation}
\tau_{exp} \bigg\{ \int_{0}^{t} \,dt'\ \mathbb{W}_{Q}(t') \bigg\} = \sum\limits_{K \geqslant 0} \int_{t_0}^{t} \,dt_{1}  \int_{t_1}^{t} \,dt_{2} \ldots \int_{t_{K-1}}^{t} \,dt_{K}\ \mathbb{W}_{Q}(t_K) \ldots \mathbb{W}_{Q}(t_1)
\end{equation}
with $K \in \mathcal{N}$. Finally, $P(C,t)$ is obtained coming back to Eq.~\eqref{eq:Q} as
\begin{equation}
\label{eq:solutionME}
\begin{split}
P(C,t) &= \sum\limits_{K \geqslant 0}\ \sum\limits_{C_{0} \ldots C_{K}}\ \int_{t_0}^{t} \,dt_{1}  \int_{t_1}^{t} \,dt_{2} \ldots \int_{t_{K-1}}^{t} \,dt_{K}  \\[5pt]
&\times r(C_0) e^{-(t_1-t_0)\ r(C_0)} \ldots r(C_{K-1}) e^{-(t_{K}-t_{K-1})\ r(C_{K-1})} \times e^{-(t-t_K)\ r(C_{K})}  \\[5pt]
&\times \frac{W(C_0 \rightarrow C_1)}{r(C_0)} \cdots \frac{W(C_{K-1} \rightarrow C_K)}{r(C_{K-1})} \times P (C_0,t=0)
\end{split}
\end{equation}
with $C_K = C$.

\subsection{Interpretation}
Equation~\eqref{eq:solutionME} allows us to have a better visualization of the process described by the master equation~\eqref{eq:1}~\cite{tagkey2007iii}. 
The dynamics occurs in jumps between configurations with transition rates $W(C\to C')$ on a time window $[0,t]$. The histories of the systems are described by 
\begin{equation}
(\vec{C},\vec{t})=(C_{0},t_0;C_{1},t_1;\ldots;C_{k},t_k;\ldots;C_{K}=C,t_K),
\end{equation}
where $K$ is the total number of actual jumps between successively distinct configurations denoted by $\vec{C} =(C_{0},C_{1},\ldots,C_{k},\ldots,C_{K}=C)$ and $\vec{t}=(t_0,t_1,\ldots,t_k,\ldots,t_K)$ is 
the increasing sequence of times at which the jumps occurs (e.g., at time $t_k$, the system jumps from configuration $C_{k-1}$ to $C_{k}$). The factors 
\begin{equation}
\frac{W(C_{k-1} \rightarrow C_k)}{r(C_{k-1})}
\end{equation}
in Eq.~\eqref{eq:solutionME} represent the probability of jumping from configuration $C_{k-1}$ to $C_k$. Hence, the probability of a history of configurations $\vec{C}$ is given by
\begin{equation}
P(\vec{C}) = \prod \limits_{k = 0}^{K-1} \frac{W(C_{k} \rightarrow C_{k+1})}{r(C_{k})}.
\end{equation}
The system rests in a configuration $C_k$ during an interval $\Delta t_k=t_{k+1}-t_k$ which is distributed with a Poisson law of parameter $r(C_k)$
\begin{equation}
\rho(r(C_k))= r(C_k)\ e^{-\Delta t_k\ r(C_k)}.
\end{equation}
Thus, the probability density of instants $\{ t_k \}_{1 \leq k \leq K}$ of the  change of configuration is
\begin{equation}
 \prod \limits_{k = 0}^{K-1} r(C_k)\ e^{-\Delta t_k\ r(C_k)}.
\end{equation}
Meanwhile the probability of not jumping betweens times $t_K$ and $t$ is 
$ e^{-(t-t_k)\ r(C_K)}$. Thus, if we fix the initial configuration $C_0$, the probability of the path is given by 
\begin{equation}
\mathbb{P}(C_{1},t_1;\ldots;C_{K},t_K \mid C_{0},t_0;t) = 
\prod \limits_{k = 0}^{K-1} \frac{W(C_{k} \rightarrow C_{k+1})}{r(C_{k-1})} \times 
\prod \limits_{k = 0}^{K-1} r(C_k) e^{-\Delta t_k r(C_k)} \times  e^{-(t-t_k)}.
\end{equation}

\section{Large Deviations of Time-Extensive Observables}
\label{sec:LDobservables}
Once we have defined our system, 
we are now interested in the distribution of history-dependent observables and its fluctuations. These dynamical observables are defined as a sum along the history of small contributions for transitions between successive configurations during a time interval $[0,t]$. In general, they are of the form
\begin{equation}
\label{eq:obs}
\mathcal O =
\sum_{k=0}^{K-1} a(C_k,C_{k+1}) + \int_0^t dt'\:b(C(t')),
\end{equation}
where $C(t')$ is the state of the system at time $t'$: when $t_k \leq t' < t_{k+1}$, $C(t')=C_k$ ($k=0,1,2,\cdots, K-1$) with $t_0=0$. The functions $a$ and~$b$ describe elementary increments: $a$ accounts for quantities associated with transitions (of state), whereas $b$ does for static quantities.  
We commonly refer to observables of the form \eqref{eq:obs} for $b=0$ to `type-A', meanwhile to those for which $a=0$ to `type-B' observables~\cite{garrahan_first-order_2009}.
A simple example of observables of this form is the dynamical activity $K$~\cite{garrahanjacklecomtepitardvanduijvendijkvanwijland,garrahan_first-order_2009,Hedges1309,ThermoCP,PhysRevE.81.011111,ChandlerGarrahan,QuantumJump,
Micromaser,Genway,Ates,Hickey,jackEast}, which is the number of configuration changes on the time interval $[0,t]$ (in this case one has $a(C,C')=1$ and $b\equiv 0$ in Eq.~\eqref{eq:obs}). Another, is the current of particles $Q$~\cite{RakosHarris,GorissenDMRG,GorissenVanderzande,GorissenLazarescu,
HurtadoSpon,EspigaresDTP,Hurtado2014,LDQ}
in a one-dimensional lattice gas, where the value of the observable $Q$ is incremented or decremented at each time a particle jumps to the left or right. This kind of observables contrasts from the static ones which depend only on the configuration of the system at a given time.

The probability density of being in configuration $C$ at time $t$ having observed a value $\mathcal O$ of observable is denoted by $P(C,\mathcal O,t)$ and is related through the probability distribution of $\mathcal O$ at time $t$, $P(\mathcal O,t)$, by
\begin{equation}
P(\mathcal O,t)= \sum_{C}^{} P(C,\mathcal O,t).
\end{equation}
This probability distribution scales as 
\begin{equation}
\label{eq:ldpO}
P(\mathcal O,t) \sim e^{t \pi(\mathcal O / t)}
\end{equation}
in the infinite-time limit. Equation~\eqref{eq:ldpO} is known as the large deviation principle for observable $\mathcal O$~\cite{touchette_large_2009}. It can be interpreted as the probability of observing an atypical value of observable $\mathcal O$ after a large-time scale.
The rate function $\pi(\mathcal O / t)$ is a dynamical equivalent of the intensive entropy in the microcanonical ensemble and it is known as the large deviation function~\cite{touchette_large_2009}. It encodes not only the Gaussian but also the non-Gaussian fluctuations (or large deviations) of the observable $\mathcal O / t$ which can be obtained by an expansion beyond the quadratic order of the function $\pi(\mathcal O / t)$. 
In the infinite-time limit the function $\pi(\mathcal O / t)$ may not be analytic which can be interpreted as a signature of dynamical heterogeneities (dynamical phase transition)~\cite{Merolle10837, jackST}.

The problem of the determination of the rate function $\pi(\mathcal O / t)$ is in general a difficult task, one thus prefers to go to the  dynamical canonical ensemble or Laplace space. Instead of fixing the value of the observable $\mathcal O $ in order to determine $\pi(\mathcal O / t)$ one introduces a parameter $s$ (intensive in time) which biases the statistical weight of histories and fixes the average value of $\mathcal O $, so that $s \neq 0$ favors its non-typical values.
In order to do that, we introduce the dynamical partition function
(or moment generating function)
\begin{equation}
  \label{eq:MCF}
  Z(s,t)=\langle e^{-s\mathcal O}\rangle,
\end{equation}
where $\langle \cdot \rangle$ is the expected value with respect to trajectories of duration~$t$. 
Since the observable $\mathcal O$ is additive and the system is described by a Markov process, 
$Z(s,t)$ satisfies at large times the scaling
\begin{equation}
\label{eq:psi}
Z(s,t) \sim e^{t\psi(s)}
\end{equation}
for $t\to\infty$. The growth rate of $Z(s,t)$ with respect to time, $\psi(s)$ is known as
the scaled cumulant generating function (CGF) which fulfils the role of a dynamical free energy.
It allows to recover the large-time limit of the cumulants of $\mathcal O$ as derivatives of $\psi(s)$ in $s=0$ from
\begin{equation}
\label{eq:cumulants}
\lim_{ t\rightarrow \infty} \frac{1}{t}\ \langle \mathcal{O}^k\rangle_c = (-1)^k\ \partial_s^k \psi (s) \mid_{s=0},
\end{equation}
where $\langle \mathcal{O}^k\rangle_c$ is the $k^{\text{th}}$ cumulant of $\mathcal{O}$.
%
The cumulative generating function $\psi(s)$ and the large deviation function $\pi(\mathcal O / t)$ are related through the Legendre transform
\begin{equation}
\psi(s) = \max_{\hat{o}} [\pi(\hat{o}) - s \hat{o} ], 
\end{equation}
where 
$\hat{o} = \mathcal O/t$. If $\pi$ is convex, i.e., $\pi''(\hat{o}) \leq 0$~\cite{Ellis}, then 
\begin{equation}
\pi(\hat{o})= \min_{s} [\psi(s) + s \hat{o} ].
\end{equation}

\section{The $s$-modified Dynamics}
\label{sec:mut}

As mentioned before, 
the parameter $s$ involves a (exponential) modification on the statistical weight of the histories of the system. Within this $s$ parametrized ensemble, averages of the observable $\mathcal O$~\eqref{eq:obs} defined as
\begin{equation}
\langle \mathcal{O} \rangle_s =  \frac{ \langle  \mathcal{O} e^{-s\mathcal O}\rangle}{\langle e^{-s\mathcal O}\rangle}
\end{equation}
for $s=0$ correspond to the steady state averages of $\mathcal O$. Meanwhile, values of $s \neq 0$ favours histories with non-typical values of observable $\mathcal O$. The $s$-modified dynamics can be obtained 
%
taking the Laplace transform of the probability distribution $P(C,\mathcal O,t)$
\begin{equation}
  \hat P(C,s,t) = \int d\mathcal O\: e^{-s\mathcal O} P(C,\mathcal O,t).
\end{equation}
This Laplace transform allows to recover the moment generating function~\eqref{eq:MCF} as 
\begin{equation}
Z(s,t)=\sum_C\hat P(C,s,t).
\end{equation}
The probability $\hat P(C,s,t)$ satisfies an $s$-modified master equation for its time-evolution~\cite{garrahan_first-order_2009}
\begin{equation}
 \label{eq:evolPhat}
 \partial_t \ket{\hat{P}(t)}  = \mathbb W_s \ket{\hat{P}(t)},
\end{equation}
where the $s$-modified master operator $\mathbb W_s$ is given by
\begin{equation}
\label{eq:defopWs}
(\mathbb W_s)_{CC'}= W_s(C' \to C) -r_s( C)\ \delta_{CC'} \ +\  \delta r_s( C)\ \delta_{CC'}, 
\end{equation}
where $\delta r_s( C)$ is defined as
\begin{equation}\label{eq:deltars}
\delta r_s( C) = r_s(C)-r(C)-s b(C),
\end{equation}
and $r(C)$ is the escape rate~\eqref{escaperate}. On the other hand, $W_s(C\to C')$ and $r_s(C)$ can be seen as $s$-modified transition and escape rates, respectively,
\begin{align}
\label{eq:Ws}
W_s(C\to C') &= e^{-s a(C,C')}\ W(C\to C'),\\
\label{eq:rs}
r_s(C) &= \sum_{C^{\prime}} W_s(C\to C'). 
\end{align}
The cumulative generating function $\psi$ in Eq.~\eqref{eq:psi} can be determined from this $s$-modified dynamics as the maximum eigenvalue of the matrix $\mathbb W_s $~\eqref{eq:defopWs} or also, by simulating Eq.~\eqref{eq:evolPhat} using a population dynamics algorithm (or cloning algorithm). Both of them are discussed below.

\subsection{ $\psi$ as the Largest Eigenvalue of $\mathbb W_s$}
\label{sec:largest}
Similarly as we saw for Eq.~\eqref{meq}, equation~\eqref{eq:evolPhat} has as solution
\begin{equation}
\label{eq:sol1s}
\ket{\hat{P}(t)}= e^{t\mathbb W_s} \ket{\hat{P}(0)}.
\end{equation}
Matrix $\mathbb W_s$ can be written in terms of its left $\bra{L_n}$ and right $\ket{R_n}$ eigenvectors and their respective eigenvalues $\lambda_n$, with $\lambda_0 > \lambda_1>\cdots$, as 
\begin{equation}
\mathbb W_s = \sum_{n} \lambda_n(s) \ket{R_n} \bra{L_n}.
\end{equation}
In the large-time limit, the exponential in Eq.~\eqref{eq:sol1s} is dominated by the largest eigenvalue $\lambda_0(s)$, so that
\begin{equation}
\label{eq:expW}
e^{t\mathbb W_s} = \ket{R_0} \bra{L_0}e^{t\lambda_0(s)}+\ldots
\end{equation}
Thus, $\ket{\hat{P}(t)}$ in the large-time limit also scales as 
\begin{equation}
\ket{\hat{P}(t)}= e^{t\mathbb W_s} \ket{\hat{P}(0)} \sim \ket{R_0} e^{t\lambda_0(s)} \braket{L_0}{P_0}+\ldots
\end{equation}
which is equivalent to 
\begin{equation}
\hat{P}(C,s,t) \sim R_0(C,s) e^{t\lambda_0(s)}.
\end{equation}
Thus, in the large-time limit 
\begin{equation}
Z(s,t)=\sum_C\hat P(C,s,t)\sim e^{t\lambda_0(s)}
\end{equation}
from which we can see that the maximum eigenvalue of matrix $\mathbb W_s$~\eqref{eq:defopWs} corresponds to the cumulative generating function $\psi(s)$.

\subsection{A Mutation-Selection Mechanism}
\label{mut}
Following the procedure used in Sec.~\ref{sec:solME} for the master equation, the solution of its $s$-modified version \eqref{eq:evolPhat} is given by 
\begin{equation}
\label{eq:solutionMEs}
\begin{split}
P(C,s, t) &= \sum\limits_{K \geqslant 0}\ \sum\limits_{C_{0} \ldots C_{K}}\ \int_{0}^{t} \,dt_{1}  \int_{t_1}^{t} \,dt_{2} \ldots \int_{t_{K-1}}^{t} \,dt_{K}  \\[5pt]
&\times r_s(C_0)\ e^{-(t_1-t_0)\ r_s(C_0)}\ \ldots r_s(C_{K-1})\ e^{-(t_{K}-t_{K-1})\ r_s(C_{K-1})} \times e^{-(t-t_K)\ r_s(C_{K})}  \\[5pt]
&\times \frac{W_s(C_0 \rightarrow C_1)}{r_s(C_0)} \cdots \frac{W_s(C_{K-1} \rightarrow C_K)}{r_s(C_{K-1})} \\[5pt]
& \times e^{(r_s(C_0)-r(C_0))\ (t_1-t_0)} \ldots e^{(r_s(C_{K-1})-r(C_{K-1}))\ (t_K-t_{K-1})} \times  e^{(r_s(C_{K})-r(C_{K}))\ (t-t_{K})}\\[5pt]
& \times P(C_0,s,0)
\end{split}
\end{equation}
which have been written conveniently in order to introduce the terms
\begin{equation}
\label{eq:Y}
Y(C_k) = e^{(r_s(C_{k})-r(C_{k}))\ \Delta t(C_k)}, 
\end{equation}
where $\Delta t(C_k)$ is the time spent in the configuration $C_k$. 
Contrarily to the original operator $\mathbb W$~\eqref{eq:defopW}, the $s$-modified operator $\mathbb W_s$~\eqref{eq:defopWs} does not conserve probability (since $\delta r_s(C)\neq 0$), implying that there is no obvious way to simulate Eq.~\eqref{eq:evolPhat}.
However, this time-evolution equation can be interpreted not as the evolution of a single system, but as a population dynamics on a large number $N_c$ of copies of the system which evolve in a coupled way~\cite{giardina_direct_2006,tailleur_probing_2007}. More precisely, reading the operator of the modified master equation~\eqref{eq:evolPhat} as in Eq.~\eqref{eq:defopWs}, we find that this equation can be seen as a stochastic process of transition rates $W_s(C\to C')$~\eqref{eq:Ws} supplemented with a selection mechanism of rates $\delta r_s( C)$~\eqref{eq:deltars}, 
where a copy of the system in configuration $C$ is copied at rate $\delta r_s( C)$ if $\delta r_s( C)>0$ or killed at rate $|\delta r_s( C)|$ if $\delta r_s( C)<0$.
As detailed below, an estimator for the CGF $\psi(s)$ may be recovered from the exponential growth (or decay) rate of a population evolving with these rules.

\section{Continuous-Time Population Dynamics}
\label{sec:cont-time-clon}
The basic idea of the population dynamics algorithm consists in preparing $N_c$ copies of the system (or clones) and in evolving them according to the transition rates $W_s(C\to C')$ given by Eq.~\eqref{eq:Ws}. During this evolution some copies are repeatedly multiplied or eliminated according to a selection mechanism of rates $\delta r_s( C)$~\eqref{eq:deltars}. 
%
%
The mutation-selection mechanism described above can be performed in a number of ways. 
One of them consists in keeping the total number of clones constant and another, in leaving this population of clones to grow (or decrease) in time (as implemented in chapter~\ref{chap:Discreteness}). Additionally, the selection mechanism can be implemented along with each evolution of the copies \textbf{(Continuous-Time)}~\cite{tailleur_simulation_2009, lecomte_numerical_2007, giardina_simulating_2011} or for each pre-fixed time-interval \textbf{(Discrete-Time)}~\cite{giardina_direct_2006}. 
This last one is implemented (in a constant population fashion) in chapter~\ref{chap:DiscreteTime}, while the continuous-time version is used in the majority of the manuscript. An explanation about important differences between continuous and discrete-time techniques can be found in  Secs.~\ref{subsubsec:differenceContinuousTime} and~\ref{Discrete-time_algorithm}. A detailed description of the continuous-time approaches is presented below.

\subsection{The Cloning Algorithm}
\label{CA}
We consider $N_{c}$ copies (or clones) of the system. 
The dynamics is continuous in time: for each copy, the actual changes of configuration occur at times (which we call `evolution times') which are separated by intervals whose duration is distributed exponentially.
At a given step of the algorithm, we denote by $ \mathbf t = \{ t^{(i)} \}_{i=1,\ldots,N_{c}}$ the set of the future evolution times of all copies and by $C = \{C_{i}\}_{i=1,\ldots,N_{c}}$ the configurations of the copies. Their initial configurations do not affect the resulting scaled cumulant generating function in the large-time limit. However, for the concreteness of the discussion, without loss of generality, we assume that these copies have the same configuration $C_0$ at ${\mathbf t}=0$. The cloning algorithm is constituted of the repetition of the following procedures:
\begin{enumerate}
\item[1.] Find the clone whose next evolution time is the smallest among all the clones, i.e., $j={\rm argmin}_i t^{(i)}$. 
\item[2.] Compute $y_j = \lfloor Y(C_j) + \epsilon  \rfloor$, 
where the cloning factor $Y(C_j)$~\eqref{eq:Y} is defined as $e^{\Delta t(C_j)\, \delta r_{s}(C_j)}$, $\Delta t(C_j)$ is the time spent by the clone $j$ in the configuration $C_j$ since its last configuration change, and
%
$\epsilon$ is a random number uniformly distributed on $[0,1]$.
\item[3.] If $y_j=0$, remove this copy from the ensemble, and if $y_{j}>0$, make $y_{j}-1$ new copies of this clone.
\item[4.] For each of these copies (if any), the state $C_{j}$ is changed independently to another state $C_{j}'$, with probability $W_{s}(C_{j} \to C_{j}')/r_{s}(C_{j})$. 
\item[5.] Choose a waiting time $\Delta t^{(j)}$ from an exponential law of parameter $r_{s}(C'_{j})$ for each of these copies. Their next change of configuration will occur at the evolution time $t^{(j)} + \Delta t^{(j)}$.
\end{enumerate}
\subsubsection{Non-Constant Population Approach}
\label{sec:NCPA}
%
The repetition of this procedure will result (after an enough time) in an exponential growth (or decay) of the number of clones. 
We can keep track of the different changes in the number of clones and the times where these changes occur and denote by $N(s,t)$ the time-dependent population. 
The CGF estimator can be computed from the slope in time of the log-population $\hat{N}(s,t) = \log N(s,t)$, which constitutes an evaluation of the population growth rate. This can be done in different ways, for example by fitting $\hat{N}(s,t)$ by $p_t = \Psi(s)t+ p_0$ where $\Psi(s)$ is the CGF estimator or also using
\begin{equation}\label{eq:6}
\Psi(s)=\frac{1}{T_{\max}-T_{\min}}\log\left(\frac{N_{\max}}{N_{\min}}\right),
\end{equation}
where $N_{\max}$ and $N_{\min} $ are the maximum and minimum values for $N(s,t)$ and $T_{\max}$ and $T_{\min}$ their respective times. This approach 
is implemented in chapter~\ref{chap:Discreteness}, where we discuss the discreteness effects in populations dynamics and their influence in the CGF estimation.
 

\subsubsection{Constant-Population Approach}
\label{sec:CPA}
In order to keep the total number of copies constant, we add to the procedure described above an additional step 
\begin{enumerate}
\item[6.] We choose randomly and uniformly: (\textit{i}) a clone $k$, $k\neq j$ and we copy it (if $y_j = 0$), or (\textit{ii}) $y_{j}-1$ clones and we erase them (if $y_j>1$).
\end{enumerate}
Thus, the CGF estimator $\Psi_{s}^{(N_{c})}$ can be obtained 
from the exponential growth rate that the population would present if it was not kept constant~\cite{giardina_simulating_2011}.
More precisely, this estimator is defined as
\begin{equation}
\Psi_{s}^{(N_{c})}=\frac{1}{t}\log \prod \limits_{i = 1}^{\mathcal K} X_{i},
\label{eq:PSI}
\end{equation}
where $X_{i} = (N_{c}+y_{i}-1)/N_{c}$ are the ``growth'' factors at each step $j$ of the procedure described above, and $\mathcal K$ is the total number of configuration changes in the full population up to time $t$ (which has not to be confused with $K$).
%
This growth rate can also be computed from a linear fit over the reconstructed log-population and the initial transient regime, where the discreteness effects are present, can be discarded in order to obtain a better estimation. This approach is used in chapters~\ref{chap:ContinuousTime},~\ref{chap:CGF} and~\ref{chap:LargeL}.
 
\section{Example Models}
\label{sec:models}

In the next chapters, the cloning algorithm is implemented in order to obtain an estimation of the CGF  $\psi(s)$. This is done 
with two specific models: 
a simple two-state annihilation-creation dynamics, and a contact process on a one-dimensional periodic lattice~\cite{CP, GRASSBERGER1979373, liggett2012interacting,lecomte_numerical_2007}.
The first system (chapters:~\ref{chap:Discreteness},~\ref{chap:DiscreteTime},~\ref{chap:ContinuousTime} and~\ref{chap:CGF}) was chosen for its simplicity and the possibility of comparing the numerical predictions with the exact values of $\psi(s)$. On the other hand, the contact process (chapters:~\ref{chap:ContinuousTime},~\ref{chap:CGF} and~\ref{chap:LargeL}) is used to extend the analysis and to verify the results on a (more complex) many body system where 
the dependence with the size of the system
can be also analyzed. 
In both cases, we consider the dynamical activity $K$~\cite{garrahanjacklecomtepitardvanduijvendijkvanwijland,garrahan_first-order_2009,Hedges1309,ThermoCP,PhysRevE.81.011111,ChandlerGarrahan,QuantumJump,
Micromaser,Genway,Ates,Hickey,jackEast} as the additive observable $\mathcal O$~\eqref{eq:obs}. The analytical expression of the CGF $\psi(s)$ is obtained (when possible) by solving the largest eigenvalue of the operator $\mathbb W_s$~\eqref{eq:defopWs} as discussed in Sec.~\ref{sec:largest}. 

\subsection{Annihilation-Creation Dynamics}
\label{sec:bdp}
The dynamics occurs in one site where the only two possible configurations $C$ are either $0$ or $1$. The transition rates between configurations are given by
%
%
%
%
\begin{align}
\label{eq:W_AC}
W(0 \rightarrow 1) &= c, \\
W(1 \rightarrow 0) &= 1-c,
\end{align}
where $c\in [0,1]$. Eq.~(\ref{eq:1}) for this process becomes
\begin{equation}\label{eq:7}
\partial_{t} \left(\begin{array}{c}
P(0,t)\\
P(1,t)
\end{array}\right) \left(\begin{array}{cc}
-c & 1-c\\
c & -1+c
\end{array}\right) \left(\begin{array}{c}
P(0,t)\\
P(1,t)
\end{array}\right).
\end{equation}
As we mentioned before, 
one advantage of considering this process for our analysis is that the large deviation function for the activity can be determined analytically. The large-time cumulant generating function $\psi_{K}(s)=\lim_{t\rightarrow\infty}\frac{1}{t}\log \langle e{}^{-sK} \rangle$ corresponds to the maximum eigenvalue of the matrix $\mathbb{W}_{s}$~\eqref{eq:defopWs} 
\begin{equation}\label{eq:WsAC}
\mathbb{W}_{s}=\left(\begin{array}{cc}
-c & (1-c)\ e^{-s}\\
c\ e^{-s} & -1+c
\end{array}\right)
\end{equation}
which results in
\vspace{-5mm}
\begin{equation} \label{eq:PSIA}
\psi_{K}(s)=-\frac{1}{2}+\frac{1}{2} \bigg( 1-4c\ (1-c)(1-e^{-2s}) \bigg)^{1/2}.
\end{equation}
Equation~\eqref{eq:PSIA} will allow us to assess the quality of our numerical results.
The inverse of the difference between the eigenvalues of $\mathbb{W}_{s}$
\begin{equation} \label{eq:tgap}
t_{\gap} = \frac{1}{\sqrt{1-4c\ (1-c)(1-e^{-2s})}}    
\end{equation}
allows us to define the typical convergence time $t_{\gap}$ for the large-time behavior for Eq.~\eqref{eq:7} which is analyzed in Sec.~\ref{Time Delay Correction}.

\subsection{Contact Process}
\label{sec:CP}
This system consists in a one-dimensional lattice with $L$ sites and periodic boundary conditions~\cite{CP, GRASSBERGER1979373, liggett2012interacting}.
Each position $i$ is occupied by a spin which is either in the state $n_{i}=0$ or $n_{i}=1$. The configuration $C$ is then constituted by the states of these spins, i.e., $C=(n_i)_{i=1}^L$.  
The transition rates for this process  are given by
\begin{eqnarray}
\label{eq:W_CP}
W(n_{i}=1 \rightarrow n_{i}=0) &=& 1, \nonumber \\
W(n_{i} = 0 \rightarrow n_{i} = 1) &=& \lambda(n_{i-1}+n_{i+1}) + h,
\end{eqnarray}
where $\lambda$ and $h$ (spontaneous rate of creation) are positive constants. This model is an example of contact processes~\cite{CP}, which have been studied in many contexts especially to model the spread of infection diseases~\cite{10.2307/2244329}. Within this context, the state $n_{i}=1$ is used to represent a sick individual and $\lambda$ can be seen as a infection rate.
The corresponding CGF develops a singularity as $L \rightarrow \infty$, showing a dynamical phase transition~\cite{lecomte_numerical_2007, Lecomte2007, ThermoCP, marro_dickman_1999}. The contact process is a model of the directed percolation universality class and its scaling properties have been discussed extensively~\cite{marro_dickman_1999, HinrichsenNoneq, OdorUniversality}.


\chapter[\quad Discreteness Effects in Population Dynamics]{Discreteness Effects \\ in Population Dynamics}
\label{chap:Discreteness}  
\section{Introduction}

In the present chapter~\cite{hidalgo_discreteness_2016}, we focus on the \textbf{non-constant population approach} of the cloning algorithm (as described in Sec.~\ref{sec:NCPA}), that we study numerically for the simple annihilation-creation (Sec.~\ref{sec:bdp}) model where its implementation and its properties can be examined in great details.
As we mentioned in the introduction, the cloning algorithm results (as time goes to infinity) in an exponential growth (for $s < 0$) or decay (for $s > 0$) of the number of clones. As we will see later, the ``discreteness effects'' in the evolution of our populations are strong at initial times. That is why the determination of the large deviation function using this algorithm is constrained not only to the parameters $(c,s)$, the initial number of clones $N_c$ and the number of realizations $R$ but also to the final time (or the maximum population) until which the process evolves in the numerical procedure.
In Sec.~\ref{sec:avepopldf} we describe issues related to the averaging of distinct runs,
that we quantify in Sec.~\ref{Parallel Behaviour in Log-Populations}.
In Sec.~\ref{sec:time_correct} we propose a new method to increase the efficiency of the population dynamics algorithm by applying a realization-dependent time delay, and
we present the results of its application in Sec.~\ref{sec:psi_timedelay}.
We characterize numerically the distribution of these time delays in Sec.~\ref{sec:time-delay-prop}.
Our conclusions and perspectives are gathered in Sec.~\ref{sec:discussionDE}.

\section{Average Population and the LDF} 
\label{sec:avepopldf}

In order to obtain an accurate estimation of~$\psi(s)$, we should average several realizations of the procedure described in Sec.~\ref{sec:NCPA}. To perform this average, we will define below a procedure that we have called \textbf{merging} which will allow us to determine in a systematic way the average population from which we can obtain this estimation that we denote $\Psi(s)$.
Noteworthy, this erroneously could be seen as obtaining~$\Psi(s)$ from the growth rate of the {average} (or equivalently the sum) of several runs of the population dynamics. 
This procedure would be incorrect since it amounts to performing a single run of the total population of the different runs, with a dynamics that would partition the total population into \emph{non-interacting} sub-populations, while, as described in~Sec.~\ref{sec:NCPA}, the population dynamics induces effective interactions among the whole set of copies inside the population.
In fact, the right way of performing this numerical estimate comes from computing~$\Psi(s)$ from the average growth rate of several runs of the population, i.e., from taking the average $\langle \log N(s,t)\rangle$ of the slopes of several $\log N(s,t)$ instead of the slope of $\log \langle N(s,t) \rangle$.
The two results differ in general since $\langle \log N(s,t)\rangle \neq \log \langle N(s,t) \rangle$. One can expect that the two results become equivalent in the large $N_c$ limit as the distribution of growth rate should become sharply concentrated around its average value; however, they are different in the finite $N_c$ regime that we are interested in. This alternative way of defining the CGF estimator is discussed deeply in Secs.~\ref{subsec:Different large deviation estimator} and 
\ref{sec:two-estimators}.

\subsection{Populations Merging}
\label{Populations Merging}
Let us consider $J$ populations: $\mathcal{N} = \{ N_{1}(s,t),N_{2}(s,t),...,N_{J}(s,t) \}$. In order to compute the average population $\langle \mathcal{N} \rangle$ defined as $\langle \mathcal{N} \rangle = \langle N_{j}(s,t) \rangle_{j = 1}^{J}$, we introduce a procedure that we have called \textbf{merging} (of populations) which is described below. 

Given $N_{i}(s,t)$ and $N_{j}(s,t)$ the result of merging these two populations $\mathcal{M}(N_{i},N_{j})$ is another population $N_{ij} = N_{i} + N_{j}$ which represents the total number of clones for each time where a change in population for $N_{i}$ and $N_{j}$ has occurred. If $\langle N_{ij} \rangle$ is the average population for $N_{i}$ and $N_{j}$, this is related to the merged population through $\langle N_{ij} \rangle = \frac{N_{ij}}{2}$. If we add, for example, to our previous result another population $N_{k}$, the result $\mathcal{M}(N_{ij},N_{k})$ is related to the average by  $\mathcal{M}(N_{ij},N_{k})=N_{ij}+N_{k}=N_{i}+N_{j}+N_{k} = N_{ijk} = 3\langle N_{ijk} \rangle$. 
These merging procedure can be repeated for each of the populations in $\mathcal{N}$ so that
 \begin{equation} \label{eq:10}
 \mathcal{M}[\mathcal{N}]=\mathcal{M}(\mathcal{M}(\mathcal{M}(\ldots(\mathcal{M}(\mathcal{M}(N_{1},N_{2}),N_{3}),N_{4})\ldots),N_{J-1}),N_{J})
 \end{equation}
is the result of systematically merging all the populations in $\mathcal{N}$. The average population $\langle \mathcal{N} \rangle$ can be recovered from $\mathcal{M}[\mathcal{N}]$ as 
\begin{equation} \label{eq:avN}
\langle \mathcal{N} \rangle = (1/J) \mathcal{M}[\mathcal{N}].
 \end{equation}
Similarly, in the case of log-populations 
($\hat{N}_{j}(s,t) = \log N_{j}(s,t)$), 
the average $\langle \hat{N} \rangle = \langle \hat{N}_{j}(s,t)\rangle_{j = 1}^{J}$ is obtained from merging all the log-populations in 
$\hat{N} = \{ \hat{N}_{1}(s,t),\hat{N}_{2}(s,t),\ldots,\hat{N}_{J}(s,t) \}$. 
The estimator $\Psi(s)$ is then computed from the slope of $\langle \hat{N} \rangle$ with $\langle \hat{N} \rangle = (1/J) \mathcal{M}[\hat{N}]$.

\subsection{Discreteness Effects at Initial Times}
\label{Discreteness Effects at Initial Times}
Issues can emerge in the determination of $\Psi(s)$~\eqref{eq:6} which are not only related to the dependence of the method in $N_{c}$ (the initial number of clones) and $J$ (the number of populations). At initial times there is a wide distribution of times at which the first series of jumps occurs. This means that fluctuations at initial times induce that some populations remain in their initial states longer than others, producing an effective delay compared to other populations that evolve faster in their initial regime. From a practical point of view, this can induce that the numerical determination of $\Psi(s)$ becomes a slow and inefficient task. One way of dealing with this issue comes from restricting the evolution of $\mathcal{N}$ up to a maximum time $T_{\max}$ or a maximum population $N_{\max}$. However, this implies that if $T_{\max}$ or $N_{\max}$ are not long enough, the determination of $\Psi(s)$ will be strongly affected by the behavior of $\mathcal{N}$ at initial times. We now discuss two issues that are encountered in the numerical evaluation of the CGF estimator: (\textit{i}) the influence of how the dynamics is halted; and (\textit{ii}) the role of initial population in the initial regime. 

\begin{figure}[t!]
        \centering
         {\includegraphics [scale=0.48] {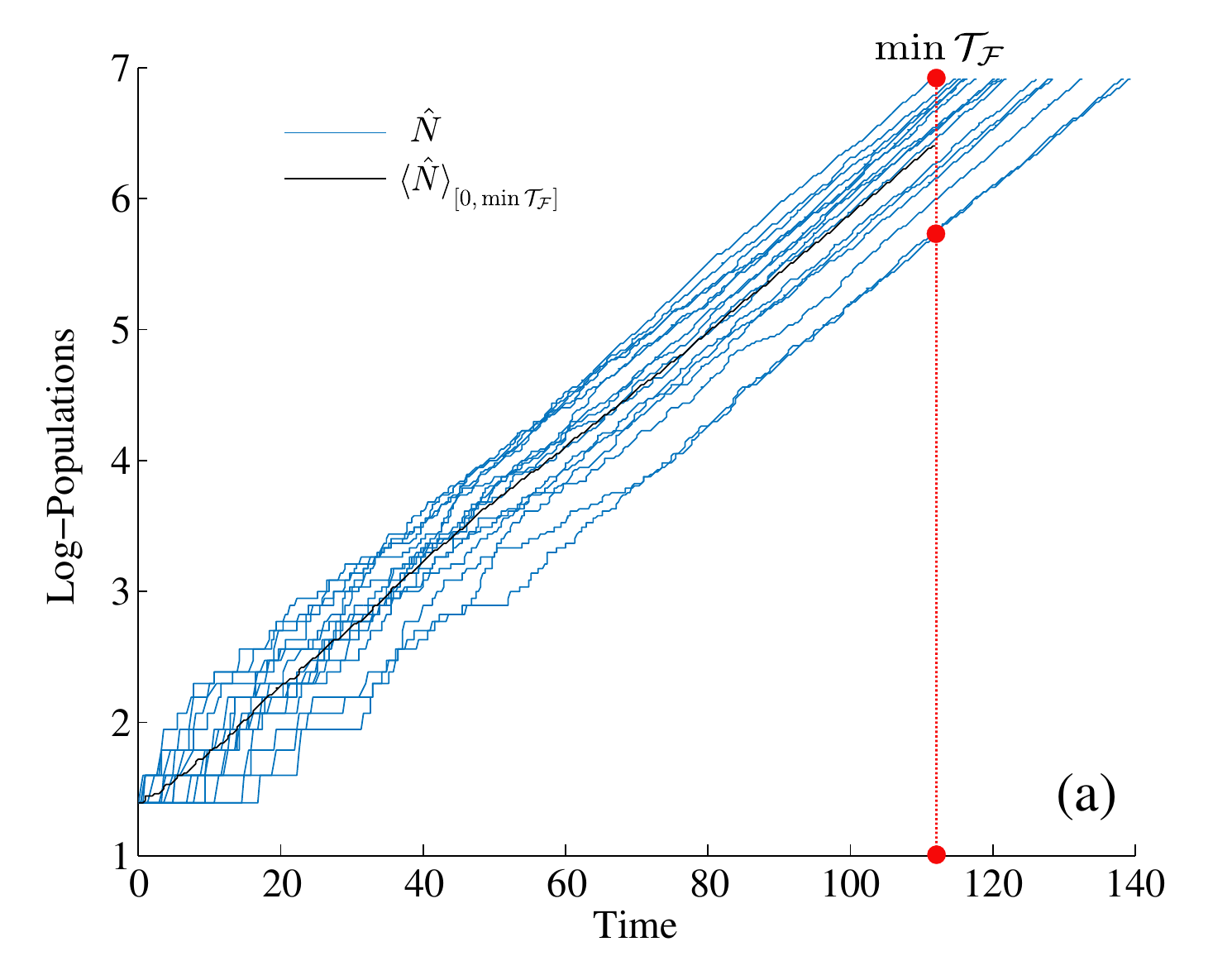}}
		{\includegraphics [scale=0.48] {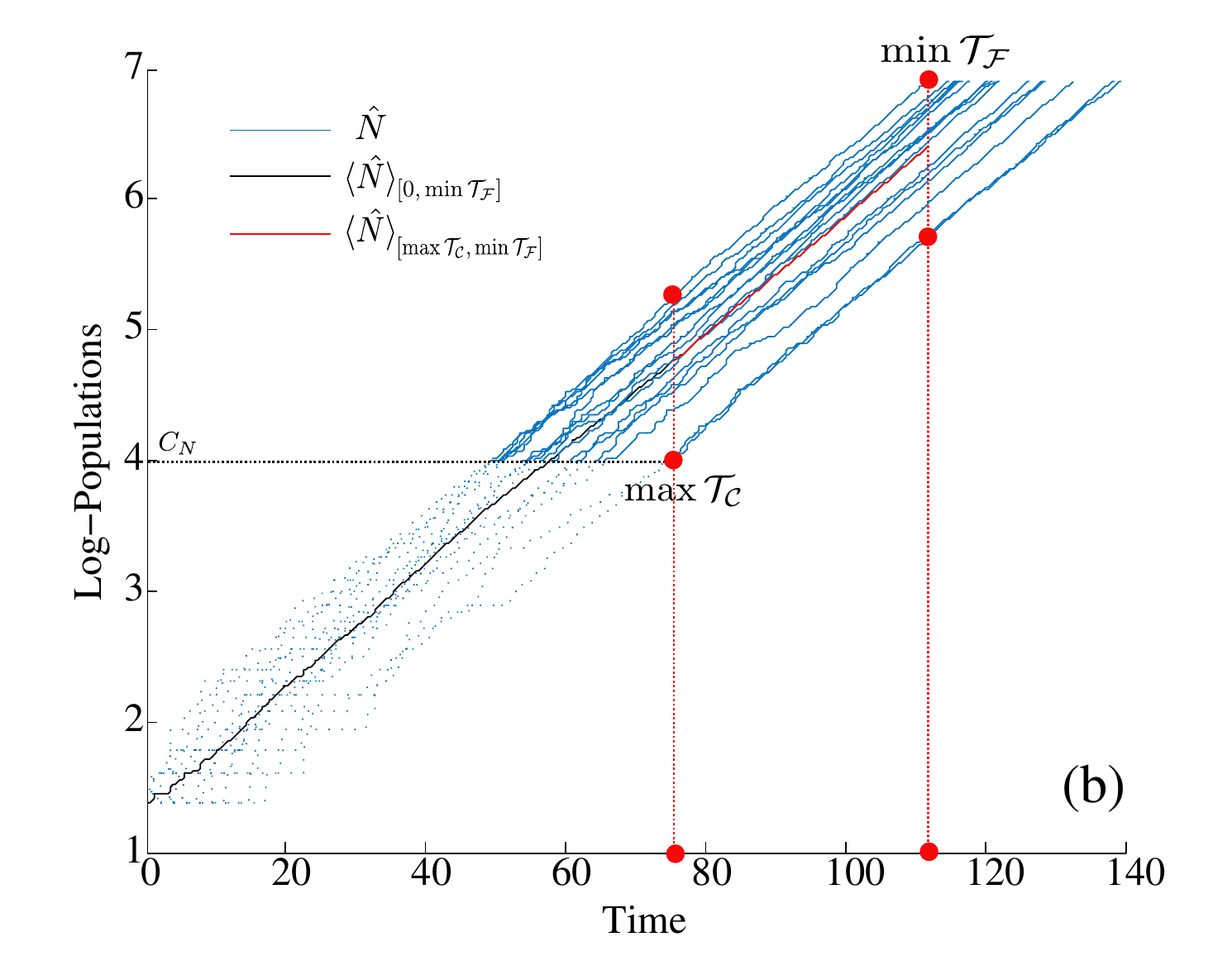}}\\ 
\centering        	
\caption[Log-Populations]{Log-populations as function of time (blue). Their evolution has been restricted up to a maximum (log) population value. \textbf{(a)} The average log-population $\langle \hat{N} \rangle$ (black) is made in the interval $[0,\min \mathcal{T_{F}}]$, where all the populations are defined. \textbf{(b)} After a cut in populations $C_{N}$ (in order to eliminate the initial discreteness effects), the average log-population (red) that represents the new $\hat{N}$ is defined only in the interval $[\max \mathcal{T_{C}},\min \mathcal{T_{F}}]$.}
\label{fig:merge}
\end{figure}

\newpage 
Let us call $\mathcal{T_{F}} = \{ t_{1}^{\mathcal{F}},\ldots,t_{J}^{\mathcal{F}} \}$ the set of final times of $\mathcal{N}$,  with $t_{j}^{\mathcal{F}} \leq T_{\max}$, $\forall j\in\{1,\ldots,J\}$. Note that $t_{j}^{\mathcal{F}}$ depends on $j$ whenever the simulation is stopped at $N_{\max}$ (as in Fig.~\ref{fig:merge}) or at $T_{\max}$. This is due to the fact that the algorithm is continuous in time and the last $\Delta t(C)$ does not exactly lead to $T_{\max}$. We say that the average population $\langle \mathcal{N} \rangle$ \textbf{represents} $\mathcal{N}$ only if the average is made in the interval $[ 0,\min \mathcal{T_{F}}]$ where all the populations are defined. In other words, the average population in this interval takes into consideration all the populations while for times $t \geq \min\mathcal{T_{F}}$ some populations have stopped evolving. This phenomenon is especially evident when considering a maximum population limit $N_{\max}$ for the evolution of the populations (Fig.~\ref{fig:merge}(a)). As a consequence, $\langle \mathcal{N} \rangle$ depends on the distribution of final times of $\mathcal{N}$ which are not necessarily equal to $T_{\max}$. 
 
An alternative that can be considered in order to overcome the influence of initial discreteness effects in the determination of $\Psi(s)$ is to get rid of the initial transient regime where  these effects are present. In other words, to cut the initial time regime of our populations. Let us call $C_{N} \geq \log N_{c}$ the initial cut in log-populations and equivalently $C_{t} \geq 0$ the initial cut in times. $\mathcal{T_{C}} = \{ t_{1}^{\mathcal{C}},\ldots,t_{J}^{\mathcal{C}} \}$ is the distribution of times at $C_{t,N}$. In that case, similarly as we analyzed before, the average population $\langle \mathcal{N} \rangle$ represents $\mathcal{N}$ only if the average is made in the interval $[\max \mathcal{T_{C}}, \min \mathcal{T_{F}}]$ which can be in fact very small and could result in a bad approximation of $\Psi(s)$ (Fig.~\ref{fig:merge}(b)).

As we will see in the next section, the log-populations after a long enough time become parallel, i.e., once the populations  have surpassed the discreteness effects regime, the distance between them is constant. We will use this fact in order to propose a method which allows us to overcome the problems we have described in this section. Throughout this chapter, we consider for our simulations $c = 0.3$, $N_{c} = 2^{2}$, $N_{\max} = 10^{3}$, $J = 2^{8}$ and $s \in \left[-0.3,0 \right]$.

\section{Parallel Behavior in Log-Populations}
\label{Parallel Behaviour in Log-Populations}
\subsection{Distance between Populations}
Given $N_{i}(s,t)$ and $N_{j}(s,t)$, we define the distance between these populations at $N^{*}$ (with $N^{*} \in N_{i}$ and $N^{*} \in N_{j}$), as
\begin{equation} \label{eq:11}
D(N_{i},N_{j})(N^{*}) = \left \vert  \left( t_{j}(N^{*}) + \frac{\Delta t_{j}(N^{*})}{2} \right) - \left( t_{i}(N^{*}) + \frac{\Delta t_{i}(N^{*})}{2} \right) \right \vert,
\end{equation}
where $\Delta t_{k}(N^{*})$ is the time interval $N_{k}(s,t)$ spent at $N^{*}$ and $t_{k}(N^{*})$ is the time where $N_{k}(s,t)$ changes to $N^{*}$. Evidently, there are cases where $N^{*} \notin N_{i}$ but $N^{*} \in N_{j}$, $N^{*} \in N_{i}$ but $N^{*} \notin N_{j}$ and $N^{*} \notin N_{i}$ and $N^{*} \notin N_{j}$. However, $D(N_{i},N_{j})(N^{*})$ for these cases can also be computed. The last analysis (and definitions) is also valid for log-populations. These distances,  $D(N_{i},N_{j})(N^{*})$ and $D(\hat{N}_{i},\hat{N}_{j})(N^{*})$ enjoy interesting properties that we discuss below.

\begin{figure}[t!]
        \centering
        {\includegraphics [scale=0.48] {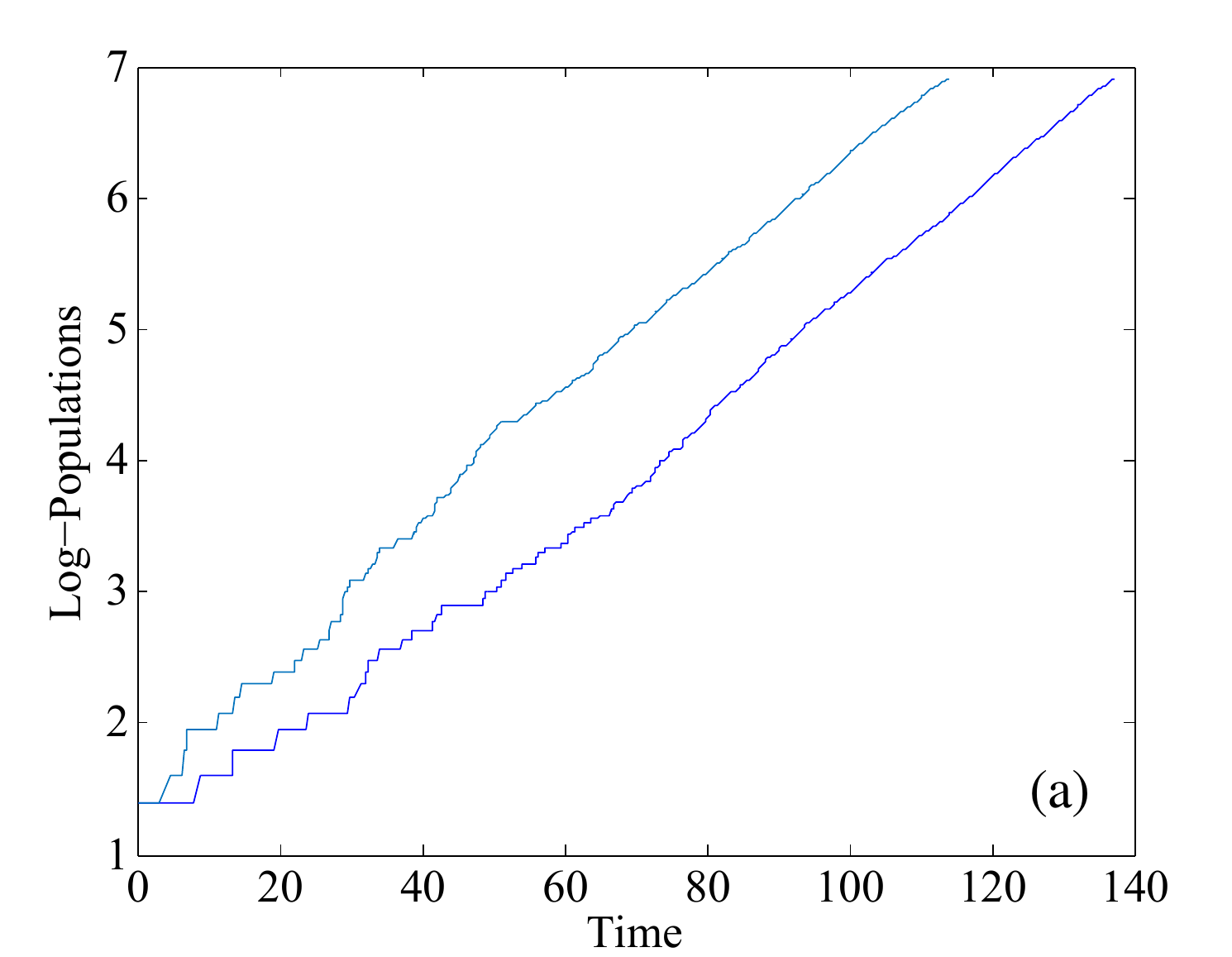}}
		{\includegraphics [scale=0.48] {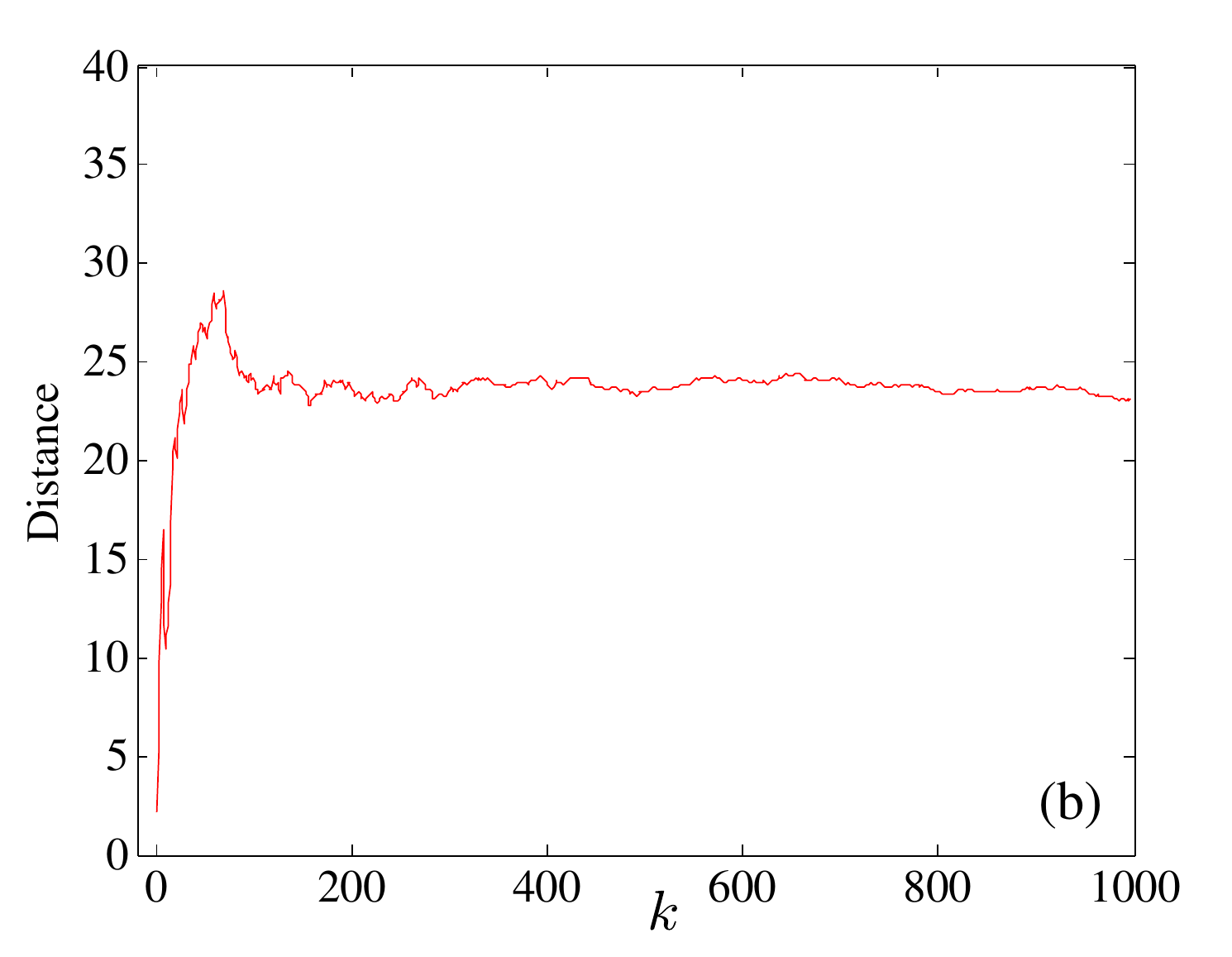}}\\ 
\centering        	
\caption[Log-Populations Distance]{Evolution of two log-populations $\hat{N}_{i},\hat{N}_{j}$ as function of time and the distance $D(\hat{N}_{i},\hat{N}_{j})$ between them as defined in Eq.~\eqref{eq:11}. \textbf{(a)} Log-populations after a long enough time become parallel. \textbf{(b)} Once the populations have overcome the initial discrete population regime, the distance between them becomes constant. $(s=-0.1)$. }
\label{fig:distance1}
\end{figure}

\subsection{Properties of $D(\hat{N}_{i},\hat{N}_{j})$}
In Fig.~\ref{fig:distance1}, we show two log-populations and the distance between them. These log-populations after a long enough time become parallel (Fig.~\ref{fig:distance1}(a)), i.e., once the populations have overcome the discreteness effects regime, the distance between them becomes constant (Fig.~\ref{fig:distance1}(b)). The region where the distance between populations is constant characterizes the exponential regime of the populations growth, i.e., the region where the discreteness effects are not strong anymore.

\newpage

\begin{figure}[h!]
        \centering
         {\includegraphics [scale=0.48] {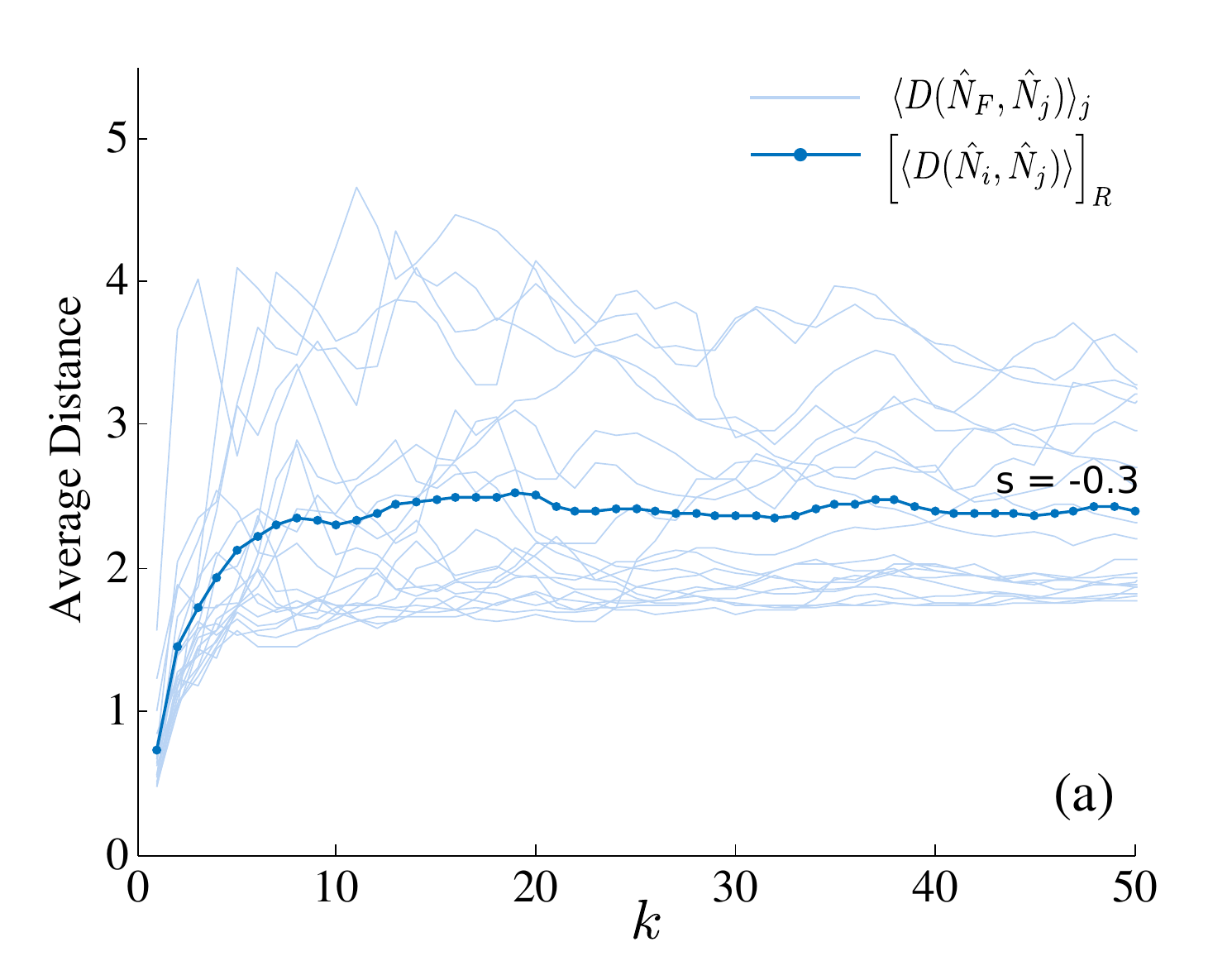}}
		{\includegraphics [scale=0.48] {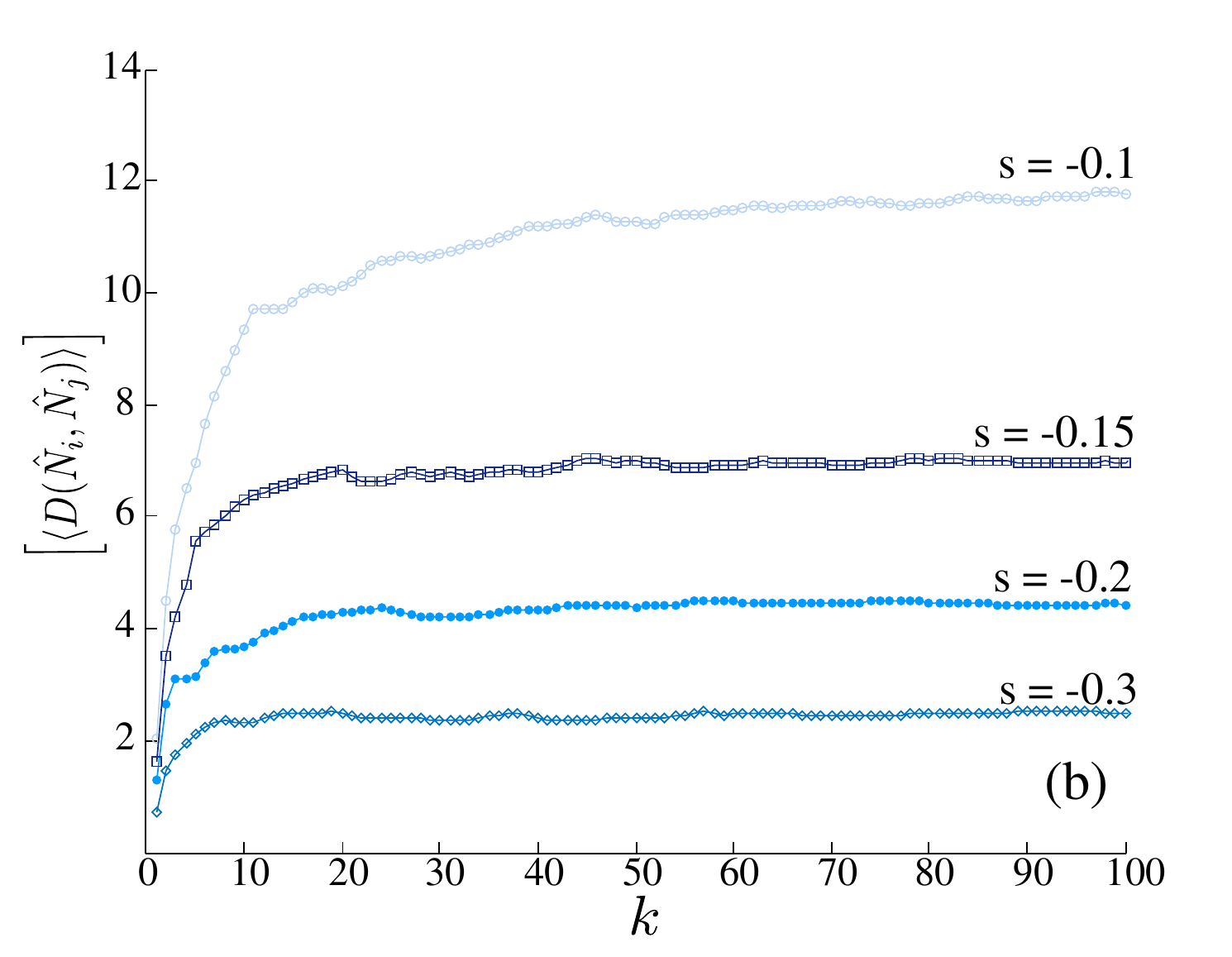}}\\ 
\centering        	
\caption[Average Distance between Log-Populations]{\textbf{(a)} Average distance of the log-populations in $\hat{N}$ with respect to a reference one for $R = 20$ realizations (light blue) and their average (dark blue). \textbf{(b)} Average distance between populations for several values of the parameter $s$. How fast a population ``exits'' from the discreteness effects regime depends on $s$: $s$ closer to zero corresponds to a slower population growth and hence a longer discreteness regime.} 
\label{fig:distance2}
\end{figure}

If we consider some population $\hat{N}_{F} \in \hat{N}$ as reference, using the definitions above, it is possible to determine the distance $D(\hat{N}_{F},\hat{N}_{j})$ between $\hat{N}_{F}$ and the rest of populations in $\hat{N} = \{ \hat{N}_{1},\hat{N}_{2},...,\hat{N}_{J} \}$. In Fig.~\ref{fig:distance2}(a) we show their average $\langle D(\hat{N}_{F},\hat{N}_{j}) \rangle_{j}$ in light blue and its average over $R = 20$ realizations $\left[\langle D(\hat{N}_{i},\hat{N}_{j}) \rangle \right]_{R}$ in dark blue. The parameter $s$ characterizes atypical behaviors of the unbiased dynamics (as we mentioned in Sec.~\ref{sec:mut}), and this induces a dependence in $s$ of the population growth. A population with a large value of $s$ corresponds to a large deviation of $K$. Also, as it is clearly illustrated in Fig.~\ref{fig:distance2}(b), the time of entrance of the system into a regime free of discreteness effects depends on $s$. 

\section{Time Correction in the Evolution of Populations}
\label{sec:time_correct}
Based on the results we just illustrated, we propose a method to improve the estimation of $\psi(s)$ and reduce the influence of the initial transient regime we described in Sec.~\ref{Discreteness Effects at Initial Times}. We aim at giving more weight to the exponential regime in the determination of $\psi(s)$. As detailed below, this can be done through a time delay in the evolution of populations. 

\subsection{Time Delay Correction}
\label{Time Delay Correction}
\begin{figure}[t!]
        \centering
        {\includegraphics [scale=0.48] {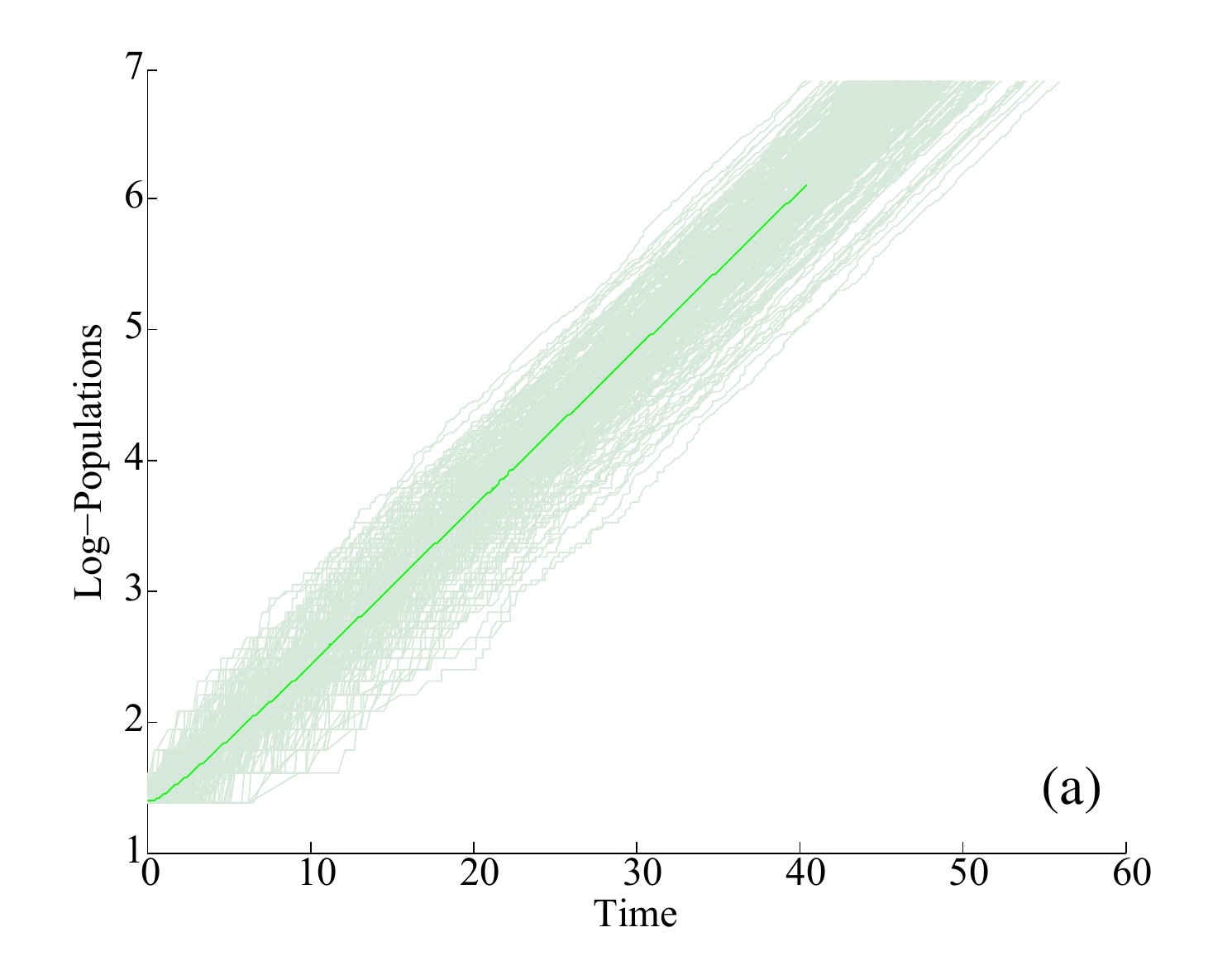}}
		{\includegraphics [scale=0.48] {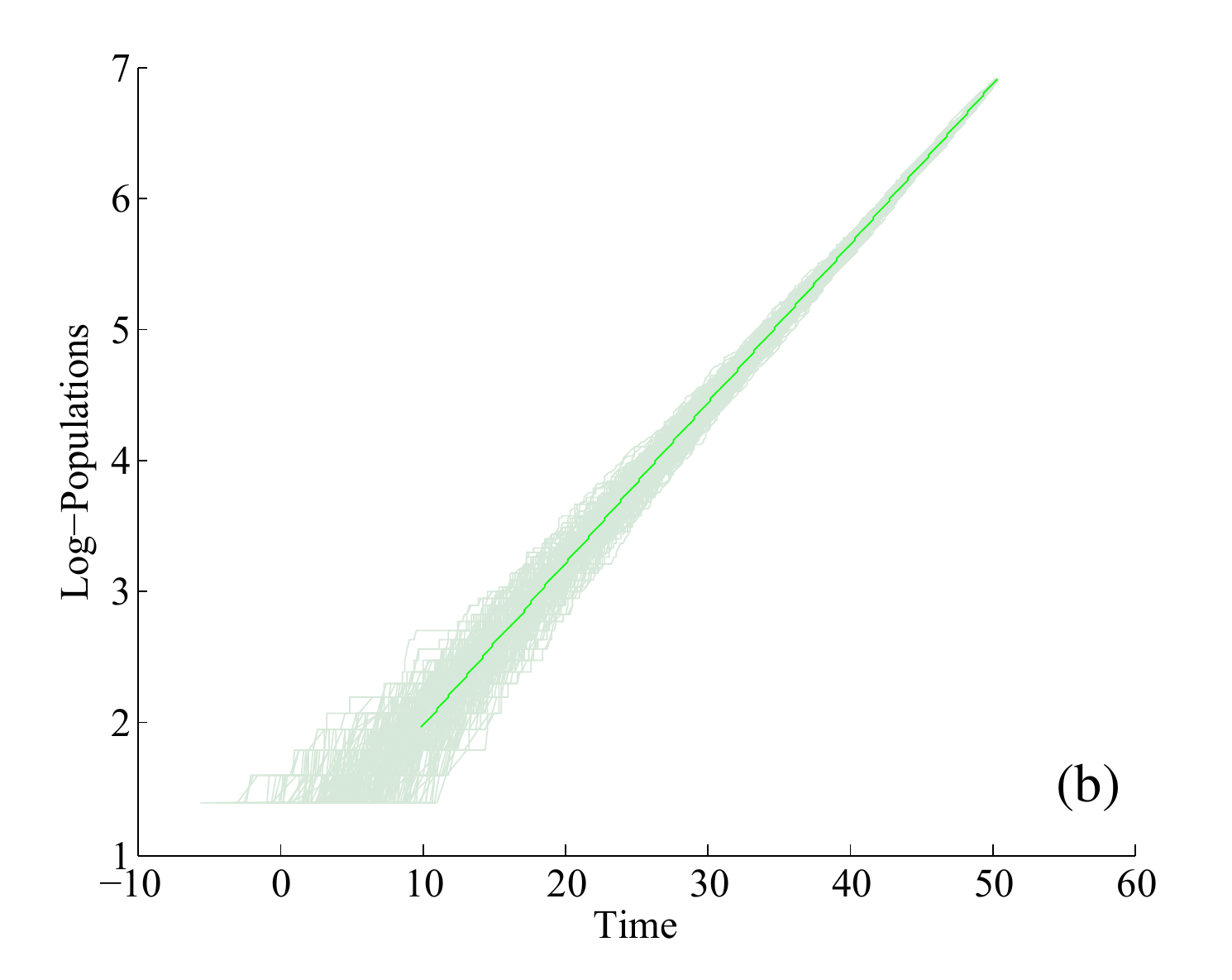}}\\ 
\centering        	
\caption[Time-Delayed Log-Populations]{\textbf{(a)} Log-populations, \textbf{(b)} time-delayed log-populations, and their average (dark green). The fluctuations at initial times produce a gap in the evolution of individual populations inducing a relative shift that lasts forever. This is compensated by delaying the populations in time, as explained in Sec.~\ref{Time Delay Correction}. $(s=-0.25)$.}
\label{fig:delayedPOP}
\end{figure}

Consider $J$ populations $\mathcal{N}$, their respective log-populations $\hat{N}$ and their distribution of final times $\mathcal{T_{F}} = \{t_{1}^{\mathcal{F}},...,t_{J}^{\mathcal{F}} \}$. We define as \textbf{delay} $\Delta\tau_{j}$ of $\hat{N}_{j}$ (with respect to a fixed reference population $\hat{N}_{F} \in \hat{N}$) the time interval

\begin{equation} \label{eq:12}
\Delta\tau_{j}=t_{F}^{\mathcal{F}} - t_{j}^{\mathcal{F}}
\end{equation}
such that, if $\Delta\tau_{j}<0$, $\hat{N}_{j}$ is ahead with respect to $\hat{N}_{F}$, and if $\Delta\tau_{j}>0$, $\hat{N}_{j}$ is delayed with respect to $\hat{N}_{F}$. This lag can be compensated by performing on $\hat{N}_{j}$ the time translation

\begin{equation} \label{eq:13}
\hat{N}_{j}^{\new} = \hat{N}_{j}(s,t+ \Delta\tau_{j})
\end{equation}
which produces that $\hat{N}_{j}^{\new}$ and $\hat{N}_{F}$ share not only the final population $N_{\max}$, but also the same final time $t_{F}^{\mathcal{F}}$. Moreover, considering also the fact that log-populations are parallel at large times, this procedure produces that $\hat{N}_{j}^{\new}$ and $\hat{N}_{F}$ overlap in a \textbf{free of discreteness effects region}. The result of performing this transformation to all the populations in $\hat{N}$ is shown in Fig.~\ref{fig:delayedPOP}.

Fig.~\ref{fig:delayedPOP} also illustrates many of the points we have discussed up to now. One of them is related to the ``wide'' distribution of final times, i.e., $\min \mathcal{T_{F}}$ and $\max \mathcal{T_{F}}$ can be very distant one from each other. This along with the fact that the average population depends on $\min \mathcal{T_{F}}$ makes that the determination of $\Psi(s)$ omits a considerable region where the populations have already entered the exponential regime. This implies precisely that more weight is given to the initial discreteness effects than to the exponential regime. These effects are in fact present up to relatively long times which means that if we  would like to get rid of the region were discreteness effects are strong by cutting the populations, the determination of $\Psi(s)$ would be restricted to the interval $[\max \mathcal{T_{C}}, \min \mathcal{T_{F}}]$. 
By applying precisely this time delay correction to $\hat{N}$ we solve these two problems. First, we give more importance precisely to the region where the population growth is exponential. Second, we omit naturally the very first initial times of the evolution of our populations. 

The inverse of the difference between the two largest eigenvalues of $\mathbb{W}_{s}$~\eqref{eq:WsAC}, $t_{\gap}$ (Eq.~\eqref{eq:tgap}) allows us to define the typical convergence time to the large time behavior for Eq.~\eqref{eq:7} (as we mentioned in Sec.~\ref{sec:bdp}). A crucial remark is that, as observed numerically, the duration before the population enters into the exponential regime is in fact larger than the time scale given by the gap: for instance, for the parameters used to obtain Fig.~\ref{fig:delayedPOP}, from Eq.~\eqref{eq:tgap} one has $t_{\gap} \approx 0.804$. The understanding of the duration of this discreteness effects regime would require a full analysis of the finite-population dynamics 
which are not fully understood. We propose in this section a numerical procedure to reduce its influence.

\subsubsection*{Log-Population Variance}
\label{Log-Population Variance}

\begin{figure}[t!]
	 \centering
        {\includegraphics [scale=0.48] {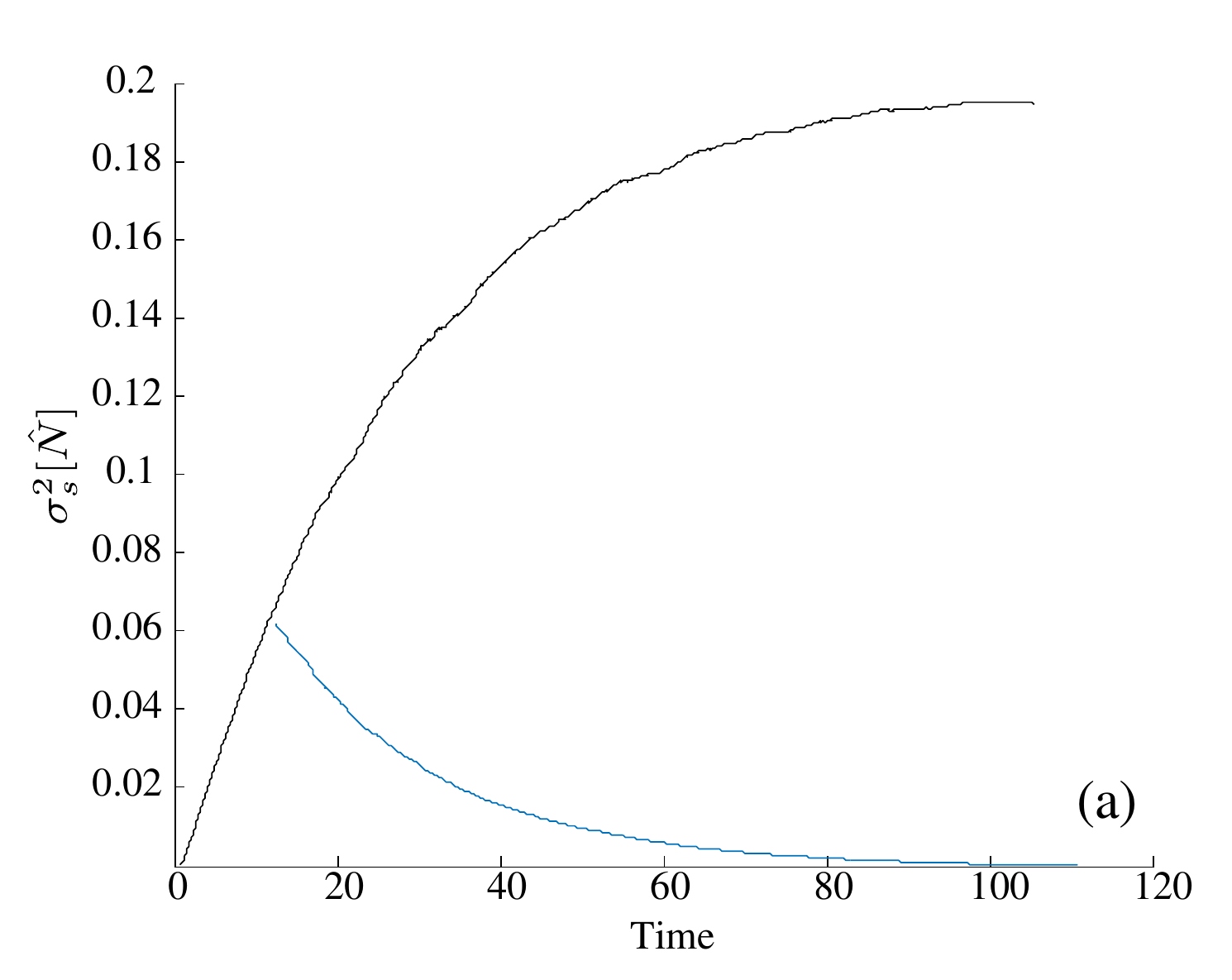}}
		{\includegraphics [scale=0.48] {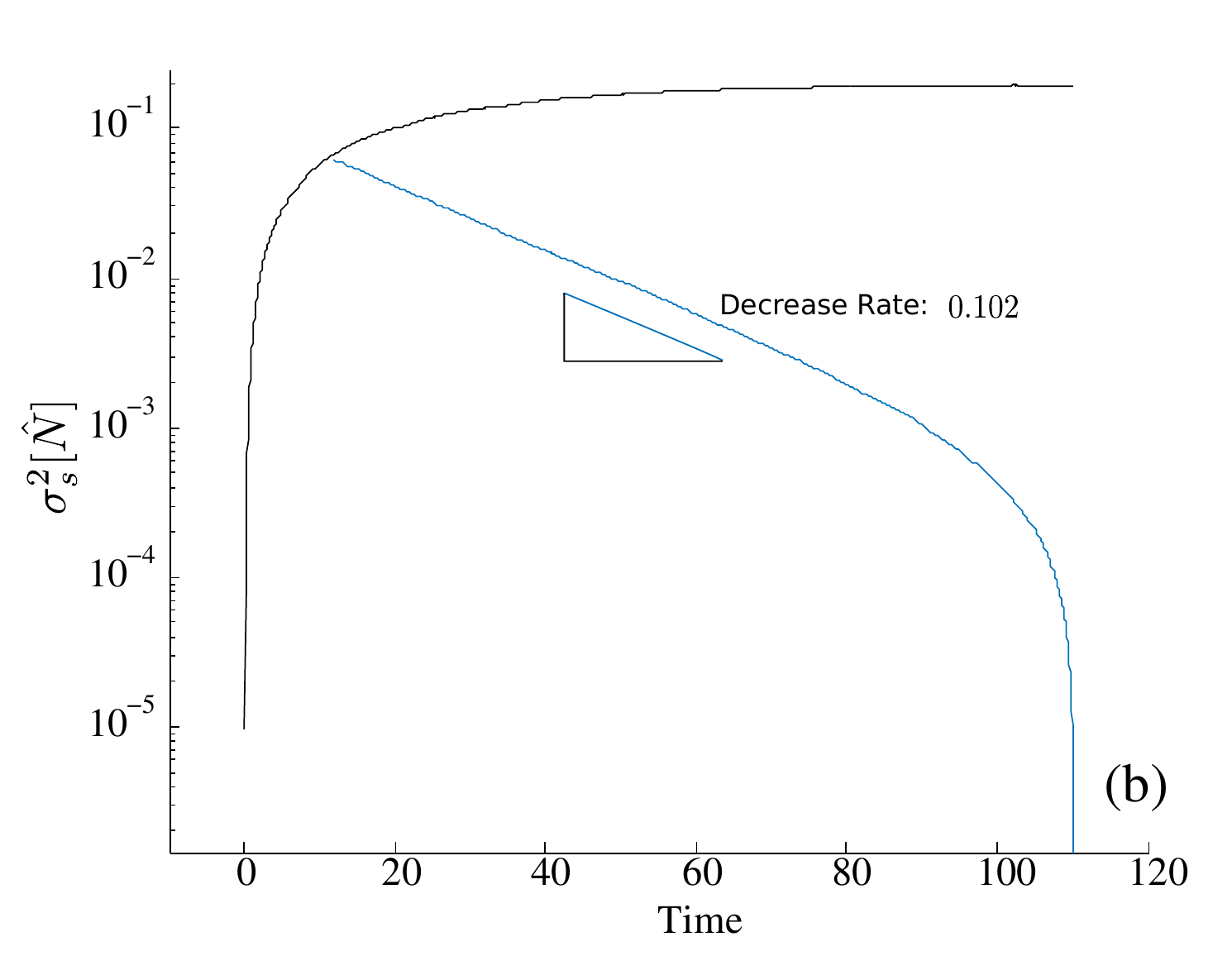}}\\ 
	\centering           	
\caption[Delayed Log-Populations Variance]{\textbf{(a)} Variance of the log-populations (black) and the delayed log-populations (blue) as a function of time. The variance of log-populations increases (or decreases, after the time transformation) as function of time. \textbf{(b)} Log-population variance in semi-log scale. $(s = -0.1)$.}
\label{fig:varlogPop}
\end{figure}

As can be seen from Fig.~\ref{fig:delayedPOP}, and as it is verified in Fig.~\ref{fig:varlogPop}, the variance of log-populations (black) increases as a function of the time, faster during the transient regime, and slower during the exponential growth regime until the variance becomes constant. After the time-delay correction, the variance of the delayed log-population (blue) decreases to zero as a function of time. The $s$-dependent decrease rate is shown in Fig.~\ref{fig:varlogPop}(b).

\section{$\Psi(s)$ Before and After the Time Delay}
\label{sec:psi_timedelay}
The CGF estimator $\Psi(s)$ can be recovered from the slope in time of the logarithm of the average population (see Sec.~\ref{Populations Merging}). We also mentioned in Sec.~\ref{Discreteness Effects at Initial Times}, that an alternative we can consider to overcome the discreteness effects would be to eliminate the initial transient regime  where these effects are strong. The improvement in the estimation of the analytical CGF $\psi(s)$~\eqref{eq:PSIA} comes precisely from these two main contributions, the time delaying of populations and the discarding of the initial transient regime of the populations.
We denote $\Psi_{\num}(s)$ the numerical estimator which is obtained from the slope of the logarithm of the average population (computed from merging several populations that have been generated using the cloning algorithm). On the other hand, $\Psi_{\tau}(s)$ is obtained through a time delay procedure over $\hat{N}$, as described above. These two numerical estimations are in fact averages over $R$ realizations and over their last $\gamma$ values. The approach followed in order to compute $\Psi_{\num}(s)$ and $\Psi_{\tau}(s)$ are computed is explained below.

\subsection{Numerical Estimators for $\psi(s)$}
\begin{figure}[t!]
	\centering
        {\includegraphics [scale=0.55]{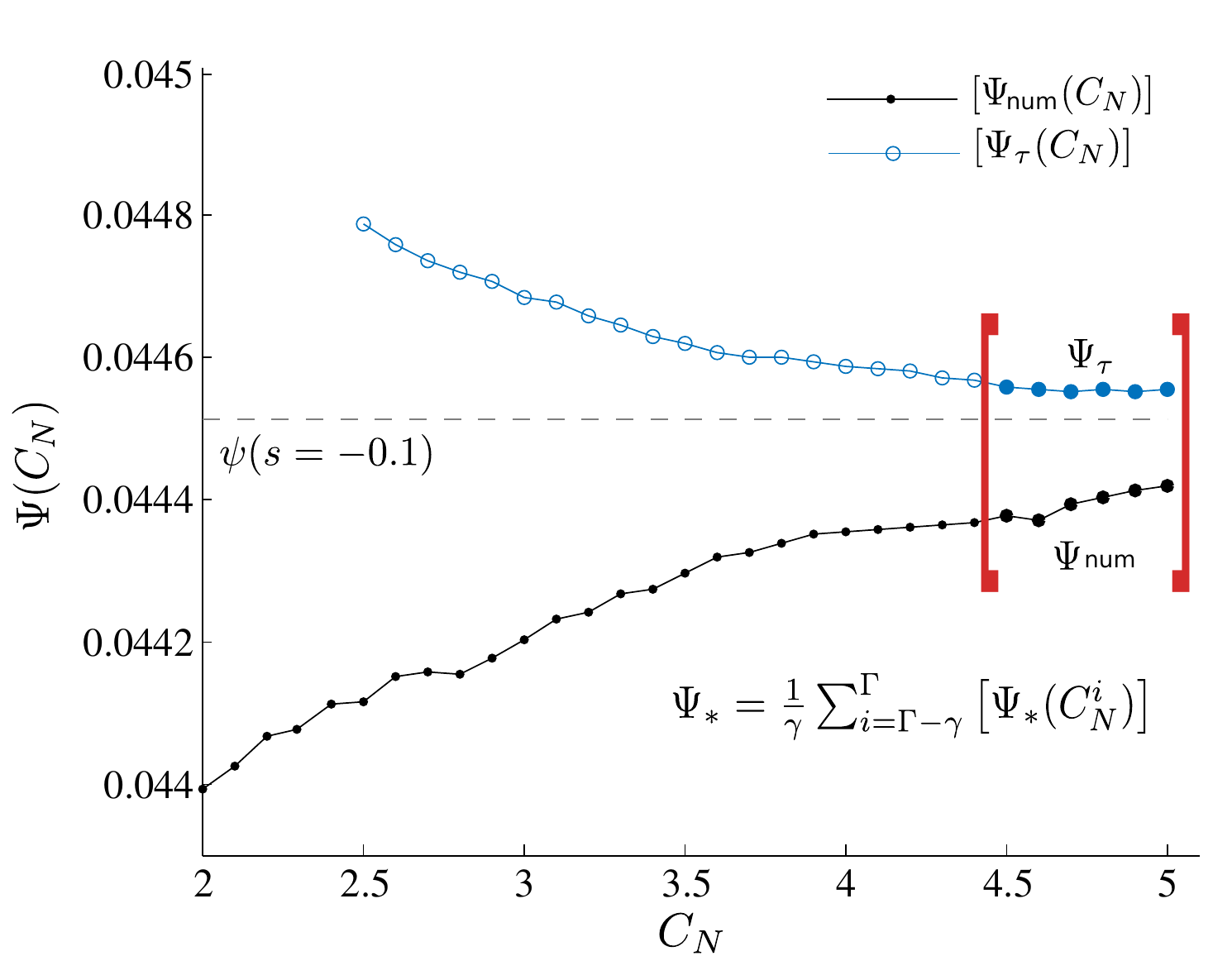}}	
	\centering        	
\caption[Numerical CGF Estimations]{Numerical estimations of $\psi(s=-0.1)$ as a function of the cut $C_{N}$ in (log) population. $\left[ \Psi_{\tau}(C_{N}) \right]$ is shown in blue and $\left[ \Psi_{\num}(C_{N}) \right]$ in black for $R = 40$. The numerical estimations $\Psi_{\num}$ and $\Psi_{\tau}$ are computed from an average of $\left[ \Psi_{*}(C_{N}^{i}) \right]$  over its last $\gamma = 6$ values. The subscript ``$*$'' denotes ``$\num$'' or ``$\tau$''.}
\label{fig:psi1}
\end{figure}

Let us call $\Psi_{*}(C_{N})$ an estimation of $\psi$ (by some method $(*)\in\{ \num, \tau \}$) as a function of the cut in log-population $C_{N}$. If we consider $C_{N}$ as $C_{N} = \{ C^{1}_{N},...,C^{\Gamma}_{N} \}$ a set of $\Gamma$ cuts, $\Psi_{*}(C_{N})$ is in fact $\Psi_{*}(C_{N}) = \{ \Psi_{*}(C^{1}_{N}),...,\Psi_{*}(C^{\Gamma}_{N}) \}$. If $\left[ \Psi_{*}(C^{i}_{N}) \right]$ is an average over $R$ realizations,
\begin{equation}
\left[ \Psi_{*}(C_{N}^{i}) \right] = \frac{1}{R} \sum_{r =1}^{R} \Psi_{*}^{r}(C_{N}^{i})
\end{equation}
our numerical estimation (for a given $s$) is then computed from an average of $\left[ \Psi_{*}(C_{N}^{i}) \right]$  over its last $\gamma$ values, i.e.,
\begin{equation}
\Psi_{*}(s)=\frac{1}{\gamma}
 \sum_{i=\Gamma-\gamma}^{\Gamma} \left[ \Psi_{*}(C_{N}^{i}) \right] =\frac{1}{\gamma R}
 \sum_{i=\Gamma-\gamma}^{\Gamma} \sum_{r =1}^{R} \Psi_{*}^{r}(C_{N}^{i})
\end{equation}
as is shown in Fig.~\ref{fig:psi1}. More details of the determination of these estimators are given in the subsection below.

\subsection{Comparison between ``Bulk'' and ``Fit'' Estimators of $\psi(s)$}
\label{``Bulk'' and ``Fit'' Slopes}
\begin{figure}[t!]
        \centering
        {\includegraphics [scale=0.48] {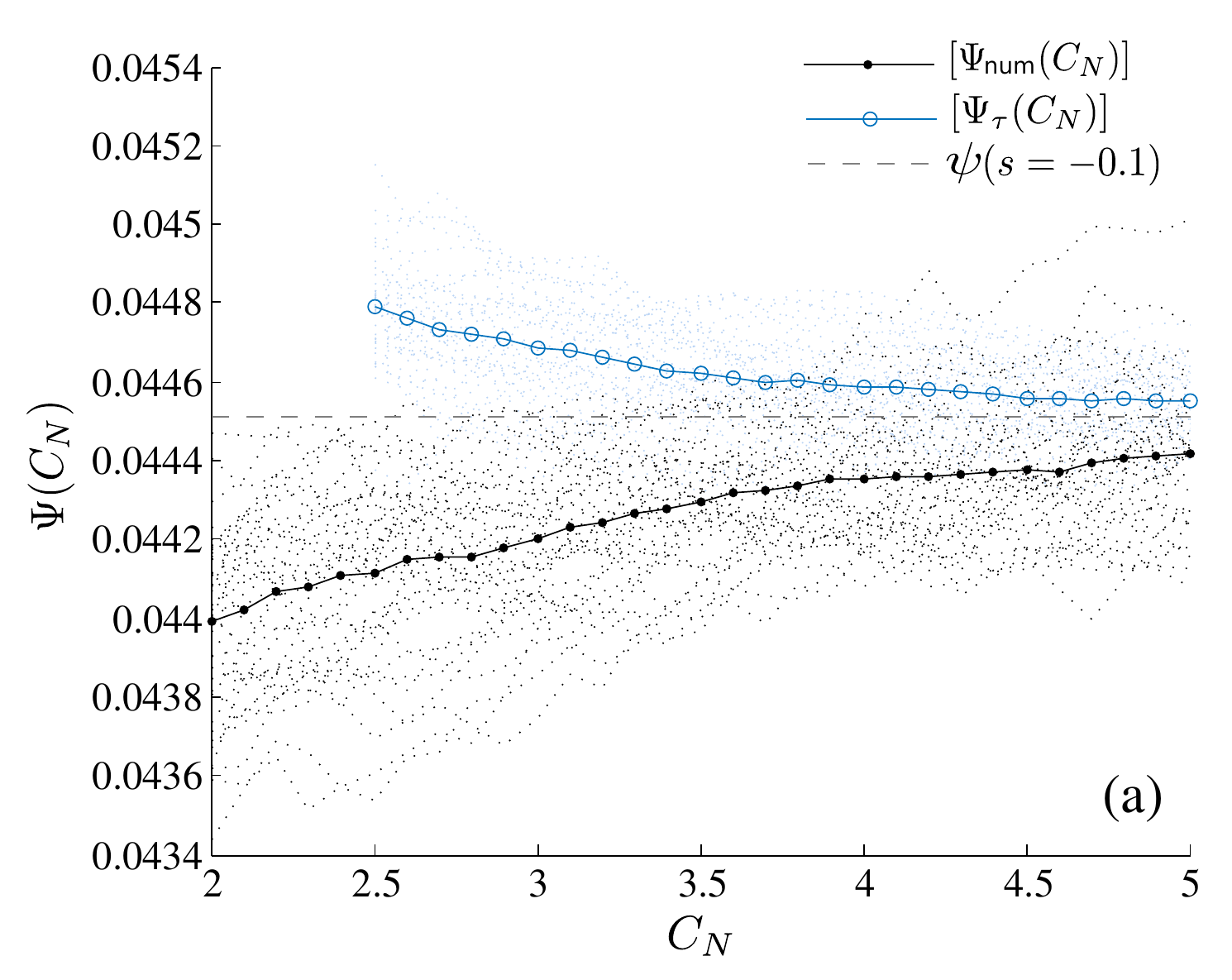}}
		{\includegraphics [scale=0.48] {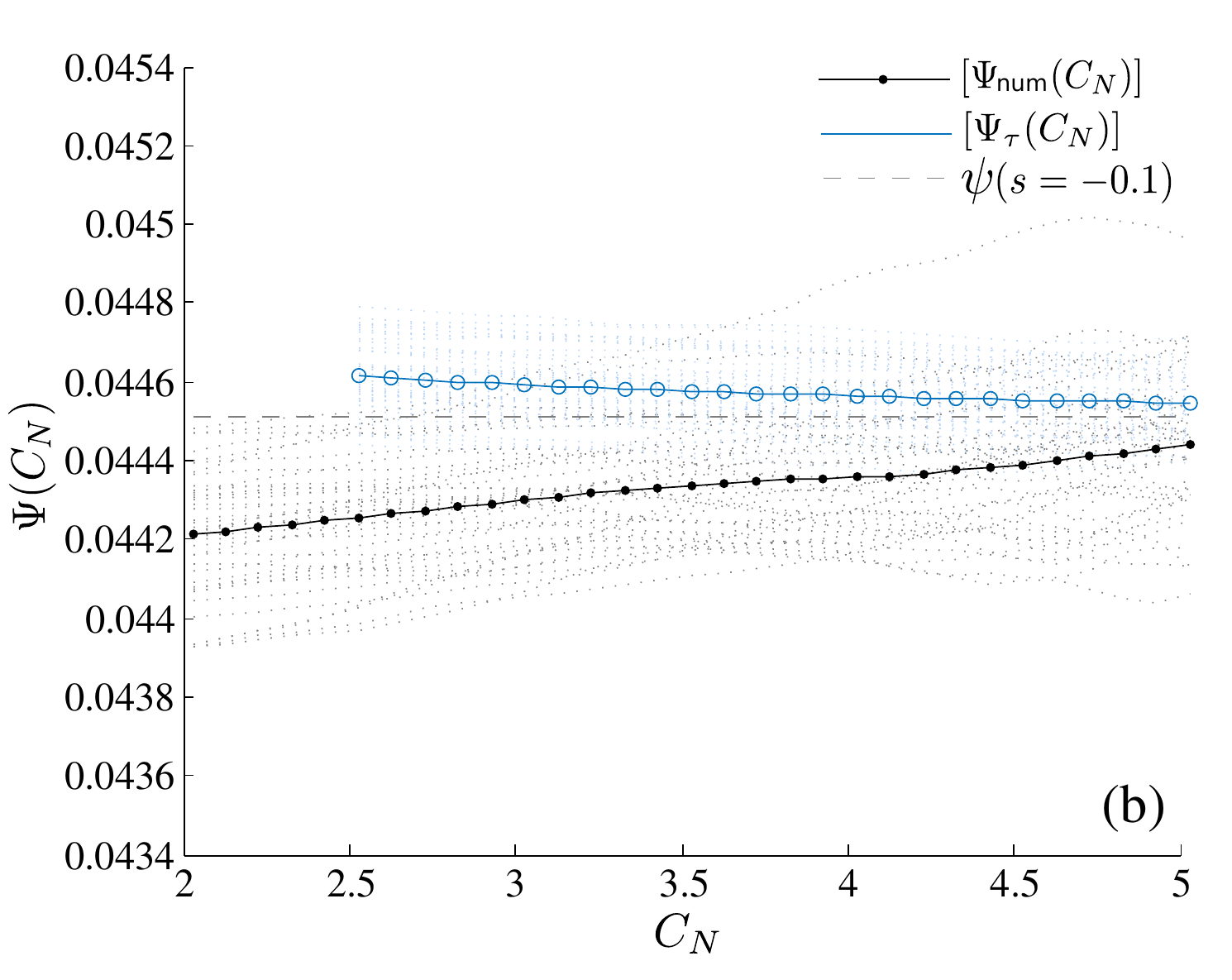}}\\ 
\centering        	
\caption[``Bulk'' and ``Fit'' Estimations]{Average over $R = 40$ realizations of the numerical estimators $\Psi_{\num}(C_{N})$ and $\Psi_{\tau}(C_{N})$ as a function of the cut in log-population for $s=-0.1$. The CGF estimations were obtained from \textbf{(a)} a ``Bulk'' slope and from \textbf{(b)} a ``Fit'' slope. The estimation for $\psi$ becomes better if we discard the initial transient regime where discreteness effects are strong.}
\label{fig:psi2}
\end{figure}

The estimators defined in the last subsection can be obtained from the ``bulk'' slope (Fig.~\ref{fig:psi2}(a)) given by Eq.~(\ref{eq:6}) and from the affine fit of the average log-population by $p_t = \Psi(s)t+ p_0$ (Fig.~\ref{fig:psi2}(b)), as explained in Sec.~\ref{sec:NCPA}. 
Fig.~\ref{fig:psi2} shows the average over $R = 40$ realizations of the numerical estimators $\Psi_{\num}(C_{N})$ and $\Psi_{\tau}(C_{N})$ as a function of the cut in log-population for $s=-0.1$. As before, $\left[ \Psi_{\tau}(C_{N}) \right]$ is shown in blue and $\left[ \Psi_{\num}(C_{N}) \right]$ (without the ``time delay'') is shown in black. As we already mentioned, the estimation for $\psi$ becomes better if we discard the initial transient regime where the discreteness effects are strong. 

The black curves in Fig.~\ref{fig:psi2} represent the standard way of estimating $\psi$ which comes from the slope of the average log-population, shown in dark green in Fig.~\ref{fig:delayedPOP}(a) for one realization. We can observe the effect of discarding the initial transient regime of these populations by cutting systematically this curve and computing $\Psi_{\num}(C_{N})$ from the growth rate $\Psi(s)$ computed on the interval $[C_{N},N_{\max}]$. Independently if $\Psi_{\num}(C_{N})$ is computed from the ``bulk'' slope or by the ``fit'' slope, for appropriate values of $C_{N}$, $\Psi_{\num}(C_{N})$ becomes closer to the theoretical value. Additionally to this result, we can add the time correction or delay proposed in Sec.~\ref{Time Delay Correction} and the estimation $\Psi_{\tau}(C_{N})$, shown by blue curves in Fig.~\ref{fig:psi2}, is closer to the theoretical value than  $\Psi_{\num}(C_{N})$ for all $C_{N}$.

Once we have proved that the estimation of $\psi$ becomes better when we discard the initial times where the discreteness effects are strong and when we perform a ``time delay'' over our populations in order to give more weight to the final regime of our populations, the question that remains is related to what we should consider as $\Psi_{\num}(s)$ and $\Psi_{\tau}(s)$. As we showed in Fig.~\ref{fig:psi1}, $\Psi_{\num}(s = -0.1)$ and $\Psi_{\tau}(s = -0.1)$ are computed from an average over the last $\gamma$ values of $\left[ \Psi_{\num}(C_{N}) \right]$ and $\left[ \Psi_{\tau}(C_{N}) \right]$. Below, we repeat this procedure and compute these estimators for several values of $s$, $s \in \left[ -0.3,-0.05 \right]$. The improvement in the determination of the CGF is measured through the relative distance of the numerical estimations with respect to the theoretical values and their errors.

\begin{figure}[t!]
        \centering
        {\includegraphics [scale=0.48] {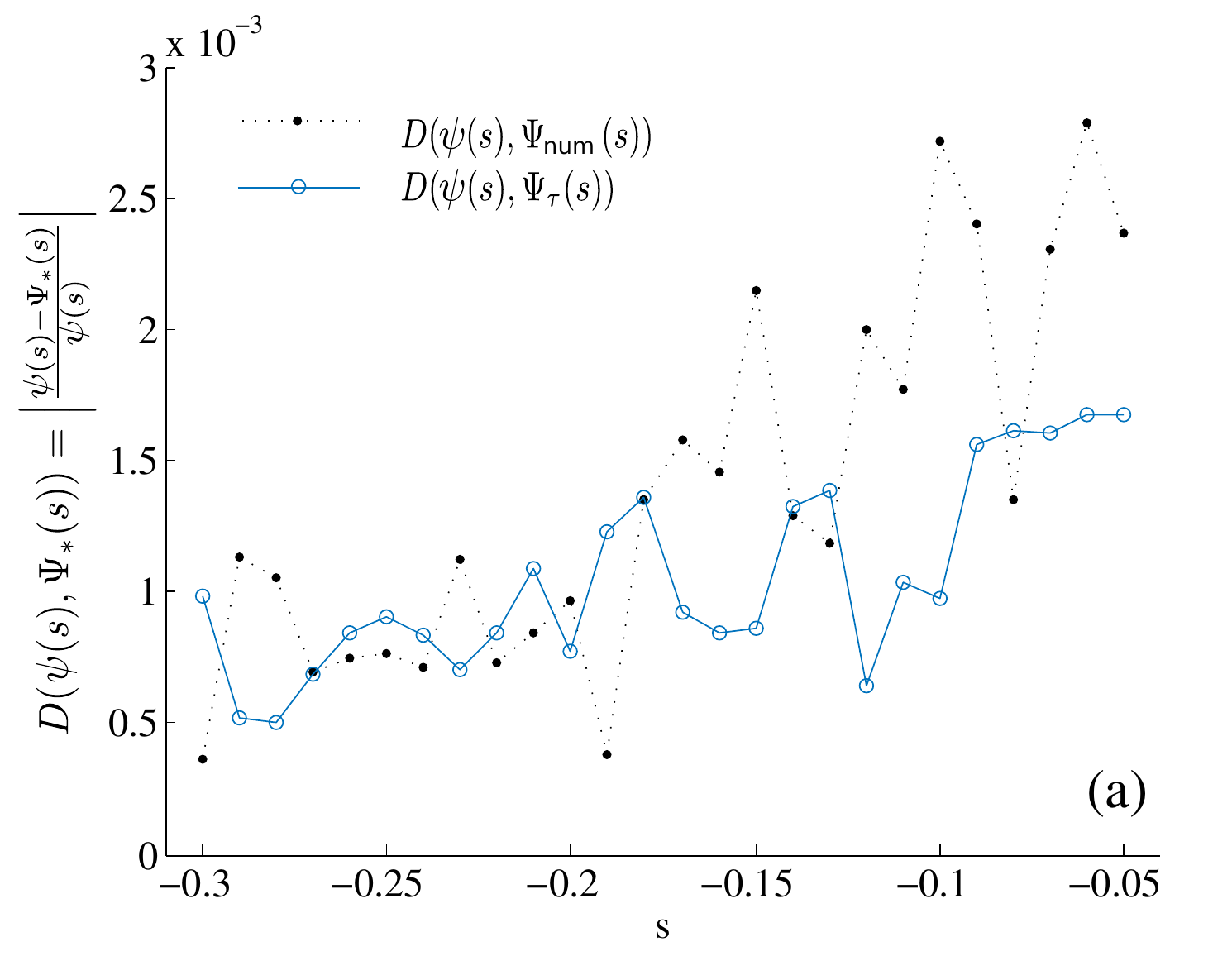}}
		{\includegraphics [scale=0.48] {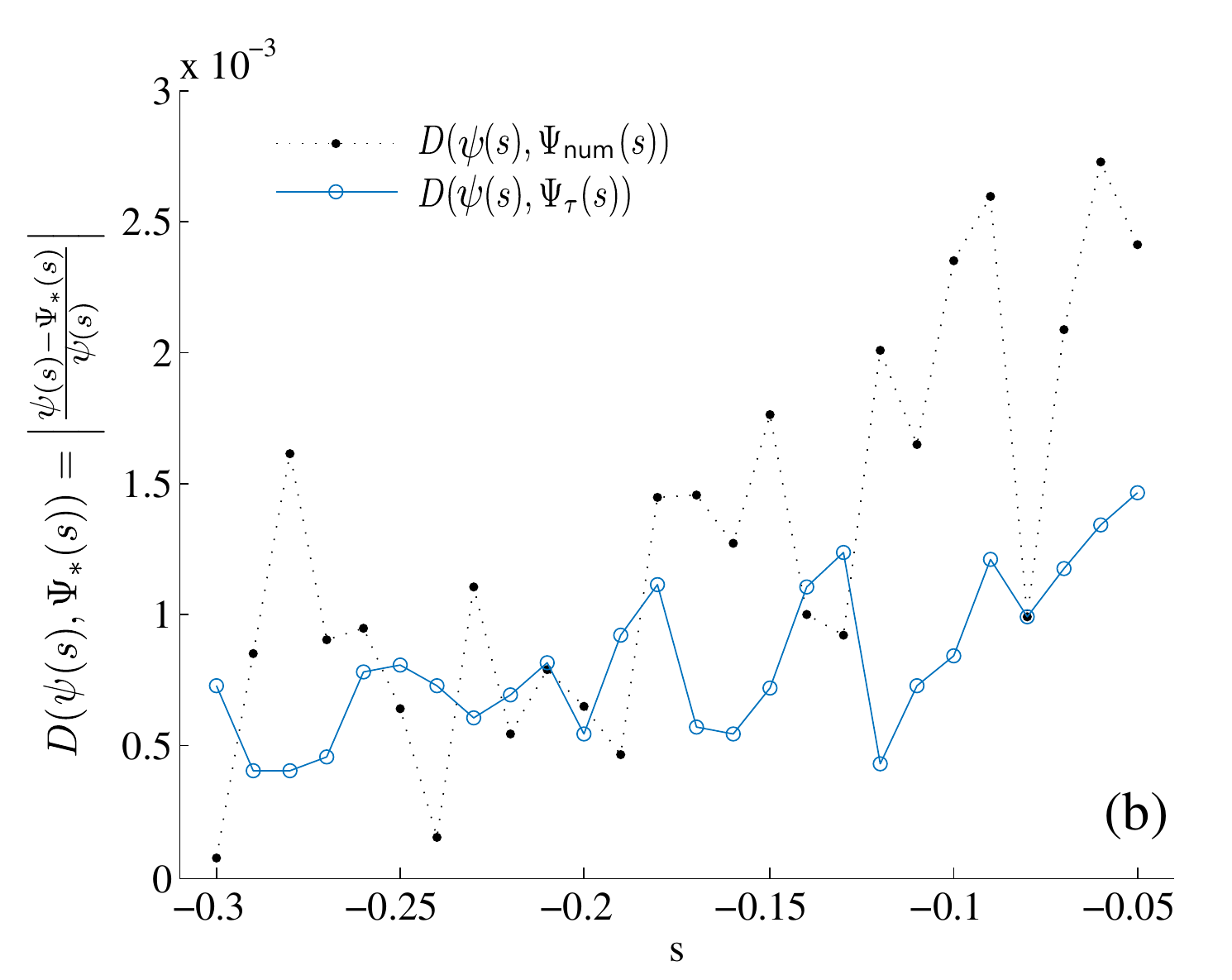}}\\ 
\centering        	
\caption[Estimator Relative Distance]{Relative distance $D(\psi(s),\Psi_{*}(s))$ between the estimator $\Psi_{*}(s)$ and its theoretical value $\psi(s)$. The deviation from the theoretical value is larger for values of $s$ close to $0$, but is smaller after the ``time delay correction'' for almost every value of $s$. The CGF estimators were obtained from \textbf{(a)} a ``Bulk'' and from \textbf{(b)} a ``Fit'' slope. }
\label{fig:relativeD}
\end{figure}
\begin{figure}[t!]
        \centering
        {\includegraphics [scale=0.48] {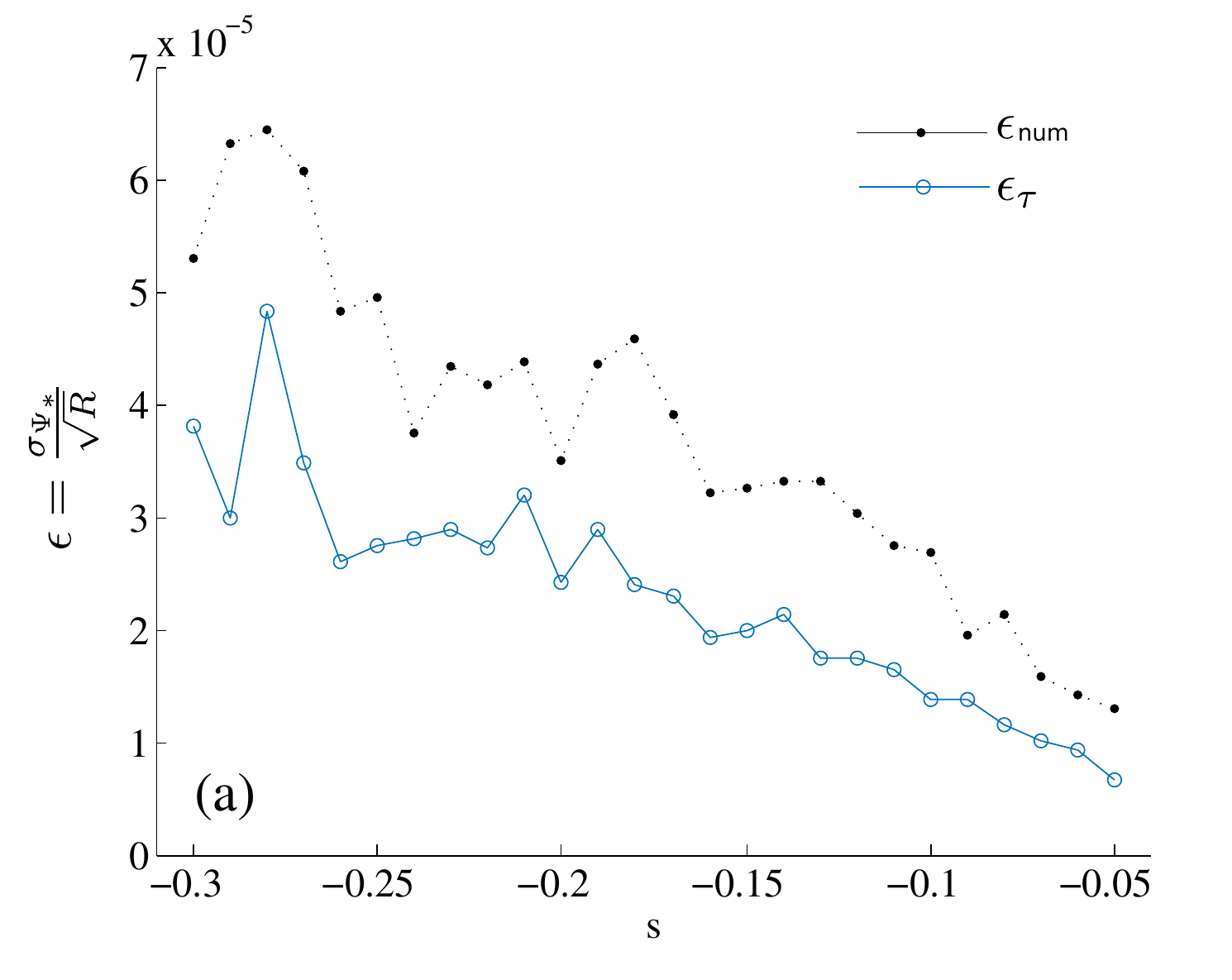}}
		{\includegraphics [scale=0.48] {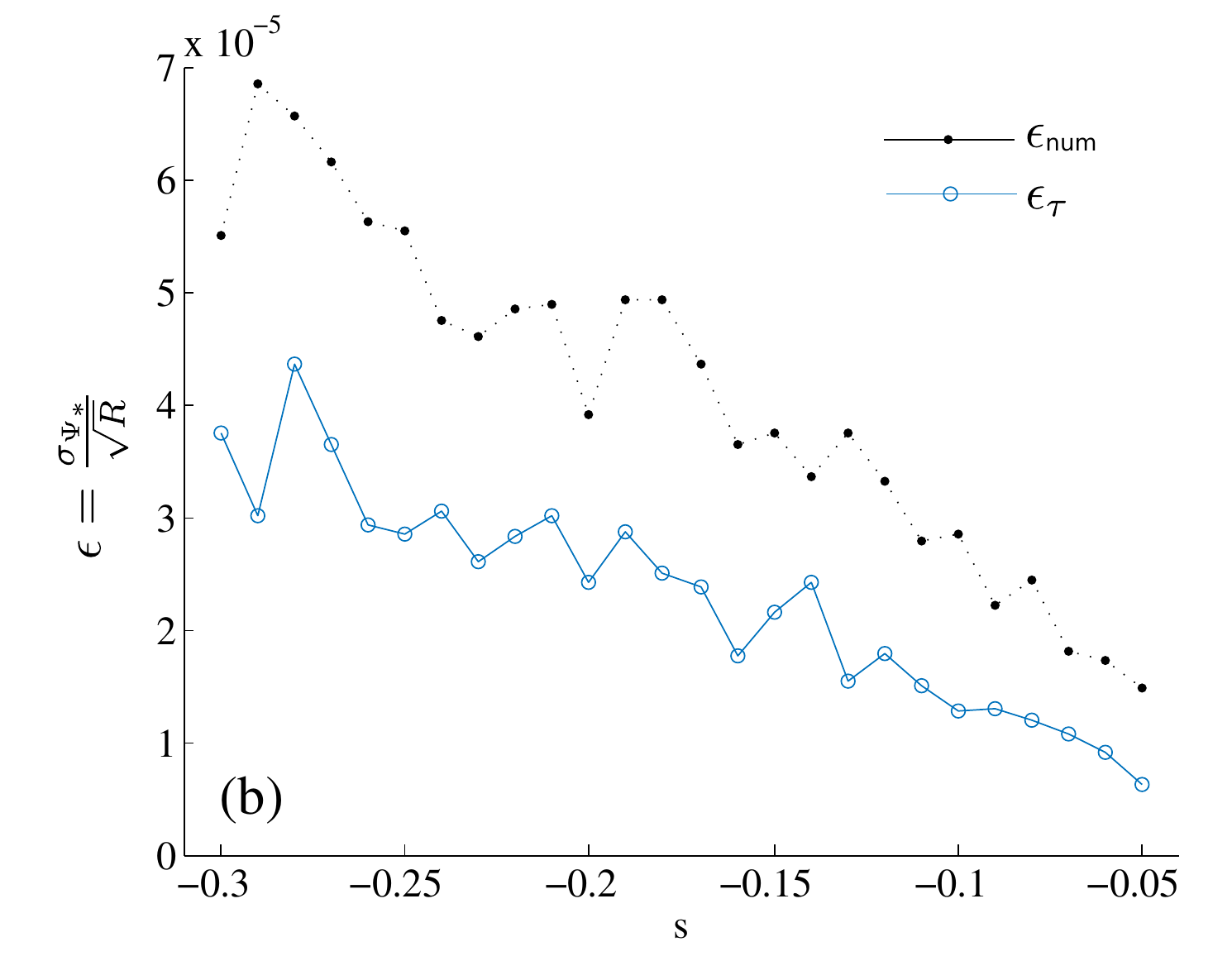}}\\ 
\centering        	
\caption[Estimator Error]{Estimator error for $\psi(s)$, $\epsilon_{num}$ (black) and $\epsilon_{\tau}$ (blue). The estimator error decreases as $s$ approaches to $0$ (for both, \textbf{(a)} ``Bulk'', \textbf{(b)} and ``Fit'' slopes) and it is always smaller for $\Psi_{\tau}(s)$ for any value of $s$.}.
\label{fig:error}
\end{figure}

\subsection{Relative Distance and Estimator Error}
\label{Relative Distance and Estimator Error}
The relative distance
\begin{equation}
D(\psi(s),\Psi_{*}(s)) = \left \vert \frac{ \psi(s) - \Psi_{*}(s)}{\psi(s)}  \right \vert
\end{equation}
between the estimator $\Psi_{*}(s)$ and its theoretical value $\psi(s)$ is shown in Fig.~\ref{fig:relativeD}.
These distances were also computed from the ``bulk'' and the ``fit'' slope and with (blue) and without time delay (black). As we can observe, the deviation from the theoretical value is larger for values of $s$ close to $0$, but is smaller after the ``time correction'' for almost every value of $s$.
Fig.~\ref{fig:error} presents the estimator error for $\psi(s)$ defined as 
\begin{equation}
\epsilon = \frac{\sigma_{\Psi_{*}}}{\sqrt{R}}, 
\end{equation}
where $R$ is the number of realizations and $\sigma_{\Psi_{*}}$ is the standard deviation of $\Psi_{*}(s)$. Similarly as in previous results, the estimator error decreases as $s$ approaches to $0$ (for both slopes) and it is always smaller for $\Psi_{\tau}(s)$ for any value of $s$.

\section{Time Delay Properties}
\label{sec:time-delay-prop}
\begin{figure}[t!]
        \centering
        {\includegraphics [scale=0.48] {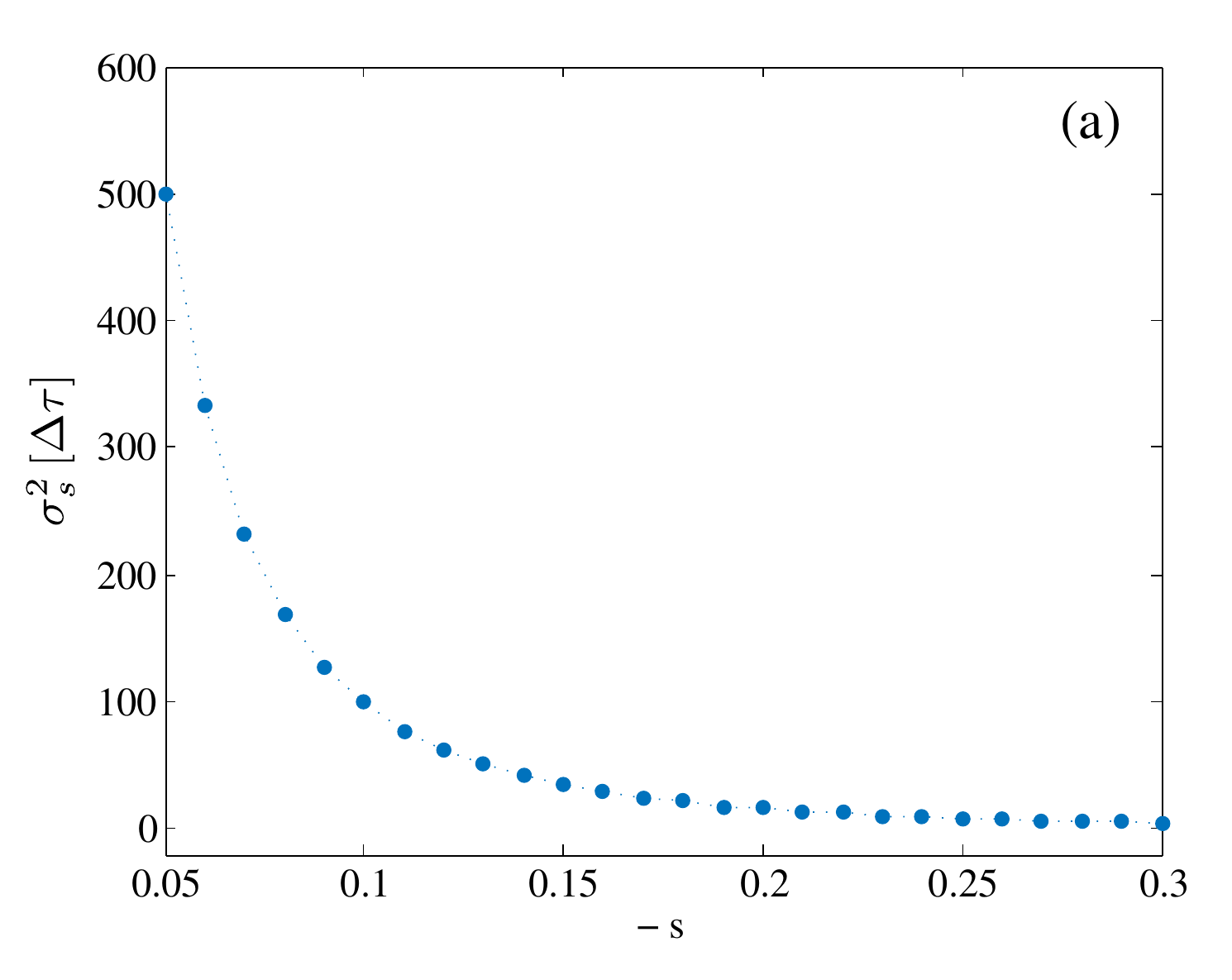}}
		{\includegraphics [scale=0.48] {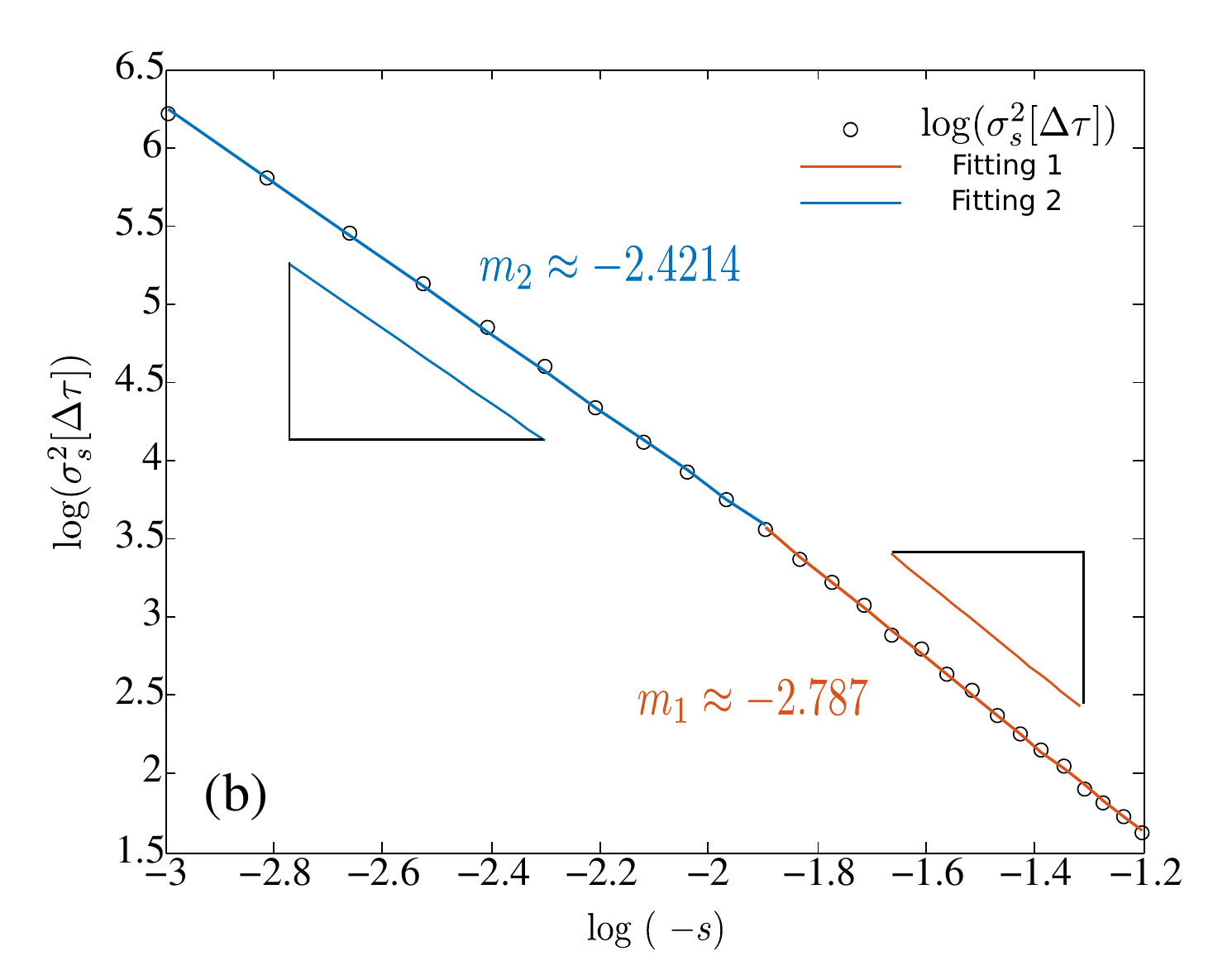}}\\ 
\centering        	
\caption[Time Delay Variance]{\textbf{(a)} Time delay variance $\sigma^{2}_{s} \left[ \Delta \tau \right]$. The dispersion of the time delays is large for values of $s$ close to $0$ and decreases rapidly as $-s$ increases. \textbf{(b)} Time delay variance regimes, one characterized with $m_{1}\approx -2.877$ ($s \in \left[ -0.15,-0.3 \right]$) and the other with $m_{2}\approx -2.4214$ ($s \in \left[ -0.05,-0.15 \right]$).}
\label{fig:timedelay}
\end{figure}

Here, we analyze the properties of the distribution of time delays $\Delta\tau(s) = \{ \Delta\tau_{1}(s),\ldots,\Delta\tau_{J}(s) \}$ which has been centered with respect to its mean. 
In Fig.~\ref{fig:timedelay}(a), we show its variance $\sigma^{2}_{s} \left[ \Delta \tau \right]$.
%
The dispersion of time delays is large for values of $s$ close to $0$ and decreases quickly as $-s$ increases. This is understood by observing that the typical growth rate $\left[ r_{s}(C) - r(C) \right] $ of the cloning algorithm goes to zero as $s \rightarrow 0 $ inducing a longer transient regime between the small and large population regimes. When we plot the variance in log-log scale, as in Fig.~\ref{fig:timedelay}(b), we can observe two linear regimes, one characterized by an exponent $m_{1}\approx -2.877$ ($s \in \left[ -0.15,-0.3 \right]$) and the other by $m_{2}\approx -2.4214$ ($s \in \left[ -0.05,-0.15 \right]$). They correspond to power-law behaviors in time of the variance of the delays, which remain to be understood.

\begin{figure}[!]
	\centering
        {\includegraphics [scale=0.48] {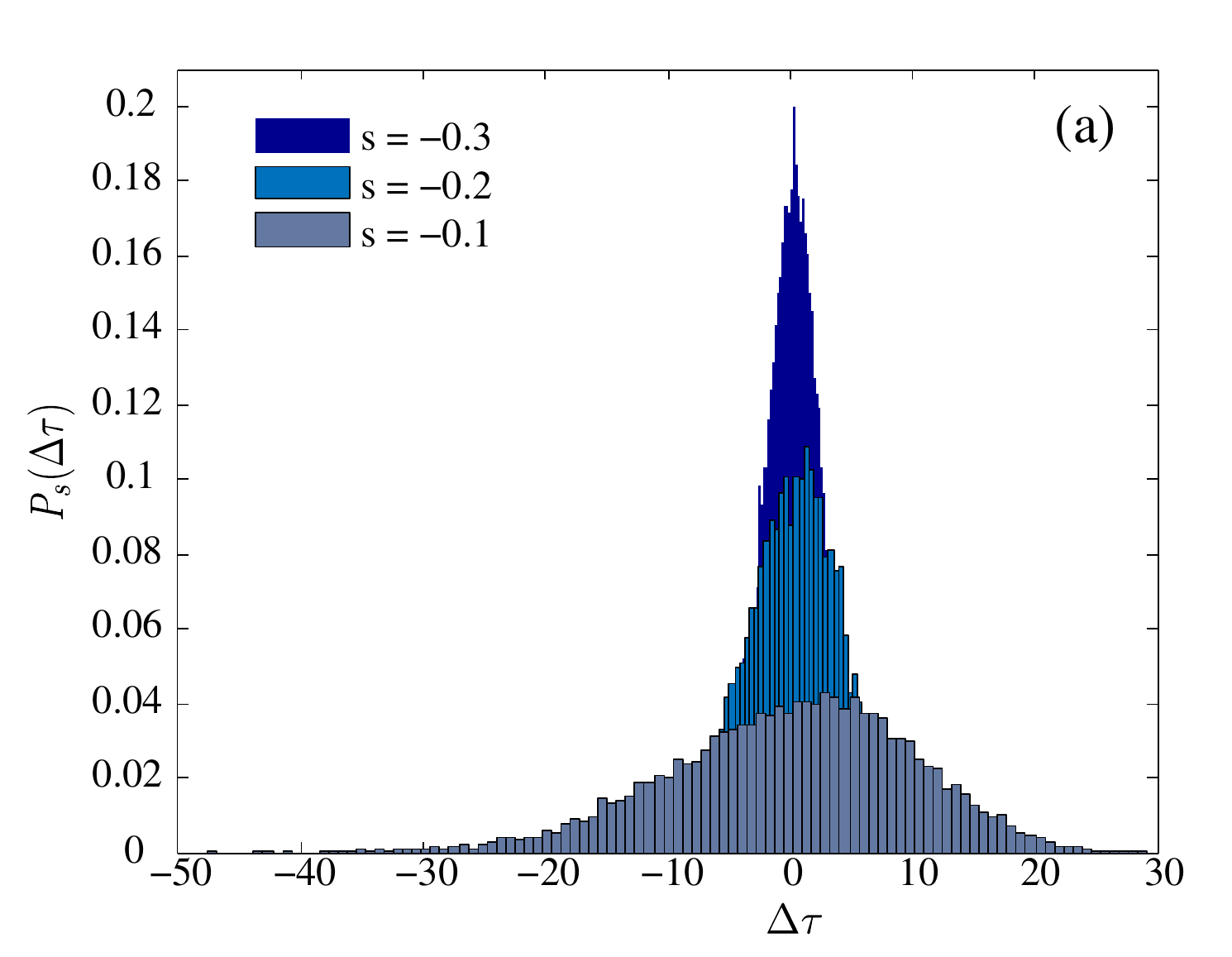}}
		{\includegraphics [scale=0.48] {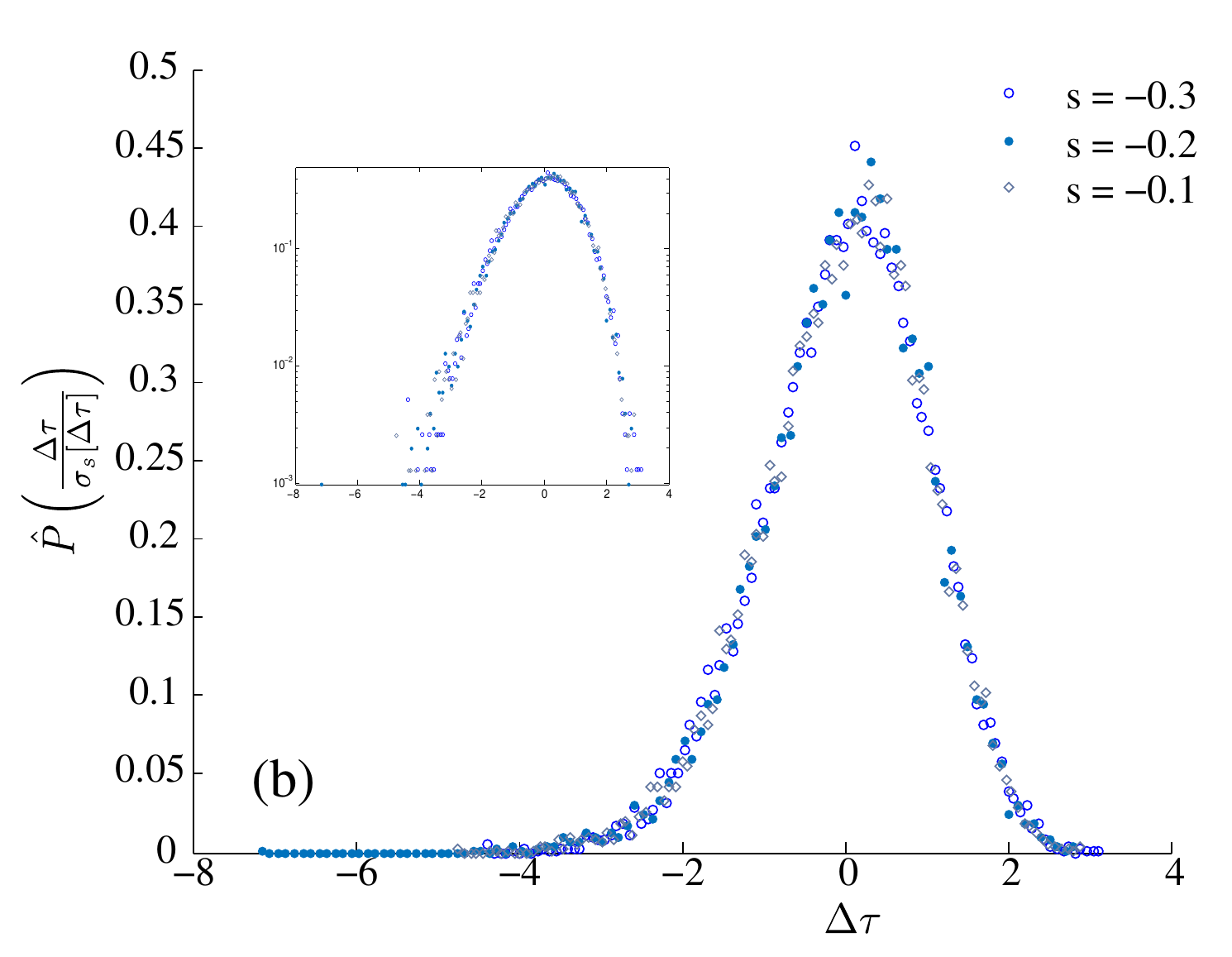}}\\ 
	\centering        	
\caption[Distribution of Time Delays ]{\textbf{(a)} Distribution of time delays for different values of $s$. The dispersion of time delays is wider for values of $s$ closer to zero. \textbf{(b)} Rescaled distribution of time delays $\hat{P} \left(\frac{\Delta \tau }{\sigma_{s} \left[\Delta\tau \right]}\right)$. The distribution of time delays depends only on their $\sigma_{s}$ as $P_{s}(\Delta\tau) = \sigma_{s} \left[\Delta\tau \right] \hat{P} \left(\frac{\Delta \tau }{\sigma_{s} \left[\Delta\tau \right]}\right)$. }
\label{fig:histogram}
\end{figure}

This dependence of the dispersion of time delays with $s$ can be better seen in the distribution of time delays $P_{s}(\Delta\tau)$ shown in Fig.~\ref{fig:histogram} for various values of $s$. This distribution is wider for values of $s$ closer to zero (Fig.~\ref{fig:histogram}(a)). However if we rescale the distributions of time delays by their respective $\sigma_{s}$, as shown in Fig.~\ref{fig:histogram}(b), the distributions become independent of $s$ as $P_{s}(\Delta\tau) = \sigma_{s} \left[\Delta\tau \right] \hat{P} \left(\frac{\Delta \tau }{\sigma_{s} \left[\Delta\tau \right]}\right)$. This provides a strong numerical evidence supporting the existence of a universal distribution $\hat P$.

\section{Discussion}
\label{sec:discussionDE}
In this chapter, we analyzed the discreteness effects at initial times in population dynamics. During the initial transient regime of the evolution of populations, there is a wide distribution of times at which the first series of jumps occurs. This means that fluctuations at initial times produce that some populations remain in their initial states for much longer than others, producing a gap in their individual evolution. This induces a relative shift that lasts forever. These effects play an important role specially for the determination of the large deviation function which may be obtained from the growth rate of the average log-population (Sec.~\ref{sec:NCPA}).

However, in Sec.~\ref{Discreteness Effects at Initial Times} we saw how by restricting the evolution of our populations up to a maximum time $T_{\max}$ (or population $N_{\max}$) which is not ``large enough'', the average population (and $\Psi(s)$) is strongly affected by the behavior of $\mathcal{N}$ at initial times. We proposed as an alternative to overcome the influence of initial discreteness effects to get rid of the regions of the populations where  these effects are present. In other words, to cut the initial transient regime of the populations. In that case, we saw that the average of populations is restricted to the interval $[\max \mathcal{T_{C}}, \min \mathcal{T_{F}}]$ which can be in fact very small and this can induce a poor estimation of $\psi(s)$ (Fig.~\ref{fig:merge}(b)).

Complementary to this, we found a way of emphasizing the effects of the exponential growth regime in the determination of $\psi(s)$ by using the fact that log-populations after a long enough time become parallel (Fig.~\ref{fig:distance1}(a)) and that once the populations have overcome the discreteness effects, the distance between them becomes constant (Fig.~\ref{fig:distance1}(b)) and the discreteness effects are not strong anymore (Sec.~\ref{Parallel Behaviour in Log-Populations}). We argued in Sec.~\ref{Time Delay Correction} that this initial discreteness effects or initial ``lag'' between populations could be compensated by performing over the populations a time translation (Eq.~\eqref{eq:13}). This time delay procedure is chosen so as to overlap the population evolutions in their large-time regime (Fig.~\ref{fig:delayedPOP}(b)). The improvement in the estimation of $\psi$ comes precisely from these two main contributions, the time delaying of populations and the discarding of the initial transient regime of the populations. 

We showed how the numerical estimations for the CGF are improved as the initial transient regime of the populations are discarded (independently of the method used to compute the growth rate of the average population, see Fig.~\ref{fig:psi2}). Also, it is was shown that if additionally, we perform the time delay procedure, the estimation of $\psi$ is improved even more and closer to the theoretical value (Sec.~\ref{``Bulk'' and ``Fit'' Slopes}). This result was confirmed later in Sec.~\ref{Relative Distance and Estimator Error} by computing the relative distance of the numerical estimators with respect to the theoretical value and their errors. As we observed in Fig.~\ref{fig:relativeD}, the deviation from the theoretical value is higher for values of $s$ close to $0$, but is smaller after the ``time correction'' for almost every value of $s$. Similarly for the error estimator (Fig.~\ref{fig:error}).
\newpage
Our numerical study was performed on a simple system, and we hope it can be extended to more complex phenomena. However, there remain open questions even for the system we have studied. The duration of the initial discrete-population regime could be understood from an analytical study of the population dynamics itself. Our numerical results also support a power-law behavior in time of the variance of the delays. Furthermore, it appeared that the distribution of the delays takes a universal form, after rescaling the variance to one. Those observations open questions for future studies.

\clearpage
\thispagestyle{empty}
\phantom{a}

\chapter[\quad Finite-Time and Finite-Size Scalings. I. Discrete Time]{Finite-Time and Finite-Size Scalings in the Evaluation of Large-Deviation Functions: \\ I. Analytical Study using \\ a Birth-Death Process}
\label{chap:DiscreteTime}      
\section[Introduction]{Introduction}
\begin{figure}[t]
\centering
\includegraphics[width=0.55\columnwidth]{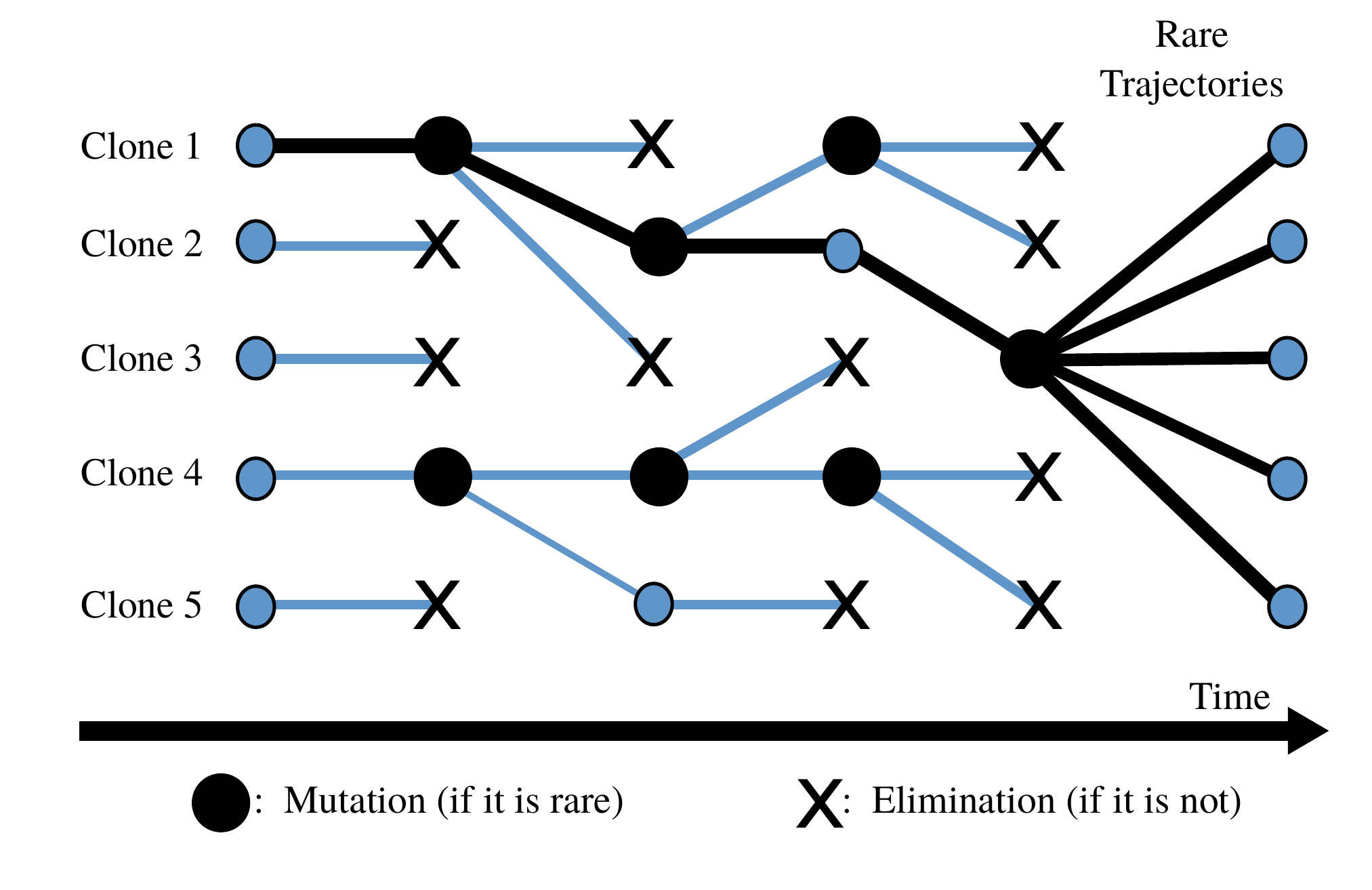}
\caption[Schematic Picture of the Population Dynamics Algorithm]{ \label{fig:traj}
Schematic picture illustrating the principle of the population dynamics algorithm. `Clones' (or copies) of the system are prepared and they evolve following a mutation-and-selection process, maintaining the total population constant.}
\end{figure}

Cloning algorithms are numerical procedures aimed at simulating rare events efficiently, using a  population dynamics scheme. In such algorithms, copies of the system are evolved in parallel and the ones showing the rare behavior of interest are multiplied iteratively \cite{DMC,Glasserman_1996,2001IbaYukito,GRASSBERGER200264,
4117599,CappeGuillinetal,Forwardinterfacesampling,PhysRevLett.94.018104,
delmoral2005,Dean2009562,lelievre,giardina_direct_2006,lecomte_numerical_2007,
tailleur_probing_2007,tailleur_simulation_2009,giardina_simulating_2011,
1751-8121-46-25-254002_2013} (See Fig.~\ref{fig:traj}). 
One of these algorithms proposed by Giardin\`a et al.~\cite{giardina_direct_2006,lecomte_numerical_2007,
tailleur_probing_2007,tailleur_simulation_2009,giardina_simulating_2011,
1751-8121-46-25-254002_2013} is used to evaluate numerically the cumulant generating function CGF (a large deviation function, LDF) of additive (or ``time-extensive'') observables in Markov processes \cite{opac-b1093895,touchette_large_2009}. 
%
%
While the method has been used widely, there have been less studies focusing on the analytical justification of the algorithm. Even though it is heuristically believed that the LDF estimator converges to the correct result as the number of copies $N_c$ increases, there is no proof of this convergence. Related to this lack of the proof, although we use the algorithm by assuming its validity, we do not have any clue how fast the estimator converges as $N_c \rightarrow \infty$.  

In this chapter~\cite{partI}, we discuss this convergence defining two types of numerical errors.
First, for a fixed finite $N_c$, averaging over a large number of realizations, the CGF estimator converges to an incorrect value, which is different from the desired large deviation result. We call this deviation from the correct value, \textbf{systematic errors}. Compared with these errors, we also consider the fluctuations of the estimated value. More precisely, for a fixed value of $N_c$, the results obtained in different realizations are distributed around this incorrect value. We call the errors associated to these fluctuations \textbf{stochastic errors}. Although both errors are important in numerical simulations, the former one can lead this algorithm to produce wrong results. For example as seen in Ref.~\cite{nemoto_population-dynamics_2016}, the systematic error grows exponentially as a temperature decreases (or generically in the weak noise limit of diffusive dynamics). 

To study these errors, we employ a birth-death process~\cite{tagkey2007iii,opac-b1079113} description of the population dynamics algorithm as explained below: We focus on physical systems described by a Markov dynamics \cite{giardina_direct_2006,giardina_simulating_2011,lecomte_numerical_2007} with a finite number of states $M$. We denote by $i$ ($i=0,1,\cdots M-1$) the states of the system. This Markov process has its own stochastic dynamics, described by the transition rates $w(i\rightarrow j)$.
In population dynamics algorithms, in order to study its rare trajectories, one prepares $N_c$ copies of the system, and simulate
these copies according to (\textit{i}) the dynamics of $w(i\rightarrow j)$ (followed independently by all copies) and (\textit{ii}) `cloning' step in which the ensemble of copies is directly manipulated, i.e., some copies are eliminated while some are multiplied (See Table~\ref{tablecorrespondence_}). 
Formally, the population dynamics represents, for a \emph{single} copy of the system, a process that does not preserve probability, as mentioned in Sec.~\ref{mut}. This fact has motivated the studies of auxiliary processes~\cite{jack_large_2010}, effective processes~\cite{1742-5468-2010-10-P10007} and  
driven processes~\cite{PhysRevLett.111.120601} to construct modified dynamics (and their approximations~\cite{PhysRevLett.112.090602}) that preserve probability.  
Different from these methods, in this chapter, we formulate explicitly the \textbf{meta-dynamics} of the copies themselves by using a stochastic birth-death process which preserves probability, and it allows us to study the numerical errors of the algorithm when evaluating LDF.  
We consider the dynamics of the copies as a stochastic birth-death process whose state is denoted by $n=(n_0,n_1,n_2,\ldots, n_{M-1})$, where $0\leq n_i\leq N_c$ represents the number of copies which are in state $i$ in the ensemble of copies.
We explicitly introduce the transition rates describing the dynamics of $n$, which we denote by $\sigma (n\rightarrow \tilde n)$. We show that the dynamics described by these transition rates lead in general to the correct LDF estimation of the original system $w(i\rightarrow j)$ in the $N_c \rightarrow \infty$ limit. 
We also show that the systematic errors are of the order $\mathcal O(1/N_c)$, whereas the numerical errors are of the order $\mathcal O(1/(\tau N_c))$ (where $\tau$ is an averaging duration). This result is in clear contrast with standard Monte-Carlo methods, where the systematic errors are always 0. 
The formulation developed in this chapter provides us the possibility to compute exactly the expressions of the convergence coefficients, as we do in Sec.~\ref{Section:Demonstrations} on a simple example. 
The analytical analysis presented here~\cite{partI} is supplemented with a thorough numerical study in the next chapter~\ref{chap:ContinuousTime}~\cite{partII}. There, we employ an intrinsically different cloning algorithm, which is the \textbf{continuous-time} population dynamics algorithm, that cannot be studied by the methods presented here (see Secs.~\ref{subsubsec:differenceContinuousTime} and~\ref{Discrete-time_algorithm}). We show in chapter~\ref{chap:ContinuousTime} that the validity of the scaling that we derive analytically here is very general, we make use of the convergence speed to propose a simple interpolation technique demonstrating in practice its efficiency in the evaluation of the LDF,
irrespectively of the details of the population dynamics algorithm. 

The present chapter is structured as follows. We first define the LDF problem in the beginning of Sec.~\ref{sec:Birth-deathprocess}, and then formulate the birth-death process used to describe the algorithm in Sec.~\ref{subsec:TransitionmatrixAlgorithm}. By using this birth-death process, we demonstrate that the  estimator of the algorithm converges to the correct large deviation function in Sec.~\ref{Subsec:DerivationLargeDeviationEstimator}. At the end of this section, in Sec.~\ref{subsec:systemsizeexpansion}, we discuss the convergence speed of this estimator (the systematic errors) and derive its scaling $\sim 1/N_c$. In Sec.~\ref{Sec:largedeviation_largedeviation}, we turn to  stochastic errors. For discussing this, we introduce the large deviation function of the estimator, from which we derive that the convergence speed of the stochastic errors is proportional to $1/(\tau N_c)$. In the next section, Sec.~\ref{Section:Demonstrations}, we introduce a simple two-state model, to which we  apply the formulations developed in the previous sections. We derive the exact expressions of the systematic errors in Sec.~\ref{subsection:systematicerrors_twostate} and of the stochastic errors in Sec.~\ref{subsection:largedeviationsInPopTwoState}. 
Then, in Sec.~\ref{subsec:Different large deviation estimator}, we propose another large deviation estimator and 
finally, in Sec.~\ref{sec:discussion}, we summarize the results obtained.
\begin{table*}[t!]
\begin{center}        
           \begin{adjustbox}{max width=\textwidth}
          \begin{tabular}{c||c|c}
          \hline	
\vphantom{\Big|}

                     	  			& Population dynamics algorithm      	&        Birth-death process describing    					 \\[-1mm]
		   	  			&								&   the population dynamics  		\\[1mm] \hline \hline
\vphantom{\Big|}
    State of the system	 	 & 		$i$              				  &        $n=(n_0,n_2,\cdots,n_{M-1})$ 					 \\
                       				 &  		($i=0,1,\cdots M-1$)		  &        ($0\leq n_i \leq N_c$ with $\sum_{i}n_i=N_c$)          \\[1mm]  \hline
\vphantom{\Big|}
 Transition rates			&  $w(i \rightarrow j)$ 	   			&	 $\sigma(n \rightarrow \tilde n)$ 					 \\[-1mm]
			 			&  Markov process on states $i$ 	   			&	 Markov process on states $n$						 \\[1mm]  \hline
\vphantom{\Big|}
         Numerical procedure     &	Prepare $N_c$ clones  and evolve those 		&   	Described by the dynamics 					\\[-1mm]
	for rare-event sampling	&    with a mutation-selection procedure  &	 	of rates $\sigma(n \rightarrow \tilde n)$			  \\[1mm] \hline
          \end{tabular}  
          \end{adjustbox}
        \end{center}
         \caption{\label{tablecorrespondence_} Correspondence between the population dynamics and the birth-death process to describe it.}
 \end{table*}

\section[Birth-Death Process and the Population Dynamics Algorithm]{ Birth-Death Process and the Population Dynamics Algorithm}
\label{sec:Birth-deathprocess}

As explained in the introduction of this chapter (also see Table~\ref{tablecorrespondence_}), the state of the population is $n=(n_0,n_1,\cdots,n_{M-1})$, where $n_i$ represents the number of clones in the state $i$. The total population is preserved: $\sum_i n_i = N_c$.  Below, we introduce the transition rates of the dynamics between the occupations $n$, $\sigma(n\rightarrow \tilde n)$ that describe corresponding large deviations of the original system whose dynamics is given by the rates $w(i\rightarrow j)$. 

As the \textbf{original system}, we consider the continuous-time Markov process in a discrete-time representation. 
By denoting by $dt$ the time step, 
the transition matrix $R_{j,i}$ for time evolution of the state $i$ is described as
\begin{equation}
R_{j,i} = \delta_{i,j} + dt \Big [ w(i\rightarrow j) -\delta_{i,j}\sum_{k}w(i\rightarrow k) \Big ],
\label{eq:ratesRij}
\end{equation}
where we set $w(i\rightarrow i)=0$. The probability distribution of the state $i$, $p_i(t)$, evolves in time as $p_i(t+dt) = \sum_{j} R_{i,j}p_j(t)$. In the $dt\rightarrow 0$ limit, one obtains the continuous-time Master equation~\eqref{eq:1} describing the evolution of $p_i(t)$ \cite{tagkey2007iii,opac-b1079113}. For simplicity, especially for the cloning part of the algorithm, we keep here a small finite $dt$. 
The reason why we use a discrete-time representation is solely for simplicity of the discussion. The main results can be derived even if we start with a continuous-time representation (see Sec.~\ref{subsubsec:dtDeltat}).
For the original dynamics described by the transition matrix~\eqref{eq:ratesRij}, we consider
an observable  $b_i$ depending on the state $i$ and we are interested in the distribution of its time-averaged value during a time interval $\tau$, defined as
\begin{equation}
B(\tau ) = \frac{1}{\tau} \sum_{t=0}^{\tau / dt} dt \ b_{i(t)}.
\label{eq:defBtau}
\end{equation}
Here $i(t)$ is a trajectory of the system generated by the Markov dynamics described by $R_{j,i}$. We note that $B(\tau)$ is a path- (or history-, or realization-) dependent quantity. Since $\tau B(\tau)$ is an additive observable, the fluctuations of $B(\tau)$ depending on the realizations are small when $\tau$ is large, but one can describe the large deviations of $B(\tau)$. Those occur with a small probability, and obey a large deviation principle \eqref{eq:ldpO}. We denote by ${\rm Prob}(B)$ the distribution function of $B(\tau)$. The large deviation principle ensures that ${\rm Prob}(B)$ takes an asymptotic form  ${\rm Prob}(B) \sim \exp (- \tau I(B))$ for large $\tau$, where $I(B)$ is a large deviation function (or `rate function') \cite{touchette_large_2009,opac-b1093895}. 
As we mentioned in the introduction, if the rate function $I(B)$ is convex, the large deviation function is expressed as a Legendre transform of a cumulant generating function $\psi(s)$ defined as
\begin{equation}
\psi(s) = \lim_{\tau \rightarrow \infty} \frac{1}{\tau} \log \left \langle e^{-s \tau B(\tau)} \right \rangle,
\label{eq:defpsi}
\end{equation}
so that $I(B) = - \inf_{s}\left [ s B +  \psi(s)  \right ]$. The large deviation function $I(B)$ and this generating function $\psi(s)$ are by definition difficult to evaluate numerically in Monte-Carlo simulations of the original system of transition rates $w(i\rightarrow j)$ (see for example Ref.~\cite{rohwer_convergence_2015}). To overcome this difficulty, population dynamics algorithms have been developed \cite{giardina_direct_2006,lecomte_numerical_2007,
tailleur_probing_2007,tailleur_simulation_2009,giardina_simulating_2011,
1751-8121-46-25-254002_2013}.
Here, we describe this population dynamics algorithm by using a birth-death process on the occupation state $n$ allowing us to study  systematically the errors in the estimation of $\psi(s)$ within the population dynamics algorithm. We mention that, without loss of generality, we restrict our study to so-called `type-B' observable (see Sec.~\ref{sec:LDobservables} in the Introduction) that do not depend on the transitions of the state~\cite{garrahan_first-order_2009}, i.e., which are time integrals of the state of the system, as in Eq.~\eqref{eq:defBtau}. Indeed, as explained for example in Refs.~\cite{giardina_simulating_2011} and~\cite{nemoto_population-dynamics_2016}, one can always reformulate the determination of the CGF of mixed-type observables into that of a type-B variable, by modifying the transition rates of the given system.   

Note that in chapter~\ref{chap:ContinuousTime}, we use a continuous-time version of the algorithm \cite{lecomte_numerical_2007} to study an observable of `type A'.
This version of the algorithm differs from the one considered here, in the sense that after its selection step, a copy in the population can have strictly more than one offspring. 
This results in an important difference: the effective interaction between copies due to the cloning/pruning procedure is unbounded (it can \emph{a priori} affect any proportion of the population), while in the discrete-time settings of the present chapter, this effective interaction is restricted to a maximum of one cloning/pruning event.
However, we observe numerically in chapter~\ref{chap:ContinuousTime} that the same finite-time and finite-population size scalings are present, illustrating their universal character.

\subsection[Transition Matrices and the Population Dynamics Algorithm]{ Transition Matrices Representing the Population Dynamics Algorithm}
\label{subsec:TransitionmatrixAlgorithm}
We denote the probability distribution of the occupation~$n$ at time~$t$ by $P_n(t)$. The time-evolution of this probability is decomposed into three parts.
 The first one is the original Monte-Carlo dynamics based on the transition rates $w(i\rightarrow j)$. The second one is the cloning procedure of the population dynamics algorithm, which favors or disfavors configurations according to a well-defined rule. The third one is a supplementary (but important) part which maintains the total number of clones to a constant $N_c$. We denote the transition matrices corresponding to these steps by $\mathcal T$, $\mathcal C$ and $\mathcal K$, respectively. By using these matrices, then, the time evolution of the distribution function is given as
\begin{equation}
\label{eq:time_Evolution}
P_{n}(t+dt) = \sum_{ \tilde n } \left (  \mathcal  K \mathcal C \mathcal T  \right )_{n,\tilde n} P_{\tilde n}(t).
\end{equation}
We derive explicit expressions of these matrices in the following sub-sections. A summary of the results obtained can be found in Table~\ref{table_Matrices}.

\begin{table*}[!]
\begin{center}
 \begin{adjustbox}{max width=\textwidth}        
          \begin{tabular}{c||c}
          \hline	
\vphantom{\bigg|}
                     	  			& Transition matrices      					 \\[2mm]  \hline \hline
\vphantom{\bigg|}
    Dynamics (``mutations'')	 	 & 		$ \mathcal T_{\tilde n, n} \equiv   \delta_{\tilde n, n} +  dt    \sum_{i=0}^{M-1} n_i  \sum_{j=0, (j\neq i)}^{M-1}w(i \rightarrow  j)    \left [  \delta _{\tilde n_i,n_i-1}   \delta _{\tilde n_j,n_j+1} \ \delta ^{i,j}_{\tilde n,n} \  -  \delta _{\tilde n,n}    \right ]  $   \\[2mm]  \hline
\vphantom{\bigg|}
  Cloning (``selection'')		&  $\mathcal C_{\tilde n, n}  =   \delta_{\tilde n, n} + s \ dt    \sum_{i=0}^{M-1}   n_i |\alpha_i|   \left [  \delta _{\tilde n_i,n_i + \alpha_i/|\alpha_i|}  \  \delta^{i}_{\tilde n, n}  - \delta_{\tilde n, n}   \right ]      + \mathcal O(dt^2)$            \\[2mm]  \hline
\vphantom{\bigg|}
  Maintaining $N_c$  	&   $\mathcal K_{\tilde n,n}  =  \delta_{\sum n_i,N_c}  \delta_{\tilde n, n}  
 + \sum_{k=-1,1} \delta_{\sum_i n_i, N_c + k}  \sum_{i=0}^{M-1}  \delta_{\tilde n_i, n_i - k} \  \delta^{i}_{\tilde n, n} \  \frac{n_i }{N_c + k}   $  \\[2mm] \hline
\vphantom{\bigg|}
	 Full process 		& $   (\mathcal K \mathcal C \mathcal T)_{\tilde n, n} 
  =   \delta_{\tilde n, n} +  dt    \sum_{i=0}^{M-1} n_i  \sum_{j=0, (j\neq i)}^{M-1} \left [ w(i \rightarrow  j) + s \tilde w_{n}(i \rightarrow  j) \right ]  \left [  \delta _{\tilde n_i,n_i-1}   \delta _{\tilde n_j,n_j+1} \ \delta ^{i,j}_{\tilde n,n} \  -  \delta _{\tilde n,n}    \right ]$  \\[2mm] 
				&  with  \ \ \  $\tilde w_n (i \rightarrow  j) = \frac{n_j  }{N_c} \left [ \alpha_j  \delta_{j\in \Omega^{(+)}} \frac{N_c}{N_c + 1} - \alpha_i \delta_{i\in \Omega^{(-)}}  \frac{N_c}{N_c - 1} \right ]$            			\\[2mm] \hline
          \end{tabular}             
          \end{adjustbox}
        \end{center}
         \caption{\label{table_Matrices} Transition matrices (Eq.~\eqref{eq:time_Evolution}) describing the birth-death process.}
 \end{table*}

\subsubsection[Original Dynamics: $\mathcal T$]{Derivation of the Original Dynamics Part: $\mathcal T$}

We first consider the transition matrix $\mathcal T$, which describes the evolution of the occupation state $n$ solely due to the dynamics based on the rates $w(i\rightarrow j)$. During an infinitesimally small time step $dt$, the occupation $n=(n_0,n_1,\cdots,n_{M-1})$ changes to $\tilde n = (n_0,n_1,\cdots, n_i-1, \cdots, n_j+1, \cdots, n_{M-1})$ where $0\leq i< M$ and $0\leq j< M$ (for all $i\neq j$). Since there are $n_i$ clones in the state $i$ before the transition, the transition probability of this change is given as $n_i w(i\rightarrow j) dt$. Thus, we obtain
\begin{equation}
 \mathcal T_{\tilde n, n} \equiv   \delta_{\tilde n, n} + dt    \sum_{i=0}^{M-1} n_i  \sum_{j=0, (j\neq i)}^{M-1}w(i \rightarrow  j) \left [  \delta _{\tilde n_i,n_i-1}   \delta _{\tilde n_j,n_j+1} \ \delta ^{i,j}_{\tilde n,n} \  -  \delta _{\tilde n,n}    \right ],
\end{equation}
where $ \delta ^{i,j}_{\tilde n,n}$ is a Kronecker $\delta$ for the indices except for $i,j$: $\delta ^{i,j}_{\tilde n,n} \equiv \prod_{k\neq i,j} \delta_{\tilde n_k,n_k}$. 
One can easily check that this matrix satisfies the conservation of the probability: $\sum_{\tilde n} \mathcal T_{\tilde n, n}=1$.
It corresponds to the evolution of $N_c$ independent copies of the original system with rates $w(i\to j)$.

\subsubsection[Cloning Part: $\mathcal C$]{Derivation of the Cloning Part: $\mathcal C$}
\label{subsubsec:CloningRatio}

In the population dynamics algorithm (for example the one described in the Appendix A of Ref.~\cite{nemoto_population-dynamics_2016}), at every certain time interval $\Delta t$, 
one evaluates the exponential factor for all clones equal to $e^{-s\int_t ^{t+\Delta t}\! dt' ~ b_{i(t')}}$ if the clone is in state $(i(t'))_{t'=t}^{t+\Delta t}$ during a time interval $t\leq t' \leq t +\Delta t$. This cloning factor  
determines whether each clone is copied or eliminated after this time interval. In the continuous-time version of the algorithm this factor was given by Eq.~\eqref{eq:Y}.
Although the details of how to determine this selection process can depend on the specific type of algorithms, the common idea is that each of the clones is copied or eliminated in such a way that a clone in state $i(t)$ has a number of descendant(s) proportional to the cloning factor
on average after this time interval.

In order to implement this idea in our birth-death process, we assume this time step $\Delta t$ to be small. For the sake of simplicity, we set this $\Delta t$ to be our smallest time interval $d t$: $\Delta t =dt$. This condition is not mandatory whenever the $\Delta t \rightarrow 0$ limit is taken at the end (see Sec.~\ref{subsubsec:dtDeltat} for the case $\Delta t > dt$). Then, noticing that the time integral $\int_t ^{t+\Delta t} dt'\ b_{i(t')}$ is expressed as $ dt\ b_{i(t)}$ for small $dt$, we introduce the following quantity for each state $i$ ($i=0,1,2,\ldots,M-1$):
\begin{equation}
\nu_i \equiv \frac{ n_i\ e^{-s\ dt\ b_i} }{\sum_{j=0}^{M-1} n_j\ e^{-s\ dt\ b_j }  }  N_c.
\label{eq:nufac}
\end{equation}
Note that there is a factor $n_i$ in front of the exponential function $e^{-s\ dt\ b_i}$
 which enumerates the number of clones that occupy the state $i$. The quantity $\nu_i$ is aimed at being the number of clones in state $i$ after the cloning process, however, since $\nu_i$ is not an integer but a real number, one needs a supplementary prescription to fix the corresponding integer number of descendants. In general, in the implementation of population dynamics, this integer is generated randomly from the factor $\nu_i$, equal either to its lower or to its upper integer part. 
The probability to choose either the lower or upper integer part is fixed by imposing that the number of descendants is equal to $\nu_i$ on average.
For instance, if $\nu_i$ is equal to $13.2$, then $13$ is chosen with probability $0.8$, and $14$ with probability $0.2$. 
Generically, $\lfloor \nu_i \rfloor$
and $\lfloor \nu_i \rfloor + 1$ are chosen with probability $1+ \lfloor \nu_i  \rfloor - \nu_i$ and $ \nu_i - \lfloor \nu_i  \rfloor $, respectively.
We note that we need to consider these two possibilities for all indices $i$.
We thus arrive at the following matrix:
\begin{equation}
\begin{split}
\mathcal C_{\tilde n, n} \equiv  &  \sum_{x_0 = 0}^1  \sum_{x_1 = 0}^1  \sum_{x_2 = 0}^1 \cdots  \sum_{x_{M-1} = 0}^1  \prod_{i=0}^{M-1}   \\
& \times \delta _{\tilde n_i, \lfloor \nu_i  \rfloor  + x_i} \left [ \left ( \nu_i - \lfloor \nu_i  \rfloor \right ) x_i +  \left (  1+ \lfloor \nu_i  \rfloor - \nu_i \right ) (1 - x_i) \right ].
\label{eq:transitionMatrixForCopyDef}
\end{split}
\end{equation}

Now, we expand $\mathcal C$ at small $dt$ and we keep only the terms proportional to $\mathcal O(1)$ and $\mathcal O(dt)$, which do not vanish in the continuous-time limit. For this purpose, we expand $\nu_i$ as
\begin{equation}
\nu_i = n_i \bigg [ 1 + s \ dt \Big ( \sum_j \frac{n_jb_j}{N_c} - b_i \Big ) \bigg ] + \mathcal O(dt^2),
\end{equation}
where we have used $\sum_i n_i =N_c$.
This expression indicates that $\lfloor \nu_i \rfloor$ is determined depending on the sign of $\sum_j n_jb_j / N_c - b_i $, where we assumed $s>0$ for simplicity without loss of generality (because when $s<0$, we can always re-define $-b$ as $b$ to make $s$ to be positive).  By denoting this factor by $\alpha_i$, i.e.,
\begin{equation}
\alpha_i(n) \equiv \sum_j \frac{n_jb_j}{  N_c} - b_i,  
\label{def:eqalphai}
\end{equation}
we thus define the following state-space $\Omega^{(\pm)}(n)$:
\begin{equation}
 \Omega^{(\pm)}(n)   = \left \{ \  i \  \big | \ 0\leq i < M  \ {\rm and} \    \pm \alpha_i(n)  > 0 \right  \}.
\end{equation}
From this definition, for sufficiently small $dt$, we obtain
%
$\left \lfloor \nu_i \right \rfloor = n_i$
for $i  \in \Omega^{(+)}$, and
$\left \lfloor \nu_i \right \rfloor = n_i - 1$
for $i  \in \Omega^{(-)}$. Substituting these results into Eq.~\eqref{eq:transitionMatrixForCopyDef} and expanding in $dt$, we obtain (denoting here and thereafter $\alpha_i=\alpha_i(n))$:
\begin{equation}
\mathcal C_{\tilde n, n}  =  \delta_{\tilde n, n} + s \ dt    \sum_{i=0}^{M-1}   n_i |\alpha_i|   \big [  \delta _{\tilde n_i,n_i + \alpha_i/|\alpha_i|}  \  \delta^{i}_{\tilde n, n}  - \delta_{\tilde n, n}   \big ]   \\
  + \mathcal O(dt^2),
\label{eq:Cdt0limit}
\end{equation}
where $\delta^{i}_{\tilde n, n}$ is a Kronecker delta for the indices except for $i$: $\delta^{i}_{\tilde n, n} = \prod_{k\neq i}\delta_{\tilde n_k,n_k}$. One can easily check that this matrix preserves probability: $\sum_{\tilde n} \mathcal C_{\tilde n, n} = 1$.

\subsubsection[Maintaining Part: $\mathcal K$]{Derivation of the Maintaining Part: $\mathcal K$}

As directly checked, the operator $\mathcal T$ preserves the total population $\sum_i n_i$.
However, the operator representing the cloning $\mathcal C$, does not.
In our birth-death implementation, 
this property originates from the rounding process $\lfloor \nu_i  \rfloor$ in the definition of $\mathcal C$:
even though $\nu_i$ itself satisfies $\sum_i \nu_i = N_c$, because of the rounding  process of $\nu_i$, the number of clones after multiplying by $\mathcal C$ (that is designed to be proportional to $\nu_i$ on average) can change.
There are several ways to keep the number  $N_c$ of copies constant without biasing the distribution of visited configurations. One of them is to choose randomly and \emph{uniformly} $\delta N_c$ clones from the ensemble, where $\delta N_c$ is equal to the number of excess (resp.~lacking) clones with respect to $N_c$, and to eliminate (resp.~multiply) them.

In our birth-death description, we implement this procedure as follows. We denote by $\mathcal K$ the transition matrix maintaining the total number of clones to be the constant~$N_c$. 
We now use a continuous-time asymptotics $dt\rightarrow 0$. In this limit, from the expression of the transition matrix elements Eq.~\eqref{eq:Cdt0limit}, we find that at each cloning step the number of copies of the cloned configuration varies by $\pm 1$ at most. Hence, the total number of clones after multiplying by $\mathcal C$, $\sum_{i} n_i$, satisfies the following inequality
\begin{equation}
N_c - 1 \leq  \sum_{i}  n_i  \leq  N_c +1.
\end{equation}
Among the configurations $n$ that satisfy this inequality, there are three possibilities, which are $\sum_{i}n_i=N_c$ and $\sum_{i}n_i=N_c\pm 1$. 
If $n$ satisfies $\sum_{i}n_i=N_c$, we do not need to adjust $n$, 
while if $n$  satisfies
$\sum_{i}n_i=N_c + 1$ (resp.~$\sum_{i}n_i=N_c - 1$), we eliminate (resp.~multiply) a clone chosen randomly and uniformly. Note that, in our formulation, we do not distinguish the clones taking the same state. This means that we can choose one of the occupations $n_i$ of a state $i$ according to a probability proportional to the number of copies $n_i$ in this state. In other words, the probability to choose the state $i$ and to copy or to eliminate a clone from this state is proportional to $ n_i  / \sum_{j=0}^{M-1}n_j $. Therefore, we obtain the expression of the matrix $\mathcal K$ as
\begin{equation}
\mathcal K_{\tilde n,n}  =  \delta_{\sum_i\! n_i,N_c}  \delta_{\tilde n, n} 
 + \sum_{k=-1,1} \delta_{\sum_i\!n_i, N_c + k}  \sum_{i=0}^{M-1}  \delta_{\tilde n_i, n_i - k} \  \delta^{i}_{\tilde n, n} \  \frac{n_i }{N_c + k}  
\label{eq:Kexpression}
\end{equation}
for $\tilde n$ that satisfies $\sum_i \tilde n_i = N_c$, and
$\mathcal K_{\tilde n,n} = 0$ otherwise.

\subsubsection[Total Transition: $\mathcal K \mathcal C \mathcal T$]{Total Transition: $\mathcal K \mathcal C \mathcal T$ }

From the obtained expressions 
we calculate the matrix $\mathcal K \mathcal C \mathcal T$, which describes the total transition of the population dynamics~(\ref{eq:time_Evolution})
\begin{equation}
  (\mathcal K \mathcal C \mathcal T)_{\tilde n, n} = \delta_{\tilde n, n} \\
   +  dt    \sum_{i=0}^{M-1} n_i  \sum_{j=0, (j\neq i)}^{M-1} \left [ w(i \rightarrow  j) + s\, \tilde w_{n}(i \rightarrow  j) \right ] \times \left [  \delta _{\tilde n_i,n_i-1}   \delta _{\tilde n_j,n_j+1} \ \delta ^{i,j}_{\tilde n,n} \  -  \delta _{\tilde n,n}    \right ],
\label{eq:KCTexpression}
\end{equation}
where the population-dependent transition rate $\tilde w_n(i\rightarrow j)$ is given as
\begin{equation}
\tilde w_n (i \rightarrow  j) = \frac{n_j  }{N_c} \left [ \alpha_j  \delta_{j\in \Omega^{(+)}} \frac{N_c}{N_c + 1} - \alpha_i \delta_{i\in \Omega^{(-)}}  \frac{N_c}{N_c - 1} \right ].
\end{equation}
The comparison of the expression~\eqref{eq:KCTexpression} with the original part $\mathcal T$ provides an insight into the result obtained. The jump ratio
$w(i\rightarrow j)$ in the original dynamics is replaced by $w(i\rightarrow j) + s\, \tilde w_n(i\rightarrow j)$ in the population dynamics algorithm. We note that this transition rate depends on the population $n$, meaning that we cannot get a closed equation for this modified dynamics at the level of the states $i$ in general.
We finally remark that the transition matrix $\sigma(n\rightarrow \tilde n)$ for the continuous-time limit is directly derived from Eq.~\eqref{eq:KCTexpression} as
\begin{equation}
 \sigma(n\rightarrow \tilde n)  =   \sum_{i=0}^{M-1} n_i  \sum_{j=0, (j\neq i)}^{M-1} \left [ w(i \rightarrow  j) + s \tilde w_{n}(i \rightarrow  j) \right ] \\ 
 \times \left [  \delta _{\tilde n_i,n_i-1}   \delta _{\tilde n_j,n_j+1} \ \delta ^{i,j}_{\tilde n,n}    \right ].
\label{eq:sigma}
\end{equation}

\subsection[Large Deviation Results in the $N_c\rightarrow \infty$ Asymptotics]{Derivation of the Large Deviation Results in the $N_c\rightarrow \infty$ Asymptotics}
\label{Subsec:DerivationLargeDeviationEstimator}

In this subsection, we study the $N_c \rightarrow \infty$ limit for the transition matrix of rates $\sigma(n\rightarrow \tilde n)$, and derive the validity of the population dynamics algorithm. 

\subsubsection{Estimator of the Large Deviation Function}
\label{subsubsec:estimator}
One of the ideal implementations of the population dynamics algorithm is as follows: We make copies of each clone at the end of simulation, where the number of copies for each realization is equal to the exponential weight $e^{-s\tau B(\tau)}$ in Eq.~(\ref{eq:defpsi}) (so that we can discuss an ensemble with this exponential weight without multiplying the probability by it, as described in the Introduction for the continuous-time case). 
In this implementation, the number of clones grows (or decays) exponentially proportionally as $\left \langle e^{-s\tau B(\tau)} \right \rangle$ by definition. In real implementations of the algorithm, however, since taking care of an exponentially large or small number of clones  can cause numerical problems, one rather keeps the total number of clones to a constant $N_c$ at every time step, as seen in Eq~\eqref{eq:nufac}. 
Within this implementation, we reconstruct the exponential change of the total number of clones as follows:
%
We compute the average of the cloning factor 
at each cloning step, and
we store the product of these ratios along the cloning steps. At final time, this product gives the empirical estimation of total (unnormalized) population during the whole duration of the simulation~\cite{hidalgo_discreteness_2016}, i.e., an estimator of  $\left \langle e^{-s\tau B(\tau)} \right \rangle$. One thus estimates the CGF $\psi(s)$ given in Eq.~(\ref{eq:defpsi})~\cite{giardina_direct_2006,lecomte_numerical_2007,
tailleur_probing_2007,tailleur_simulation_2009,giardina_simulating_2011,
1751-8121-46-25-254002_2013} as the logarithm of this reconstructed population, divided by the total time.

In this formulation, the average cloning ratio is given as $\sum _i n_i e^{- s\ dt\ b_i}/N_c$, and thus the multiplication over whole time interval reads $ \prod_{t=0}^{\tau/dt}    
 \{ n_i(t) e^{- s\ dt\ b_i}/N_c\}$.
Given that we empirically assume that the CGF estimator converges to $\psi(s)$ in the $N_c, \tau \rightarrow \infty$ limit, 
the following equality is expected to hold in probability 1:
\begin{equation}
\psi(s)  
\stackrel{?}{=}
 \lim_{N_c \rightarrow \infty} \lim_{\tau \rightarrow \infty} \frac{1}{\tau} \sum_{t=0}^{\tau/dt}  \ \log   
\sum _i \frac{n_i(t) e^{- s\ dt\ b_i}}{N_c}
 + O(dt). 
\label{eq:expected1}
\end{equation}
Since the dynamics of the population $n$ is described by a Markov process, ergodicity is satisfied, i.e., time averages can be replaced by the expected value with respect to the stationary distribution function which applied to the right-hand side of Eq~\eqref{eq:expected1}, we obtain
\begin{equation}
 \lim_{\tau \rightarrow \infty}  \frac{1}{\tau} \sum_{t=0}^{\tau/dt} 
\log \sum _i \frac{n_i(t) e^{- s dt b_i}}{N_c}  
 = \frac{1}{dt}
\sum_{n}P_{n}^{\rm st} \ \log   
\sum _i \frac{n_i e^{-s\ dt\ b_i}}{N_c} 
+ \mathcal O(dt), 
\end{equation}
where $P_{n}^{\rm st}$ is the stationary distribution function of the population $n$ in the $dt \rightarrow 0$ limit, (namely, $P_{n}^{\rm st}$ is  the stationary distribution of the dynamics of transition rates $\sigma(n\rightarrow \tilde n)$). 
By expanding this right-hand side with respect to $dt$, we rewrite the expected equality~\eqref{eq:expected1} as
\begin{equation}
\psi(s)  \stackrel{?}{=} - s \lim_{N_c \rightarrow \infty} \sum_{n}P_{n}^{\rm st} \sum_{i}\frac{n_i b_i}{N_c} + O(dt).
\label{eq:expected2}
\end{equation}
where we used that $\sum_i n_i=N_c$ is a conserved quantity.
Below we demonstrate that this latter equality~\eqref{eq:expected2} is satisfied by analyzing the stationary distribution function $P_n^{\rm st}$.

\subsubsection[Connection between the Distribution Functions of the Population and of the Original System]{Connection between the Distribution Functions of the Population \\ and of the Original System}

From the definition of the stationary distribution function $P_n^{\rm st}$, we have
\begin{equation}
\sum_{\tilde n} P_{\tilde n}^{\rm st} \sigma(\tilde n\rightarrow  n) - \sum_{\tilde n}P_n^{\rm st} \sigma(n\rightarrow \tilde n) = 0,
\label{eq:stationaryMaster}
\end{equation}
which is a stationary Master equation. 
In this equation, we use the explicit expression of $\sigma$ shown in Eq.~\eqref{eq:sigma}. By
denoting by $n^{j \rightarrow i}$ the configuration where one clone in the state $j$ moves to the state $i$: $n^{j \rightarrow i} \equiv (n_0, n_1, \cdots, n_i + 1, \cdots, n_j -1, \cdots, n_{M-1})$, the stationary Master equation~\eqref{eq:stationaryMaster} is rewritten as
\begin{equation}
\sum_{i,j (i\neq j)}\left [  f_{i \rightarrow j}(n^{j\rightarrow i}) - f_{i \rightarrow j}(n) \right ] = 0,
\label{eq:derivation1}
\end{equation}
where we defined $f_{i \rightarrow j}(n)$ as
\begin{equation}
f_{i \rightarrow j}(n) = P_{n}^{\rm st} n_i \left [w(i \rightarrow j) + s \tilde w_n(i \rightarrow j) \right ].
\end{equation}
Now we multiply expression~\eqref{eq:derivation1} by $n_k$ ($k$ is arbitrary from $k=0,1,2,\cdots, M-1$), and sum it over all configurations $n$:
\begin{equation}
\sum_{n}\sum_{i,j (i\neq j)} n_k \left [  f_{i \rightarrow j}(n^{j\rightarrow i}) - f_{i \rightarrow j}(n) \right ] = 0.
\label{eq:derivation2}
\end{equation}
We can change the dummy summation variable $n$ in the first term to $n^{i\rightarrow j}$, which leads to
$\sum_{n}\sum_{i,j (i\neq j)} (n^{i\rightarrow j})_k    f_{i \rightarrow j}(n) $. Since the second term has almost the same expression as the first one except for the factor $n_k$, the sum in Eq.~\eqref{eq:derivation2} over the indices $(i,j)$, where none of $i$ nor $j$ is equal to $k$, becomes 0. The remaining term in Eq.~\eqref{eq:derivation2} is thus
\begin{equation}
0  = \sum_{n} \sum_{j(j\neq k)} \left ( (n^{k\rightarrow j})_k -  n_k\right ) f_{k \rightarrow j}(n) 
 + \sum_{n} \sum_{i(i\neq k)} \left ( (n^{i\rightarrow k})_k -  n_k\right ) f_{i \rightarrow k}(n).
\end{equation}
Using the definition of $n^{i \rightarrow j}$ in this equation, we arrive at
\begin{equation}
0  = \sum_{n} \bigg [  \sum_{i(i\neq k)}  f_{i \rightarrow k}(n) - f_{k \rightarrow i}(n)  \bigg].
\label{eq:derivation3}
\end{equation}
This equation~\eqref{eq:derivation3} connects the stationary property of the population dynamics (described by the occupation states $n$) and the one in the original system (described by the states $i$). 


The easiest case where we can see this connection is when $s=0$. By defining the empirical occupation probability of the original system as $p_i \equiv \sum_{n}P_{n}^{\rm st}n_i/N_c$, Eq.~\eqref{eq:derivation3} leads to the following (stationary) master equation for $w(i\rightarrow j)$:
\begin{equation}
0 = \sum _{j}p_j w(j\rightarrow i)  - \sum_{j}p_i w(i\rightarrow j)  \quad\textnormal{(for $s=0$)}
\label{eq:mastereqpiemp}
\end{equation}
This is valid for any $N_c$, meaning that, for original Monte-Carlo simulations in $s=0$, the empirical probability~$p_i$ is exactly equal to the steady-state probability, as being the unique solution of Eq.~\eqref{eq:mastereqpiemp}.
It means that there are no systematic errors in the evaluation of $p_i $. However, in the generic case $s\neq 0$, this property is not satisfied. One thus needs to understand the $N_c \rightarrow \infty$ limit to connect the population dynamics result with the large deviation property of the original system. 

\subsubsection{Justification of the Convergence of the Large Deviation Estimator as Population Size becomes Large}
We define a scaled variable $x_i$ as $n_i/N_c$. While keeping this occupation fractions $x_i$ to be $\mathcal O(1)$, 
we take the $N_c \rightarrow \infty$ limit 
in Eq.~(\ref{eq:derivation3}), which leads to 
\begin{equation}
\begin{split}
0 =  \sum_n P_n^{\rm st} & \Bigg [ 
 \sum _{j}x_j w(j\rightarrow i)  - \sum_{j}x_i w(i\rightarrow j)  \\
 & - s ~ x_i \left ( b_i  -  \sum_k x_k b_ k  \right )    \Bigg ]
 +  \mathcal O(1/N_c).
\end{split}
\label{eq:derivation4}
\end{equation}
Inspired by this expression, we define a matrix $L_{i,j}^s$ as
\begin{equation}
L_{i,j}^s = w(j\rightarrow i) - \delta _{i,j} \left (\sum_{k} w(i \rightarrow k)  + s ~ b_i\right ),
\end{equation}
and a correlation function between $x_i$ and $x_j$ as
\begin{equation}
c_{i,j} = \sum_{n}x_ix_j P_{n}^{\rm st} - p_i p_j,
\end{equation}
(where we recall $p_i \equiv \sum_n x_i P_{n}^{\rm st}$). 
From these definitions, Eq.\eqref{eq:derivation4} is rewritten as
\begin{equation}
\sum_{j}p_j L_{i,j}^{s}  = - s p_i \sum_{k}  p_kb_k - s \sum_{k}c_{i,k} b_k  + \mathcal O \left ( \frac{1}{N_c} \right ).
\label{eq:generaleigenvalue}
\end{equation}
Since $x_i\equiv n_i/N_c$ is an averaged quantity (an arithmetic mean) with respect to the total number of clones, we can safely assume that the correlation $c_{i,j}$ becomes 0 in $N_c \rightarrow \infty$ limit:
\begin{equation}
\lim_{N_c\rightarrow \infty}c_{i,k}=0.
\label{eq:Assumption}
\end{equation}
For a more detailed discussion of why this is valid, see the description after Eq.~\eqref{eq:systemsizeexpansion}.
Thus, by defining $p_i^{\infty}\equiv \lim_{N_c \rightarrow \infty} p_i$, we obtain
\begin{equation}
\sum_{j}p_j^{\infty} L_{i,j}^{s}  = - s p^{\infty}_i \sum_{k}  p^{\infty}_kb_k.
\end{equation}
From the Perron-Frobenius theory, the positive eigenvector of the matrix $L_{i,j}^{s}$ is unique  and corresponds to its eigenvector of largest eigenvalue (in real part). This means that $-s \sum_{k}p_{k}^{\infty } b_k$ is the largest eigenvalue of the matrix $L_{i,j}^{s}$. Finally, by recalling that
the largest eigenvalue of this matrix $L_{i,j}^{s}$ is equal to the generating function $\psi(s)$ (see Ref.~\cite{garrahan_first-order_2009} for example), we have justified that the CGF estimator~\eqref{eq:expected2} is valid in the large-$N_c$ limit. This is equivalent to what we saw in Sec.~\ref{sec:largest} for the continuous-time version.

\subsection[Systematic Errors due to Finite $N_c$: Convergence Speed of the Large Deviation Estimator as $N_c\to\infty$]{Systematic Errors due to Finite $N_c$: Convergence Speed of the Large Deviation Estimator as $N_c\to\infty$}
\label{subsec:systemsizeexpansion}

In the introduction of this chapter, 
we defined the \textbf{systematic errors} as the deviations of the large deviation estimator from the correct value due to a finite number of clones~$N_c$. 
From Eq.~(\ref{eq:expected2}), we quantitatively define this systematic error $\epsilon_{\rm sys} $ as
\begin{equation}
\epsilon_{\rm sys} \equiv \left |  \psi(s) + s  \sum_i p_i b_i \right |.
\label{eq:systematicdef}
\end{equation}
From a simple argument based on a system size expansion, we show below that this $\epsilon_{\rm sys}$ is of order $\mathcal O(1/N_c)$. 
We first show that one can perform a system size expansion (as for example in Ref.~\cite{tagkey2007iii}) for the population dynamics. In Eq.~(\ref{eq:derivation1}), by recalling the definition of the vector $x$ as $x=n/N_c$, and by denoting $\tilde P^{\rm st}(x) = P^{\rm st}_{x N_c} $, we obtain
\begin{equation}
\begin{split}
0 = & \sum_{i,j (i\neq j)}\sum_{r=1}^{\infty}\frac{1}{r !} \frac{1}{N_{c}^r} \left (\frac{ \partial}{\partial x_i} - \frac{\partial}{\partial x_j} \right )^r x_i  \tilde P^{\rm st}(x) \\
 & \times \left [ w(i\rightarrow j) + s \tilde w_n(i\rightarrow j)|_{n=xN_c}  \right ].
\label{eq:systemsizeexpansion}
\end{split}
\end{equation}
This indicates that the stochastic process governing the evolution of $x$ becomes deterministic in the $N_c \rightarrow \infty$ limit.  The deterministic trajectory for $x$ is governed by a differential equation derived from the sole term  $r=1$ in the expansion~(\ref{eq:systemsizeexpansion}) (see~Sec.~3.5.3 in ref.~\cite{opac-b1079113} for the detail of how to derive this property). Thus if $x$ converges to a fixed point as $N_c$ increases, which is normally observed in implementations of cloning algorithms, the assumption~\eqref{eq:Assumption} is satisfied. 

From the expression of $\epsilon_{\rm sys}$, we see that the dependence in~$N_c$ comes solely from $p_i$, which can be calculated from the first order correction of $P^{\rm st}_n$ (at large $N_c$). The equation to determine $P^{\rm st}_n$ is the stationary Master equation (\ref{eq:stationaryMaster}) or equivalently, the system-size expansion formula (\ref{eq:systemsizeexpansion}). We expand the  jump ratio  $w(i\rightarrow j) + s \tilde w_n(i\rightarrow j)$ in Eq.~(\ref{eq:systemsizeexpansion}) with respect to $1/N_c$ as:
\begin{equation}
w(i\rightarrow j) + s \tilde w_n(i\rightarrow j)  
 = w(i\rightarrow j) + s \tilde w_x^\infty (i \rightarrow j) + \frac{s}{N_c} \delta w_x(i\rightarrow j) + \mathcal O(1/N_c^2),
\label{eq:wexpansion}
\end{equation}
where $\tilde w_x^\infty (i \rightarrow j)$ and $\delta w_x (i \rightarrow j)$ are defined as
\begin{equation}
\tilde w_x^\infty (i \rightarrow j)  = x_j \left [\alpha_j \delta_{j\in \Omega^{(+)}} - \alpha_i \delta_{i\in \Omega^{(-)}}  \right ]
\end{equation}
and
\begin{equation}
\delta w_x (i \rightarrow j)  = - x_j \left [\alpha_j \delta_{j\in \Omega^{(+)}} + \alpha_i \delta_{i\in \Omega^{(-)}}  \right ].
\end{equation}
By substituting Eq.~\eqref{eq:wexpansion} into the system-size expansion formula~\eqref{eq:systemsizeexpansion} and performing a perturbation expansion, we find that a first-order correction of $p$ is naturally of order $\mathcal O(1/N_c)$, i.e., $\epsilon_{\rm sys}=\mathcal O(1/N_c)$. For a practical scheme of how to implement this perturbation on a specific example, see Sec.~\ref{subsection:systematicerrors_twostate}.
In chapter~\ref{chap:ContinuousTime}~\cite{partII}, the scaling analysis of the $1/N_c$ correction is shown to hold numerically with the continuous-time cloning algorithm. 
We also show that the $1/N_c$ correction behavior remains in fact valid at finite time, an open question that remains to be investigated analytically.

\subsection[Remarks]{Remarks}
Here, we discuss some remarks on the formulation presented in this section. 

\subsubsection{Relaxing the Condition $dt=\Delta t$}
\label{subsubsec:dtDeltat}
In Sec.~\ref{subsubsec:CloningRatio}, we set the discretization time of the process $dt$ to be equal to the time interval for cloning $\Delta t$, and we took the $dt=\Delta t \rightarrow 0$ limit at the end. We note that the condition $\Delta t = dt $ is not necessary if both limits $\Delta t \rightarrow 0$ and $d t \rightarrow 0$ (with $dt<\Delta t$) are taken at the end. 
This is practically important, because we can use the continuous-time  process to perform the algorithm presented here by setting $dt=0$ first, and $\Delta t \rightarrow 0$ limit afterwards. More precisely, replacing $dt$ by $\Delta t$ in the matrix $\mathcal C$ and $\mathcal K$, we build a new matrix $\mathcal K \mathcal C (\mathcal T^{\Delta t/dt})$.
Taking the $dt \rightarrow 0$ limit in this matrix while keeping $\Delta t$ non-infinitesimal (but small), this matrix represents the population dynamics algorithm of a continuous-time process with a finite cloning time interval $\Delta t$. The arguments presented in this section can  then be applied in the same way, replacing $dt$ by $\Delta t$.
We note that the deviation due to a non-infinitesimal $\Delta t$ should thus appear as $O(\Delta t)$ (see Eq.~(\ref{eq:expected1})  for example).

\subsubsection{A Continuous-Time Cloning Algorithm }
\label{subsubsec:differenceContinuousTime}
The $\Delta t \rightarrow 0$ limit is the key point in the formulation developed in this section (and in this chapter).
Thanks to this limit, upon each cloning step, the total number of clones $\sum_{j=0}^{M-1}n_j$ always varies only by $\pm 1$, which makes the expression of the matrices $\mathcal C$ and $\mathcal K$ simple enough to develop the arguments presented in Secs.~\ref{Subsec:DerivationLargeDeviationEstimator} and~\ref{subsec:systemsizeexpansion}.
Furthermore, during the time interval $\Delta t$ separating  two cloning steps, the configuration is changing at most once. 
The process between cloning steps is thus simple, which allows us to represent the corresponding time-evolution matrix as $\mathcal T$ (by replacing $dt$ by $\Delta t$ as just explained above). 
Generalizing our analytical study to a cloning dynamics in which the limit $\Delta t\to 0$ is not taken is therefore a very challenging task, which is out of the scope of this chapter. 

However, interestingly, in chapter~\ref{chap:ContinuousTime}~\cite{partII} we observe numerically that our predictions for the finite-time and finite-population scalings are still valid in a different version of algorithm
for which $\sum_{j=0}^{M-1}n_j$ can vary by an arbitrary amount --~supporting the hypothesis that the analytical arguments that we present here could be extended to more general algorithms.
More precisely, in chapter~\ref{chap:ContinuousTime}~\cite{partII}, we use a continuous-time version of the algorithm~\cite{lecomte_numerical_2007} 
to study numerically an observable of `type~A'~\cite{garrahan_first-order_2009} (See Sec.~\ref{sec:LDobservables}).
This version of the algorithm differs from that considered in this chapter, in the sense that 
the cloning steps are separated by non-fixed non-infinitesimal time intervals. These time intervals are distributed exponentially, in contrast to the fixed ones taken here where $\Delta t$ is a constant.
%
This results in an important difference: the effective interaction between copies due to the cloning/pruning procedure is unbounded (it can \emph{a~priori} affect any proportion of the population), while in the algorithm of the present here, this effective interaction is restricted to a maximum of one cloning/pruning event in the $\Delta t \to 0$ limit.
We stress that the $dt\to 0$ limit of the cloning algorithm studied in here with a fixed $\Delta t$ \textbf{does not yield the continuous-time cloning algorithm}, stressing that these two versions of the population dynamics present essential differences.
%

\section[Stochastic Errors: Large Deviations of the Population Dynamics]{\large{Stochastic Errors: Large Deviations of the Population Dynamics}}
\label{Sec:largedeviation_largedeviation}

In the previous section, we formulated the population dynamics algorithm as a birth-death process and evaluated the systematic errors (which are the deviation of the large deviation estimator from the correct value) due to a finite number of clones (Table~\ref{Table:Numericalerrors}). In this section, we focus on \textbf{stochastic errors} corresponding to the run-to-run fluctuations of the large deviation estimator within the algorithm, at fixed $N_c$.

In order to study stochastic errors, we formulate the large deviation principle of the large deviation estimator. In the population dynamics algorithm, the CGF estimator is the time-average of the average cloning ratio of the population (see Sec.~\ref{subsubsec:estimator}):
\begin{equation}
\psi_{N_c,\tau}(s) \equiv - s \frac{1}{\tau} \int_{0}^{\tau} dt \sum_{i=0}^{M-1}   \frac{n_i(t) b_i}{N_c}.
\label{eq:estimetor2}
\end{equation}
As $\tau$ increases, this quantity converges to the expected value (which depends on $N_c$) with probability 1. However whenever we consider a finite $\tau$, dynamical fluctuations are present, and there is a probability that this estimator deviates from its expected value. Since the population dynamics in the occupation states $n$ is described by a Markov process, the probability of these deviations are themselves described by a large deviation principle~\eqref{eq:ldpO}~\cite{touchette_large_2009,opac-b1093895}: By denoting by $\rm Prob (\psi)$ the probability of $\psi_{N_c,\tau}(s)$, one has:
\begin{equation}
{\rm Prob} (\psi) \sim \exp \left (  - \tau I_{N_c,s}(\psi) \right ),
\end{equation}
where $I_{N_c,s}(\psi)$ is a large deviation ``rate function'' (of the large deviation estimator). 
To study these large deviations, we can apply a standard technique using a biased evolution operator for our population dynamics: For a given Markov system, to calculate large deviations of additive quantities such as Eq.~\eqref{eq:estimetor2}, one biases the time-evolution matrix with an exponential factor~\cite{opac-b1093895}.  
Specifically, by defining the following matrix
\begin{equation}
L^{h}_{\tilde n,n} = \sigma(n \rightarrow \tilde n) - \delta_{\tilde n, n} \sum_{n^{\prime}}\sigma(n \rightarrow n^{\prime}) -  h s \sum_{i=0}^{M-1}   \frac{n_i b_i}{N_c}.
\label{eq:DefinitionLh}
\end{equation}
and by denoting the largest eigenvalue of this matrix $G(h,s)$ (corresponding, as a function of $h$, to a scaled cumulant generating function for the observable~\eqref{eq:estimetor2}), the large deviation function $I_{N_c,s}(\psi)$ is obtained as the Legendre transform $\sup_{h} \left [h \psi -  G(h,s) \right ]$.
In chapter~\ref{chap:ContinuousTime}~\cite{partII}, we show that a quadratic approximation of the rate function~$I_{N_c,s}(\psi)$ (i.e., a Gaussian approximation) can be estimated directly from the cloning algorithm.

We consider the scaling properties of $I_{N_c,s}$ in the large-$N_c$ limit. For this, we define a scaled variable $\tilde h \equiv h /N_c$ and a scaled function $\tilde G(\tilde h,s) \equiv G(\tilde h N_c,s)/N_c$. If this scaled function $\tilde G(\tilde h,s) \equiv G(\tilde h N_c,s)/N_c$ is well-defined in the $N_c \rightarrow \infty$ limit (which is natural as checked in the next paragraph), then we can derive that $I_{N_c,s}$ has the following scaling:
\begin{equation}
I_{N_c,s}(\psi) = N_c I_{s}(\psi) + o(N_c)
\label{eq:scalingNcINcs}
\end{equation}
or equivalently,
\begin{equation}
{\rm Prob}(\psi) \sim e^{-\tau N_c  I_{s}(\psi)},
\label{eq:largedeviation_largedeviation}
\end{equation}
where $I_{s}(\psi) = \max_{\tilde h} \left [ \tilde h \psi -  \tilde G(\tilde h,s)   \right ]$.
The scaling form~\eqref{eq:scalingNcINcs} is validated numerically in chapter~\ref{chap:ContinuousTime}~\cite{partII}. From this large deviation principle, we can see that the stochastic errors of the large deviation estimator is of $\mathcal O(1/(N_c \tau))$ as shown in Table~\ref{Table:Numericalerrors}.

In the largest eigenvalue problem for the transition matrix~\eqref{eq:DefinitionLh}, by performing a system size expansion (see Sec.~\ref{subsec:systemsizeexpansion}), we obtain
\begin{equation}
\begin{split}
\tilde G(\tilde h,s) = & \sum_{i,j (i\neq j)}  \left (\frac{ \partial}{\partial x_i} - \frac{\partial}{\partial x_j} \right ) x_i  q(x) \times \left [ w(i\rightarrow j) + s \tilde w_x^{\infty}(i\rightarrow j)  \right ] \\
& \quad \quad - \frac{\tilde h}{s} s \sum_i x_i b_i q(x) + \mathcal O(1/N_c),
\label{eq:systemsizeexpansion2}
\end{split}
\end{equation}
where $q(x)$ is the right-eigenvector associated to the largest eigenvalue of $L_{\tilde n,n}^h$ (represented as a function of $x\equiv n/N_c$). 
The first order of the right-hand side is of order $\mathcal O(N_c^0)$, so that $\tilde G(\tilde h,s)$ is also of order $\mathcal O(N_c^0)$ in $N_c \rightarrow \infty$. (For an analytical example of the function $\tilde G(\tilde h, s)$, see Sec.~\ref{subsection:largedeviationsInPopTwoState}).
\begin{table}[t]
\begin{center}          
          \begin{tabular}{c||c}
          \hline
                     	  			& Magnitude of errors				 \\  \hline \hline
    Systematic errors	 	 & 		$\mathcal O(1/N_c)$   \\  \hline
  Numerical errors			&              $\mathcal O(1/(\tau N_c))$  \\ \hline
          \end{tabular}  
          \caption{\label{Table:Numericalerrors} Magnitudes of the numerical errors}
        \end{center}
 \end{table}

\section[Example: A Simple Two-State Model]{Example: A Simple Two-State Model}
\label{Section:Demonstrations}

In order to illustrate the formulation that we developed in the previous sections, here we consider a simple two state model. 
In this system, the dimension of the state $i$ is two ($M=2$) and the transition rates $w( i \rightarrow j)$ are
\begin{eqnarray}
w(0  \rightarrow 1) =& c,\\
w(1  \rightarrow 0) =& d
\end{eqnarray}
%
with $c,d>0$	 and $w(i  \rightarrow  i) = 0$.
In this model, the quantity
$\alpha_{i}$ defined in Eq.~\eqref{def:eqalphai} becomes
\begin{equation}
\alpha_{i} = \delta_{i,0} \frac{n_1}{N_c}\left (b_1 - b_0 \right ) +  \delta_{i,1} \frac{n_0}{N_c} (b_0 - b_1).
\end{equation}
Hereafter, we assume that $b_1>b_0$ without loss of generality. From this, the space $\Omega^{(\pm)}$ is determined as $\Omega^{(+)}=\{ 0 \}$ and $\Omega^{(-)}=\{ 1 \}$, which leads to the jump ratio $\tilde w_{n}(i \rightarrow j)$ as
\begin{equation}
\tilde w_{n}(i \rightarrow j) =\delta_{i,1}\delta_{j,0}  \frac{n_0}{N_c}(b_1 - b_0) \left [\frac{n_1}{N_c+1} + \frac{n_0}{N_c -1} \right ].
\end{equation}
Finally, from the conservation of the total population: $n_0+n_1=N_c$, we find that the state of the population $n$ can be
uniquely determined by specifying only the variable $n_0$.
Thus the transition rate for the population dynamics is a function of $n_0$ (and $\tilde n_0$), $\sigma (n_0 \rightarrow \tilde n_0)$, which is derived as
\begin{equation}
\begin{split}
 \sigma(n_0\rightarrow \tilde n_0) = \delta_{\tilde n_0, n_0 + 1 } \bigg [ (N_c - n_0) d 
&+ k(n_0,N_c - n_0) \\
&\times \Big( \frac{n_0}{N_c -1} + \frac{N_c - n_0}{N_c + 1} \Big) \bigg ] + \delta_{\tilde n_0, n_0 - 1 } \ n_0 \,c,
\end{split}
\end{equation}
where we have defined
\begin{equation}
k(n_0,n_1) = \frac{n_0 n_1}{N_c} s \left [b_1 - b_0 \right ].
\end{equation}

\subsection[Systematic Errors]{Systematic Errors}
\label{subsection:systematicerrors_twostate}
In order to evaluate the systematic errors (see Sec.~\ref{subsec:systemsizeexpansion}), we consider the distribution function $P_{n}^{\rm st}$. 
Since the system is described by a one dimensional variable $n_0$ restricted to $0\leq n_0 \leq N_c$, the transition rates $\sigma(n_0 \rightarrow \tilde n_0)$ satisfy the detailed balance condition:
\begin{equation}
P_{n_0}^{\rm st}  \sigma(n_0 \rightarrow n_0 + 1)  = P_{n_0+1}^{\rm st} \sigma(n_0 + 1 \rightarrow n_0).
\end{equation}
We can solve this equation exactly, but to illustrate the large-$N_c$ limit, it is in fact sufficient to study the solution in an expansion $1/N_c\ll 1$. The result is
\begin{equation}
P^{\rm st}_{x N_c} = C \exp \left [ - N_c I_{\rm conf}(x) + \delta I(x) + \mathcal O(1/N_c)  \right ]
\end{equation}
with  $x\equiv n_0 / N_c$ and explicitly
\begin{equation}
\begin{split}
 I_{\rm conf}(x) = x &+ \log(1-x) - \frac{d \log \left [d+(b_1-b_0) s x\right ]}{(b_1-b_0) s} \\
&- x \log \left [  \frac{1}{cx}(1-x)\left (d + (b_1 - b_0) s x \right ) \right ]
\end{split}
\end{equation}
and
\begin{equation}
\begin{split}
\delta I(x) =  &- x  - \frac{2 d x}{(b_1-b_0) s} + x^2 - \log x  \\
& + \frac{2 d^2 \log \left [ d + (b_1 - b_0)s x\right ] }{(b_1-b_0)^2 s^2} + \frac{d \log \left[ d + (b_1-b_0) s x \right ]}{(b_1-b_0)s}.
\end{split}
\end{equation}

We now determine the value of $x$ that minimizes $ - N_c I_s(x) + \delta I(x)$, which leads to a finite-size correction (i.e., the systematic errors) of the population dynamics estimator. Indeed, denoting this optimal value of $x$ by $x_{N_c}^*$, the large deviation estimator is obtained as
\begin{equation}
\psi_{N_c}(s) = - s \left [ x^*_{N_c} b_0 +(1-x^*_{N_c})b_1  \right ]
\end{equation}
(see Sec.~\ref{subsubsec:estimator}).
From a straightforward calculation based on the expressions $I_{\rm conf}(x)$ and $\delta I(x)$, we obtain
the expression of $x_{N_c}^*$ as
\begin{equation}
x^*_{N_c} = x^{*} + \frac{1}{N_c} \delta x^* + \mathcal O((1/N_c)^2),
\end{equation}
with
\begin{equation}
x^{*} = \frac{-c - d + (b_1-b_0) s }{2 (b_1 - b_0) s} 
 + \frac{ \sqrt{4 d (b_1-b_0) s + \left [-c-d  + (b_1-b_0)s \right ]^2}}{2 (b_1 - b_0) s} 
\label{eq:xstar}
\end{equation}
and
\begin{equation}
\delta x ^*  =  \left (2 d + 2 (b_1 - b_0) s x^* \right )^{-1} \times    \frac{2 c \left [ - d - (b_1 - b_0) s x^*  \left ( 1 +  x^*  - 2 (x^*)^2 \right ) \right ]}{\sqrt{4 d (b_1-b_0)s + [c+d-(b_1-b_0)s ]^2)}  }.
\end{equation}
We thus arrive at
\begin{equation}
 \psi(s)  =  \frac{-c - d - (b_1 + b_0) s }{2 } + \frac{ \sqrt{4 d (b_1-b_0) s + \left [-c-d  + (b_1-b_0)s \right ]^2}}{2 } 
\label{eq:PsiInfinitetime_InfiniteCopy}
\end{equation}
and
\begin{equation}
\epsilon _{\rm sys} =  \frac{1}{N_c}   \frac{1}{\left | d+(b_1-b_0)sx^*\right |} 
 \times \left |  \frac{s c(b_0-b_1) \left ( d + (b_0 - b_1) s (x^* - 1)  x^* (1+2x^*) \right ) }{ \sqrt{4(b_1-b_0)d s + [c+d +(b_0 - b_1)s]^2}} \right |
\end{equation}
(see Eq.~\eqref{eq:systematicdef} for the definition of the systematic error $\epsilon_{\rm sys}$).
We check easily that the expression of $\psi(s)$ is the same as the one obtained from a standard method by solving the largest eigenvalue problem of a biased time-evolution operator (as explained in Sec.~\ref{sec:largest}, and implemented in chapter \ref{chap:Discreteness}~\cite{hidalgo_discreteness_2016}).

\subsection[Stochastic Errors]{Stochastic Errors}
\label{subsection:largedeviationsInPopTwoState}

We now turn our attention to the stochastic errors. The scaled cumulant generating function $N_c \tilde G(\tilde h, s)$ is the largest eigenvalue of a matrix $L_{\tilde n,n}^{h}$~(\ref{eq:DefinitionLh}). We then recall 
a formula to calculate this largest eigenvalue problem from the following variational principle:
\begin{equation}
\begin{split}
\tilde G(\tilde h,s) 
 = &\sup_{\phi>0} \sum_{n}  p_{\rm st}(n_0)\phi(n_0)^2  \Bigg [ \frac{\sigma(n\rightarrow n+1)}{N_c} \left ( \frac{\phi(n_0 + 1)}{\phi(n_0)} - 1 \right ) \\
& + \frac{\sigma(n\rightarrow n-1)}{N_c } \left (   \frac{\phi(n_0 - 1)}{\phi(n_0)} - 1 \right ) - s \tilde h \frac{\sum_{i}n_i b_i}{N_c^2}  \Bigg ].
\end{split}
\end{equation}
(See e.g., the appendix G of~\cite{PhysRevE.84.061113} or~\cite{garrahan_first-order_2009} for the derivation of this variational principle). By following the usual route to solve such equations (see e.g., the Sec.~2.5 of Ref.~\cite{1742-5468-2014-10-P10001}), we obtain
\begin{equation}
\tilde G(\tilde h,s) =   \sup_{x}  \Bigg [ - \left ( \sqrt{(1-x)(d+(b_1 - b_0)s x)} - \sqrt{c x} \right )^2 
  - s \tilde h \left [  x b_0  + (1-x) b_1 \right ]    \Bigg ].
\end{equation}
Thus, $\tilde G(\tilde h,s)$ is well-defined, justifying that the large deviation principle~\eqref{eq:largedeviation_largedeviation} is satisfied. 
Furthermore, by expanding this variational principle with respect to $\tilde h $, we obtain
\begin{equation}
\tilde G(\tilde h,s) =\psi (s)  \tilde h  + \frac{\kappa_s}{2} \tilde h ^2  + \mathcal O(\tilde h^3),
\label{eq:expansionGtilde}
\end{equation}
where $ \psi (s)$ is given in Eq.~\eqref{eq:PsiInfinitetime_InfiniteCopy}, and the variance $\kappa_s$ is given as
\begin{equation}
\begin{split}
\kappa_s &= c + \frac{c s(b_1 - b_0)}{\sqrt{4(b_1-b_0) s d + (c+d+(b_0 - b_1)s)^2}} \\
 &-  \frac{ c(c+d)^2 + c (b_0 - b_1)(c-3d)s }{c^2 + 2 c \left [ d + (b_0 - b_1) s \right ] + (d+(b_1-b_0)s)^2}.
\end{split}
\end{equation}
We note that the expansion~\eqref{eq:expansionGtilde} is equivalent to the following expansion of the large deviation function
 $I_{s}(\psi)$~\eqref{eq:largedeviation_largedeviation} around the expected value $\psi (s)$:
\begin{equation}
I_{s}(\psi)  = \frac{(\psi - \psi(s))^2}{2\kappa_s}  + \mathcal O ((\psi - \psi_s)^3).
\end{equation}
The variance of the obtained large deviation estimator is thus $\kappa_s/(N_c \tau)$.

\subsection[A Different Large Deviation Estimator]{A Different Large Deviation Estimator}
\label{subsec:Different large deviation estimator}
As an application of these exact expressions, we expand the systematic error $\epsilon _{\rm sys}$ and the stochastic error (variance) $\kappa_s$ with respect to $s$. 
A straightforward calculation leads to
\begin{equation}
\begin{split}
\epsilon _{\rm sys} N_c = &\Bigg |   \frac{2c(b_0 - b_1)}{c+d} s \Bigg | + \mathcal O(s^2) 
\label{eq:epsilon_expansion}
\end{split}
\end{equation}
and 
\begin{equation}
\kappa_s = \frac{2 (b_0 - b_1)^2 c d }{(c+d)^3} s^2 + \mathcal O(s^3).
\label{eq:kappa_expansion}
\end{equation}
We thus find that the first-order of the error $\epsilon _{\rm sys}$ scales as $\mathcal O(s)$ at small $s$, but that the variance $\kappa_s$ is of order $\mathcal O(s^2)$. From this scaling, as we explain below,  
one can argue that 
the following large deviation estimator can be better than the standard one for small $s$:
\begin{equation}
\tilde \Psi(s) \equiv   \frac{1}{\tau}   \log  \overline{ \prod_{t=0}^{\tau / dt}  
\sum _i \frac{n_i(t) e^{- s dt b_i}}{N_c} },
\label{Definition_Psi_s}
\end{equation}
where the overline represents the averaging with respect to the realizations of the algorithm. Normally, this realization-average is taken \emph{after} calculating the logarithm, which corresponds to the estimator \eqref{eq:expected1}.
Mathematically, this average (Eq.\eqref{Definition_Psi_s}, before taking the logarithm) corresponds to a bias of the time-evolution matrix $\sigma$ as seen in Eq.~(\ref{eq:DefinitionLh}) for $h=1$. This means that, in the limit $\tau \rightarrow \infty$ with a sufficiently large number of realizations, this averaged value behaves as $\tilde \Psi(s) \sim e^{\tau G(1,s)}$. By combining this result with the expansion (\ref{eq:expansionGtilde}), we thus obtain
\begin{equation}
\begin{split}
\lim_{\tau \rightarrow \infty} \lim_{\substack{ \text{many} \\ \text{realizations}}}
\tilde \Psi(s)
= \psi(s) + \frac{\kappa_s}{2} N_c^{-1} + \mathcal O(N_c^{-2})
\label{eq:G(1s)}
\end{split}
\end{equation}
(recalling $\tilde G = G/N_c$ and $\tilde h = h/N_c$).
When we consider small $s$, by recalling $\epsilon_{\rm sys}N_c=\mathcal O(s)$ and $\kappa_s=\mathcal O(s^2)$, we thus find that the deviations from the correct value are smaller in the estimator $\tilde \Psi(s)$ than in the normal estimator given in Eq.~(\ref{eq:expected1}), which comes as a surprise because in Eq.~\eqref{Definition_Psi_s} the average and the logarithm are inverted with respect to a natural definition of the CGF estimator. 

To use this estimator, we need to discuss the two following points. First, since the scaled cumulant generating function $G(1,s)$ has small fluctuations, one needs a very large number of realizations in order to attain the equality~\eqref{eq:G(1s)}. The difficulty of this measurement is the same level as the one of direct observations of a large deviation function, see for example Ref.~\cite{rohwer_convergence_2015}. However, we stress that this point may not be fatal in this estimator, because we do not need to attain completely this equality, i.e., our aim is the zero-th order coefficient, $\psi(s)$, in Eq.~\eqref{eq:G(1s)}. Second, we have not proved yet the scaling properties with respect to $s$, which are $\epsilon_{\rm sys}N_c=\mathcal O(s)$ and $\kappa_s=\mathcal O(s^2)$, in a general set-up aside from this simple two state model.  
We show in practice in the next chapter that for small values of~$s$, the estimator~\eqref{Definition_Psi_s} is affected by smaller systematic errors, in the numerical study of the creation-annihilation process studied in this section. This alternative way of defining the CGF estimator is studied again for the continuous-time version of the algorithm in Sec~\ref{sec:two-estimators}. 

\section[Discussion]{Discussion}
\label{sec:discussion}

In this chapter, we formulated a birth-death process that describes population dynamics algorithms and evaluated numerically large deviation functions. 
We showed that this birth-death process leads generically to the correct large deviation results in the large limit of the number of clones $N_c \rightarrow \infty$. 
We also showed that the deviation of large deviation estimator from the desired value (which we called systematic errors) is small and proportional with $\mathcal O(N_c^{-1})$.
%
%
%
%
%
%
In the next chapter, we verify and use the $1/\tau$- and $1/N_c$-scalings of the CGF estimator 
in order to 
interpolate its large-$\tau$ and large-$N_c$ asymptotic value from the measured values for finite $\tau$ and $N_c$. 
We demonstrate numerically that the interpolation technique is very efficient, by a direct comparison of the resulting CGF estimation to its analytical value, which can be determined in the studied system.
We also underline that this is done for a different version of the algorithm, a continuous in time population dynamics~\cite{lecomte_numerical_2007}. For a description of their conceptual difference refer to Secs.~\ref{subsubsec:differenceContinuousTime} and~\ref{Discrete-time_algorithm}.
%

\chapter[\quad Finite-Time and Finite-Size Scalings. II. Continuous Time]{Finite-Time and Finite-Size Scalings in the Evaluation of Large-Deviation Functions: \\ II. Numerical Approach \\ in Continuous Time}
\label{chap:ContinuousTime}
\section[Introduction]{\quad Introduction}

In chapter~\ref{chap:DiscreteTime}~\cite{partI}, we performed an analytical study of a {discrete-time} version of the population dynamics algorithm. We derived the finite-$N_c$ and finite-$t$ scalings of the systematic errors of the LDF estimator, showing that these behave as $1/N_c$ and $1/t$ in the large-$N_c$ and large-$t$ asymptotics respectively. 
In principle, knowing the scaling {\it a priori} means that the asymptotic limit of the estimator in the $t \to \infty$ and $N_c \to \infty$ limits may be interpolated from the data at finite $t$ and $N_c$. 
However, whether this idea is actually useful or not is a non-trivial question, as there is always a possibility that onset values of $N_c$- and $t$-scalings are too large to use these scalings. 
In the present chapter, we consider a {continuous-time} version of the population dynamics algorithms~\cite{lecomte_numerical_2007,tailleur_simulation_2009}. We show numerically that one can indeed make use of these scaling properties in order to improve the estimation of CGF, in an application to a system with many-body interactions (a contact process). 
%
We emphasize that the two versions of the algorithm differ on a crucial point which makes that an extension of the analysis developed in chapter~\ref{chap:DiscreteTime}~\cite{partI} cannot be done straightforwardly in order to comprehend the continuous-time case (see section~\ref{Discrete-time_algorithm}). We thus stress that the observation of these scalings themselves is also non-trivial.

This chapter~\ref{chap:ContinuousTime}~\cite{partII} is organized as follows. 
%
In Sec.~\ref{sec:LDFT} we study the behavior of the CGF estimator as a function of the duration of the observation time (for a fixed population~$N_{c}$) and we see how its infinite-time limit can be extracted for the numerical data.  In Sec.~\ref{sec:LDFN} we analyze the behavior of the estimator as we increase the number of clones (for a given final simulation time) and the infinite-size limit of the LDF estimator. Based on these results, we present in Sec.~\ref{sec:LDFinfTinfN} a method which allows us to extract the infinite-time, infinite-size limit of the large deviation function estimator from a finite-time, finite-size scaling analysis. 
%
%
In Sec.~\ref{Discrete-time_algorithm}, we discuss the difficulty of an analytical approach to the continuous-time algorithm. Finally, our conclusions are made in Sec.~\ref{sec:conclusionsP3}.

\section[CGF Estimator: Constant-Population Approach]{\quad CGF Estimator: Constant-Population Approach}
\label{sec:CGF-CPA}
In practice, in order to obtain a good estimation of the CGF, it is normal to launch the simulation several times (where we denote  by $R$ the number of realizations of the same simulation), and
%
%
to estimate the arithmetic mean of the obtained values of Eq.~\eqref{eq:PSI} over these $R$ simulations. Strictly speaking (as discussed in Sec.~\ref{Discreteness Effects at Initial Times}), as the simulation does not stop exactly at the final simulation time $T$ but at some time $t_{r}^{\mathcal{F}} \leq T$ (which is different for every $ r\in\{1,...,R\}$), the average over $R$ realizations of $\Psi_{s}^{(N_{c})}$ is then correctly defined as
\begin{equation} \label{eq:PSI1}
\overline { \Psi_{s}^{(N_{c})} }  = \frac{1}{R} \sum\limits_{r = 1}^{R}  \frac{1}{t_{r}^{\mathcal{F}}}\log\prod \limits_{i = 1}^{\mathcal K_{r}} X_{i}^{r}.
\end{equation}
However, we have observed that for not too short simulation times, $ \big \vert \overline { \Psi_{s}^{(N_{c})}(T) }  - \overline { \Psi_{s}^{(N_{c})} (t_{r}^{\mathcal{F}}) } \big \vert$ is small. By assuming $t_{r}^{\mathcal{F}} \approx T $, Eq.~\eqref{eq:PSI1} can be approximated by replacing $t_{r}^{\mathcal{F}}$ by $T$ (which is what we do in practice)
 \begin{equation} \label{eq:PSI2}
 \overline { \Psi_{s}^{(N_{c})} }  \simeq \frac{1}{R} \frac{1}{T} \sum\limits_{r = 1}^{R}  \log\prod \limits_{i = 1}^{\mathcal K_{r}} X_{i}^{r}.
 \end{equation}
The CGF estimator can be defined differently from Eq.~\eqref{eq:PSI1} by using an alternative way of computing the average over $R$ realizations (as in Sec.~\ref{subsec:Different large deviation estimator} in discrete-time and as in Sec.~\ref{sec:two-estimators} in continuous-time).
 Equations~\eqref{eq:PSI1} and~\eqref{eq:PSI2} allow us to estimate the CGF using the constant-population approach of the continuous-time cloning algorithm for a $s$-biased Markov process, given a fixed number of clones $N_{c}$, a simulation time $T$ and $R$ realizations of the algorithm. 

\section[Finite-Time and Finite-$N_c$ Behavior]{\quad Finite-Time and Finite-$N_c$ Behavior of CGF Estimator}

In this section, we focus on the annihilation-creation process~(Sec.~\ref{sec:bdp}) for a particular value of parameter $s$ ($s=-0.2$), which is representative of the full range of $s$ on which we study large deviations. 
\begin{figure} [ht]
\centering
\includegraphics[width=0.6\textwidth]{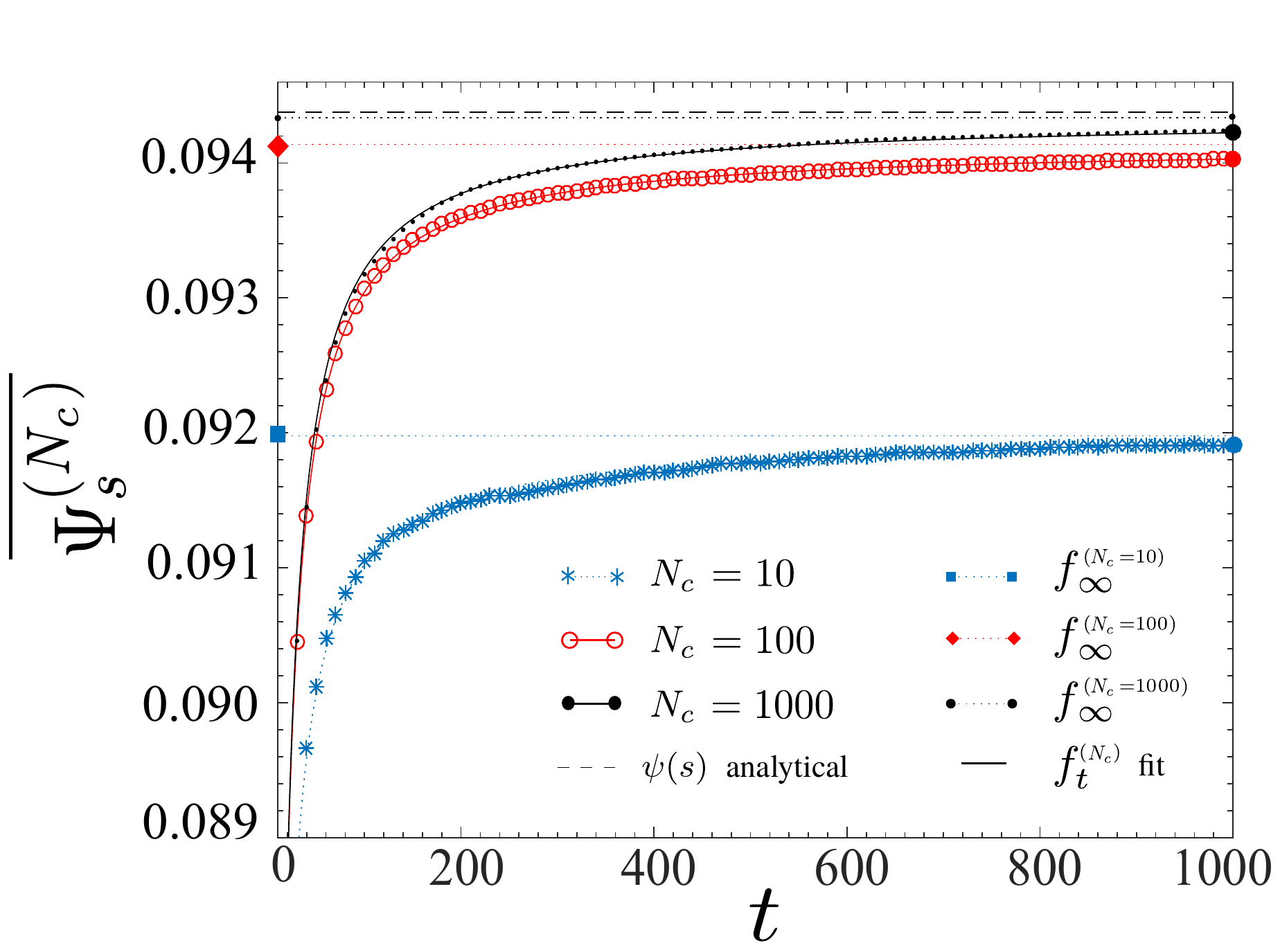}
\caption[Time Evolution of the CGF Estimator]{\label{fig:averagePsi_1} 
Average over  $R = 10^{4}$ realizations of the CGF estimator $\Psi_{s}^{(N_{c})}$~\eqref{eq:PSI1} as a function of duration $t$ of the observation window, for $N_{c} \in \{10,100,1000 \}$ clones, for the annihilation-creation dynamics with $c=0.3$. The analytical expression for the large deviation function $\psi(s)$~\eqref{eq:PSIA} is shown with a black dashed line and the fitting functions $f_{t}^{(N_{c})}$ encoding the finite-$t$ scaling (Eq.~(\ref{eq:fitdef})) are shown with continuous curves. The (\emph{a priori}) best estimation of the large deviation function (to which we refer as standard estimator) is given by $\overline{ \Psi_{s}^{(N_{c})}(t)}$ at the largest simulation time $T = 1000$, which are shown with solid circles (at the right end of the figure). The extracted infinite-time limits $f_{\infty}^{(N_{c})}$ are shown as dotted lines and squares ($N_{c} = 10$), diamonds ($N_{c} = 100$) and circles ($N_{c} = 1000$). }
\end{figure}

\subsection[Finite-Time Scaling] {\quad Finite-Time Scaling}
\label{sec:LDFT}

Here, we study the large-time behavior of the CGF estimator, at fixed number of clones $N_c$. 
Fig.~\ref{fig:averagePsi_1} presents the average over $R=10^4$ realizations of the CGF estimator $\overline{ \Psi_{s} ^{(N_{c})} }$~\eqref{eq:PSI1} as a function of the (simulation) time for given numbers of clones $N_{c} = \{10,100,1000 \}$. It is compared with the analytical value $\psi(s)$~\eqref{eq:PSIA} which is shown with a black dashed line. 
%
%
%
As can be seen in Fig.~\ref{fig:averagePsi_1} for a small number of clones ($N_{c}=10$),
the CGF estimator $\overline{ \Psi_{s} ^{(N_{c})} }$ highly deviates from the analytical value $\psi(s)$. However, as $N_c$ and the simulation time $t$ become larger, the CGF estimator gets closer to the analytical value $\psi(s)$. 
One can expect that in the $t \to \infty$ and $N_c \to \infty$ limits, $\psi(s)$ will be obtained from the estimator as
\begin{equation}
\lim_{N_c \to \infty} \lim_{t \to \infty} \overline{\Psi_{s}^{(N_{c})}(t)} = \psi(s),
\label{eq:limitNcT}
\end{equation}
as it was derived in chapter~\ref{chap:DiscreteTime}~\cite{partI}. However, in a practical implementation of the algorithm, this infinite-time and -size limits are not achievable and we use large but {\it finite} simulation time $t$ and number of clones $N_{c}$. 
This fact motivates our analysis of the actual dependence of the estimator with $t$ and $N_{c}$.
The standard estimator of the large deviation function is the value of $\overline{ \Psi_{s} ^{(N_{c})} }$ at the largest simulation time $T$ and for the largest number of clones $N_c$, ($\overline{ \Psi_{s} ^{(N_{c})}(T) }$ for $N_{c}=1000$ and $T=1000$), i.e., the black solid circle~{\small$\bullet$} in Fig.~\ref{fig:averagePsi_1}. This value provides the (\emph{a priori}) best estimation of the large deviation function that we can obtain from the continuous-time cloning algorithm. 
However encouragingly, as we detail later, this estimation can be improved by taking into account the convergence speed of the CGF estimator. 


The result of fitting $\overline{ \Psi_{s} ^{(N_{c})}(t)}$ with the curve $f_{t}^{(N_{c})}$ is shown with solid lines in Fig.~\ref{fig:averagePsi_1}. This is defined as
\begin{equation}
f_{t}^{(N_{c})}  \equiv f_{\infty}^{(N_{c})} + b_{t}^{(N_{c})}t^{-1},
\label{eq:fitdef}
\end{equation}
where the fitting parameters $f_{\infty}^{(N_{c})}$ and $b_{t}^{(N_{c})}$ can be determined from the least squares method by minimizing the deviation from $\overline{ \Psi_{s} ^{(N_{c})}(t)}$. 
The clear coincidence between $\overline{ \Psi_{s} ^{(N_{c})}(t)}$ and the fitting lines indicates the existence of a $1/t$-convergence of $\overline{ \Psi_{s} ^{(N_{c})}(t)}$ to $\lim_{t\rightarrow \infty}\overline{ \Psi_{s} ^{(N_{c})}(t)}$ (that we call $\mathbf{t^{-1}}$ \textbf{-scaling}). 
%
This property can be derived from the assumption that the cloning algorithm itself is described by a Markov process: in chapter~\ref{chap:DiscreteTime}~\cite{partI} with a different version of the algorithm, we constructed a meta-Markov process to describe the cloning algorithm by expressing the number of clones by a birth-death process. 
Once such meta process is constructed, the CGF estimator~\eqref{eq:PSI} is regarded as the time-average of the observable $X_i$ within such meta-Markov process.
In other words, $t \Psi_{s}^{(N_{c})}$ is an additive observable of the meta-process describing the cloning algorithm. 
We now recall that time-averaged quantities converge to their infinite-time limit with an error proportional to $1/t$ when the distribution function of the variable converges exponentially (as in Markov processes). This leads to the $t^{-1}$-scaling of CGF estimator (\ref{eq:fitdef}). 
We note that constructing such a meta-Markov process explicitly is not a trivial task, and for the algorithm discussed here, such a construction 
remains as an open problem. 

By assuming the validity of the scaling form~(\ref{eq:fitdef}), it is possible to extract the infinite-time limit of the CGF estimator from finite-time simulations. We denote this infinite-time limit as $f_{\infty}^{(N_{c})}$ and it is expected to be a 
the better estimator of CGF than $\overline{ \Psi_s^{(N_c)}(T)}$ at finite $T$, provided that
\begin{equation}
f_\infty^{(N_c)} = \lim _{t\rightarrow \infty }\overline{ \Psi_s^{(N_c)}(t)}.
\label{eq:infT}
\end{equation}
%
In Fig.~\ref{fig:averagePsi_1},
we show $f_{\infty}^{(N_{c})}$ with dotted lines and circles ($N_{c} = 10$), diamonds ($N_{c} = 100$) and squares ($N_{c} = 1000$).
As can be seen, 
this parameter indeed provides a better numerical estimate of $\psi({s})$ than $\overline{ \Psi_{s} ^{(N_{c})}(T)}$. 

\subsection[Finite-$N_c$ Scaling]{\quad Finite-$N_c$ Scaling \label{sec:LDFN}}
\label{sec:LDFT_Nc}

\begin{figure}[t]
\centering
\includegraphics[width=0.6\textwidth]{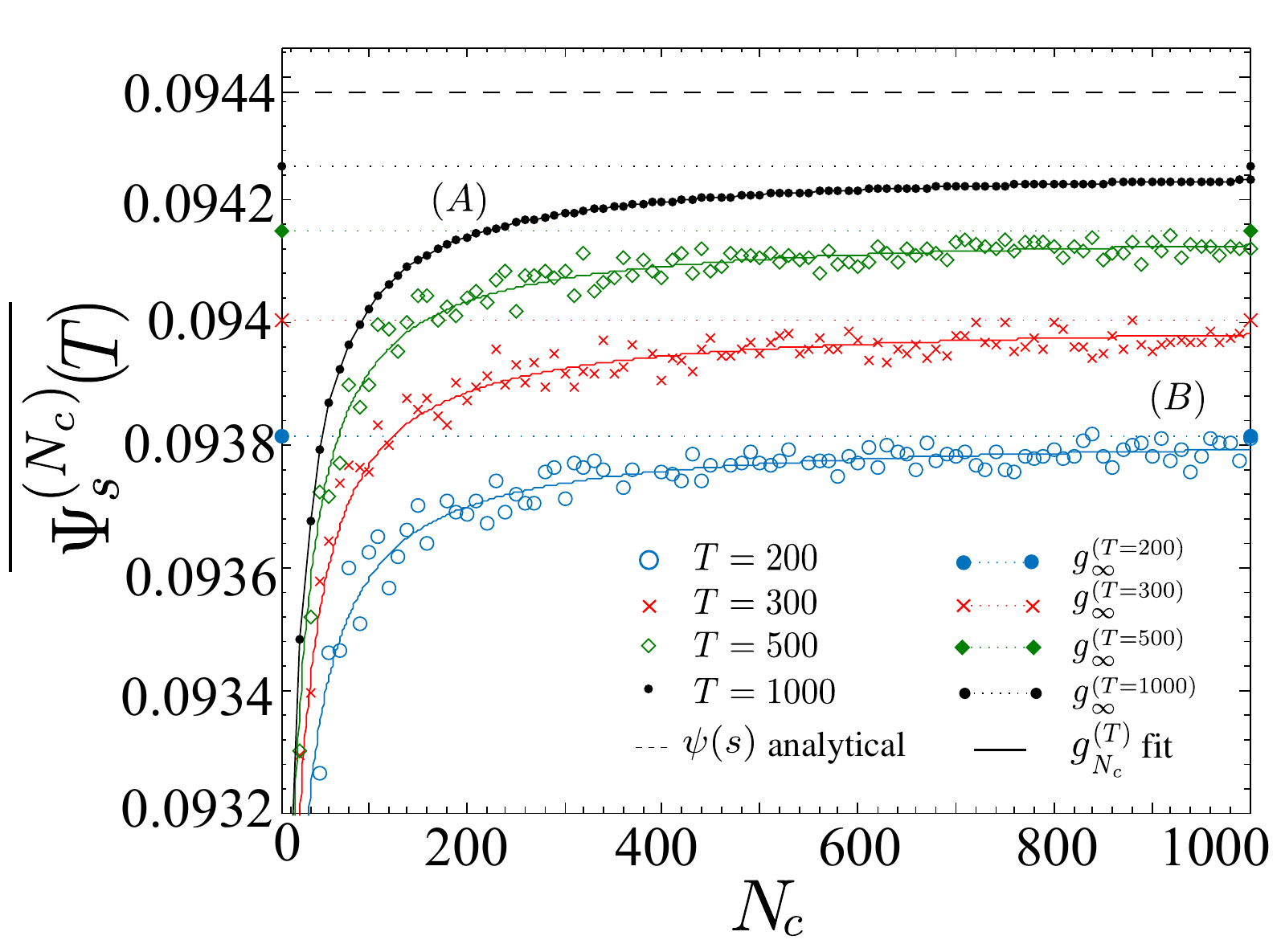}
\caption[Population-Size Evolution of the CGF Estimator]{\label{fig:PsiN} CGF estimator $\overline{ \Psi_{s}^{(N_{c})} (T) }$~\eqref{eq:PSI1} for given final (simulation) times $T = \{200,300,500, 1000 \}$ as a function of the number of clones $N_c$ (on the range $10\leq N_{c}\leq 1000$). The analytical value $\psi(s)$~\eqref{eq:PSIA} is shown with a dashed line and the fits $g_{N_c}^{(T)}$~\eqref{eq:PSI4N} with continuous curves. A large simulation time for a small number of clones, shown in~(\emph{A}), produces a better estimation compared to the one given by the largest number of clones with a relatively short simulation time, which is shown in~(\emph{B}). The best CGF estimation we can naively obtain would be given by $\overline{ \Psi_{s}^{(N_{c})}(T) }$ at largest simulation time $T$ and largest number of clones $N_{c}$. However, the extracted infinite-size limits $g_{\infty}^{(T)}$ provide a better estimation in comparison. These limits are shown with dotted lines and circles ($T=200$), crosses ($T=300$), diamonds ($T=500$) and dots ($T=1000$). Additionally, $c=0.3$ and $s=-0.2$.
} 
\end{figure}

Here, we study the behavior of the (standard) CGF estimator $\overline{ \Psi_{s}^{(N_{c})} (T) }$ as we increase the number of clones $N_c$, for a given final (simulation) time $T$. 
%
%
%
Similar to what we did in Sec.~\ref{sec:LDFT}, we consider a curve in the form
\begin{equation} \label{eq:PSI4N}
g_{N_{c}}^{(T)} = g_{\infty}^{(T)} + \tilde b_{N_{c}}^{(T)}N_{c}^{-1},
\end{equation}
where $g_{\infty}^{(T)}$ and $\tilde b_{N_{c}}^{(T)}$ are fitting parameters which are determined by the least squares fitting to 
$\overline{ \Psi_{s}^{(N_{c})} (T) }$. The obtained $g_{N_{c}}^{(T)}$ as a function of $N_{c}$ are shown in Fig.~\ref{fig:PsiN} as solid lines.   
We considered four values of final simulation time $T = \{200,300,500, 1000 \}$ and population sizes in the range $10\leq N_{c}\leq 1000$. 
As can be seen, these curves describe well the dependence in $N_c$ of $\overline{ \Psi_{s}^{(N_{c})} (T) }$, indicating that $\overline{ \Psi_{s}^{(N_{c})} (T) }$ converges to its infinite-$N_c$ limit with an error proportional to $1/N_c$ (that we call $\mathbf{N_c^{-1}}$\textbf{-scaling}). 
This scaling could be proved under general assumptions in 
chapter~\ref{chap:DiscreteTime}~\cite{partI}, 
(\textit{i}) however 
%
without covering the continuous-time algorithm discussed here,
and (\textit{ii}) for the CGF estimator $\overline{ \Psi_{s}^{(N_{c})} (T)}$ considered the $T \to \infty$ limit, instead of finite $T$. The generalization of the argument presented in chapter~\ref{chap:DiscreteTime}~\cite{partI} in order to cover the general cases (\textit{i}) and (\textit{ii}) is an important open direction of research.

By assuming the validity of such $N_c^{-1}$-scaling,
we can evaluate
the $N_c \to \infty$ limit of $\overline{ \Psi_{s}^{(N_{c})} (T) }$ as
the fitting parameter $g_{\infty}^{(T)}$ obtained from finite $N_c$ simulations as
\begin{equation}
g_{\rm \infty}^{(T)} =  \lim_{N_c\to \infty}  \overline{ \Psi_{s}^{(N_{c})} (T) }.
\end{equation} 
These parameters $g_{\infty}^{(T)}$ (to which we refer as infinite-size limit) are shown in Fig.~\ref{fig:PsiN} as dotted lines and provide 
better estimations of $\psi({s})$ than the standard estimator $\overline{ \Psi_{s}^{(N_{c})} (T) }$. 
%


\section[Finite-Time and Finite-$N_c$ Scaling Method]{\quad Finite-Time and Finite-$N_c$ Scaling Method to estimate Large Deviation Functions  \label{sec:LDFinfTinfN}}

In the previous section, we have shown how
it is possible to extract $f^{(N_c)}_\infty$ and $g^{(T)}_\infty$ from finite $T$- and finite $N_c$- simulations respectively. In this section, we combine both of these $1/t$- and $1/N_c$- scaling methods in order to extract the infinite-time and -size limit of the CGF estimator. This limit gives a better evaluation of the large deviation function within the cloning algorithm than the standard estimator. 

\begin{figure}[!t]
\centering
\includegraphics[width=0.55\textwidth]{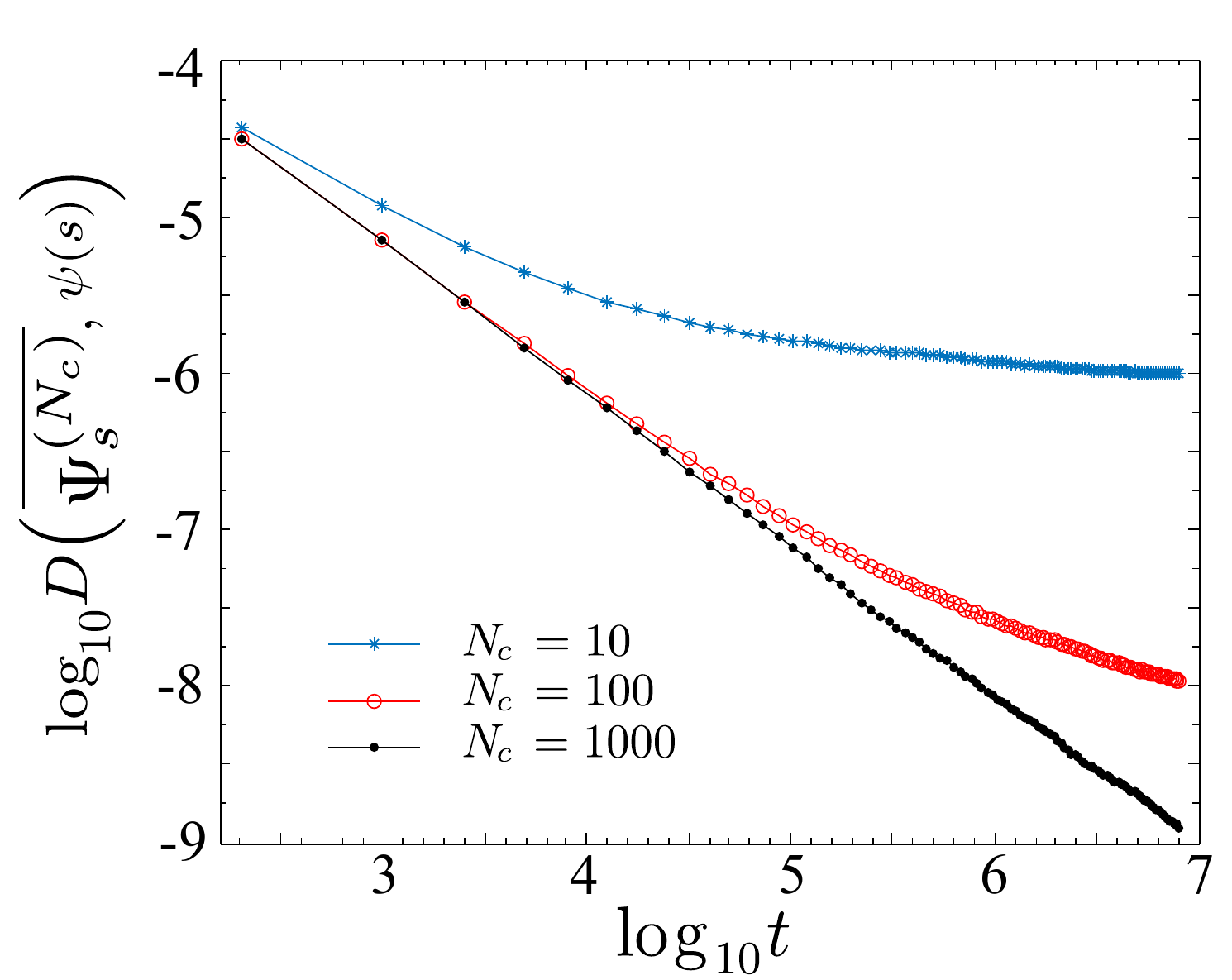}
\caption[CGF Estimator Distance]{\label{fig:averagePsi_2} 
Distance $D$ (Eq.~\eqref{eq:D1}) between the analytical CGF $\psi(s)$ and its numerical estimator $\overline{ \Psi_{s}^{(N_{c})} }$, as a function of time~$t$ in log-log scale. The distances are computed from the values in Fig.~\ref{fig:averagePsi_1}. This distance behaves as a power law of exponent $-1$ on a time window, where the size of the time window increases as $N_{c}$ increases. This illustrates the scaling~\eqref{eq:D2}. The parameters of the model are $c = 0.3$, $s=-0.2$.}
\end{figure}


We first note that either of $f_{\infty}^{(N_c)}$ or $g_{\rm \infty}^{(T)}$
%
is expected to converge to $\psi(s)$ as $N_c \to \infty$ or as $T \to \infty$. 
We checked numerically this property by defining the distance $D$ between $\psi({s})$ and its numerical estimator $\overline{ \Psi_{s}^{(N_{c})}}$, 
\begin{equation} \label{eq:D1}
D \big(  \overline{\Psi_{s}^{(N_{c})} },\psi(s) \big) = \big \vert \overline{ \Psi_{s}^{(N_{c})} } - \psi(s) \big \vert.
\end{equation}
This quantity is shown in Fig.~\ref{fig:averagePsi_2} as a function of $t$ in log-log scale. As we can see, as $N_{c}$ increases, $\log D$ behaves as straight line with slope $-1$ on a time window which grows with $N_c$. In other words, when $N_{c}\to\infty$, 
\begin{equation} \label{eq:D2}
\big \vert \overline{ \Psi_{s}^{(N_{c})} } - \psi(s) \big \vert \sim t^{-1}.   
\end{equation}

Inspired by this observation, we assume the following scaling
for the fitting parameter $f_{\infty}^{N_{c}}$. If we consider a set of simulations performed at population sizes $\vec{N}_{c}= \{ N_{c}^{(1)},...,N_{c}^{(j)} \}$, the obtained infinite-time limit of the CGF estimator $f_{\infty}^{N_{c}}$ behaves as a function of $N_{c}$ as
\begin{equation} \label{eq:PSIinfinf}
 f_{\infty}^{(N_c)} \simeq f_{\infty}^{\infty} + b_{\infty}^{(N_c)} N_c^{-1},
\end{equation}
which means that $f_{\infty}^{(N_c)}$ itself exhibits $1/N_c$ corrections for large but finite $N_c$. 
By using this scaling, we detail below in Sec.~\ref{ssec:SM} the method to extract the infinite-time infinite-$N_c$ limit of the CGF estimator $\overline{\Psi_{s}^{(N_{c})}(T)}$ from finite-time and finite-$N_c$ data. We note that this method can be used for a relatively short simulation time and a relatively small number of clones (see Fig.~\ref{fig:PSIT-PSIN}). In Sec.~\ref{ssec:exCP}, we 
present numerical examples of the application of this method to the contact process. 

\subsection[The Scaling Method]{\quad The Scaling Method}
\label{ssec:SM}
The procedure is summarized as follows:
\begin{enumerate}
\item[1.] Determine the average over $R$ realizations $\overline{ \Psi_{s}^{(N_{c})}(t) }$~\eqref{eq:PSI1} up to a final simulation time $T$ for each $N_c \in \vec{N}_{c}$.
\item[2.] Determine the fitting parameters $f_{\infty}^{(N_{c})}$'s defined in the form $f_{t}^{(N_{c})} = f_{\infty}^{(N_{c})} + b_{t}^{(N_{c})}t^{-1}$~\eqref{eq:fitdef} from each of the obtained $\overline{ \Psi_{s}^{(N_{c})}(t) }$'s.  
\item[3.] Determine $f_{\infty}^{\infty}$ from a fit in size $f_{\infty}^{(N_c)} = f_{\infty}^{\infty} + b_{\infty}^{(N_c)}N_c^{-1}$~\eqref{eq:PSIinfinf} on the extracted $f_{\infty}^{(N_{c})}$'s.
\end{enumerate}
The result obtained for $f_{\infty}^{\infty}$ renders a better estimation of $\psi(s)$ than the standard estimator $\overline{ \Psi_{s}^{(N_{c})}(t) }$ evaluated for $N_c = \max \vec{N_c}$ and for $t = T$. 

\subsection[Application to the Contact Process]{\quad Application to the Contact Process}
\label{ssec:exCP}
\begin{figure}[t!]
\centering
\includegraphics[width=0.55\textwidth]{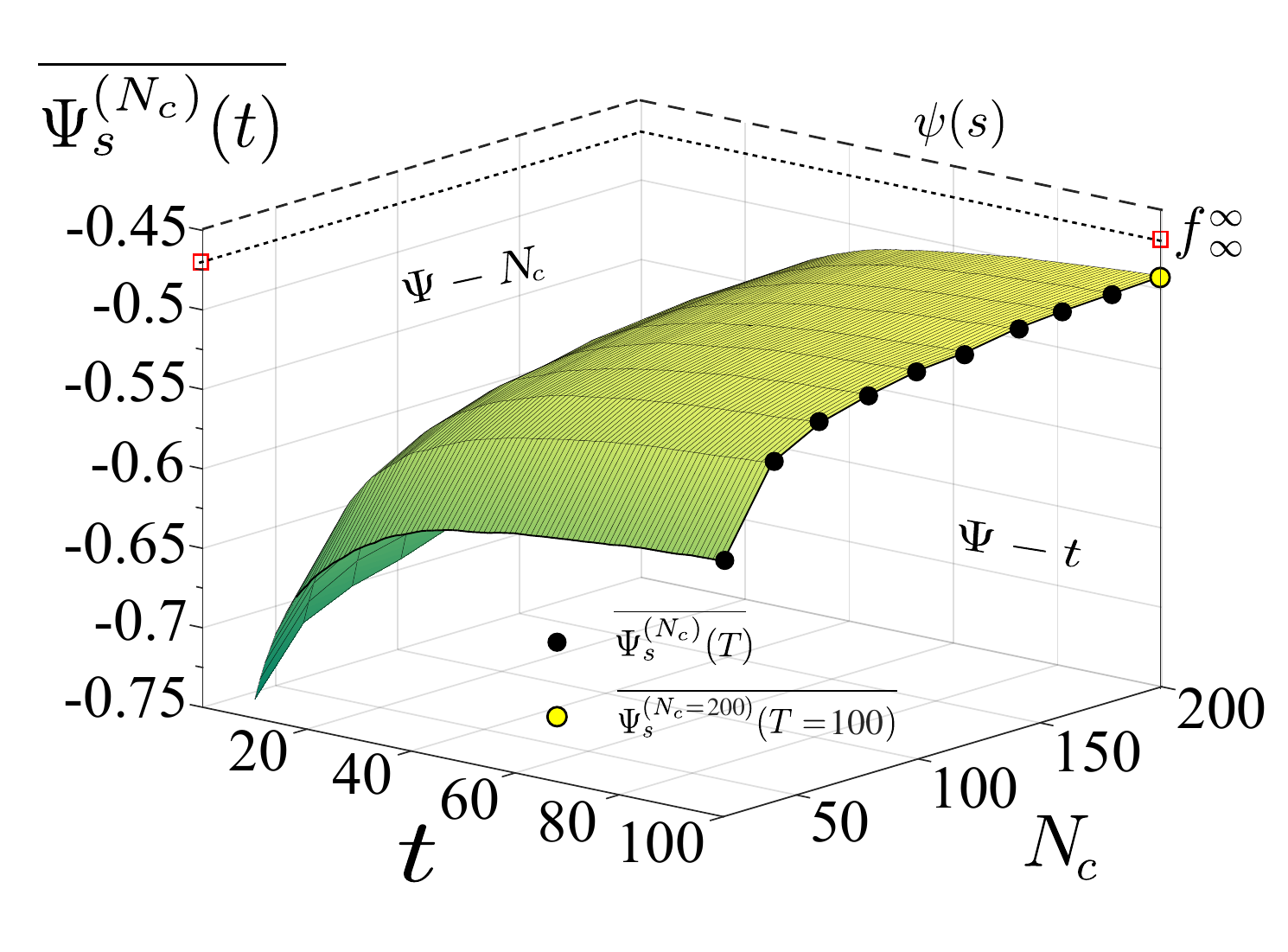}
\caption[Continuous-Time CGF Estimator]{\label{fig:surfPSI} Estimator of the large deviation function $\overline{ \Psi_{s}^{(N_{c})} (t) }$ as a function of time and the number of clones. The estimator $\overline{ \Psi_{s}^{(N_{c})} (T) }$ at final simulation time $T = 100$ as a function of the number of clones (up to $N_{c}=200$) is shown as black circles. The best CGF estimation under this configuration given by the standard estimator, i.e., $\overline{ \Psi_{s}^{(N_{c}=200)} (T=100) }$ is shown as a yellow circle. The analytical value of the CGF $\psi({s})$ is obtained from the largest eigenvalue of the matrix~\eqref{eq:defopWs} and shown as a black dashed line. The extracted limit $f_\infty^\infty$ is shown with red squares. Additionally, $L=6$, $s=0.15$, $h = 0.1$, $\lambda=1.75$ and $R=10^{3}$.
} 
\end{figure}

We apply the scaling method to the one-dimensional contact process (see Sec.~\ref{sec:CP}). We set $L = 6$, $h = 0.1$, $\lambda=1.75$, $T=100$ and $s=0.15$. 
%
%
As we detail below, we compare the improved estimator $f_{\infty}^{\infty}$ obtained from 
the application of the scaling method (for $\vec{N}_{c} = \{20,40,...,180,200 \}$) with the standard estimator $\overline{ \Psi_{s}^{(N_{c})} (T) }$ (for $N_{c}=\max \vec{N}_{c} = 200$).
Fig.~\ref{fig:surfPSI} represents the behavior of the estimator $\overline{ \Psi_{s}^{(N_{c})}(t) }$ as a function of the simulation time~$t$ and of the number of clones $N_{c}$. 
%
%
%
The values of the estimator at the final simulation time~$T$ are represented with black circles for each $N_{c} \in \vec{N}_{c}$ and with a yellow circle for $N_{c} = \max \vec{N}_{c}$. 
The analytical expression for the large deviation function $\psi({s})$ is shown in a black dashed line.

\begin{figure*} [!t]
\includegraphics[width=0.48\textwidth]{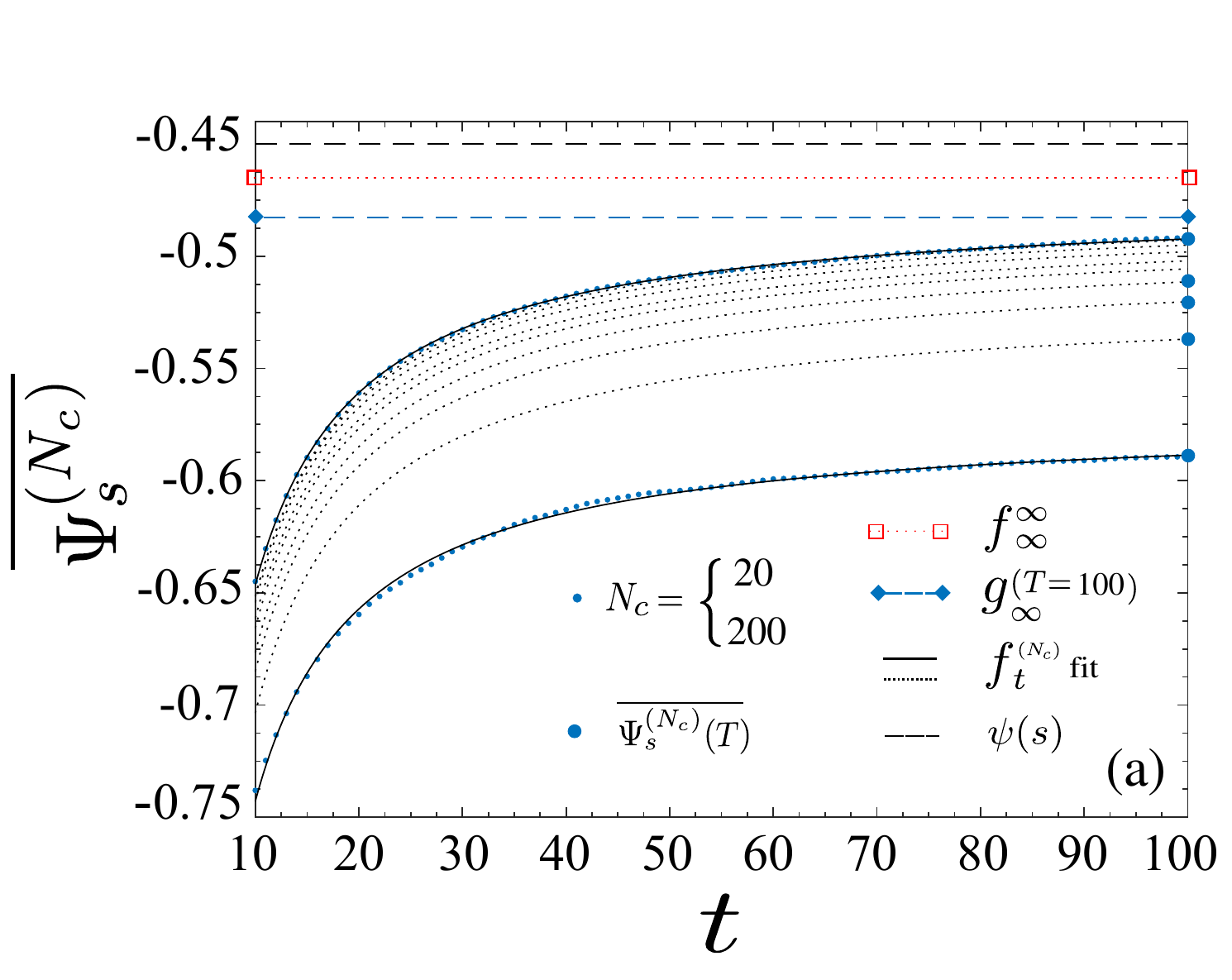}
\includegraphics[width=0.48\textwidth]{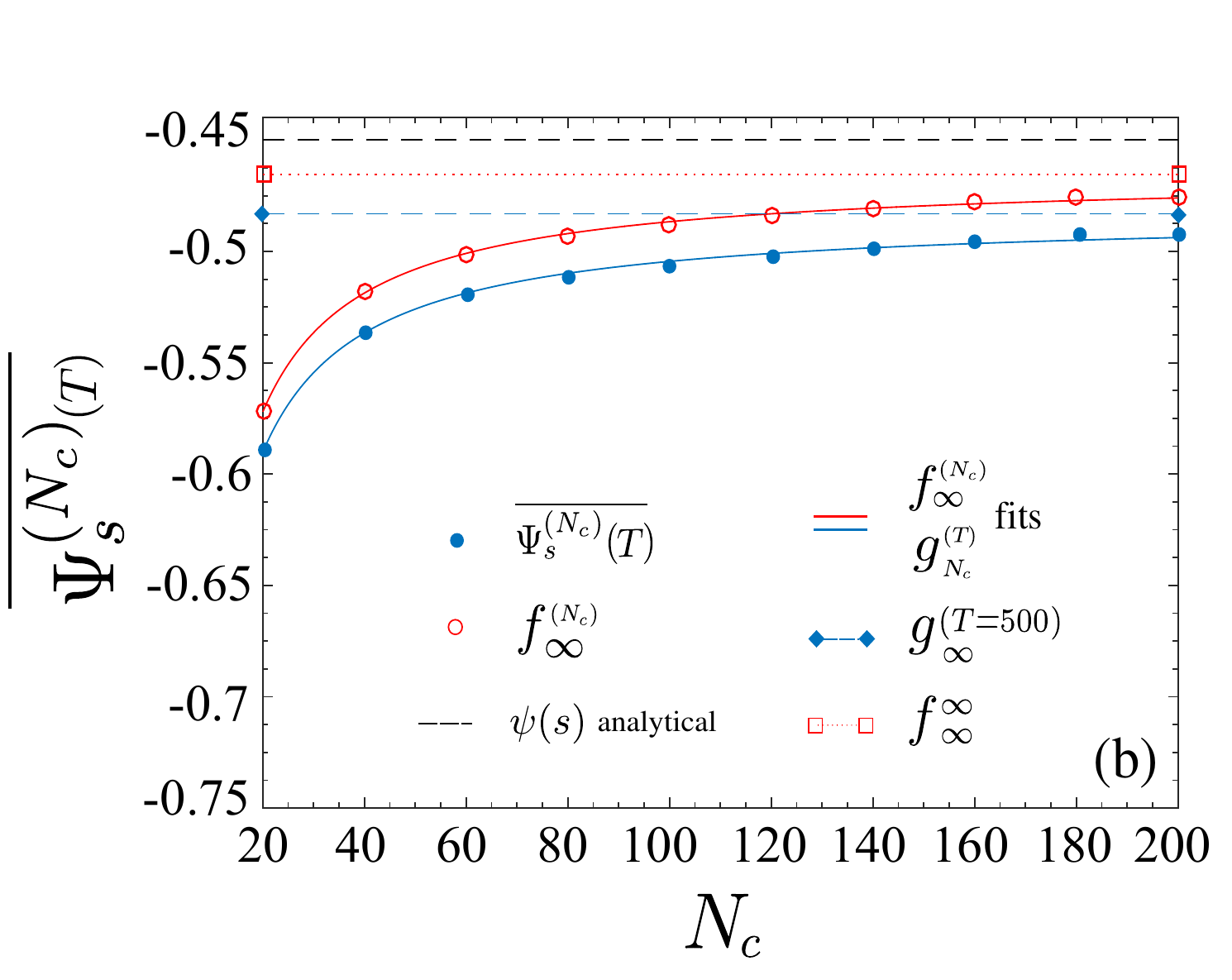}
\caption[Projection of the CGF Estimator]{\label{fig:PSIT-PSIN} \textbf{(a)} Projection of the surface represented in Fig.~\ref{fig:surfPSI} over the plane $\Psi-t$. $\overline{ \Psi_{s}^{(N_{c})}(t) }$ is represented for $N_{c} = 20$ and $N_{c}= 200$ with blue dots. The estimations $\overline{ \Psi_{s}^{(N_{c})}(T) }$ of the large deviation (at the final simulation time $T=100$) are shown in large blue dots for all the values of $N_{c}$ considered. 
The fit in time (Eq.~\eqref{eq:fitdef}) over $\overline{ \Psi_{s}^{(N_{c})}(t)}$ is shown as black solid lines (for $N_{c} = 20$ and $N_{c}= 200$) and dotted lines (for other values of $N_c$). 
\textbf{(b)} Projection at the final simulation time $T=100$ on the plane $\Psi-N_{c}$, $\overline{ \Psi_{s}^{(N_{c})}(T) }$ is shown in large blue dots.
The infinite-time limit $f_{\infty}^{(N_{c})}$ as a function of $N_c$ (see Eq.~(\ref{eq:fitdef})) is represented in red circles.
The results of fitting $\overline{ \Psi_{s}^{(N_{c})}(T) }$ (Eq.~(\ref{eq:PSI4N})) and  $f_{\infty}^{(N_{c})}$ (Eq.~(\ref{eq:PSIinfinf})) are shown with blue and red solid curves respectively. The infinite-$N_c$ limit $g_{\infty}^{(T)}$ is shown with blue dashed line and diamonds meanwhile the infinite-size and time limit $f_{\infty}^{\infty}$ is shown with a red dotted line in both of \textbf{(a)} and \textbf{(b)}. 
The extracted limit $f_{\infty}^{\infty}$ renders a better estimation of the large deviation function than $\overline { \Psi_{s}^{(N_{c} = 200)}(T = 100) }$ (and also than $g_{\infty}^{(T)}$) demonstrating the efficacy of the method proposed.
}
\end{figure*}

\begin{figure}[!t]
\centering
\includegraphics[width=0.55\textwidth]{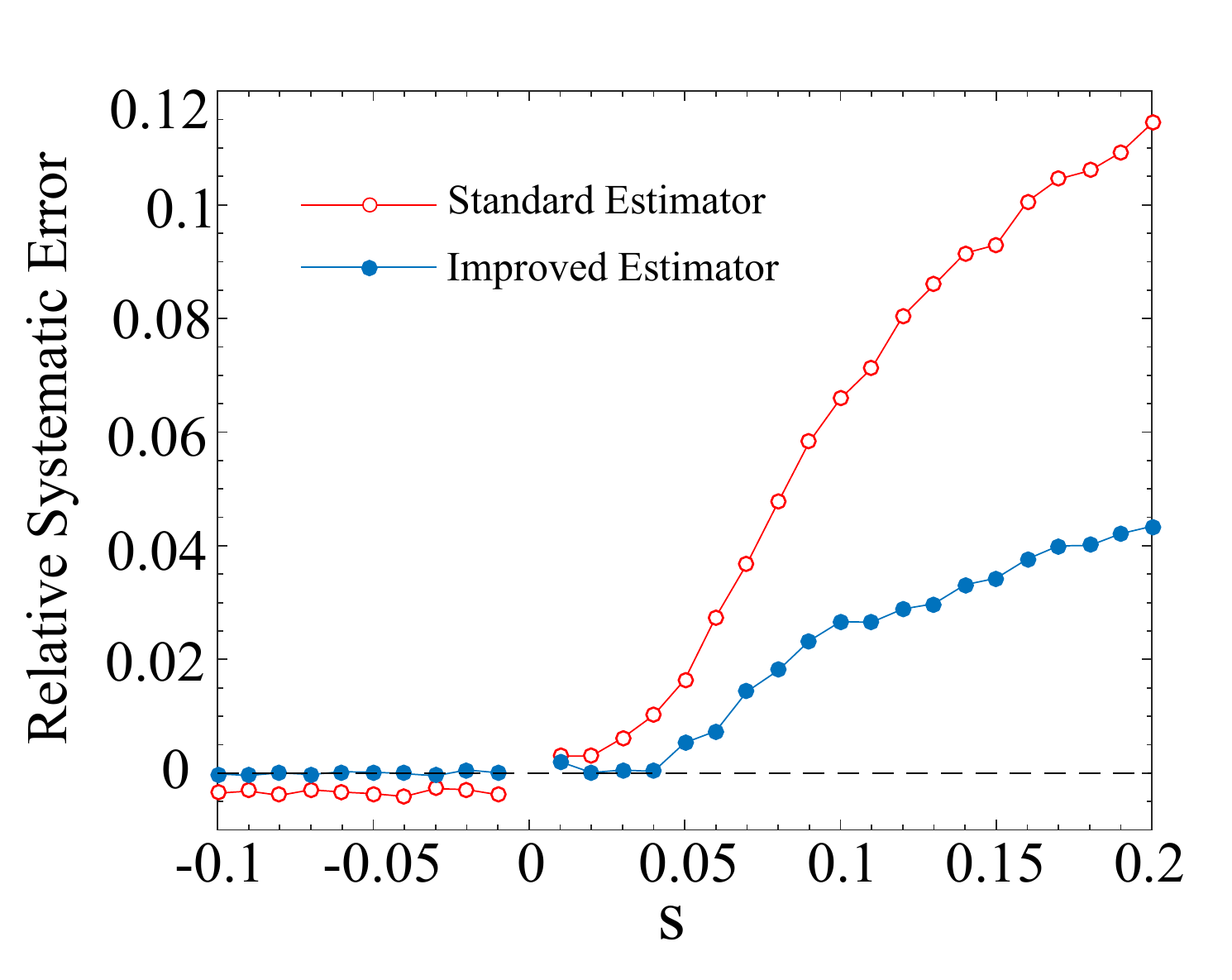}
\caption[Relative Systematic Error]{\label{fig:illustration-results} Relative systematic error $\left [ \Psi(s)-\psi(s) \right ]/\psi(s)$ between the numerical estimators $\Psi(s)$ and the analytical CGF $\psi(s)$. The error for the standard estimator $\overline {\Psi_s^{(N_c)}(T)}$ is shown in blue and for the improved one, $f_\infty^{\infty}$ (Eq.~(\ref{eq:PSIinfinf})) in red. The scaling method proposed in this chapter was tested on the contact process (with $L = 6$, $h = 0.1$, and $\lambda=1.75$) 
for a set of populations $\vec{N}_{c} = \{20,...,200 \}$, a simulation time $T=100$, and $R =1000$ realizations. As can be seen, the errors due to finite-size and -time effects can be reduced through the improved estimator.
}
%
\end{figure}

On Fig.~\ref{fig:PSIT-PSIN}(a) we show the projection of the surface of Fig.~\ref{fig:surfPSI} on the plane $\Psi-t$. The behavior in~$t$ of the estimator $\overline{ \Psi_{s}^{(N_{c})}(t) }$ is shown for $N_{c} = 20$ and $N_{c} = 200$, in blue dots in Fig.~\ref{fig:PSIT-PSIN}(a). The standard CGF estimators, $\overline{ \Psi_{s}^{(N_{c})}(T) }$, are shown in large blue dots in Fig.~\ref{fig:PSIT-PSIN}(a) (on the axis for $T=100$).
The fitting curves $f_{t}^{(N_{c})}$ (Eq.~(\ref{eq:fitdef})) are shown in black continuous lines (for $N_c=20$ and $N_c=200$) and black dotted lines (for other intermediate values of $N_c$). 
Next, we show in Fig.~\ref{fig:PSIT-PSIN}(b) the projection of the surface of Fig.~\ref{fig:surfPSI} on the plane $\Psi-N_c$ where the time has been set to the largest $t=T$. 
The standard CGF estimators, $\overline{ \Psi_{s}^{(N_{c})}(T)}$ are plotted as blue filled circles, and the fitting curve $g_{N_c}^{(T)}$ (Eq.~\eqref{eq:PSI4N}) on $\overline{ \Psi_{s}^{(N_{c})}(T)}$ is shown as a blue solid line. From these curves, we determine $g_{\infty}^{(T)}$ (see Sec.~\ref{sec:LDFT_Nc}), which is shown as a blue dashed line and diamonds. 
Finally, the parameter $f_{\infty}^{(N_{c})}$ extracted from the fitting on $\overline{ \Psi_{s}^{(N_{c})}(t)}$ (for each value of $N_{c}$) is shown as red circles in Fig.~\ref{fig:PSIT-PSIN}(b). These values also scale as $1/N_c$ (Eq.~\eqref{eq:PSIinfinf}) and their fit is shown as a red solid curve. 
The scaling parameter $f_{\infty}^{\infty}$ obtained from this last step  provides a better estimation of the large deviation function than the standard estimator $\overline { \Psi_{s}^{(N_{c} = 200)}(T = 100) }$ that is widely used in the application of cloning algorithms. 
This improvement is valid on a wide range of values of the parameter $s$ as can be visualized in Fig~\ref{fig:illustration-results},  where we represented the relative systematic error $\left [ \Psi(s)-\psi(s) \right ]/\psi(s)$ between the standard and improved estimators $\Psi(s)$ and the analytical CGF~$\psi(s)$.

\newpage
\section[Issues on an Analytical Approach]{\quad Issues on an Analytical Approach \label{Discrete-time_algorithm}}

In chapter~\ref{chap:DiscreteTime}~\cite{partI}, we considered a \emph{discrete-time} version of the population dynamics algorithm, where a cloning procedure is performed every small time interval $\Delta t$. 
We have proved the convergence of the algorithm in the large-$N_c$, -$t$ limits,
and we also derived that the systematic error of the LDF estimator (i.e., the deviation of the estimator from the desired LDF) decayed proportionally to $1/t$ and $1/N_c$. 
From a practical point of view, however, the formulation used there had one problem. In order to prove the result, we took the large frequency limit of cloning procedure or, in other words, we took the $\Delta t \rightarrow 0$ limit. 
A rough estimate of the error due to non-infinitesimal $\Delta t$ proves to be $O(\Delta t)$. For a faster algorithm, it is better to take this value to be larger, and indeed empirically, we expect that this error to be very small (or rather disappearing in the large $t, N_c$ limits). However, the detailed analytical estimation of this error 
is still an open problem. 

In the main part of this chapter, from a different point of view, we consider the \emph{continuous-time} version of the population dynamics algorithm 
%
%
\cite{lecomte_numerical_2007,tailleur_simulation_2009}.
Here, the cloning is performed at each change of state of a copy.
The time intervals $\Delta t$ which separate those changes of state are non-infinitesimal, which
means that the formulation used in chapter~\ref{chap:DiscreteTime}~\cite{partI} cannot be applied to understand its convergence. 
Furthermore, because these time intervals are of non-constant duration and stochastically distributed, the continuous-time algorithm is more difficult to handle analytically than the discrete-time version. 
Instead of pursuing the analytical study within the continuous-time algorithm, we performed a numerical study, and we have shown that the $1/N_c$ and $1/t$ scalings are also observed for the continuous-time algorithm. 
%
%
%
Although the proof of these scalings are beyond the scope of this chapter,
these numerical observations support a conjecture that such scaling in large $t$ and in large $N_c$ limits are generally valid in cloning algorithms to calculate large deviation functions. 

\section[Conclusions]{\quad Conclusions}
\label{sec:conclusionsP3}

Direct sampling of the distribution of rare trajectories is a rather difficult numerical issue (see for instance Ref.~\cite{rohwer_convergence_2015}) because of the scarcity of the non-typical trajectories. We have shown how to increase the efficiency of a commonly used numerical method (the so-called cloning algorithm) in order to improve the evaluation of large deviation functions which quantify the distribution of such rare trajectories, in the large time limit.
We used the finite-size and finite-time scaling behavior of CGF estimators in order to propose an improved version of the continuous-time cloning algorithm which provides more reliable results, 
%
less affected by finite-time and -$N_c$ effects.
We verified the results observed for the discrete-time version of the cloning algorithm in chapter~\ref{chap:DiscreteTime}~\cite{partI} and we showed their validity also for the continuous case~\cite{partII}. Importantly, we showed how these results can be applied to more complex systems.

%

We note that the scalings which rule the convergence to the infinite-size infinite-time limits (with corrections in $1/N_c$ and  in $1/t$) have to be taken into account properly: indeed, as power laws, they present no characteristic size and time above which the corrections would be negligible.
The situation is very similar to the study of the critical depinning force in driven random manifolds: the critical force presents a corrections in one over the system size~\cite{kolton_uniqueness_2013} which has to be considered properly in order to extract its actual value.
Generically, such scalings also provide a convergence criterion to the asymptotic regimes of the algorithm: one has to confirm that the CGF estimator does present corrections (first) in $1/t$ and (second) in $1/N_c$ with respect to an asymptotic value in order to ensure that such value does represent a correct evaluation of the CGF.

It would be interesting to extend our study of these scalings to systems presenting dynamical phase transitions (in the form of a non-analyticity of the CGF), where it is known that the finite-time and the finite-size scalings of the CGF estimator can be very hard to overcome~\cite{lecomte_numerical_2007}. 
In particular, in this context, it would be useful to understand how the dynamical phase transition of the original system translates into anomalous features of the distribution of the CGF estimator in the cloning algorithm. 
These phase transitions are normally accompanied with an infinite system-size  limit (although there was a report of dynamical phase transitions without taking a such limit \cite{0295-5075-116-5-50009}). To overcome these difficulties (caused by a large system size and/or by the presence of a phase transition), it may be useful to use the adaptive version of the cloning algorithm \cite{PhysRevLett.118.115702}, which has been recently developed to study such phase transitions, with the scaling method presented in this chapter.

\chapter[\quad Fluctuations of CGF Estimator]{Fluctuations of CGF Estimator}
\label{chap:CGF}
In order to complement the main discussion done in the previous chapter, 
here we study the fluctuations of the CGF estimator~\cite{partII} as defined in Eq.~\eqref{eq:PSI1}. This is done by studying its distribution and its dependence with the simulation time and the number of clones. Compatible with the central limit theorem, we show how  a proper rescaling of the CGF estimator produces a collapse of the distributions into a normal standard distribution for different values of $N_c$  and simulation times. 
Additionally, we discuss in Sec.~\ref{sec:two-estimators} an alternative way of defining it which was already introduced in Sec.~\ref{subsec:Different large deviation estimator} for the discrete-time version.

\section[Central Limit Theorem]{\quad Central Limit Theorem}
\label{sec:CLT}
\begin{figure*}[b!]
\includegraphics[width=0.32\textwidth]{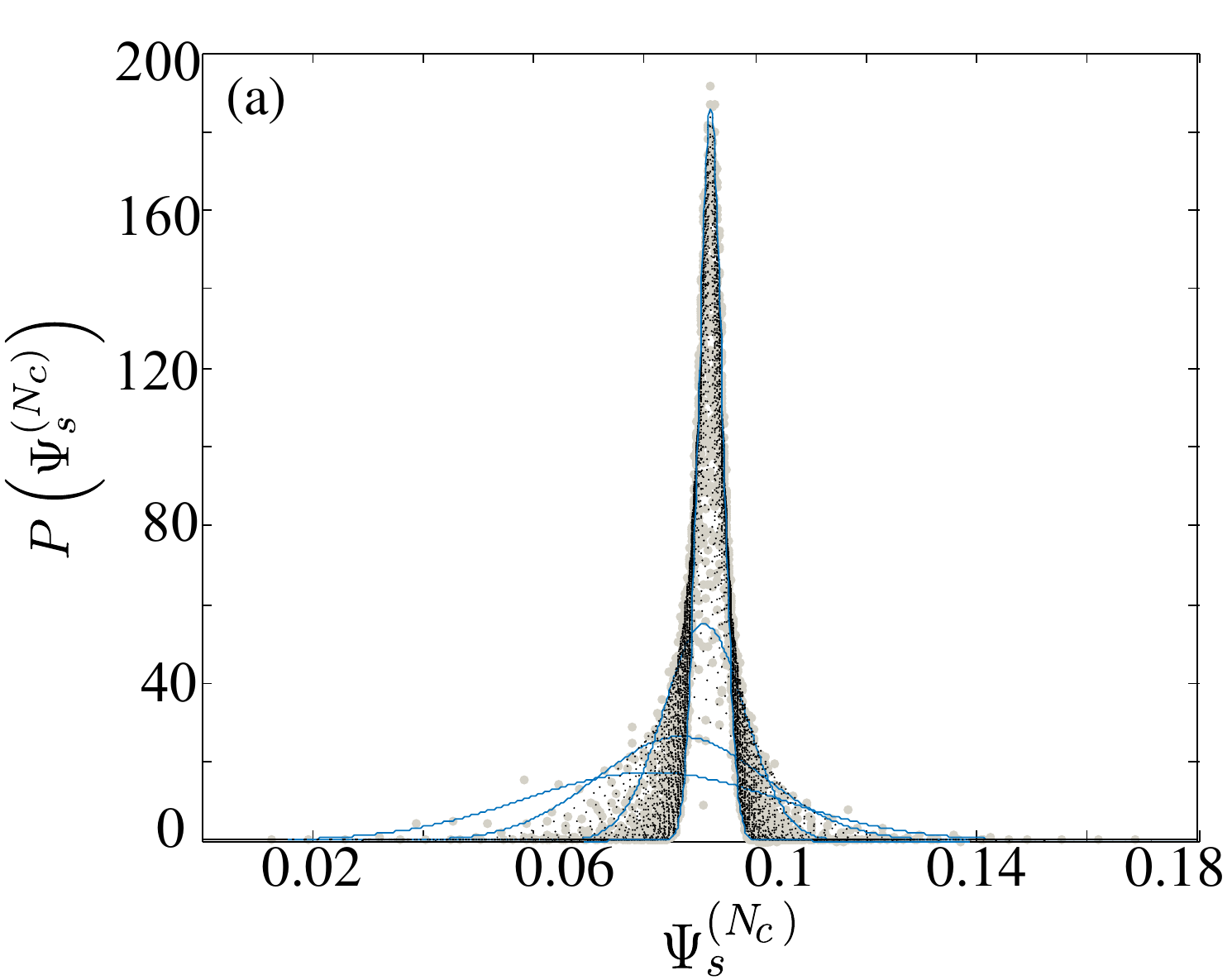}
\includegraphics[width=0.32\textwidth]{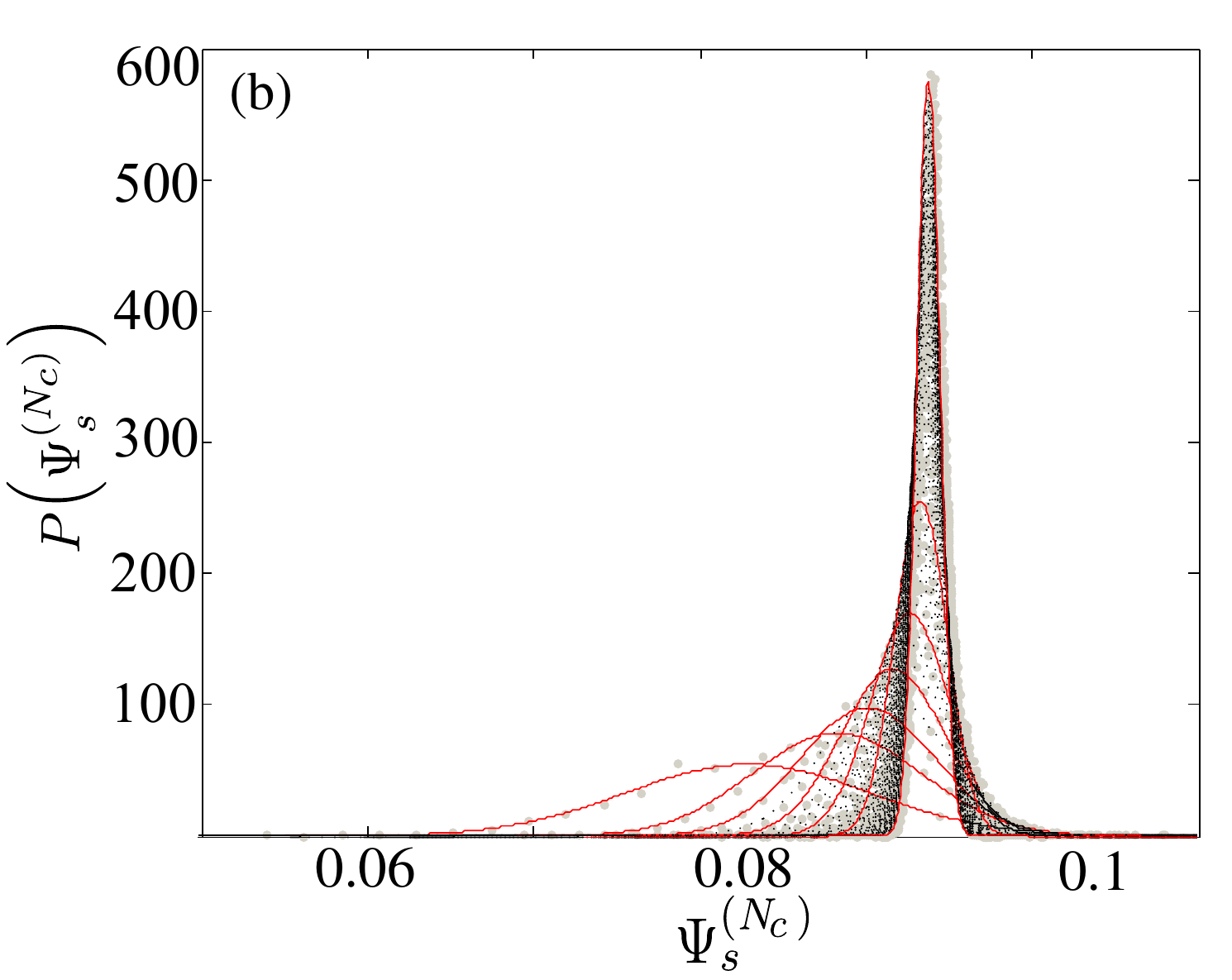}
\includegraphics[width=0.32\textwidth]{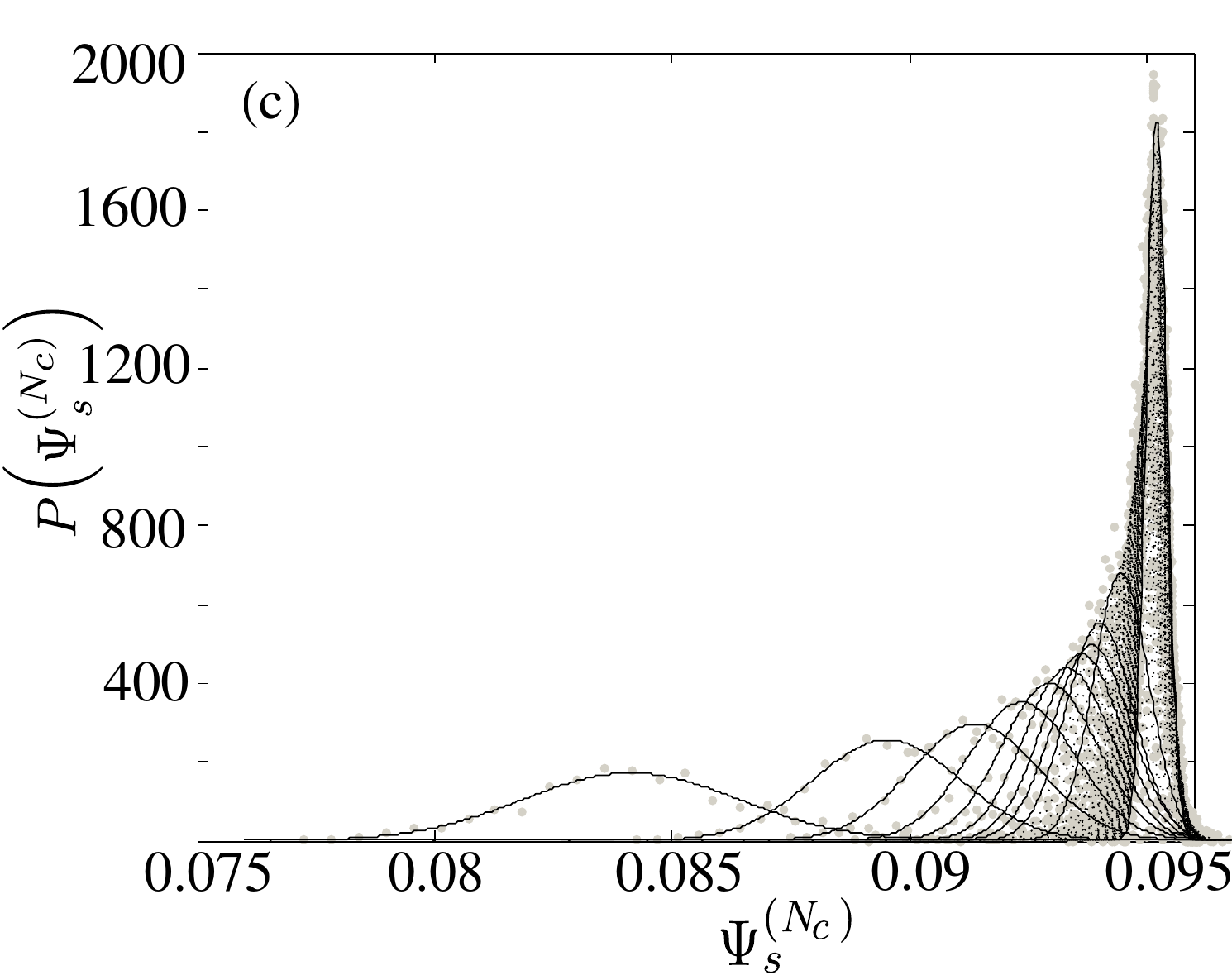}
\caption[Distribution of the CGF estimator]{\label{fig:PsiDistribution2} Distribution $P \left(  \Psi_s^{(N_{c})}  \right)$ of the CGF estimator $\Psi_s^{(N_{c})}$ for \textbf{(a)} $N_{c}=10$, \textbf{(b)} $N_{c}=100$ and \textbf{(c)} $N_{c}=1000$ and for simulation times $t \in \left[10,1000 \right]$. Each realization ($R=10^{4}$ for each simulation time) is shown with gray dots meanwhile its respective Gaussian fit (Eq.~\eqref{eq:DPSI1}) is shown with a dotted or a continuous curve. The dispersion of  $\Psi_s^{(N_{c})}$ is wider for shorter simulation times and small $N_{c}$. The mean value of the distribution converges to the theoretical value as the simulation time and the number of clones increase.}
\end{figure*}

\begin{figure}[t!]
\centering
\includegraphics[width=0.55\textwidth]{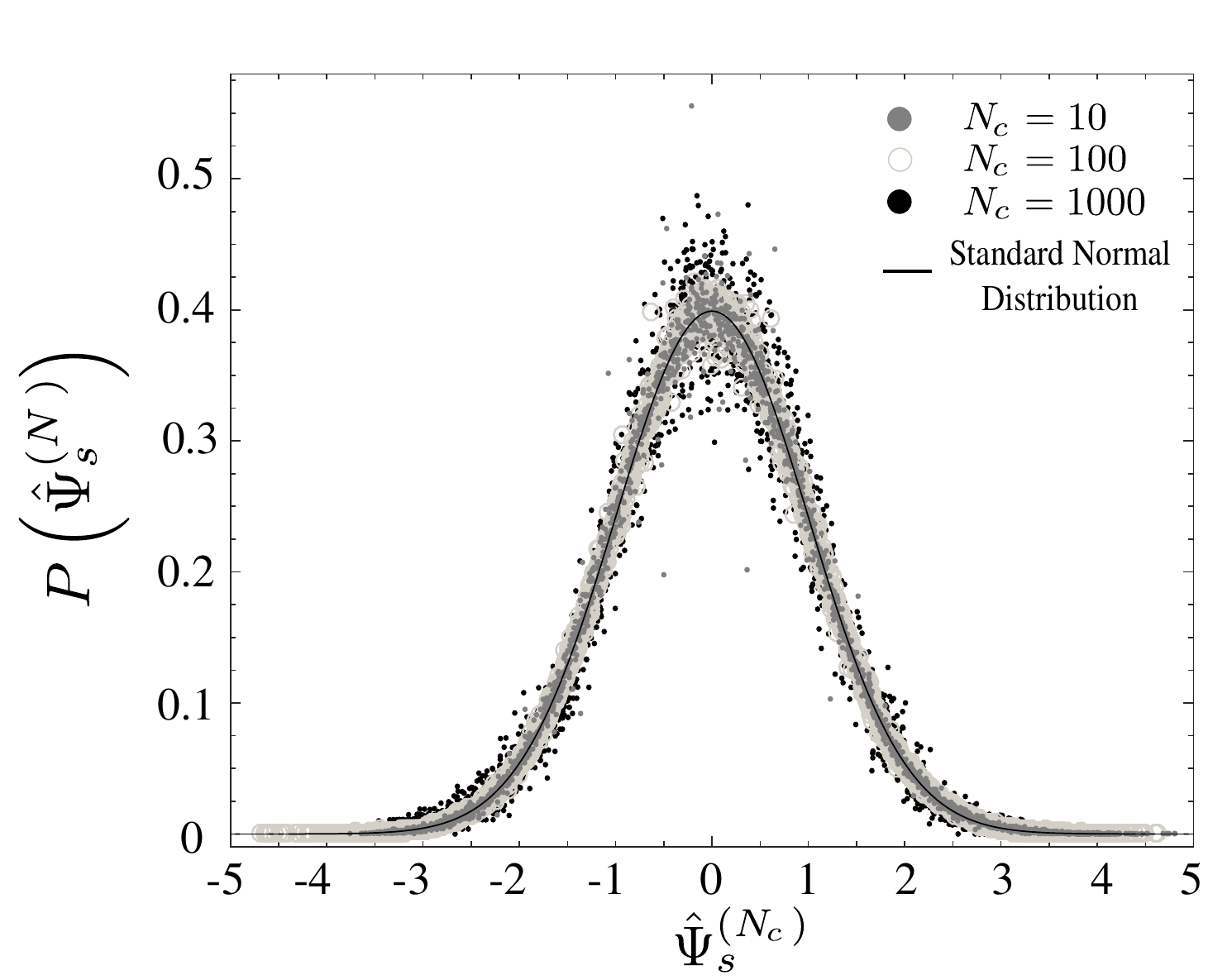}
\caption[Collapse of the CGF Estimator Distribution ]{\label{fig:NormDistribution} The distribution function of the rescaled variable $\hat{\Psi}_s^{(N_{c})}$ (Eq.~(\ref{eq:PSInorm})). Compatible with the central limit theorem, a collapse of the distribution function into a standard normal distribution for different number of clones is observed.}
\end{figure}

From relation~\eqref{eq:PSI1}, one can infer that the dispersion of the distribution of $\Psi_{s}^{(N_{c})}$ depends on the simulation time~$t$. This determines whether or not a large number of realizations $R$ is required in order to minimize the statistical error. In fact, as seen in Fig.~\ref{fig:PsiDistribution2}, the dispersion of $\Psi_{s}^{(N_{c})}$  is concentrated around its mean value, which approaches the analytical value $\psi(s)$ as the simulation time and the number of clones increase. 
We numerically confirm that these distributions are well-approximated by a Gaussian distribution
\begin{equation} \label{eq:DPSI1}
P \left(  \Psi_s^{(N_{c})}  \right) \sim A\, e^{-\frac{1}{C^{2}} \left( \Psi_s^{(N_{c})} -B \right) ^2}
\end{equation}
where the parameter $B$ is equal to $\overline{\Psi_{s}^{(N_{c})}(T)}$ and the parameters $A$ and $1/C^2$ are respectively of the order of $N_{c}^{1/2}$ and $N_{c}$. 
%
%
A mathematical argument to explain this obtained Gaussian distribution is given as follows: At any given time (not necessarily at $T$), let us perform the following rescaling
\begin{equation} \label{eq:PSInorm}
 \hat{\Psi}_s^{(N_{c})} = \frac{\Psi_{s}^{(N_{c})} - \overline{ \Psi_{s}^{(N_{c})} }}{\sigma_{\Psi_{s}^{(N_{c})}}},
\end{equation}
where 
\begin{equation} \label{eq:sigmaPsi1}
\sigma_{\Psi_{s}^{(N_{c})}}^{2} = \frac{1}{R-1} \sum\limits_{r = 1}^{R} \left \vert \big( \Psi_{s}^{(N_{c})} \big)_{r} - \overline{ \Psi_{s}^{(N_{c})} }    \right \vert^{2}
\end{equation}
is the variance of the $R$ realizations of $\Psi_{s}^{(N_{c})}$.
Then, this rescaling produces a collapse of the distributions $P \big(  \hat{\Psi}_s^{(N_{c})} \big)$, for any  $t$ and any $N_{c}$ (Fig.~\ref{fig:NormDistribution}). We remark then that the CGF estimator~\eqref{eq:PSI1} is an additive observable of the history of the population, which follows a Markov dynamics. Hence, the rescaled estimator $\hat{\Psi}_s^{(N_{c})} $ follows a standard normal distribution in the large time limit, according to the central limit theorem (CLT):
\begin{equation} \label{eq:DPSI2}
P \big(  \hat{\Psi}_s^{(N_{c})} \big) = \frac{1}{\sqrt{2 \pi}} e^{-\frac{1}{2} \big( \hat{\Psi}_s^{(N_{c})} \big) ^2}.
\end{equation}
%
The verification of the CLT allows us to ensure if the steady-state of the population dynamics has been reached. Note that in general the typical convergence time to the steady state is larger than the inverse of the spectral gap of the biased evolution operator as discussed in Secs.~\ref{sec:bdp} and~\ref{Time Delay Correction}.

By considering the scaling~\eqref{eq:PSInorm} we focus only on the small fluctuations of $\Psi_{s}^{(N_{c})}$ around $\overline{ \Psi_{s}^{(N_{c})} }$. But in general, the distribution function is not Gaussian, and in that case we need to consider a large deviation principle as below.

\section[Logarithmic Distribution of CGF Estimator]{\quad Logarithmic Distribution of CGF Estimator}
\begin{figure*}[t]
\centering
\includegraphics[width=0.48\textwidth]{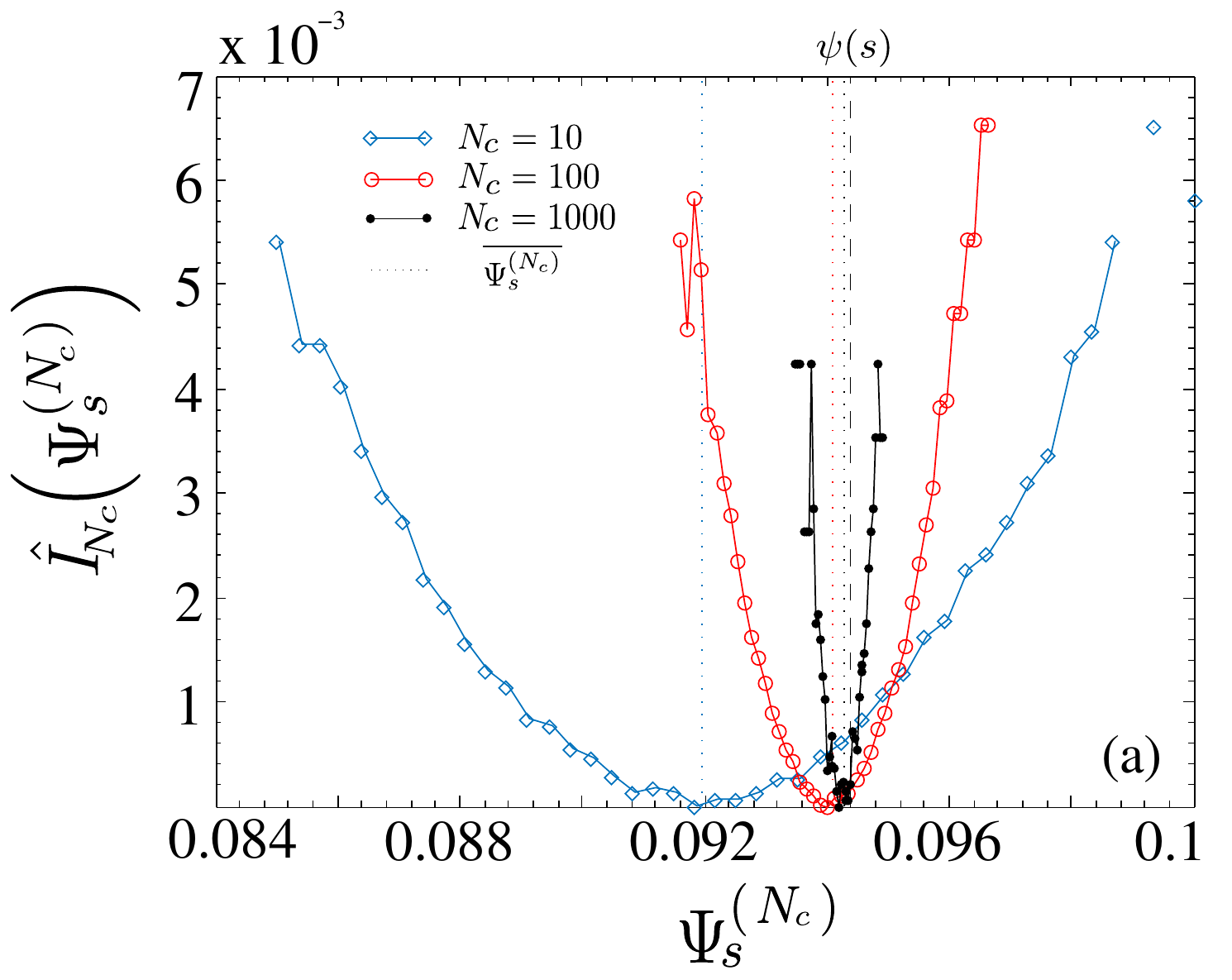}
\includegraphics[width=0.48\textwidth]{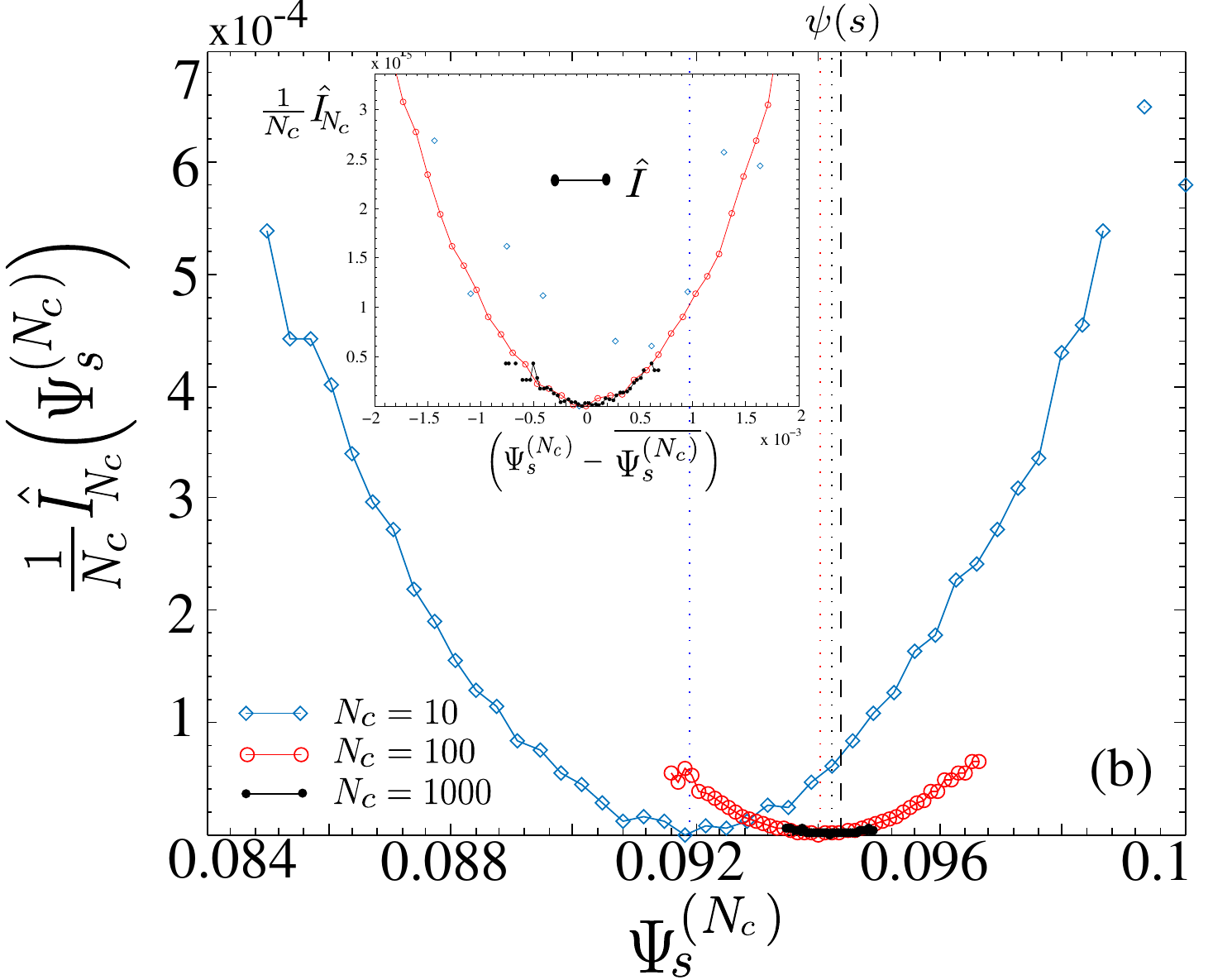}
\caption[Logarithmic Distribution]{\label{fig:I1} \textbf{(a)} Logarithmic distribution $\hat{I}_{N_c} \big( \Psi_s^{(N_{c})} \big)$ (Eq.~\eqref{eq:Ihat}).
Numerical evaluations were made for three fixed population sizes $N_{c} \in \{10,100,1000 \}$ with a fixed simulation time $T=1000$.
The logarithmic distribution presents a smaller width as $N_c$ increases.
The average over $R$ realizations of the CGF estimator $\overline{ \Psi_{s}^{(N_{c})} (T) }$ corresponds to the minimum of $\hat{I}_{N_c} \big( \Psi_s^{(N_{c})} \big)$ (dotted lines) and converges to the analytical value $\psi(s)$  (dashed lines) as $N_c\to\infty$. 
\textbf{(b)} Rescaled logarithmic distribution $\frac{1}{N_{c}}\hat{I}_{N_c} \big( \Psi_s^{(N_{c})} \big)$ as a function of $\Psi_s^{(N_{c})}$ and as a function of $\check{\Psi}_{s}^{(N_{c})} = \big( \Psi_s^{(N_{c})} - \overline{ \Psi_{s}^{(N_{c})}} \big)$ (inset) for a final simulation time $T=1000$.} 
\end{figure*}

Since $\Psi_{s}^{(N_{c})}$ is itself an additive observable of the dynamics of the ensemble of clones (chapter~\ref{chap:DiscreteTime}~\cite{partI}), the distribution of the CGF estimator $\Psi_{s}^{(N_{c})}$ satisfies itself a large deviation principle
\begin{equation} \label{eq:LDP1}
P \big(  \Psi_s^{(N_{c})}  \big) \sim e^{-t\: I_{N_c}\big( \Psi_s^{(N_{c})}  \big)},
\end{equation}
where $I_{N_c}\big( \Psi_s^{(N_{c})}  \big)$ is the rate function. This 
rate function could be evaluated in principle from the empirical distribution $P \big(\Psi_s^{(N_{c})} \big)$ as
\begin{equation} \label{eq:I1}
I_{N_c} \big( \Psi_s^{(N_{c})} \big) \approx -\frac{1}{t} \log P \big(  \Psi_s^{(N_{c})}  \big)  
\end{equation}
for a large $t$. 
Here we try to estimate the rate function from this equation. 
The numerical estimation of the right-hand side of the last expression at final simulation time $T$ is shown in Fig.~\ref{fig:I1}(a), where we have defined
\begin{equation} \label{eq:Ihat}
\hat{I}_{N_c} \big( \Psi_s^{(N_{c})} \big) \equiv -\frac{1}{t} \log P \big(  \Psi_s^{(N_{c})}  \big)   + \frac{1}{t} \log P \big(  \overline{ \Psi_{s}^{(N_{c})} }   \big)  
\end{equation}
so that $\hat{I}_{N_c} \big( \overline{ \Psi_{s}^{(N_{c})} } \big) = 0$. 
In the same figure, we also show $\overline{ \Psi_{s}^{(N_{c})} (T) }$ as vertical dotted lines which correspond to the minima of the logarithmic distribution $\hat{I}_{N_c} \big( \Psi_s^{(N_{c})} \big)$. As can be seen, these minima are displaced towards the analytical value $\psi(s)$ (shown with a dashed line) as \mbox{$N_{c}\to\infty$}. The logarithmic distribution $\hat I_{N_c}$ also becomes more concentrated as $N_{c}$ increases. 


Next, in order to study this decreasing of the width, 
we show
%
%
%
%
a rescaled logarithmic distribution
function $(1/N_c)\hat{I}_{N_c} \big( \Psi_s^{(N_{c})} \big)$
in Fig.~\ref{fig:I1}(b).
The minimum converges to the analytical value $\psi(s)$ (black dashed line) as $N_{c}\to\infty$. 
In the infinite-time infinite-size limit of $\Psi_s^{(N_{c})}$, it would be thus compatible with a logarithmic distribution function given by
\begin{equation} \label{eq:Int}
I \big( \Psi_s^{(N_{c})} \big) = - \lim_{N_c\to\infty} \:\frac{1}{N_c}\: \lim_{t\to\infty} \frac{1}{t} \log P \big(  \Psi_s^{(N_{c})} (t) \big) 
\end{equation}
which is shown (rescaled) with black dots in Fig.~\ref{fig:I1}(b). 
By performing the shift $\check{\Psi}_{s}^{(N_{c})} = \big( \Psi_s^{(N_{c})} - \overline{ \Psi_{s}^{(N_{c})} }  \big)$ we can see in the inset of Fig.~\ref{fig:I1}(b) the superposition of quadratic deviations of the numerical estimator $\Psi_{s}^{(N_{c})}$ around the minimum of $\hat{I_{N_c}}$ (especially for $N_c=100, 1000$). This indicates the decreasing of the fluctuation of CGF estimator proportional with both of $T$ and $N_c$ (see chapter~\ref{chap:DiscreteTime}~\cite{partI} for more detailed explanation). 
%


The obtained logarithmic distribution is well-approximated by a quadratic form, although these large deviations are in general not quadratic (chapter~\ref{chap:DiscreteTime}~\cite{partI}). 
This means that the direct observation discussed here cannot capture the large deviations of the CGF estimator (see also Ref.~\cite{rohwer_convergence_2015} for more detailed study of the direct estimation of rate functions).  
However we note that, for practical usage of the algorithm, we only consider small fluctuations described by central limit theorem,  although these large fluctuations might play an important role in more complicated systems, such as the ones presenting dynamical phase transitions.

\section[A Different CGF Estimator]{\quad A Different CGF Estimator}
\label{sec:two-estimators}
Normally, the CGF estimator is defined as an arithmetic mean over many realizations, as seen in Eq.~\eqref{eq:PSI1}. 
Here we show that another definition of the CGF estimator can be used, which
indeed provides better results than the ones from the standard estimator (in some parameter ranges). We define a new estimator as
\begin{equation} \label{eq:Lav}
 \Phi_{s}^{(N_{c})}  = \frac{1}{T} \log \overline{ \prod \limits_{i = 1}^{K_{r}} X_{i}^{r} },
\end{equation}
%
where we note that the average with respect to realizations are taken {\it inside} the logarithm.
As we discussed in Sec.~\ref{subsec:Different large deviation estimator}, this estimator provides a correct value of CGF $\psi(s)$ in the infinite-time infinite-$N_c$ limits. This is thanks to the fact that the distribution of $\Psi_{s}^{(N_{c})}$ concentrates around $\psi(s)$ in those limits (the so-called ``self-averaging'' property).
At any finite population, one can rewrite $\Phi_{s}^{(N_{c})} $ using the large-time LDF principle~\eqref{eq:LDP1} as follows:
\begin{align} \label{eq:Lav2}
 \Phi_{s}^{(N_{c})} 
& = \frac{1}{T} \log \overline{ e^{T \Psi_{s}^{(N_{c})}}}
\\
& = \frac{1}{T} \log 
\int d\Psi
\
e^{-T \left[I_{N_c}(\Psi)+\Psi\vphantom{|^I}\right]}
\end{align}
which proves that in the large-$T$ limit, 
\begin{equation} \label{eq:Lav3}
 \Phi_{s}^{(N_{c})} = \min_{\Psi} \left[I_{N_c}(\Psi)+\Psi\vphantom{|^I}\right],
\end{equation}
to be compared to
\begin{equation} \label{eq:Lav4}
\overline{ \Psi_{s}^{(N_{c})} } = \underset{\Psi}{\operatorname{argmin}}\ I_{N_c}(\Psi) .
\end{equation}

\begin{figure}[t]
\centering
\includegraphics[width=0.55\textwidth]{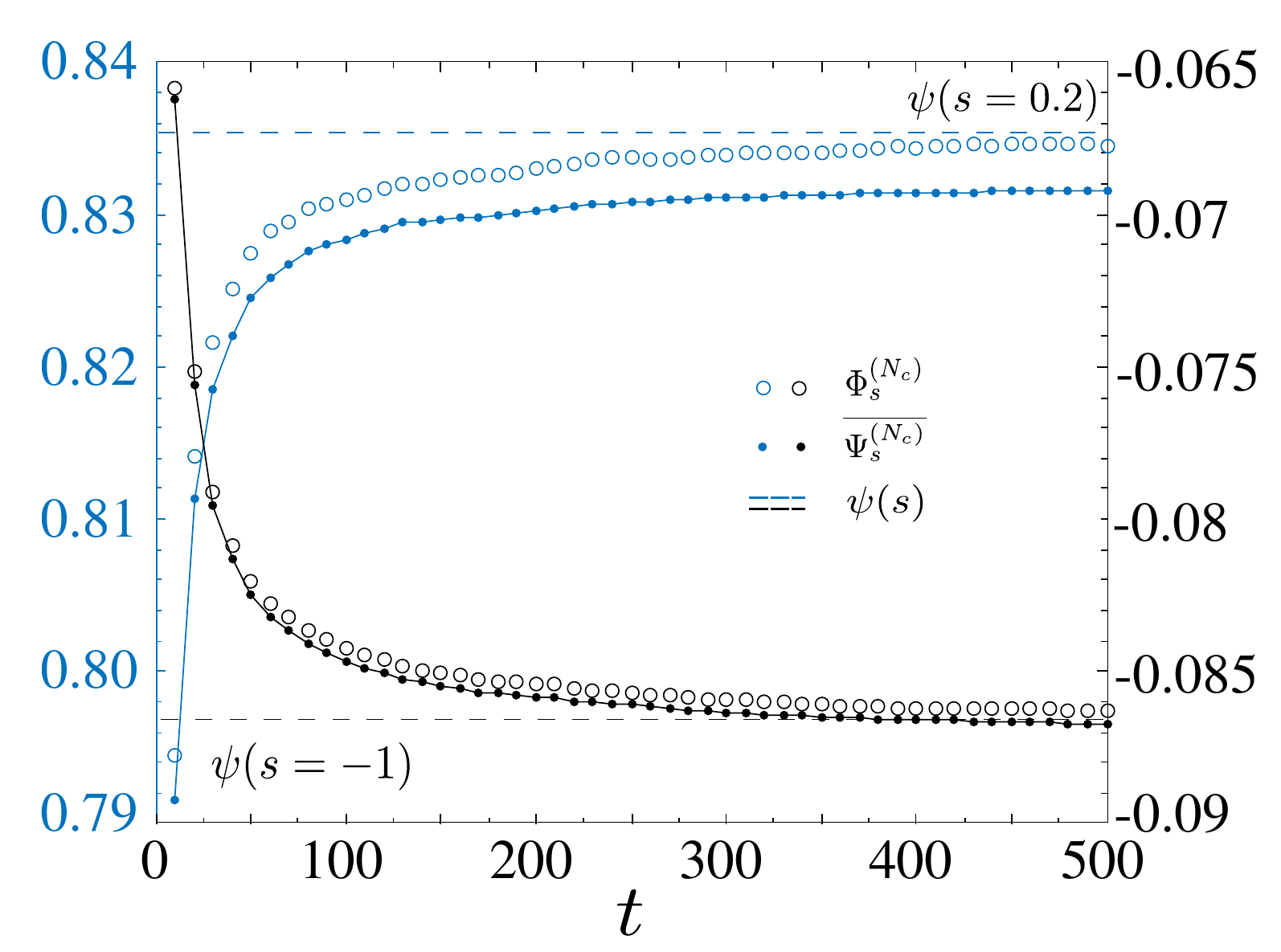}
\caption[Two Different CGF Estimators]{\label{fig:avLLav} Comparison between two different estimators of the large deviation function, $\overline{ \Psi_{s}^{(N_{c})} }$ (Eq.~\eqref{eq:PSI1}) shown in dots and $\Phi_{s}^{(N_{c})}$ (Eq.~\eqref{eq:Lav}) in circles, for the annihilation-creation dynamics (Sec.~\ref{sec:bdp}). The analytical value  $\psi({s})$ (Eq~\eqref{eq:PSIA}) is shown with a dashed line. Here we have also compared two different values of parameter $s=0.2$ (blue) and $s=-1$ (black). Additionally, $N_{c}=100$, $c=0.4$, $T = 500$ and $R=500$.
As discussed in the text, $\Phi_{s}^{(N_{c})}$ provides a better numerical evaluation of the CGF at small $s$.
}
\end{figure}

On one hand, the definition~\eqref{eq:Lav} amounts to estimate $\psi$ from the exponential growth rate of the average of the final-$T$ population of many small (non-interacting) ``\mbox{islands}'', where the cloning algorithm would be operated. 
On the other hand, the estimator~\eqref{eq:PSI1} amounts to estimate $\psi$ from growth rate of a large ``island'' gathering the full set of the $R$ populations. The latter is thus expected to be a better estimator of $\psi(s)$ than the former because it corresponds to a large population, where finite-size effects are less important. As a consequence, the estimator $\Phi_{s}^{(N_{c})}$ appears \emph{a priori} to be worse estimator than $\overline{ \Psi_{s}^{(N_{c})} }$ of $\psi(s)$.
However, as shown in Sec.~\ref{subsec:Different large deviation estimator}, at small~$|s|$ and finite-$N_c$, a supplementary bias introduced by taking Eq.~\eqref{eq:Lav} in fact \emph{compensates} the finite-$N_c$ systematic error presented by Eq.~\eqref{eq:PSI1}, for a simple two state model.
Namely, the error is $O(sN_c^{-1})$ for Eq.~\eqref{eq:PSI1} while it is $O(s^2N_c^{-1})$ for Eq.~\eqref{eq:Lav}.
This fact is illustrated on Fig.~\ref{fig:avLLav}, where we show that at small $s=0.2$, $\Phi_{s}^{(N_{c})}$ provides a better estimation of $\psi(s)$ than $\overline{ \Psi_{s}^{(N_{c})} }$, while at larger $|s|$ ($s=-1$) the two estimators yield a comparable error.

\clearpage
\thispagestyle{empty}
\phantom{a}

\chapter[\quad Finite-Time and Finite-$N_c$ Scalings in the Large-$L$ Limit]{Breakdown of the Finite-Time \\ and Finite-$N_c$ Scalings \\ in the Large-$L$ Limit}
\label{chap:LargeL}
\section[Introduction]{\quad Introduction}

The analysis of the finite-$t$ and finite-$N_c$ scalings in the evaluation of the LDF 
was performed following two different approaches: an analytical one, in chapter~\ref{chap:DiscreteTime}~\cite{partI}, using a discrete-time version of the population dynamics algorithm~\cite{giardina_direct_2006}, and a numerical one, in chapter~\ref{chap:ContinuousTime}~\cite{partII}, using a continuous-time version~\cite{lecomte_numerical_2007,tailleur_simulation_2009}. 
In both cases, the systematic errors of these scalings were found to behave as $1/t$ and $1/N_c$ in the large-$t$ and large-$N_c$ asymptotics respectively. 
Moreover, it was shown how these scaling properties can be used in order to improve the LDF estimation by the implementation of a scaling method (Sec.~\ref{ssec:SM}). 
This was done considering that the asymptotic behavior of the estimator in the $t \to \infty$ and $N_c \to \infty$ limits may be interpolated from the data obtained from simulations at \textbf{finite and relative small} simulation time and number of clones. 
%

However, the validity of these scalings and the method efficiency were proved
only in cases for which the number of sites $L$ (where the dynamics occurs) 
was small: a simple two-states annihilation-creation dynamics~(Sec.~\ref{sec:bdp}) (in one site) and a one-dimensional contact process~(Sec.~\ref{sec:CP}) (with $L=6$ sites).
Here, we complement the results presented in chapters~\ref{chap:DiscreteTime}~\cite{partI} and~\ref{chap:ContinuousTime}~\cite{partII} by extending the analysis to a large-$L$ contact process. 
%
This is done by introducing the exponents $\gamma_{t}$ and $\gamma_{N_c}$ such that the generalized $\mathbf{t^{-\gamma_{t}}}$\textbf{-} and $\mathbf{N_{c}^{-\gamma_{N_c}}}$\textbf{-scalings} allow to characterize the scaling behavior in the large-$L$ limit where we verify that $t^{-1}$ and $N_c^{-1}$-scalings are no longer valid.

This chapter~\ref{chap:LargeL}~\cite{largeLCP} is organized as follows:
%
%
The generalization to large-$L$ systems of the finite-time and finite-$N_c$ scalings of the LDF 
%
is done in Sec.~\ref{subsec: tnScalingL}. 
We make use of these results in Sec.~\ref{sec: CPL100} where we check the validity of the $t^{-1}$- and $N_c^{-1}$-scalings (Sec.~\ref{sec: CGFL100}), their behavior in the $s$-modified dynamics (Sec.~\ref{sec: gamma_tn}) as well as the applicability of the scaling method (Sec.~\ref{sec: SML100}) for a contact process with $L=100$ sites.
This analysis is generalized in Sec.~\ref{sec: planeSL} where we characterize the finite-$t$ and finite-$N_c$ scalings of the LDF in the plane $s-L$. Before presenting our conclusions in Sec.~\ref{sec:conclusion}, we discuss about the effects of the dynamical phase transition 
in the scalings in Sec.~\ref{sec:DPTcp}. 
\section[Finite Scalings of the Large Deviation Function Estimator]{\quad Finite Scalings of the Large Deviation Function Estimator}
\label{sec: Scaling}
Below, we summarize the finite-time and finite-$N_c$ scalings of the CGF estimator and its generalization to large-$L$ systems.

\subsection[Large-Time and Large-$N_{c}$ Limit]{\quad Large-Time and Large-$N_{c}$ Limit}
\label{subsec: tnScaling}

When we analyze the time behavior of the CGF estimator~\eqref{eq:PSI2} 
for a fixed number of clones $N_c$,
we observe 
this can be well described by a curve $f_{t}^{(N_{c})}$~\eqref{eq:fitdef} indicating the existence of a $t^{-1}$-convergence to the value $f_{\infty}^{(N_{c})}$. We call this $\mathbf{t^{-1}}$\textbf{-scaling} and is valid independently if $N_c$ is small or large. 
The curve $f_{t}^{(N_{c})}$ is determined from a fit in time over $\overline{ \Psi_s^{(N_c)}(t)}$ up to (the final simulation) time $T$ and allows the extraction of the infinite-time limit of the CGF estimator $f_{\infty}^{(N_{c})}=\lim_{t\rightarrow \infty}\overline{ \Psi_{s} ^{(N_{c})}(t)}$ which provides a better CGF estimation 
\footnote{Additionally, the behavior of the standard CGF estimator $\overline{ \Psi_{s}^{(N_{c})} (T) }$ as a function of the population size $N_c$ is well described by a behavior of the form~\eqref{eq:PSI4N}
\begin{equation*}
g_{N_{c}}^{(T)} = g_{\infty}^{(T)} + \tilde b_{N_{c}}^{(T)}N_{c}^{-1}
\end{equation*}
indicating that $\overline{ \Psi_{s}^{(N_{c})} (T) }$ also converges to its infinite-$N_c$ limit $g_{\infty}^{(T)}=  \lim_{N_c\to \infty}  \overline{ \Psi_{s}^{(N_{c})} (T) }$ with an error proportional to $1/N_c$.}.

When we repeat this procedure for different values of population size $N_c \in \vec{N}_{c}= \{ N_{c}^{(1)},...,N_{c}^{(j)} \}$, extracting in each case the corresponding 
$f_{\infty}^{(N_{c})}$, we observe they exhibit $1/N_c$ corrections in $N_c$ ($\mathbf{N_{c}^{-1}}$ \textbf{-scaling}).
In other words, the $f_{\infty}^{(N_c)}$'s satisfy a equation of the form~\eqref{eq:PSIinfinf} which can be obtained from a fit in $N_{c}$ over the extracted $f_{\infty}^{(N_{c})}$'s. 
Thus, the $t^{-1}$- and $N_{c}^{-1}$-scalings of the CGF estimator are given by Eqs.~\eqref{eq:fitdef} and~\eqref{eq:PSIinfinf}, i.e., 
\begin{align}
f_{t}^{(N_{c})}  = f_{\infty}^{(N_{c})} + b_{t}^{(N_{c})} t^{-1}, \\
f_{\infty}^{(N_c)} = f_{\infty}^{\infty} + b_{\infty}^{(N_c)} N_c^{-1}.
\end{align}
%
%
These equations imply that $\overline{ \Psi_s^{(N_c)}(t)}$ converges to its infinite-$t$ and infinite-$N_c$ limit, $f_{\infty}^{\infty} = \lim_{N_{c}\to\infty} f_\infty^{(N_c)}$, proportionally to $1/t$ and $1/N_c$. 
Importantly, this limit can be obtained using a small number of clones and simulation time by making use of the \textbf{scaling method} (Sec.~\ref{ssec:SM}~\cite{partII}). The results obtained for $f_{\infty}^{\infty}$ rendered a better estimation of $\psi(s)$ than the standard estimator $\overline{ \Psi_{s}^{(N_{c})}(t) }$ evaluated for $N_c = \max \vec{N_c}$ and for $t = T$.
%
%
\subsection[Scalings in the Large-$L$ Limit]{\quad Scalings in the Large-$L$ Limit}
\label{subsec: tnScalingL}
In order to verify whether the 
scalings observed in small 
systems are also valid in the large-$L$ limit, 
%
%
we assume that the CGF estimator 
can be described by equations of the form
\begin{align}
\label{eq:tScal2}
\chi_{t}^{(N_{c})} &\equiv 
 \chi_{\infty}^{(N_{c})} + \kappa_{t}^{(N_{c})} t^{-\gamma_{t}}, \\
\label{eq:nScal2}
\chi_{\infty}^{(N_c)} &\equiv 
 \chi_{\infty}^{\infty} + \kappa_{\infty}^{(N_c)} N_c^{-\gamma_{N_{c}}},
\end{align}
redefining in a more general way the scalings~\eqref{eq:fitdef} and~\eqref{eq:PSIinfinf}.
We will refer to Eq.~\eqref{eq:tScal2} as
$\mathbf{t^{-\gamma_{t}}}$\textbf{-scaling} whereas Eq.~\eqref{eq:nScal2} as 
$\mathbf{N_{c}^{-\gamma_{N_{c}}}}$ \textbf{-scaling}.
The problem reduces in determining the exponents $\gamma_{t}$ and $\gamma_{N_{c}}$ in order to verify if effectively $\gamma_{t} \approx 1$ and $\gamma_{N_{c}} \approx 1$ and
whether the terms $\chi_{\infty}^{(N_{c})}$ and $\chi_{\infty}^{\infty}$ represent the limits in $t \to \infty$ and $N_{c} \to \infty $ of the CGF estimator.
Thus, a value of the exponent $\gamma_{t} \approx 1$, verifies $\chi_{\infty}^{(N_{c})} \approx f_{\infty}^{(N_{c})} $ and $\gamma_{N_{c}} \approx 1$, verifies $\chi_{\infty}^{\infty} \approx f_{\infty}^{\infty}$.
This is done in Sec.~\ref{sec: CPL100} on a contact process with $L=100$ sites. Below we describe the procedure followed in order to obtain these exponents \footnote{\label{foot:Gamma_n} Additionally to Eqs.~\eqref{eq:tScal2} and~\eqref{eq:nScal2}, the $N_c$-behavior of $\overline{ \Psi_{s}^{(N_{c})} (T) }$ can be described by the equation
\begin{equation}
\label{eq:nScal3}
\chi_{N_{c}}^{(T)} = \chi_{\infty}^{(T)} + \tilde \kappa_{N_{c}}^{(T)}N_{c}^{-\gamma_{N_{c}}^{T}},
\end{equation}
where $\chi_{\infty}^{(T)} = \lim_{N_{c}\to\infty} \overline{ \Psi_{s}^{(N_{c})} (T) }$. Here it is important to remark that both $\chi_{\infty}^{(N_{c})}$ and $\overline{ \Psi_{s}^{(N_{c})} (T) }$ scale in the same way in $N_c$. In other words,
 $\gamma_{N_{c}} \approx \gamma_{N_{c}}^{T}$.}.

\subsubsection[Determination of the Exponents $\gamma_{t}$ \& $\gamma_{N_{c}}$]{\quad Determination of the Exponents $\gamma_{t}$ \& $\gamma_{N_{c}}$}
\label{sec:exp}
\begin{figure*} [t!]
\centering
\includegraphics[width=0.48\textwidth]{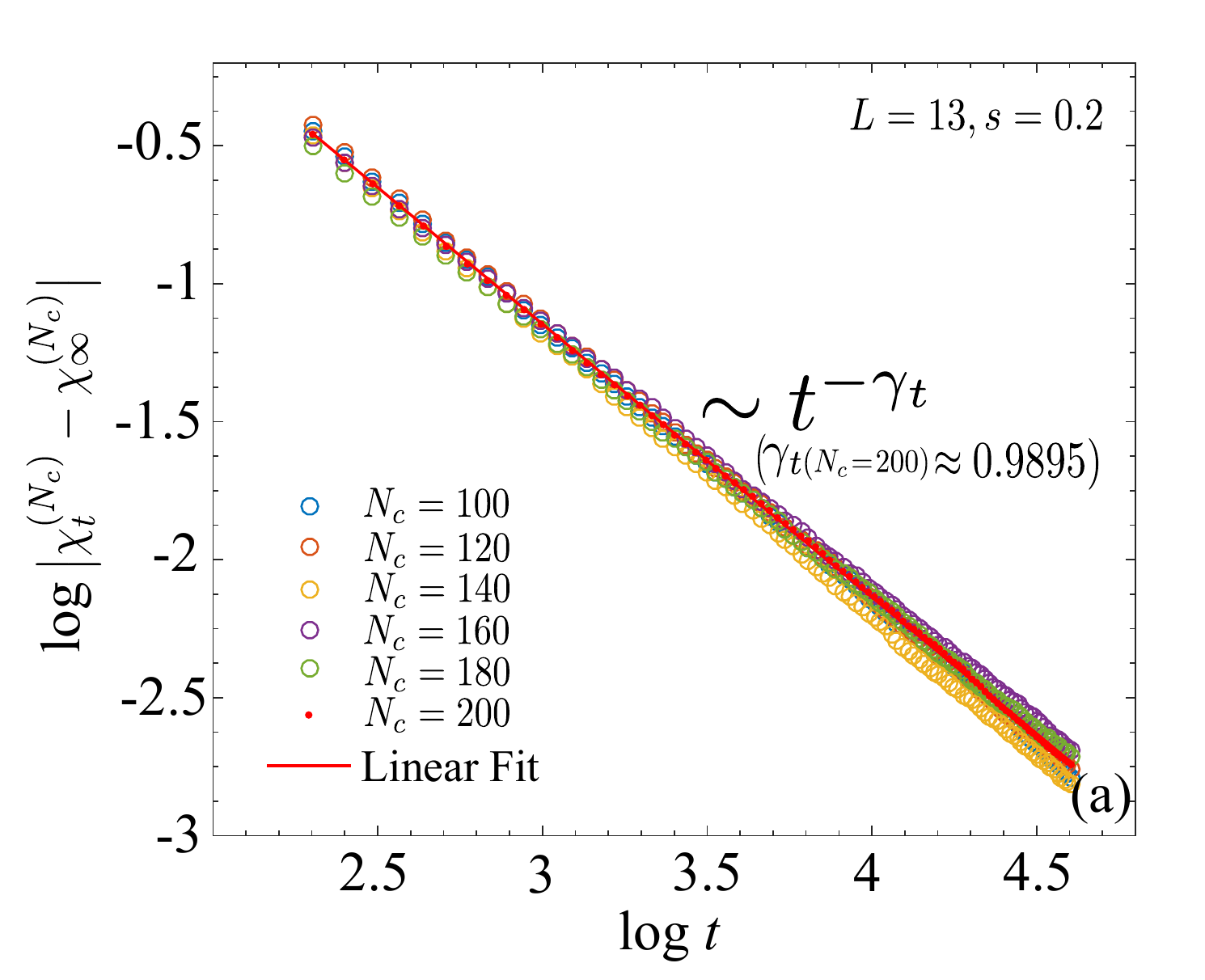}
\includegraphics[width=0.48\textwidth]{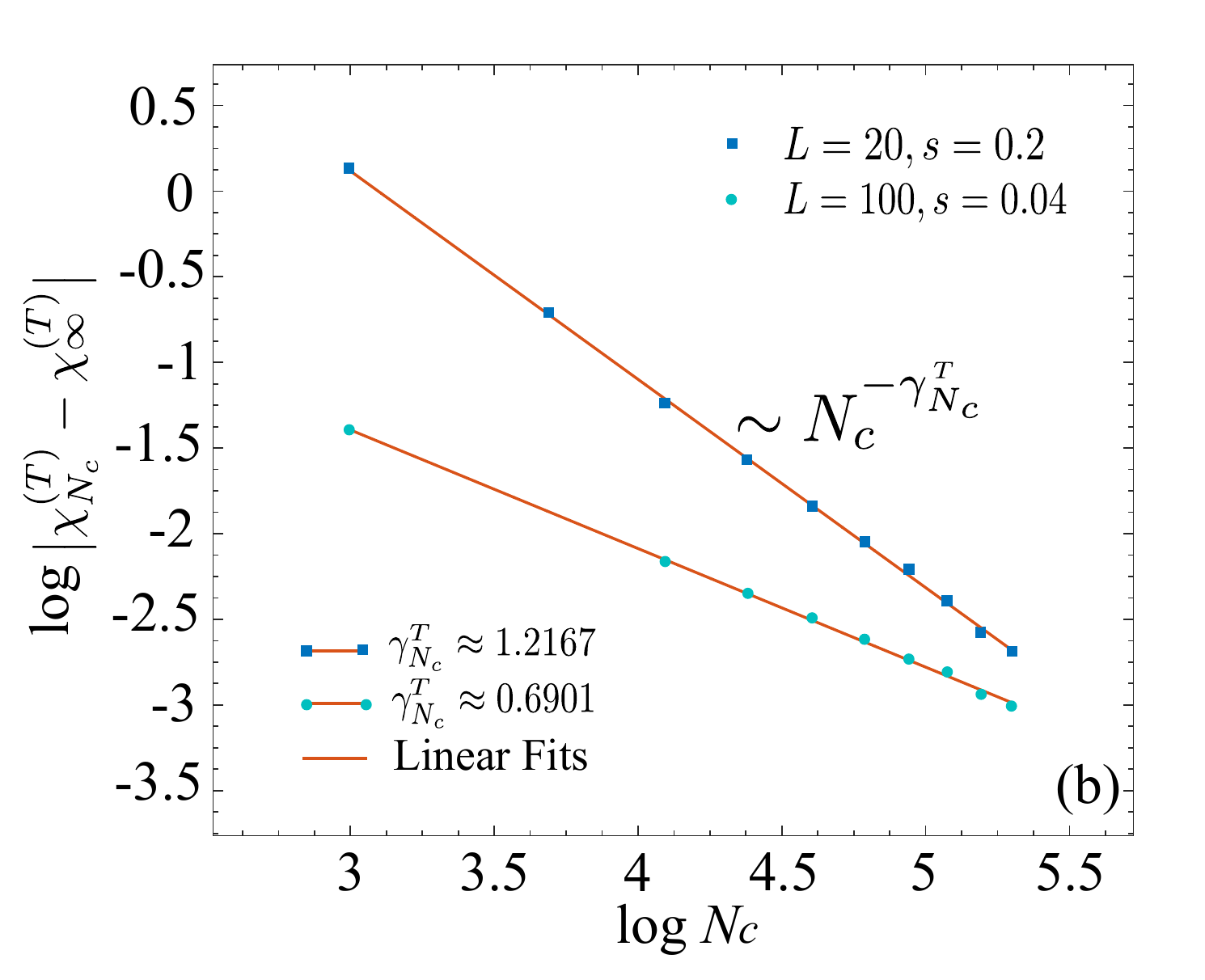}
\centering
\caption[Finite Scalings in the Large-L Limit]{\label{fig:powerlaw} Exponents \textbf{(a)} $\gamma_{t}$ \& \textbf{(b)} $\gamma_{N_{c}}$ which characterize the finite-time and -size scalings of the CGF estimator of the activity for a contact process with $\lambda=1.75$ and $h = 0.1$. These exponents were determined from the slope of a linear fit in log-log scale over Eqs.~\eqref{eq:tScal2PL} and~\eqref{eq:nScal3}, respectively. For $\gamma_{t}$, we used $L=13$, $s=0.2$ and $N_c \in \vec{N_c} = \{100,120,...,200 \}$. Meanwhile,  $\gamma_{N_{c}}$ was computed for $L=20$ and $s=0.2$, and for $L=100$ and $s=0.04$. Additionally, in all the cases $T=100$ and $R = 500$. 
}
\end{figure*}

From Eqs.~\eqref{eq:tScal2} and~\eqref{eq:nScal2} we expect that, independently of $N_c$, $T$, $L$ or $s$, a power law behavior of the form
\begin{align}
\label{eq:tScal2PL}
\big \vert \chi_{t}^{(N_{c})} - \chi_{\infty}^{(N_{c})} \big \vert &\sim  t^{-\gamma_{t}}, \\
\label{eq:nScal2PL}
\big \vert \chi_{\infty}^{(N_c)} - \chi_{\infty}^{\infty} \big \vert &\sim  N_c^{-\gamma_{N_{c}}},
\end{align}
be observed. Thus, the exponents $\gamma_{t}$ \& $\gamma_{N_{c}}$ can be obtained from the slope of a straight curve in log-log scale of Eqs.~\eqref{eq:tScal2PL} and~\eqref{eq:nScal2PL}. This can be seen in Fig.~\ref{fig:powerlaw}. Despite only some representative configurations have been considered, we confirm this power law behavior independently of the parameters chosen.
%
%

\section[Finite Scalings for a Large-$L$ Contact Process]{\quad Finite Scalings for a Large-$L$ Contact Process}
\label{sec: CPL100}
\begin{figure*} [t]
\centering
\includegraphics[width=0.48\textwidth]
{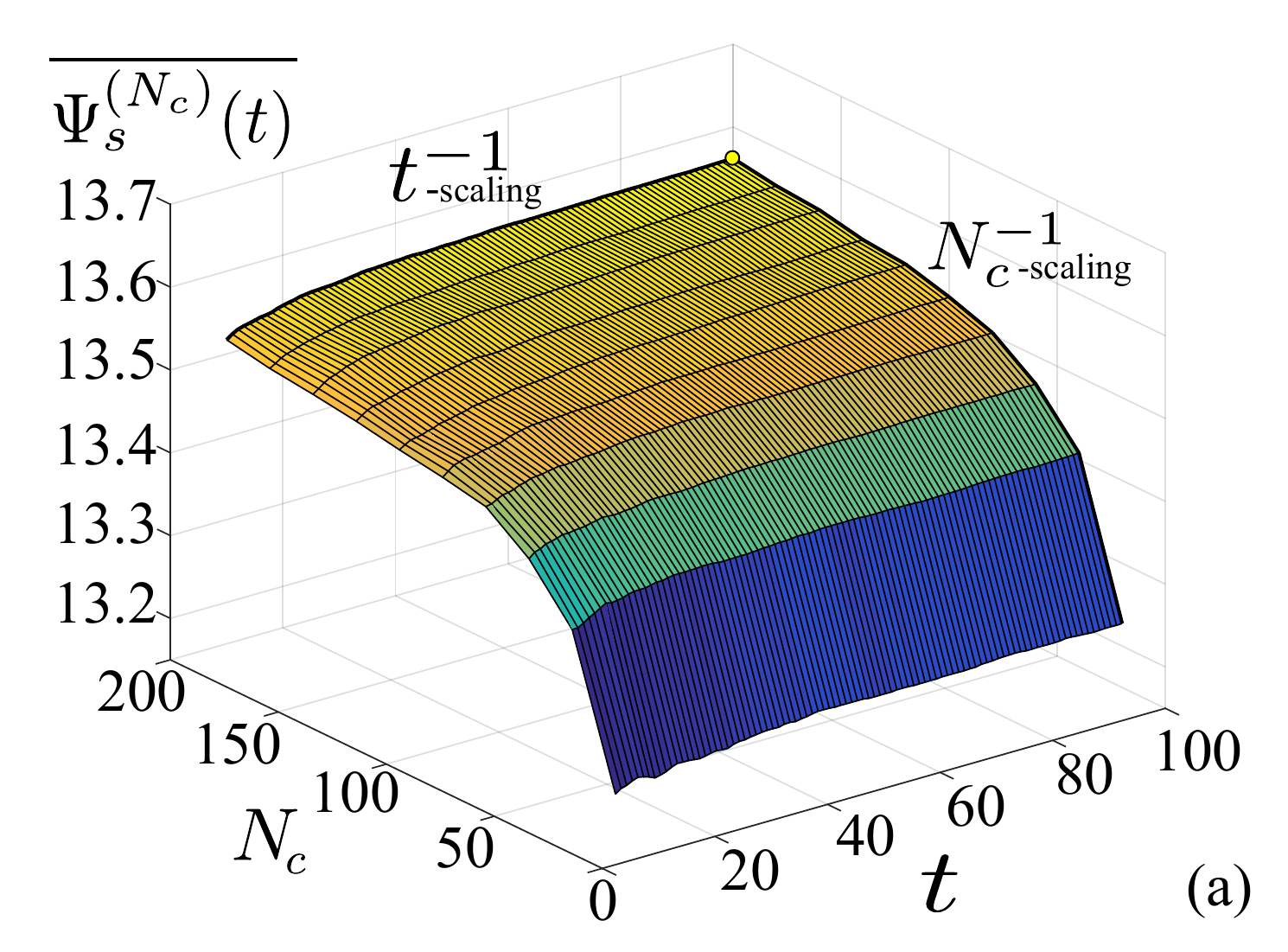}
\includegraphics[width=0.48\textwidth]
{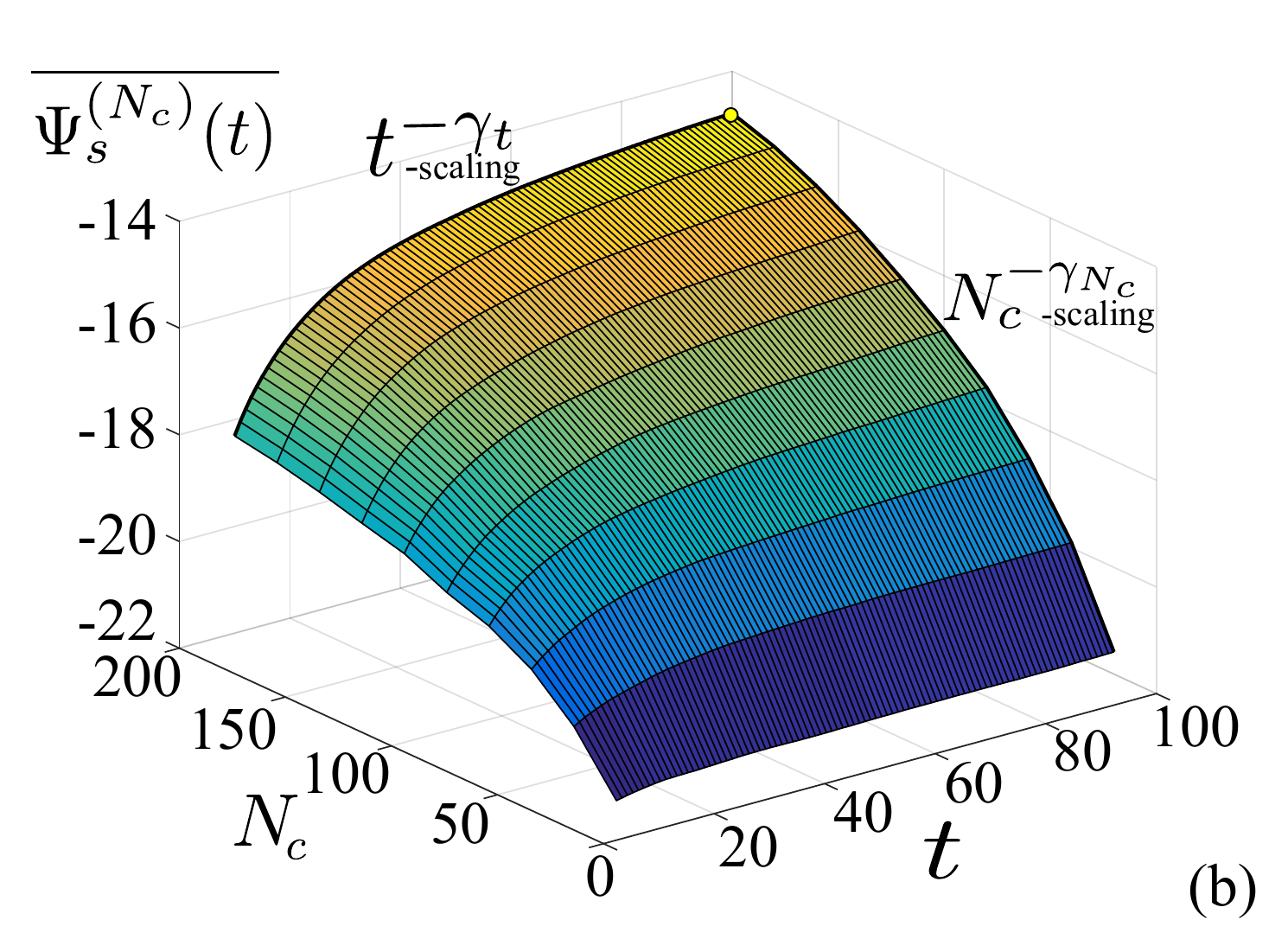}
\caption[CGF Estimator in the Large-L Limit]{\label{fig:surfPSIL100} Large deviations of the activity for a contact process with $L=100$, $\lambda=1.75$  and $h = 0.1$. The CGF estimator  $\overline{ \Psi_{s}^{(N_{c})} (t) }$ (Eq.~\eqref{eq:PSI1}) is presented as a function of time $t$ and the number of clones $N_c$ for \textbf{(a)} $s = -0.1$ and \textbf{(b)} $s = 0.2$. These surfaces were computed using the continuous-time cloning algorithm up a final simulation time $T=100$, $\vec{N}_{c} = \{20,...,200 \}$ and $R =500$ realizations. The $N_{c}^{-1}$-scaling observed in small-$L$ systems holds only for $s =-0.1$ whereas for $s=0.2$ a $N_{c}^{-\gamma_{N_{c}}}$-scaling is observed ($\gamma_{N_{c}}(s=0.2) \approx -0.16$). Similarly, for the time-scaling for which $\gamma_{t}(s=0.2) \approx 0.7$.
}
\end{figure*}

In Fig.~\ref{fig:surfPSIL100}, we compare the behavior of $\overline{\Psi_{s}^{(N_{c})}(t)}$ as function of $t$ and $N_{c}$, for two representative values of the parameter $s$, $s=-0.1$ (left) and $s=0.2$ (right). The size of the system is $L=100$ sites. Each point of these surfaces was obtained using the cloning algorithm (Eq.~\eqref{eq:PSI1}) up to time $T=100$, for $\vec{N_c} = \{20,40,...,180,200 \}$ and for $R=500$ realizations. 
The best possible CGF estimation 
(i.e., at largest $T$ and $N_c$)
in both cases is shown with solid circles
which, according to 
the results presented in 
chapter~\ref{chap:ContinuousTime}~\cite{partII}, could be improved by using the $t^{-1}$ and $N_c^{-1}$-scalings (if still valid for large-$L$).

\subsection[Finite-Time and Finite-$N_c$ Scalings]{\quad Finite-Time and Finite-$N_c$ Scalings}
\label{sec: CGFL100}

Although the exponents $\gamma_{t}$ and $\gamma_{N_{c}}$ can be  computed in principle for any value of $N_c \in \vec{N_c}$ and for any $t \leq T$, as we saw above, 
%
from now on, we will consider these exponents defined at the highest number of clones and at final simulation time, i.e.,
\begin{align}
\label{eq:GammaT}
\gamma_{t} &:= \gamma_{t}(N_{c}=\max \vec{N_{c}}), \\
\label{eq:GammaN}
\gamma_{N_c} &:= \gamma_{N_c}(t = T).
\end{align}
Thus, the exponent $\gamma_{t}$ is obtained as described in Sec.~\ref{sec:exp} after adjusting Eq.~\eqref{eq:tScal2} to $\overline{ \Psi_{s}^{(N_c)}(t)}$ for $N_c=\max \vec{N_{c}}=200$. 
On the other hand, $\gamma_{N_{c}}$ is determined after fitting $\chi_{\infty}^{(N_c)}$ with Eq.~\eqref{eq:nScal2} at $T=100$ or, as $\gamma_{N_{c}} \approx \gamma_{N_{c}}^{T}$, after fitting $\overline{ \Psi_{s}^{(N_c)}(t=T)}$ using Eq.~\eqref{eq:nScal3}.
In simple words, these exponents can be obtained from an adequate fit over the thick curves in Fig.~(\ref{fig:surfPSIL100}). They characterize the finite-$t$ and finite-$N_c$ scalings of the large deviations of the dynamical activity $K$. 

Following this approach, we found that the $t^{-1}$-scaling~\eqref{eq:fitdef} is satisfied only for $s=-0.1$, 
%
meaning that the exponent $\gamma_{t}$ was found to be $\gamma_{t} \approx 1$. As a consequence, the parameter $\chi_{\infty}^{(N_{c})}$ obtained from Eq.~\eqref{eq:tScal2} effectively represents the limit in $t \to \infty$ of the CGF estimator, i.e., $\chi_{\infty}^{(N_{c})} \approx f_{\infty}^{(N_{c})}$. This is not the case for $s=0.2$ for which $\gamma_{t}(s=0.2) \approx 0.7$.
Similarly, 
a $N_{c}^{-\gamma_{N_{c}}}$-scaling is observed for $s = 0.2$, whereas for $s=-0.1$, the $N_c^{-1}$-scaling~\eqref{eq:PSIinfinf} holds. 
%
It is important to remark that a value of exponent $\gamma_{N_{c}}>0$ could still guaranty the convergence of the CGF estimator in the infinite-$N_{c}$ limit. However, even though $\gamma_{N_{c}}(t=10)>0$ at initial times, at final time $T$, the exponent is negative ($\gamma_{N_{c}}(t=T) \approx -0.16$), which would imply that $\chi_{\infty}^{\infty}$ does not corresponds to the $t,N_c \to \infty$ 
limit 
of the CGF estimator. This fact will be addressed later in Sec.~\ref{sec:DPTcp}.
%
%
Below, we present how the change in the scalings is produced depending on $s$.

\subsection[Exponents Characterization \& $s$-Dependence]{\quad Exponents Characterization \& $s$-Dependence}
\label{sec: gamma_tn}
%
Here, we consider values of $s$ ranging in the interval $s \in [-0.1,0.2]$. 
%
%
%
For $s<0$, the exponent $\gamma_{t}(s)$ varies around $1$. 
However for $s>0$, 
$\gamma_{t}$ deviates slightly from $1$ decreasing with $s$ up to $\gamma_{t} \approx 0.7$ at $s=0.2$. 
In order to describe the behavior of this exponent, results convenient to define 
$s'$ as the value of the parameter $s \in [s_a,s_b]$ such that $\gamma_{t}(s < s') \approx 1$, i.e., until which the $t^{-1}$-scaling holds. Thus,
\begin{equation}
\label{eq:Gammat_s}
\gamma_{t}(L=100):
    \begin{cases}
           \gamma_{t}(s) \approx 1, & \text{for}\ s<s'  \\         
           0<\gamma_{t}(s) <1, & \text{otherwise.}
    \end{cases}
\end{equation}
%
%
If the scaling holds $\forall s \in [s_a,s_b]$ (given some system size $L$), then $s' = s_b$. 

\begin{figure} [t]
\centering
\includegraphics[width=0.55\textwidth]{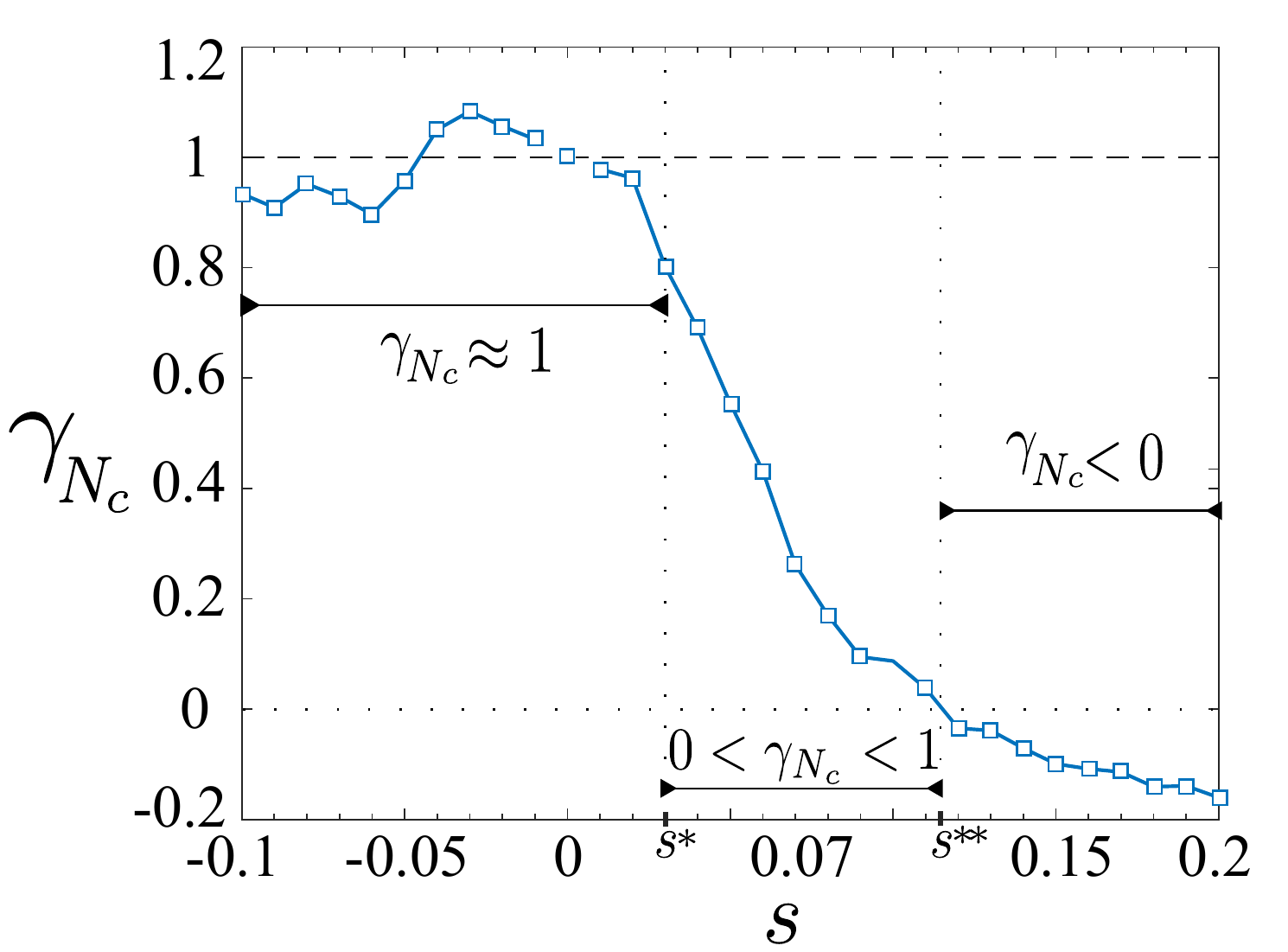}
\centering
\caption[Finite Population-Size Scaling in the Large-L Limit]{\label{fig:Gamma_tn_s} Dependence of the $N_c^{-\gamma_{N_c}}$-scaling with the parameter $s \in [-0.1,0.2]$. 
The exponent $\gamma_{N_{c}}(s)$ is obtained by fitting $\overline{\Psi_{s}^{(N_{c})}(T=100)}$ as function of $N_c \in \vec{N_c} = \{20,40,...,180,200 \}$ by Eq.~\eqref{eq:nScal3} in log-log scale as described in Sec.~\ref{sec:exp}.
Three stages of $\gamma_{N_{c}}(s)$ can be clearly distinguish for $L=100$: (\textit{i}) $\gamma_{N_{c}}(s<s^{*})\approx 1$, (\textit{ii}) $0 < \gamma_{N_{c}}(s^{*} < s < s^{**})<1$, and (\textit{iii}) $\gamma_{N_{c}}(s>s^{**}) < 0$. The exponent $\gamma_{N_{c}}$ for $s=0$ was set to $\gamma_{N_{c}}(s=0) = 1$.
}
\end{figure}

On the other hand, the value of $s \in [s_a,s_b]$ which signals the validity of the $N_{c}^{-1}$-scaling is denoted by $s^{*}$.
From this point, $\gamma_{N_{c}}$ decreases until eventually it becomes negative, as can be seen in Fig.~\ref{fig:Gamma_tn_s}. Here, we introduce $s^{**}$ such that $\gamma_{N_{c}}(s = s^{**}) = 0$. Thus, $\gamma_{N_{c}}<0$ for $s > s^{**}$. 
This behavior was not observed in chapter~\ref{chap:ContinuousTime}~\cite{partII} for $L=6$ for which the $N_{c}^{-1}$-scaling was valid independently of $s$. In those cases,  $s^{*}=s_b$ and $\nexists s^{**}$.
%
Instead of confirming for $L=100$ the $N_c^{-1}$-scaling of the CGF estimator presented in chapter~\ref{chap:ContinuousTime}~\cite{partII}, here we have been able to distinguish clearly three stages for the exponent $\gamma_{N_{c}}(s)$:
\begin{equation}
\label{eq:GammaN_s}
\gamma_{N_{c}}(L=100):
    \begin{cases}
          \gamma_{N_{c}}(s)\approx 1, & \text{for}\ s<s^{*}  \\   
           0<\gamma_{N_{c}}(s)<1,     & \text{for}\ s^{*}<s<s^{**}\\  
           \gamma_{N_{c}}(s) < 0,     & \text{for}\ s>s^{**}.
    \end{cases}
\end{equation}
%
%
%
The possibility of extracting the infinite-$t$ and infinite-$N_c$ limit of the CGF estimator relied on the validity of the $t^{-1}$- and $N_{c}^{-1}$-scalings. How the results obtained from the application of the scaling method are affected by $\gamma_{t}$ and $\gamma_{N_{c}}$ are presented below.

\subsection[Implementation of the Scaling Method]{\quad Implementation of the Scaling Method}
\label{sec: SML100}
\begin{figure} [t]
 \centering
\includegraphics[width=0.55\textwidth]{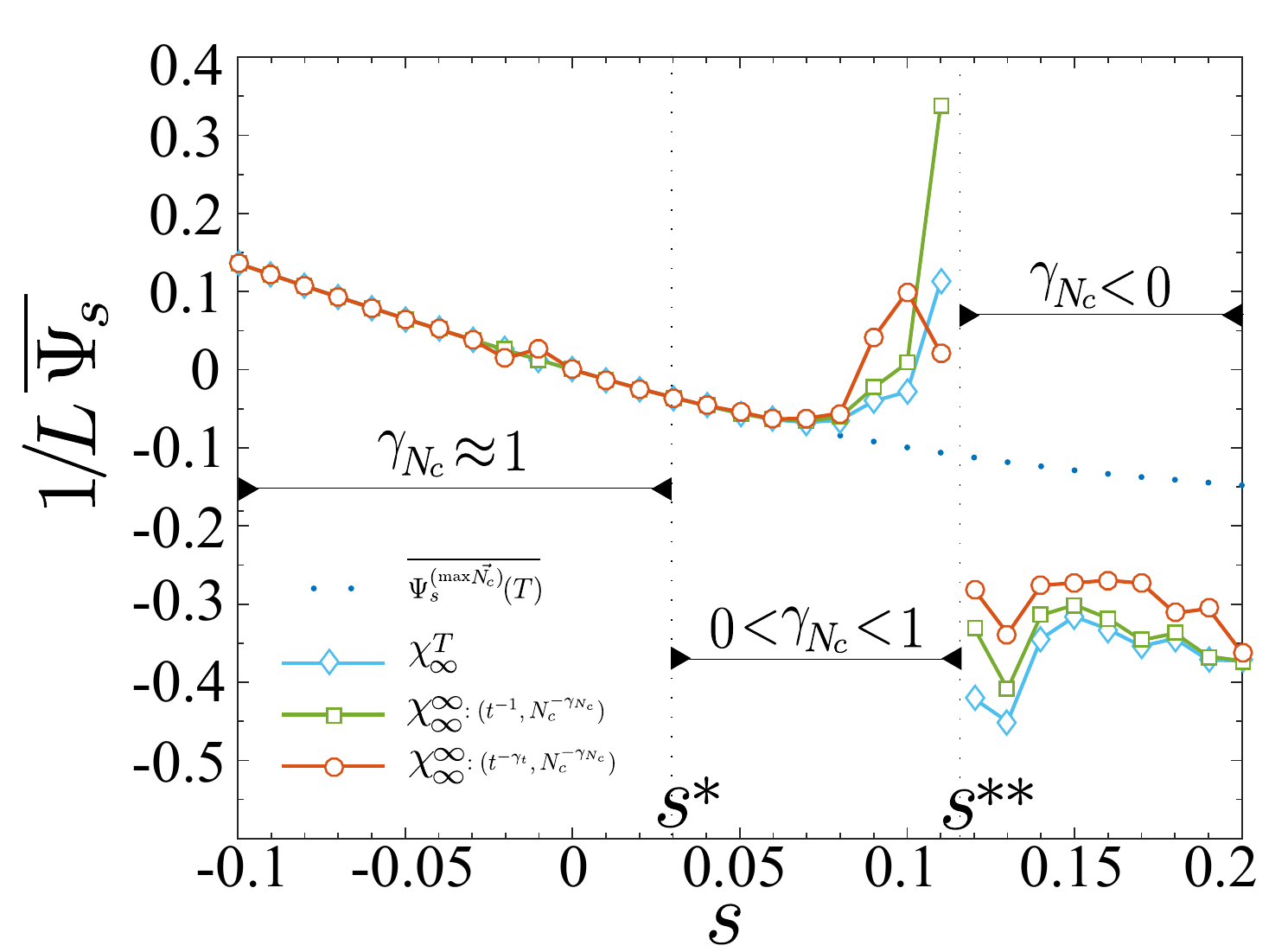}
 \centering
\caption[CGF Estimators in the Large-L Limit]{\label{fig:Psi_s} Different estimators of the large deviations of the activity as function of the parameter $s \in [-0.1,0.2]$ for a contact process with $L=100$ sites. The standard CGF estimator $\overline{\Psi_{s}^{(N_{c})}(t)}$ (evaluated at $N_c = \max \vec{N_c}=200$, $t = T=100$ and for $R=500$ realizations) is shown with dots meanwhile the ones obtained from the scaling method $\chi_{\infty}^{\infty}$ are presented in squares and circles, and $\chi_{\infty}^{T}$ in diamonds. The legend $(t^{-1},N_{c}^{-\gamma_{N_c}})$ refers to the assumption of a $t^{-1}$-scaling for $\overline{\Psi_{s}^{(N_{c})}(t)}$ (setting $\gamma_{t}=1$ in Eq.~\eqref{eq:tScal2}) and a $N_{c}^{-\gamma_{N_c}}$-scaling~\eqref{eq:nScal2} for the $\chi_{\infty}^{(N_c)}$'s. On the other hand, $(t^{-\gamma_{t}},N_{c}^{-\gamma_{N_c}})$ refers to the fact that we have left $\gamma_{t}$ and $\gamma_{N_c}$ as free parameters.
The different estimators correspond to each others up to $s=s^{*}$ from which they diverge up to $s = s^{**}$  where there is a discontinuity and they become negative for $s > s^{**}$. This is directly related with the behavior of the exponent $\gamma_{N_c}$ observed in Fig.~\ref{fig:Gamma_tn_s}. 
}
\end{figure}

The scaling method allows to determine the asymptotic limit to which the CGF estimator~\eqref{eq:PSI1} converges in the $t \to \infty$ and $N_c\to \infty$ limits. This limit, that we have denoted $f_{\infty}^{\infty}$ (Eq.~\eqref{eq:PSIinfinf}), was proved to render a better estimation of the analytical CGF $\psi(s)$ than the standard estimator $\overline{ \Psi_{s}^{(\max \vec{N_c})}(T)}$, at least for the cases analyzed in chapter~\ref{chap:ContinuousTime}~\cite{partII}.
However, the results we just presented 
%
%
would suggest that the determination of $f_{\infty}^{\infty}$ could be affected depending whether $\gamma_{t} \approx \gamma_{N_{c}} \approx 1$ or not. If this holds, the scaling method could render valid results in our example only for $s<0$. Solely in this region the extracted $\chi_{\infty}^{\infty}$ (obtained from Eq.~\eqref{eq:nScal2}) would represent the infinite-$t$ and -$N_c$ limit of the CGF estimator. Indeed, this can be observed in Fig.~\ref{fig:Psi_s} where we have applied the scaling method to our example. 

The method can be performed following two different approaches: 
(\textit{i}) $(t^{-1},N_{c}^{-\gamma_{N_c}})$: First, imposing a $t^{-1}$-scaling for $\overline{\Psi_{s}^{(N_{c})}(t)}$ (setting $\gamma_{t}=1$ in Eq.~\eqref{eq:tScal2}) and then, considering a $N_{c}^{-\gamma_{N_c}}$-scaling~\eqref{eq:nScal2} for the extracted $\chi_{\infty}^{(N_c)}$'s. Alternatively,
(\textit{ii}) $(t^{-\gamma_{t}},N_{c}^{-\gamma_{N_c}})$: Leaving $\gamma_{t}$ and $\gamma_{N_c}$ as free parameters in Eqs.~\eqref{eq:tScal2} and~\eqref{eq:nScal2}. 
Both resulting estimators $\chi_{\infty}^{\infty}(i)$ and $\chi_{\infty}^{\infty}(ii)$ are shown in Fig.~\ref{fig:Psi_s} with squares and circles, respectively. Additionally, the infinite-$N_c$ limit $\chi_{\infty}^{T}$~\eqref{eq:nScal3} is also presented with diamonds. 
The standard CGF estimator $\overline{\Psi_{s}^{(\max \vec{N_c})}(T)}$ (in dots) serves as reference. 

As can be seen in Fig.~\ref{fig:Psi_s}, the different estimators correspond to each other up to $s=s^{*}$.
%
%
From this point, their distance with respect to $\overline{\Psi_{s}^{(\max \vec{N_c})}(T)}$ increases rapidly with $s$ up to $s = s^{**}$ 
where a discontinuity occurs. 
%
In fact, the behavior observed in Fig.~\ref{fig:Psi_s} keeps correspondence with the $N_c^{-\gamma_{N_c}}$-scaling of the CGF estimator. Specifically, with the stages of the exponent $\gamma_{N_c}$ that were presented in Sec.~\ref{sec: gamma_tn} and Fig.~\ref{fig:Gamma_tn_s}. 
Thus, the discontinuity in $\chi_{\infty}^{\infty}$ is related precisely with the change in sign of $\gamma_{N_c}$ in $s = s^{**}$ in the same way as the divergence of the estimators from the standard one at $s=s^{*}$ is related with the fact that from this point, $\gamma_{N_c} \neq  1$.

The example presented through this section related the effectiveness of the scaling method (as proposed in chapter~\ref{chap:ContinuousTime}~\cite{partII}) with the actual scaling of the CGF estimator in large-$L$ systems. Depending on the value of the exponents $(\gamma_{t}, \gamma_{N_c})$ it is possible to extract or not the infinite-$t$ and infinite-$N_c$ limit of the CGF estimator. Moreover, we also presented the dependence of these exponents with the parameter $s$. Below we extend our analysis by considering the scaling behavior on a wider range of values of $L$. This will provide a complete overview of how the CGF estimator behaves and how the change in scaling is given.

\section[$L$-Dependence of the Finite Scalings]{\quad $L$-Dependence of the Finite Scalings}
\label{sec: planeSL}

In this section, we detail the behavior of the finite-$t$ and -$N_c$ scalings of the CGF estimator 
for $s>0$ and $L$ ranging in the interval $L\in [3,100]$. For each pair $(s,L)$, the exponents $\gamma_{t}$ and $\gamma_{N_c}$ were computed as described in Sec.~\ref{sec:exp} for $T=100$ and $\vec{N_c} = \{20,40,...,180,200 \}$. 
\begin{figure} [t]
\centering
\includegraphics[width=0.55\textwidth]{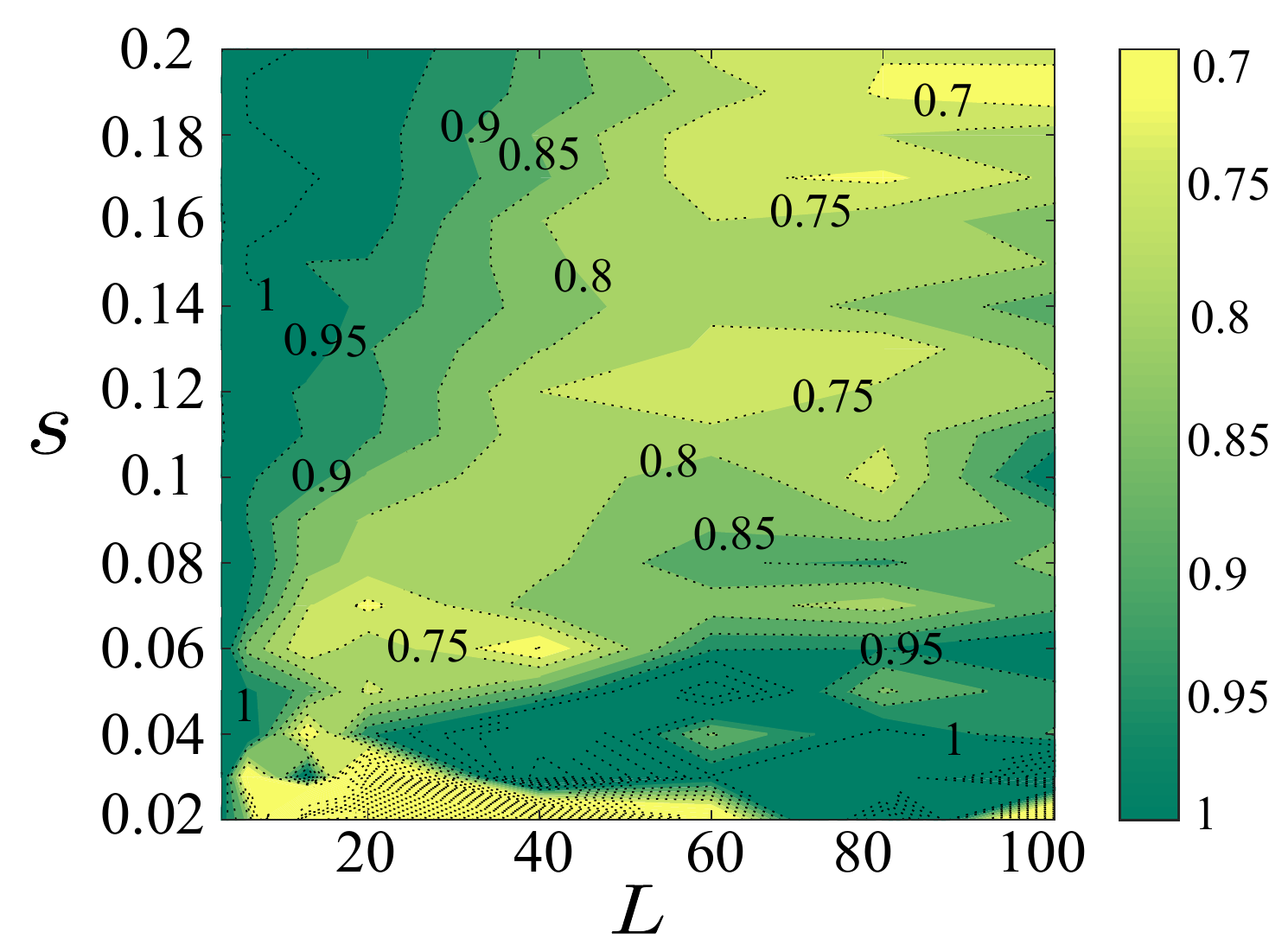}
\centering
\caption[Finite-Time Scaling on the Plane s-L]{\label{fig:Gammat_LS} Behavior of the exponent $\gamma_{t}$ in a region of the plane $s-L$ given by the parameters $s \in \big[0.02,0.2\big]$ and $L\in \big[ 3,100 \big]$. This exponent characterizes the finite-time scaling of the large deviations of the activity in the contact process ($\mathbf{t^{-\gamma_{t}}}$\textbf{-scaling}). 
The values of $\gamma_{t}$ closest to $1$ are presented with the darkest tones whereas smaller values are shown with clearer ones.
Although for small values of $L$ the $t^{-1}$-scaling holds independently of $s$, in general the exponent $\gamma_{t}$ decreases gradually as $s$ and $L$ increase.
}
\end{figure}

\subsection[Characterization of the Exponent $\gamma_{t}(s,L)$]{\quad Characterization of the Exponent $\gamma_{t}(s,L)$}
The contour plot in Fig.~\ref{fig:Gammat_LS} shows the value of the exponent $\gamma_{t}$ as it changes depending on the parameters $s$ and $L$. 
%
%
We have focused in the region for $s\in \big[0.02,0.2\big]$ as for $s<0$, $\gamma_{t}\approx 1$ and thus, the $t^{-1}$-scaling~\eqref{eq:fitdef} holds. 
The values closest to $1$ are presented with the darkest tone while smaller values are shown with clearer tones. As can be seen, the exponent $\gamma_{t}$ decreases gradually as $L$ and $s$ increase.

For a given system size $L$, we can describe qualitatively the behavior of $\gamma_{t}$ with respect to $s$ is similar way as we did for $L=100$ in Sec.~\ref{sec: gamma_tn}.  
%
In order to extend that description into the plane $s-L$, we introduce a number of sites dependency of the bound $s'$. We denote by $s'(L)$ the value of $s$ until which the $t^{-1}$-scaling is valid given a particular $L$. Similarly, $\gamma_{t}^{\circ}(L)$ is the lower bound of $\gamma_{t}^{(L)}(s)$.
%
%
%
Thus, the exponent $\gamma_{t}$ which characterizes the $t^{-\gamma_{t}}$-scaling~\eqref{eq:tScal2} of the CGF estimator 
%
%
is given by
%
%
%
%
\begin{equation}
\label{eq:Gammat_sL}
\gamma_{t}:
    \begin{cases}
\gamma_{t}^{(L)}(s) \approx 1,                           &\text{for}\ s<s'(L) \\
\gamma_{t}^{\circ}(L) \leq \gamma_{t}^{(L)}(s) \lesssim 1,  &\text{otherwise,}\    
    \end{cases}
\end{equation}
%
%
where $s'(L)>0$, $\gamma_{t}^{\circ}(L)>0$ and $L$ is large. In fact, for this case, $\gamma_{t}^{\circ}(L)>1/2$, for all $L$.

\subsection[Characterization of the Exponent $\gamma_{N_{c}}(s,L)$]{\quad Characterization of the Exponent $\gamma_{N_{c}}(s,L)$}

\begin{figure*} [t]
\centering
\includegraphics[width=0.48\textwidth]{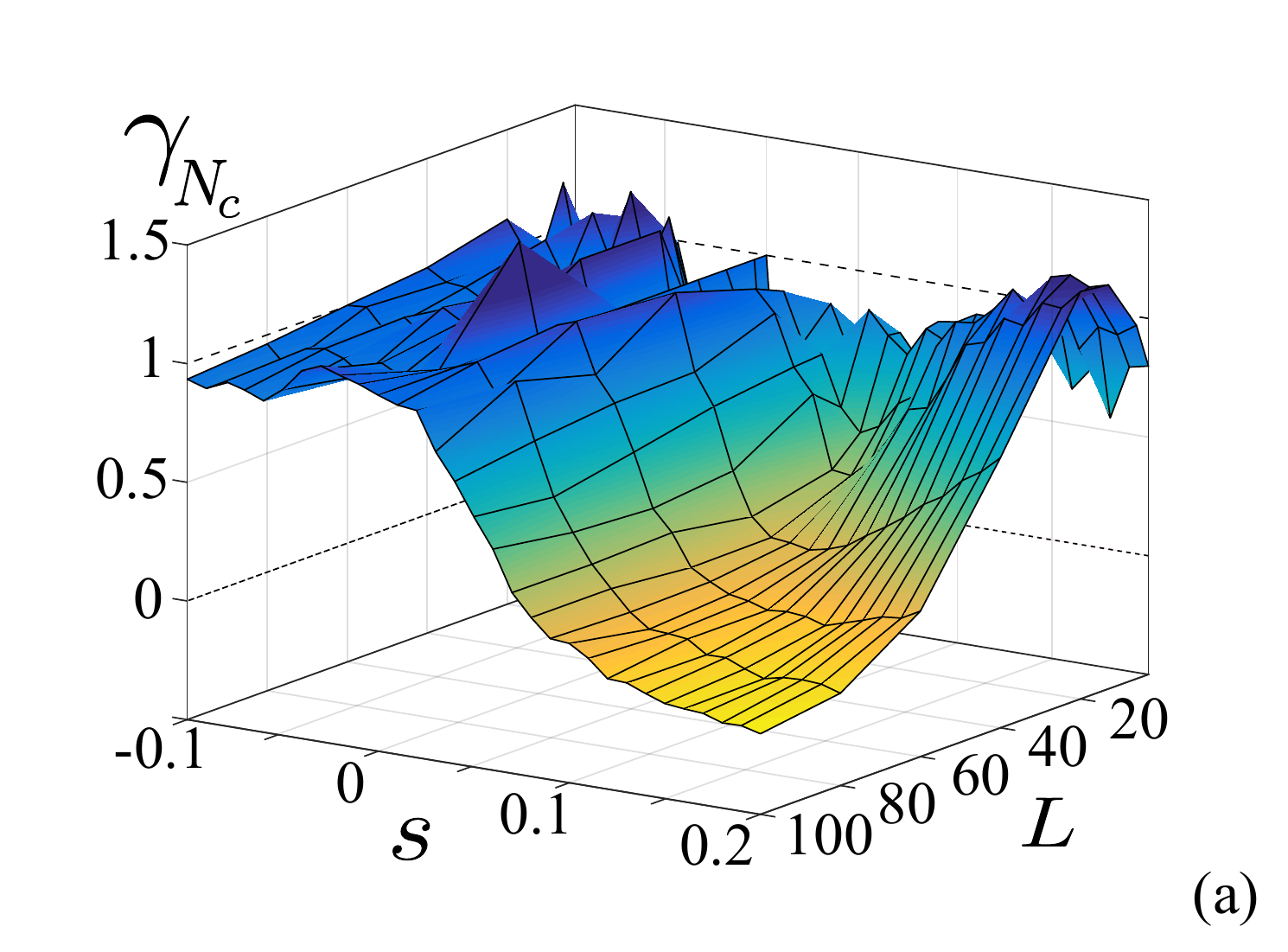}
\includegraphics[width=0.48\textwidth]{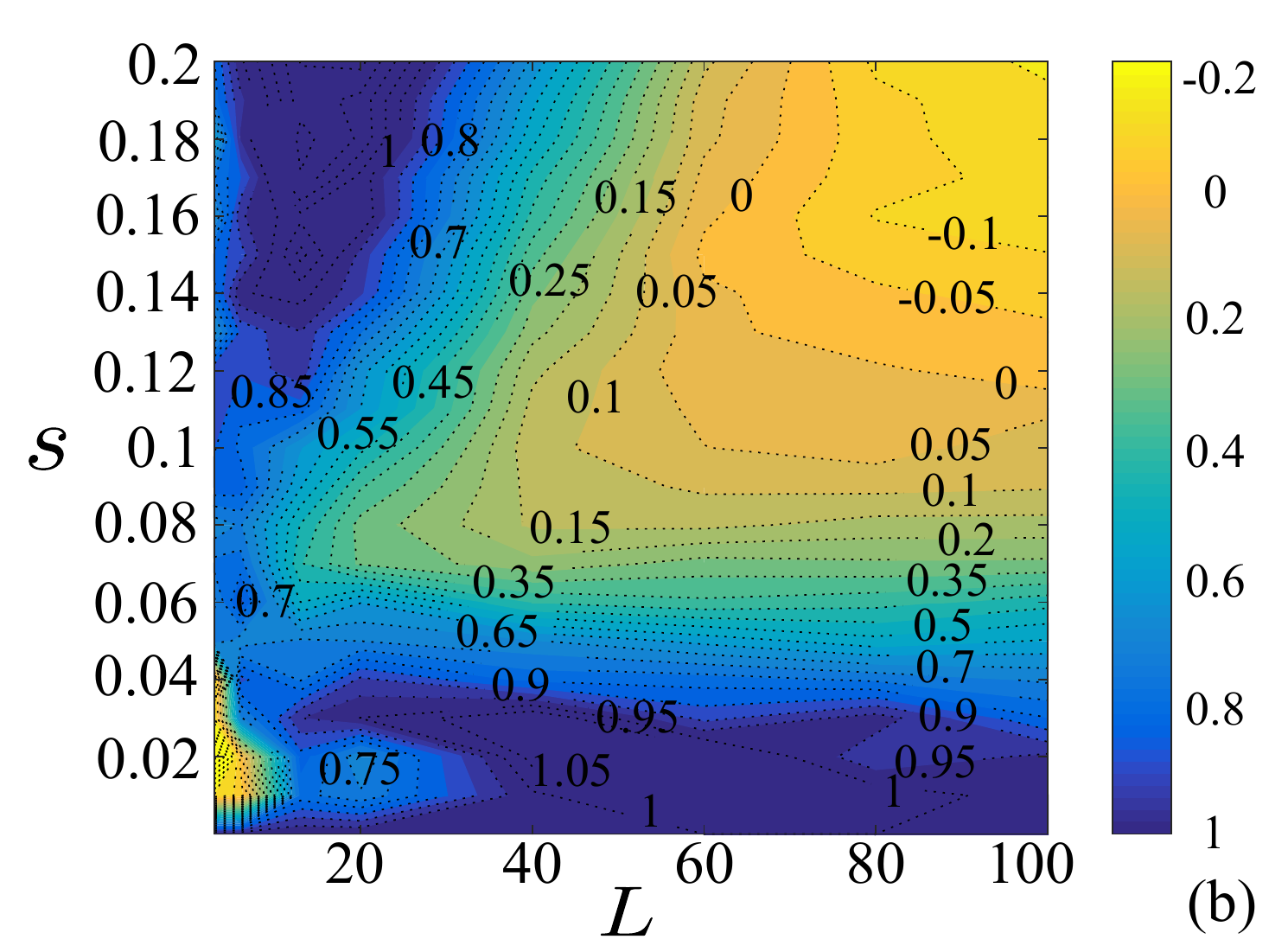}
\caption[Finite Population-Size Scaling on the Plane s-L]{\label{fig:Gamman_LS} Behavior of the exponent $\gamma_{N_c}$ in a region of the plane $s-L$ given by the parameters $s \in \big[-0.1,0.2\big]$ and $L\in \big[ 3,100 \big]$. This exponent characterizes the finite-$N_c$ scaling of the large deviations of the activity in the contact process ($\mathbf{N_{c}^{-\gamma_{N_{c}}}}$ \textbf{-scaling}). \textbf{(a)} The surface $\gamma_{N_c}(s,L)$ illustrates the change in the scaling of the CGF estimator. The exponent $\gamma_{N_{c}}(s,L) \approx 1$  up to some value of $s$ (which depends of $L$), from which it decreases as $s$ and $L$ increases, even becoming negative for some values of $L$.  \textbf{(b)} Projection of the surface in (a) on the plane $s-L$. The values of $\gamma_{N_{c}}(s,L)$ closer to $1$ are shown in dark tones while the smaller tones with clearer ones. The $N_c^{-\gamma_{N_c}}$-scaling can be characterized depending on the number of zeros of the exponent $\gamma_{N_{c}}(s)$ for a given $L$ in three regions: $\mathcal{L}_{I}$ if has no zeros, $\mathcal{L}_{II}$: if have two zeros and $\mathcal{L}_{III}$: if have one zero. 
}
\centering
\end{figure*}

Similarly as above, in Fig.~\ref{fig:Gamman_LS} we present the exponent $\gamma_{N_{c}}$ as it changes depending of some particular choice of the parameters $(s,L)$ within the intervals considered. 
The surface in Fig.~\ref{fig:Gamman_LS}(a) illustrates clearly the change in the $N_c$-scaling of the CGF estimator.
For every value of $L$ considered, the exponent $\gamma_{N_{c}}$ is approximately $1$ up to some value of $s$, denoted as $s^{*}(L)$ (Sec.~\ref{sec: gamma_tn}). 
%
However, from this point, its value decreases as $s$ and $L$ increases, becoming, in some cases, negative.
This change in the $N_c^{-\gamma_{N_c}}$-scaling is also shown in the contour plot in Fig.~\ref{fig:Gamman_LS}(b) where we have focus in the region for $s>0$. The values of $\gamma_{N_{c}}$ closer to $1$ are shown in dark tones. 

In Sec.~\ref{sec: gamma_tn}, we also defined $s^{**}$ such that $\gamma_{N_{c}}^{(L)}(s^{**}) = 0$. This value of course depends on $L$ and in some cases it does not even exists.
However, for some particular values of $L$ (large), the exponent $\gamma_{N_{c}}^{(L)}$ changes sign twice (as can be seen in Fig.~\ref{fig:Gamman_LS}(b)).
We will use this fact in order to characterize the $N_c^{-\gamma_{N_c}}$-scaling depending on the number of zeros of the exponent $\gamma_{N_{c}}^{(L)}(s)$ for a given $L$.

We define $\mathcal{L}_{I}$ as the set of values of $L$, for which the exponent $\gamma_{N_{c}}^{(L)}(s)$ has no zeros, $\mathcal{L}_{II}$: if has two zeros ($s_{1}^{**}(L)$ and $s_{2}^{**}(L)$, with $s_{2}^{**}(L) > s_{1}^{**}(L)$) and $\mathcal{L}_{III}$: if has one zero  ($s^{**}(L)$). 
These regions are bounded by $L_{inf}$ and/or by $L_{sup}$, where $L_{inf}$ is the smallest value of $L$ such that the curve $L =L_{inf}$ is tangent to $\gamma_{N_{c}}(s,L)=0$ in one single point. On the other hand, $L_{sup}$ is the largest $L$ such that the curve $L = L_{sup}$ cuts $\gamma_{N_{c}}(s,L)=0$ in two points. 
Thus, the region $\mathcal{L}_{I}$ groups the values of $L$ such that  $L<L_{inf}$, $\mathcal{L}_{II}$ the values of $L$ within the interval $L_{inf} < L < L_{sup}$ and $\mathcal{L}_{III}$, the values of $L$ such that $L > L_{sup}$.
%
Thus, the exponent $\gamma_{N_{c}}$ which characterizes the $N_c^{-\gamma_{N_c}}$-scaling~\eqref{eq:nScal2} of the CGF estimator is given by
%
%
%
%
\begin{equation}
\label{eq:GammaN_sL}
\gamma_{N_{c}}:
    \begin{cases}    
    \mathcal{L}_{I}: 
    \begin{cases}
    		\gamma_{N_{c}}^{(L)}(s)\approx 1, & \text{for}\ s<s^{*}(L) \\
     	0 < \gamma_{N_{c}}^{(L)}(s) \lesssim 1, & \text{otherwise.}\\
    	\end{cases}\\ \\
    	             
    \mathcal{L}_{II}: 
    \begin{cases}
    			\gamma_{N_{c}}^{(L)}(s)\approx 1, & \text{for}\ s<s^{*}(L) \\         
                 0 < \gamma_{N_{c}}^{(L)}(s) < 1, & \begin{aligned}
            & \text{for}\ s^{*}(L)<s<s^{**}_{1}(L) \\ 	 												                   	        & \text{and}\ s>s^{**}_{2}(L) \\
            \end{aligned}\\
    \gamma_{N_{c}}^{(L)}(s)<0, & \text{for}\ s^{**}_{1}(L)<s<s^{**}_{2}(L)
    \end{cases}\\ \\
    
    \mathcal{L}_{III}: 
    \begin{cases} 
            \gamma_{N_{c}}^{(L)}(s)\approx 1, & \text{for}\ s<s^{*}(L) \\  
            0 < \gamma_{N_{c}}^{(L)}(s) < 1, & \text{for}\ s^{*}(L)<s<s^{**}(L) \\
           \gamma_{N_{c}}^{(L)}(s)<0, & \text{for}\ s>s^{**}(L) 
    \end{cases}   
   \end{cases}
\end{equation}

%
%

\section[Dynamical Phase Transition and Finite-Scalings ]{\quad Dynamical Phase Transition and Finite-Scalings}
\label{sec:DPTcp}

In Sec.~\ref{sec:LDobservables}, we introduced the biasing parameter (or field) $s$ (conjugated to an observable $\mathcal O$) to characterize a non-equilibrium ensemble of trajectories. 
%
%
Within this ``$s$-ensemble'', space-time or \textbf{dynamical phase transitions} 
manifest themselves as singularities in the CGF and, in our case, express a dynamical coexistence of histories with high and low activity $K$~\cite{Lecomte2007}.

The contact process~\cite{CP, GRASSBERGER1979373, liggett2012interacting} is well know to exhibit a dynamical phase transition in the $L \rightarrow \infty$ limit~\cite{lecomte_numerical_2007, Lecomte2007, ThermoCP, marro_dickman_1999} even in one-dimension~\cite{marro_dickman_1999}. 
However in Ref.~\cite{lecomte_numerical_2007} evidence of the presence of a phase transition (in the active phase of $\lambda$) was reported to occur at $s_c\approx0.057$ for finite-$L$. There, the authors used the same version of the contact process and the same approach we used throughout this thesis (i.e., the cloning algorithm).
On the other hand, in Ref.~\cite{ThermoCP}, using a 
density matrix re-normalization group approach (DMRG)~\cite{WhiteDMRG1, WhiteDMRG2, reviewDMRG, Kaulke1998, Carlon1999}, 
it was showed that for every value of infection rate $\lambda$, either if this belong to the absorbing or to the active phase, there exists a phase transition as a function of $s$. 
For the case of the active phase, this transition was found to occur at $s_c = 0$. 
It is important to remark that even if the versions of the contact process used in Refs.~\cite{lecomte_numerical_2007} and~\cite{ThermoCP} are different, both present a dynamical phase transition.  Meanwhile in the later case the particles are created just at the boundaries, in Ref.~\cite{lecomte_numerical_2007} (and here) they are created at every site
and also, the spontaneous rate of creation $h$ is considered different from $0$ (in order to circumvent the absorbing state in finite size~\cite{Lecomte2007}).

Despite our main interest 
is not the study of the dynamical phase transition in the contact process, 
what does concern us is how this could affect the finite scalings of the CGF. Importantly, the relation that $s^{*}$ (but also $s'$ or $s^{**}$) could have with $s_c$ where this transition occurs. 
In Sec.~\ref{sec: Scaling} we showed how the scaling behavior given by Eqs.~\eqref{eq:tScal2} and~\eqref{eq:nScal2} was robust independently of $T$, $N_c$, $s$ or $L$ (Fig.~\ref{fig:powerlaw}), but not the exponents $\gamma_{t}$ and $\gamma_{N_c}$ whose behavior change depending on $s$ and $L$ specially in $s\geq0$ as $L$ becomes larger.
We remark that even if the infinite-$L$ limit is not achievable numerically,
the effects induced by a
dynamical phase transition should become more evident as $L$ increases  (which could explain many of the behavior observed throughout this chapter).
This was clearly illustrated for $L=100$ for which $\gamma_{N_c}$ has an abrupt change for $s\geq0$, where we know the dynamical phase transition occurs, even taking negative values and inducing a divergence of the infinite-$t$ and infinite-$N_c$ limit of the CGF estimator (Fig.~\ref{fig:Psi_s}).

We recall here that our purpose was to verify the validity of the scalings (and the scaling method) presented in chapters~\ref{chap:DiscreteTime}~\cite{partI} and~\ref{chap:ContinuousTime}~\cite{partII} (for small size systems) in the large-$L$ limit
where
a main result was
the possibility of
extracting the infinite-$N_c$ infinite-$t$ limit of the CGF estimator from finite and small number of clones and time. 
An analysis of the dynamical phase transition, on the other hand,
would require
a large-$N_c$ and -$t$ configuration which under our approach
is a task difficult to fulfill.
%
This however does not represent any surprise 
given that is well know that 
the existing methods
~\cite{giardina_direct_2006, Hedges1309, PitardDT, SpeckDPT, Speck}
perform poorly in the vicinity of a dynamical phase transition, 
or they are 
numerically expensive in order to obtain accurate estimations ~\cite{SpeckDPT, LimmerIce, DPTpath} developing if not important finite-size effects~\cite{hurtado_current_2009}.
However, recently has been proposed a promising method~\cite{nemoto_population-dynamics_2016, PhysRevLett.118.115702} which combines the existing cloning algorithm~\cite{giardina_direct_2006, giardina_simulating_2011, tailleur_simulation_2009, lecomte_numerical_2007, Hedges1309, PitardDT, SpeckDPT, Speck, partI, partII} with a modification of
the dynamics~\cite{jack_large_2010, 1742-5468-2010-10-P10007, PhysRevLett.111.120601, PhysRevLett.112.090602} resulting in a significant improvement of its computational efficiency. The method was successfully applied to the study of the dynamical phase transition of 1D FA model~\cite{FAmodel} using a relatively small $N_c$ and $L$. 
The implementation of this method will 
provide 
in a next stage
a clear contrast between the results obtained following the two different approaches and a correct relation between $s_c$ and $s^{*}$. 

\section[Conclusion]{\quad Conclusion}
\label{sec:conclusion}
%
%
The dependence of the CGF estimator (and of its accuracy) with the simulation time $T$ and number of clones $N_c$ was studied in chapters~\ref{chap:DiscreteTime}~\cite{partI} and~\ref{chap:ContinuousTime}~\cite{partII} where the finite-$t$ and finite-$N_c$ scalings of the systematic errors of the LDF were found to behave as $1/N_c$ and $1/t$ in the large-$N_c$ and large-$t$ asymptotics, respectively. By making use of these convergence-speeds, a scaling method was proposed which allowed to extract the asymptotic behavior of the CGF estimator in the $t \to \infty$ and $N_c \to \infty$ limits. At least for the cases analyzed in chapters~\ref{chap:DiscreteTime}~\cite{partI} and~\ref{chap:ContinuousTime}~\cite{partII}, this infinite-time and infinite-$N_c$ limit resulted to render a better LDF estimation in comparison with the standard estimator. 
However, the validity of the method and of these scalings were proved only for a simple one-site annihilation-creation dynamics and for a contact process with $L=6$ sites, leaving an analysis of the dependence 
with the number of sites $L$ pending.
In order to do so, 
in this chapter, 
%
we redefined these scalings in a more general way.
We assume the behavior of the CGF estimator described by a
$t^{-\gamma_{t}}$-scaling (Eq.~\eqref{eq:tScal2}) and a $N_{c}^{-\gamma_{N_{c}}}$-scaling (Eq.~\eqref{eq:nScal2}). This redefinition allowed us to verify in large-$L$ systems if effectively $\gamma_{t} \approx 1$ and $\gamma_{N_{c}} \approx 1$ and
whether the terms $\chi_{\infty}^{(N_{c})}$ and $\chi_{\infty}^{\infty}$ represent the limits in $t \to \infty$ and $N_{c} \to \infty $ of the CGF estimator.

This was done at first in Sec.~\ref{sec: CGFL100} where we considered a contact process with $L=100$ sites and two representative values of the parameter $s$.
Although the $t^{-1}$-scaling and $N_{c}^{-1}$-scaling were proved to hold for $s = 0.1$, this was not the case for $s=0.2$. 
How this change in the scaling is produced depending on the parameter $s$ was presented in detail in Sec.~\ref{sec: gamma_tn} where the exponents $\gamma_{t}(s)$ and $\gamma_{N_{c}}(s)$ were characterized.
Particularly, for $\gamma_{N_{c}}(s)$, we were able to distinguish three stages in its behavior, where, the $N_{c}^{-1}$-scaling was valid up to $s=s^{*}$, then $\gamma_{N_{c}}$ decreases to $0$ at $s=s^{**}$ and finally, it becomes negative for $s>s^{**}$.
In Sec.~\ref{sec: SML100} we showed how these scalings affect the determination of the infinite-$t$ and infinite-$N_c$ limit of the CGF estimator. This occurs because the scaling method relied on the validity of the $t^{-1}$- and $N_{c}^{-1}$-scalings. As for $L=100$ this was not the case, it was possible to see how the different estimators corresponded to each others up to $s=s^{*}$ from which they diverge up to $s = s^{**}$  where there is a discontinuity.

This analysis was extended to the plane $s-L$ in Sec.~\ref{sec: planeSL} where the exponents $\gamma_{t}$ and $\gamma_{N_c}$ were computed for a grid of values of the parameters $(s,L)$.
Their characterization was done introducing a number-of-sites dependency of the bounds $s'$, $s^{*}$ and $s^{**}$ previously defined in Sec.~\ref{sec: CPL100} as well as the use of the number of zeros of the exponent $\gamma_{N_{c}}^{(L)}(s)$ in order to characterize the different groups of $L$.
Whether the results presented through this chapter are restricted only to the contact process or not is left as a pending problem and a possible direction for future research.

\clearpage
\thispagestyle{empty}
\phantom{a}

\chapter[\quad Intra-day Seasonalities in Financial Time Series]{Intra-day Seasonalities in High Frequency Financial Time Series} 
\label{chap:Intraday}
\section[\quad Introduction]{\quad Introduction}
From the statistical study of financial time series have arisen a set of  properties or empirical laws sometimes called ``stylized facts'' or seasonalities. These properties have the characteristic of being  common and persistent across different markets, time periods and assets~\cite{1, 2, 3, 4, 5, 6, 7}.
As it has been suggested in Ref.~\cite{7}, the reason why these ``patterns''  appear could be because markets operate in synchronization with human activities which leave a trace in the financial time series.
However using the ``right clock'' might be of primary importance when dealing with statistical properties and the patterns could vary depending if we use daily data or intra-day data and event time, trade time or arbitrary intervals of time (e.g. $T = 1$, $5$, $15$ minutes, etc). 
For example, it is a well-known fact that empirical distributions of  financial returns and log-returns are fat tailed~\cite{8, 9}, however as one increases the time scale the fat-tail property becomes less pronounced  and the distribution approach the Gaussian form~\cite{10}. As was stated in Ref.~\cite{4}, the fact that the shape of the distribution changes with time makes it clear that the random process underlying prices must have a non-trivial temporal structure.
In a previous work, Allez et al.~\cite{7} established several new stylized facts concerning the intra-day seasonalities of single and cross-sectional stock dynamics. This dynamics is characterized by the evolution of the moments of its returns  during a typical day.  Following the same approach, we show the bin size dependence of these patterns for the case of returns and, motivated by the work of Kaisoji~\cite{11}, we extend the analysis to relative prices and show how in this case, these patterns are independent of the size of the bin, also independent of the index we consider but characteristic for each index. These facts could be used in order to detect an anomalous behavior during the day, like market crashes or intra-day bubbles~\cite{11, 12}. The work presented in this chapter~\ref{chap:Intraday}~\cite{binsize} is completely empirical but it could offer signs of the  underlying stochastic process that governs the financial time series. 

\section[\quad Definitions]{\quad Definitions}
The data consists in two sets of intra-day high frequency time series, the CAC $40$ and the S\&P $500$. For each of the $D = 22$ days of our period of analysis (March $2011$), we dispose with the evolution of the prices of each of the stocks that composes our indexes during a specific day from $10:00$ a.m. to $16:00$ p.m.  The
main reasons why we chose to work with these two indexes are: The number of stocks that
compose them ($N_{1}=40$ and $N_{2}=500$), the time gap between their respective markets and the
different range of stock prices (between $5$ and $600$ USD for the S\&P $500$
and between $5$ and $145$ EU for the CAC $40$). 

\begin{figure}[h!]
\centering
\includegraphics [width=0.55\textwidth] {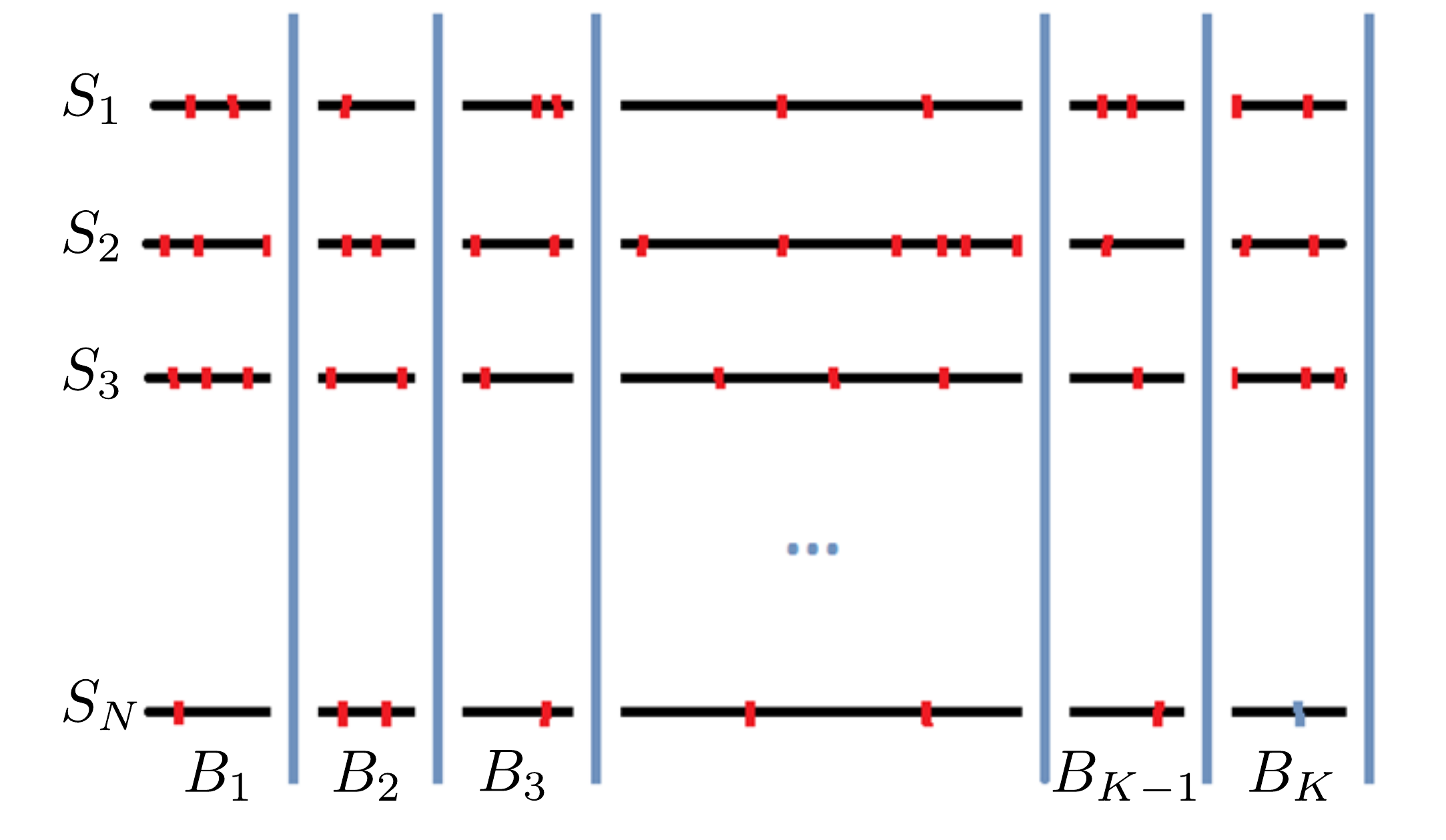}
\caption[\quad Intra-Day Asynchronous Financial Time Series]{Intra-day asynchronous financial time series. $S_{i}$ are the stocks and $B_{k}$ are bins. The asynchronous prices are show in red and the bin limit in blue.}
\label{fig:Fig1E}
\end{figure}

As the changes in prices are not synchronous between different stocks (Fig.~\ref{fig:Fig1E}), we manipulated our original data in order to construct a new homogeneous matrix $P_{D}^{(j)}$ of bin prices. In order to do this, we divided our daily time interval $[10:00,16:00]$ in $K$ bins of size $T$ (minutes), i.e., $B_{1} = [10:00,10:00 + T]$, $B_{2} = [10:00 + T,10:00 + 2T],\ldots, B_{K} = [16:00 - T,16:00]$, where the right endpoints of these intervals are called bin limits. For a particular day $j$, the prices that conform the matrix $P_{D}^{(j)}$ are given by the last prices that reaches that stock $i$ just before a specific bin limit.

Each row in the matrix below represents the evolution of the prices of a
particular stock as function of the bins. For example, the element $%
(P_{D})_{ik}^{(j)}$, represents the price for a particular day $j$ of the stock $i$ and just before the bin limit of the bin $B_{k}$.
\begin{equation}
P_{D}^{(j)}=%
\begin{pmatrix}
P_{11}^{(j)} & P_{12}^{(j)} & ... & ... & ... & P_{1K}^{(j)} \\ 
P_{21}^{(j)} & P_{22}^{(j)} & ... & ... & ... & P_{2K}^{(j)} \\ 
... & ... & ... & ... & ... & ... \\ 
... & ... & ... & P_{ik}^{(j)} & ... & ... \\ 
... & ... & ... & ... & ... & ... \\ 
P_{N1}^{(j)} & P_{N2}^{(j)} & ... & ... & .. & P_{NK}^{(j)}%
\end{pmatrix}
\label{A1}
\end{equation}%
In a similar way, we can construct the matrix $P_{S}^{(i)}$ for each of the $%
i=1,...,N_{1,2}$ stocks. $(P_{S})_{jk}^{(i)}$ is the price of the stock $(i)$ in the day $j$ and just before the bin limit of the bin $B_{k}$.%
\begin{equation}
P_{S}^{(i)}=%
\begin{pmatrix}
P_{11}^{(i)} & P_{12}^{(i)} & ... & ... & ... & P_{1K}^{(i)} \\ 
P_{21}^{(i)} & P_{22}^{(i)} & ... & ... & ... & P_{2K}^{(i)} \\ 
... & ... & ... & ... & ... & ... \\ 
... & ... & ... & P_{jk}^{(i)} & ... & ... \\ 
... & ... & ... & ... & ... & ... \\ 
P_{D1}^{(i)} & P_{D2}^{(i)} & ... & ... & .. & P_{DK}^{(i)}%
\end{pmatrix}
\label{A2}
\end{equation}

\bigskip 

In the following and for simplicity, we will refer to the price $P$ of a particular stock $i = \alpha $ during a particular day $j = t$ and just before the bin limit of the bin $B_{k}$ as $P_{\alpha }(k,t)$ where $P_{\alpha }(k,t) = P_{tk}^{(\alpha)}=P_{\alpha k}^{(t)}$. 
We will perform our statistical analysis over the variable $x_{\alpha }(k,t)$ that can be computed from the matrices above. For our interests we will be working with returns 
\begin{equation}
x_{\alpha }^{(1)}(k,t)=\frac{P_{\alpha }(k+1,t)-P_{\alpha }(k,t)}{P_{\alpha}(k,t)},  
\label{1}
\end{equation}%
and relative prices~\cite{11, 12}%
\begin{equation}
x_{\alpha }^{(2)}(k,t)=\frac{P_{\alpha }(k,t) - P_{\alpha }(1,t)}{P_{\alpha }(1,t)}.
\label{2}
\end{equation}
The single or collective stock dynamics is characterized by the evolution of the moments of the returns (or relative prices). Below, we show how we computed these
moments~\cite{7}.  

\subsection[\quad Single Stock Properties]{\quad Single Stock Properties}
The distribution of the stock\ $\alpha $\ in bin $k$ is characterized by its
four first moments: mean $\mu _{\alpha }(k)$, standard deviation
(volatility) $\sigma _{\alpha }(k),$ skewness $\zeta _{\alpha }(k)$ and
kurtosis $\kappa _{\alpha }(k)$ defined as%
\begin{eqnarray}
\mu _{\alpha }(k) &=&\left\langle x_{\alpha }(k,t)\right\rangle,   \label{3}
\\
\sigma _{\alpha }^{2}(k) &=&\left\langle x_{\alpha }^{2}(k,t)\right\rangle
-\mu _{\alpha }^{2}(k),  \label{4} \\
\zeta _{\alpha }(k) &=&\frac{6}{\sigma _{\alpha }(k)}(\mu _{\alpha
}(k)-m_{\alpha }(k)),  \label{5} \\
\kappa _{\alpha }(k) &=&24\left( 1-\sqrt{\frac{\pi }{2}}\frac{\left\langle
|x_{\alpha }(k,t)-\mu _{\alpha }(k)|\right\rangle }{\sigma _{\alpha }(k)}%
\right) +\zeta _{\alpha }^{2}(k),  \label{6}
\end{eqnarray}%
where $m_{\alpha }(k)$ is the median of all values of $x_{\alpha }(k,t)$ and time averages for a given stock in a given bin are expressed with angled brackets $\langle...\rangle$.

\subsection[\quad Cross-Sectional Stock Properties]{\quad Cross-Sectional Stock Properties}
The cross-sectional distributions (i.e., the dispersion of the values of the
variable $x$ of the $N$ stocks for a given bin $k$ in a given day $t$) are
also characterized by the four first moments%
\begin{eqnarray}
\mu _{d}(k,t) &=&\left[ x_{\alpha }(k,t)\right],   \label{7} \\
\sigma _{d}^{2}(k,t) &=&\left[ x_{\alpha }^{2}(k,t)\right] -\mu _{d}^{2}(k,t),
\label{8} \\
\zeta _{d}(k,t) &=&\frac{6}{\sigma _{d}(k,t)}(\mu _{d}(k,t)-m_{d}(k,t)),
\label{9} \\
\kappa _{d}(k) &=&24\left( 1-\sqrt{\frac{\pi }{2}}\frac{\left[ |x_{\alpha
}(k,t)-\mu _{\alpha }(k)|\right] }{\sigma _{d}(k)}\right),   \label{10}
\end{eqnarray}%
where $m_{d}(k,t)$ is the median of all the $N$ values of the variable $x$
for a given $(k,t)$ and the square brackets $\left[...\right]$ represent averages over the ensemble
of stocks in a given bin and day. If $x_{\alpha }(k,t)$ are the returns, $\mu _{d}(k,t)$ can be seen as the return of an index
equi-weighted on all stocks.

\section[\quad Intra-day Seasonalities for Returns]{\quad Intra-day Seasonalities for Returns}
The following results are in complete agreement with the ones presented in~\cite{5, 6, 7}.
\begin{figure}[t!]
        \begin{center}
        \subfigure[MEAN]{\includegraphics [width=0.48\textwidth] {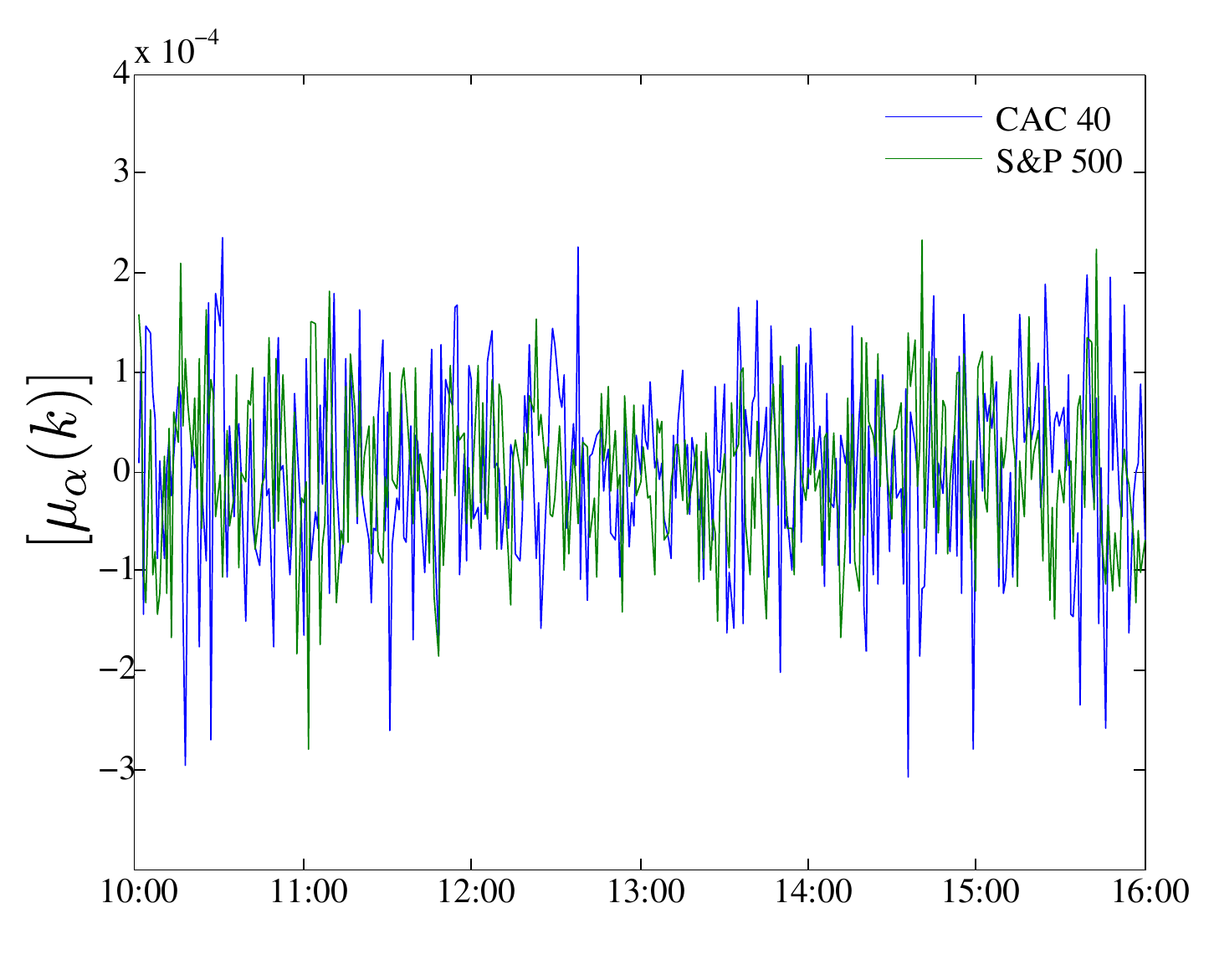}}
		\subfigure[VOLATILITY]{\includegraphics [width=0.48\textwidth] {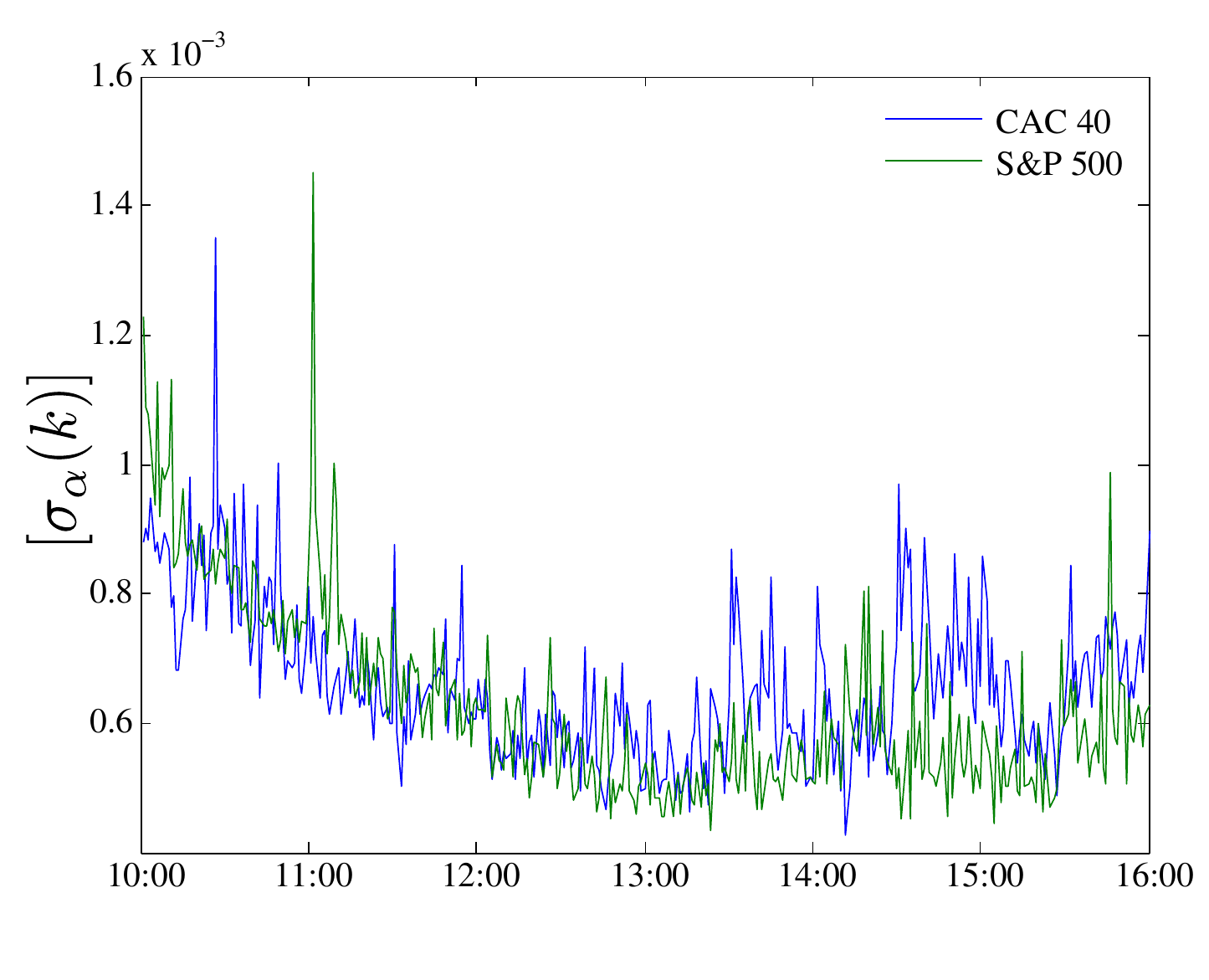}}\\ 
        \subfigure[SKEWNESS]{\includegraphics [width=0.48\textwidth] {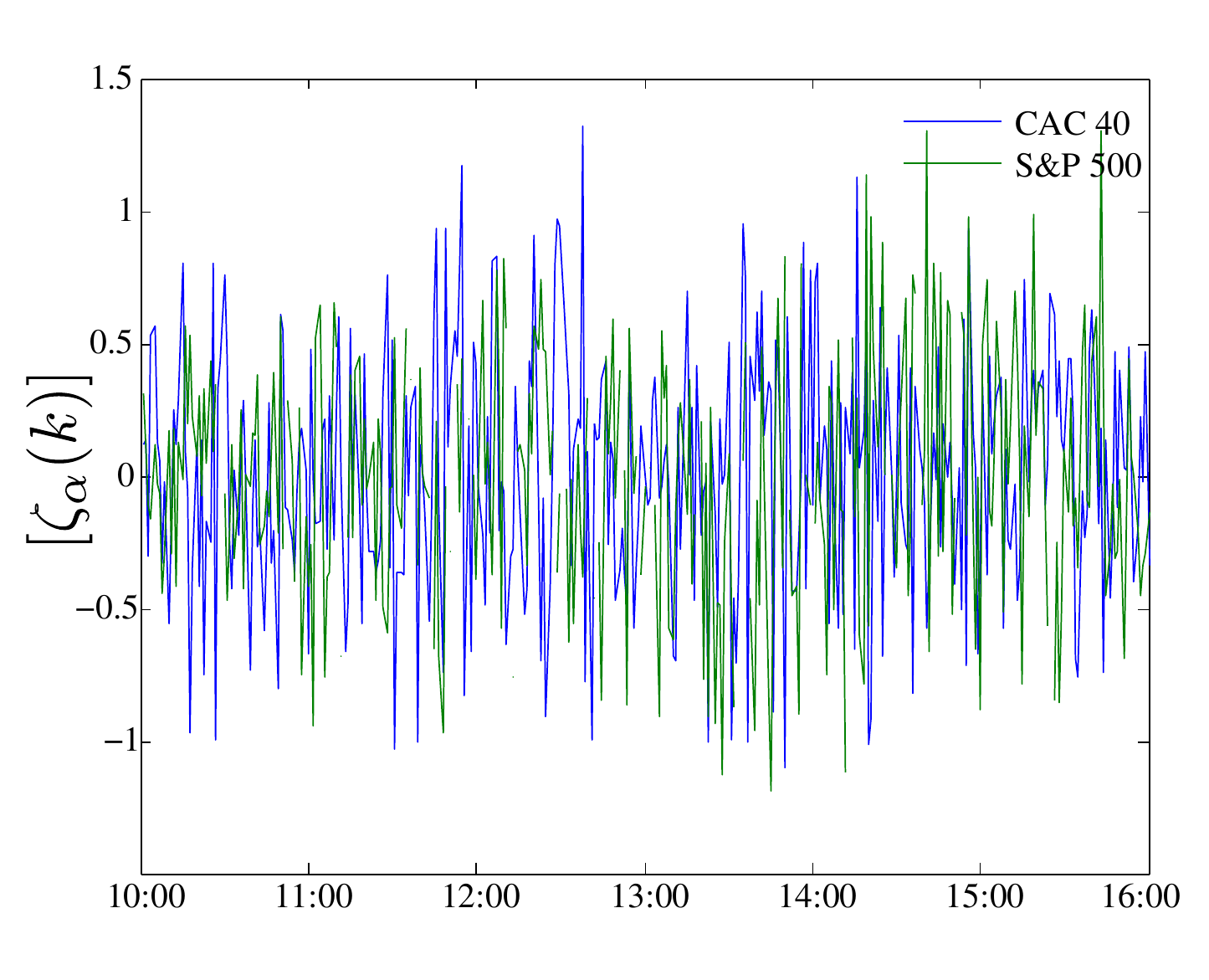}}
        \subfigure[KURTOSIS]{\includegraphics [width=0.48\textwidth] {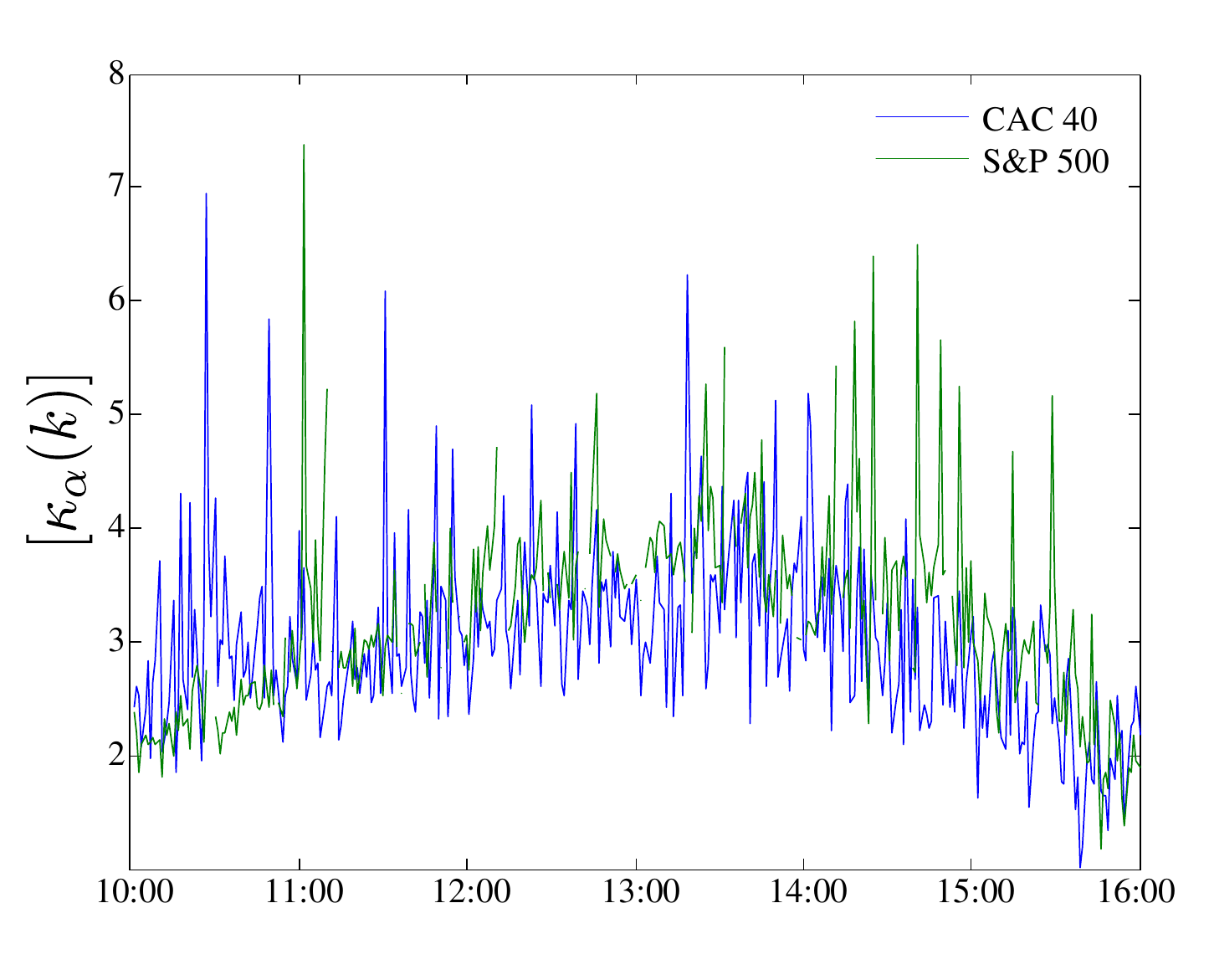}}                  
    	\end{center}
    \caption[\quad Single Stock Intra-Day Seasonalities: Returns]{Single Stock Intra-day Seasonalities: Stock average of the single stock mean, volatility, skewness and kurtosis for the CAC $40$ (blue) and the S\&P $500$ (green). $T=1$.}
\label{fig:Fig2E}
\end{figure}

\subsection[\quad Single Stock Intra-day Seasonalities]{\quad Single Stock Intra-day Seasonalities}
Figure~\ref{fig:Fig2E} shows the stock average of the single stock mean $\left[\mu_{\alpha }(k)\right]$, volatility $\left[\sigma _{\alpha }(k)\right]$, skewness $\left[\zeta_{\alpha }(k)\right]$ and kurtosis $\left[\kappa_{\alpha }(k)\right]$ for the CAC\ $40$ (blue) and the S\&P $500$ (green), and $T=1$ minute bin. As can be seen in Fig.~\ref{fig:Fig2E}(a), the mean tends to be small (in the order of $10^{-4}$) and noisy around zero. The average volatility reveals the well known U-shaped pattern (Fig.~\ref{fig:Fig2E}(b)), high at the opening of the day, decreases during the day and increases again at the end of the day. The average skewness (Fig.~\ref{fig:Fig2E}(c)) is also noisy around zero. The average kurtosis exhibits an inverted U-pattern (Fig.~\ref{fig:Fig2E}(d)), it increases from around  $2$ at the beginning of the day to around $4$ at mid day, and decreases again during the rest of the day.

\begin{figure}[t!]
        \begin{center}
        \subfigure[MEAN]{\includegraphics [width=0.48\textwidth] {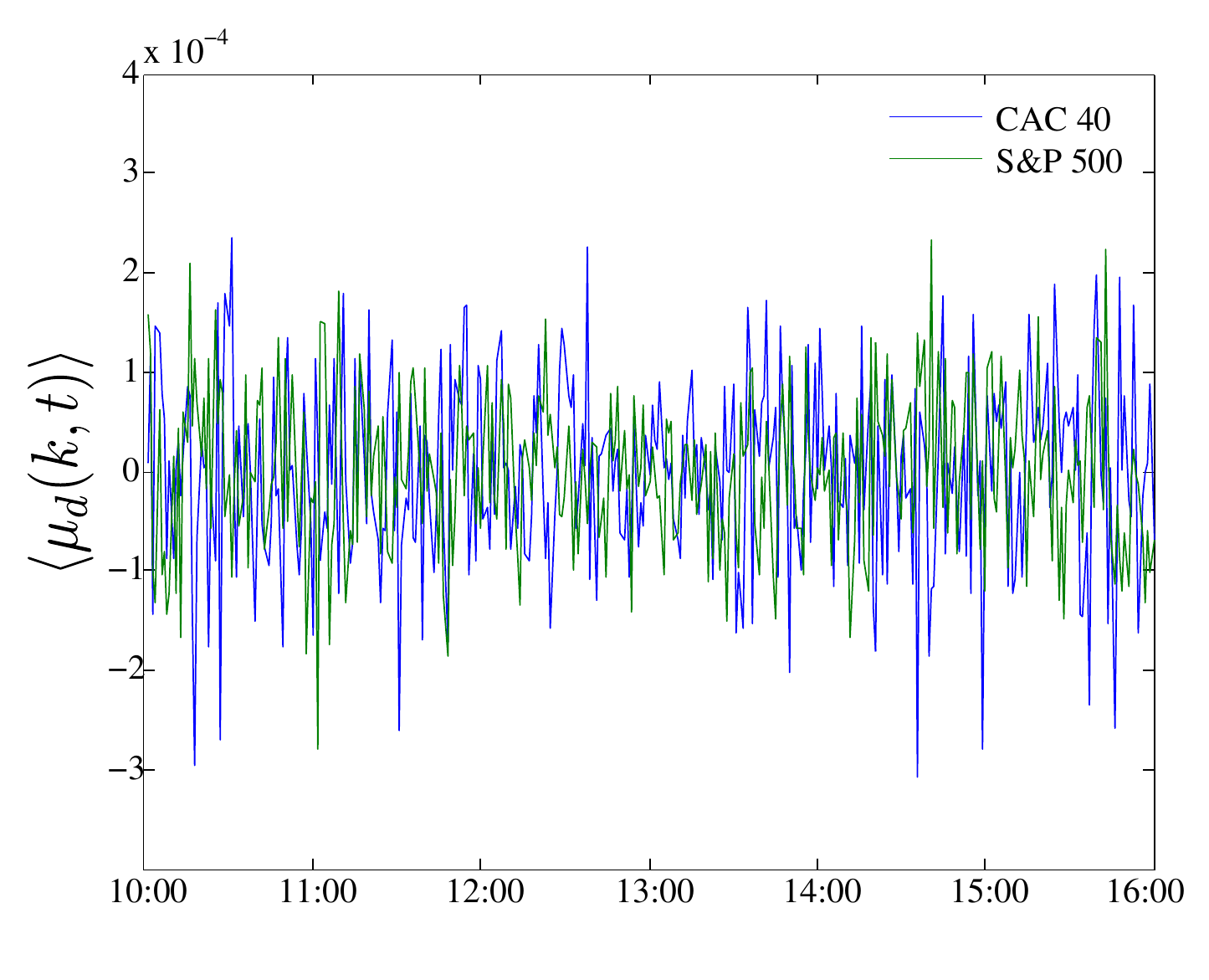}}
		\subfigure[VOLATILITY]{\includegraphics [width=0.48\textwidth] {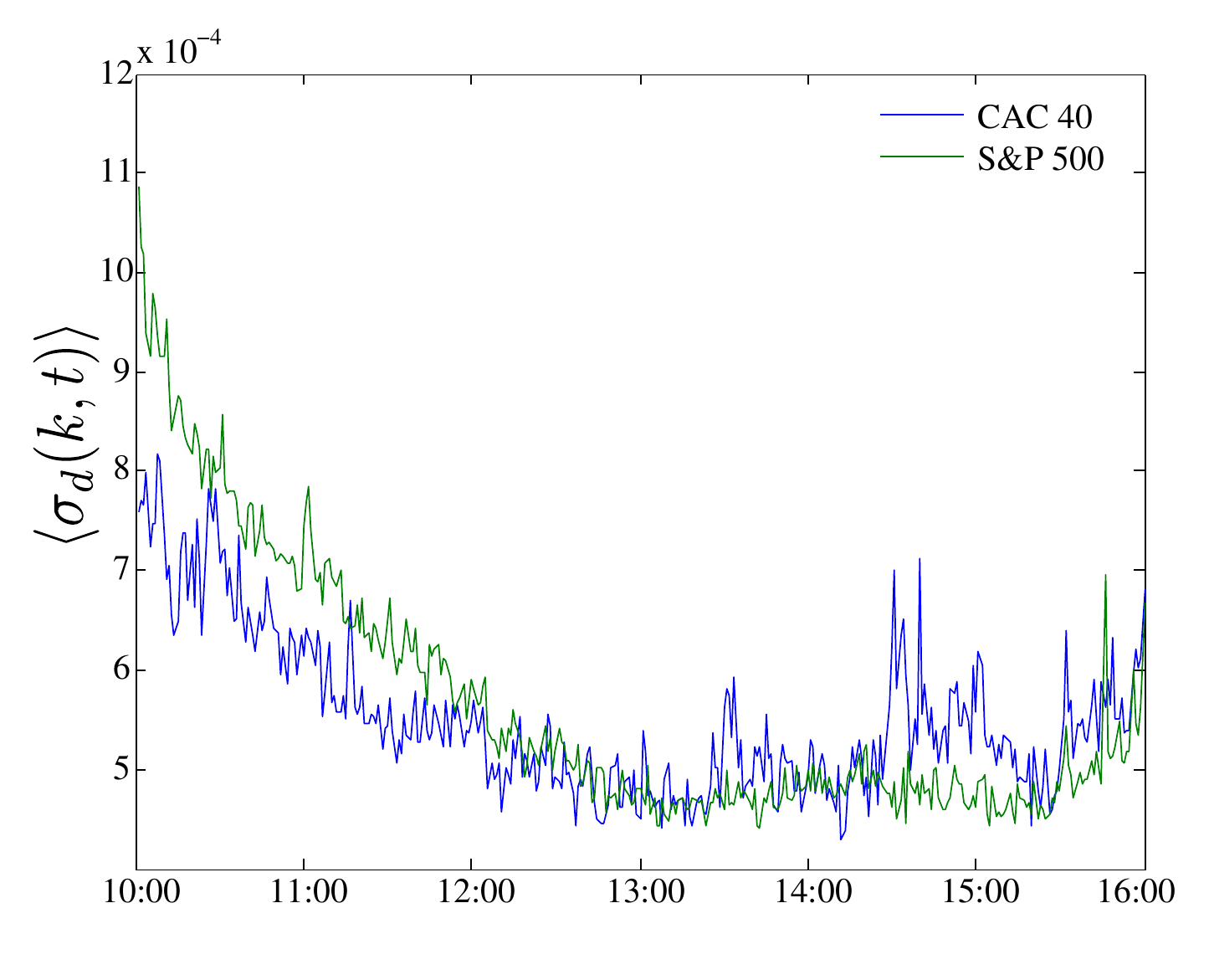}}\\ 
        \subfigure[SKEWNESS]{\includegraphics [width=0.48\textwidth] {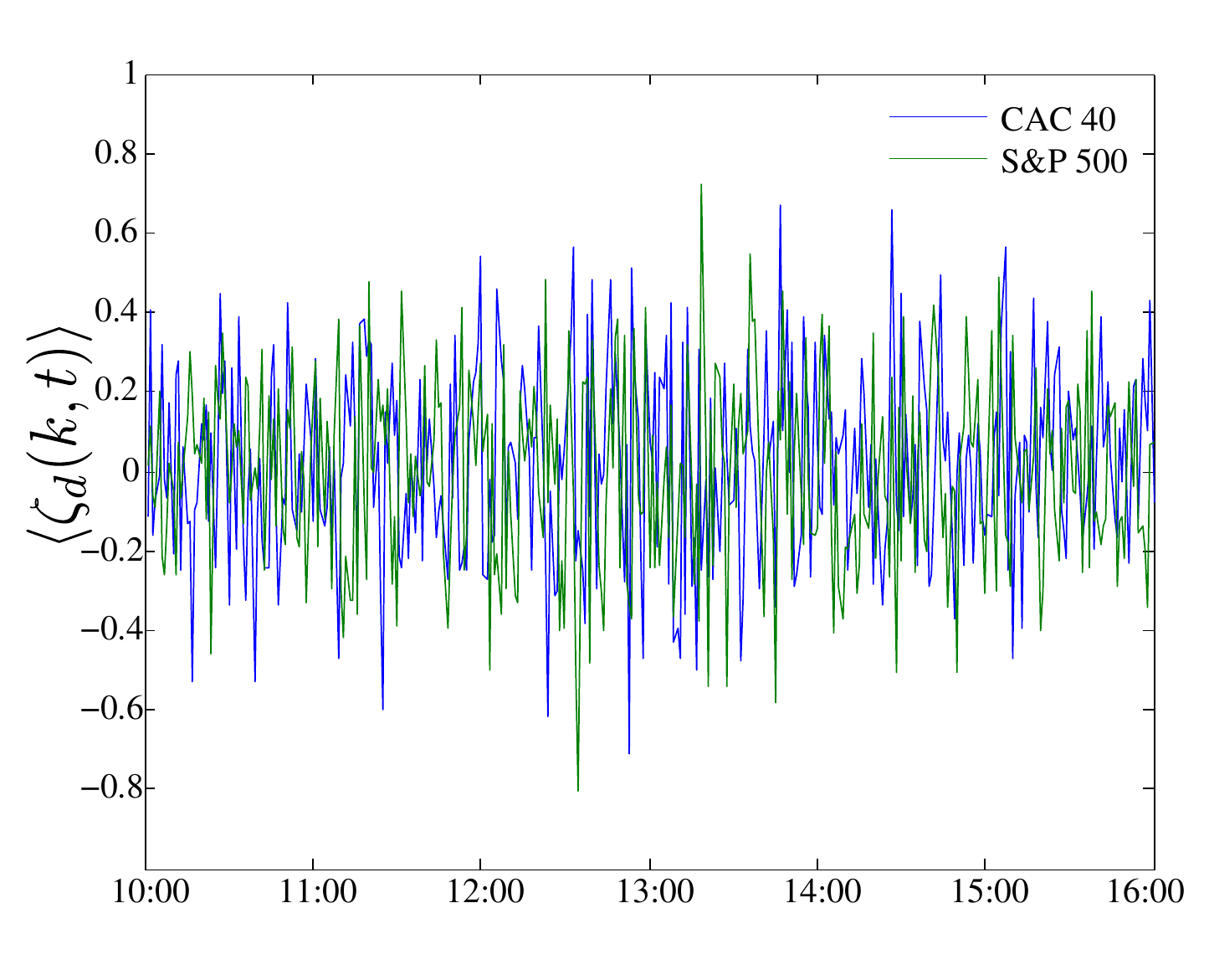}}
        \subfigure[KURTOSIS]{\includegraphics [width=0.48\textwidth] {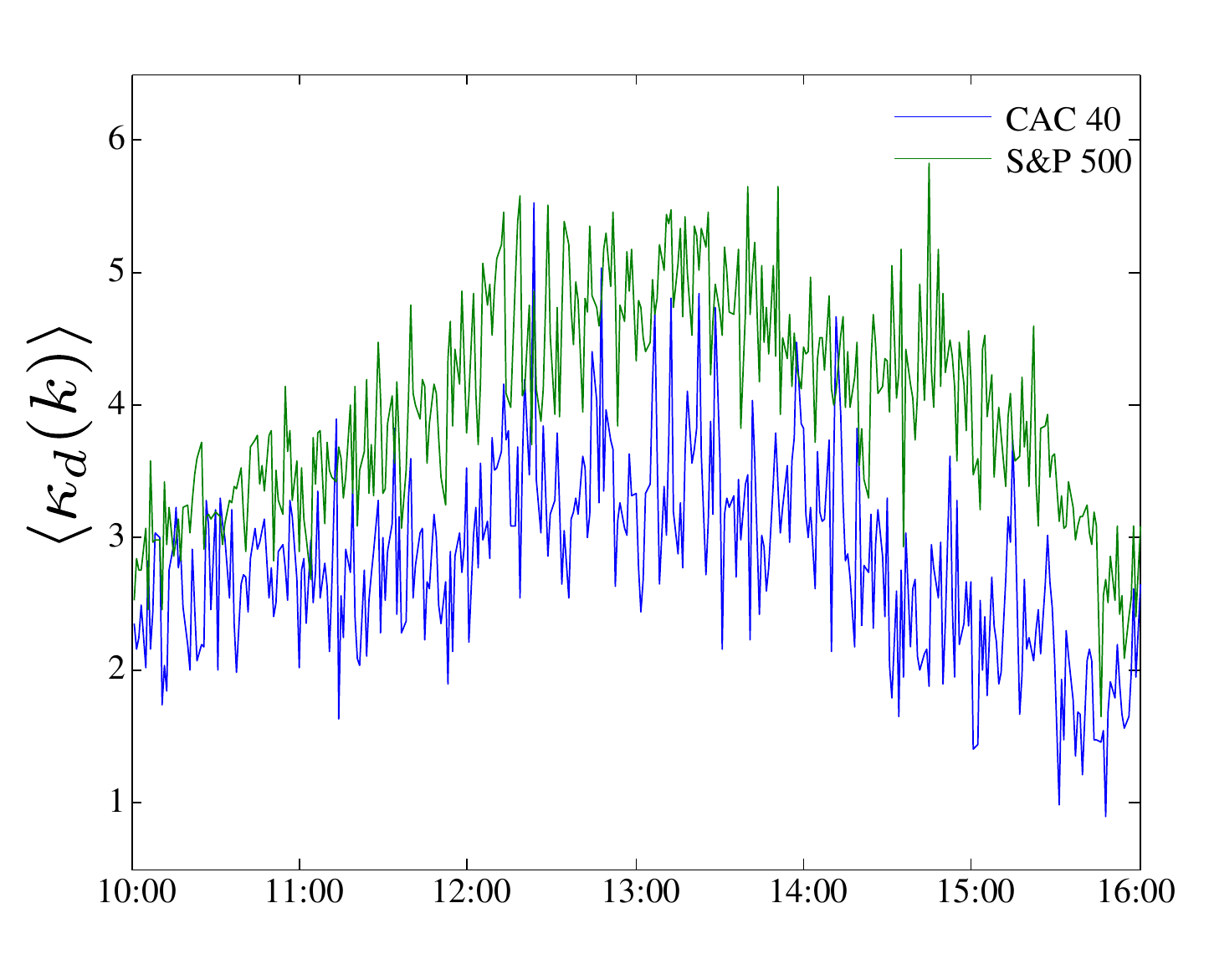}}                  
    	\end{center}
    \caption[\quad Cross-Sectional Intra-day Seasonalities: Returns]{Cross-Sectional Intra-day Seasonalities: Time average of the cross-sectional mean, volatility, skewness and kurtosis for the CAC $40$ (blue) and the S\&P $500$ (green), and $T=1$ minute bin.}
\label{fig:Fig3E}
\end{figure}

\subsection[\quad Cross-Sectional Intra-day Seasonalities]{\quad Cross-Sectional Intra-day Seasonalities}
As the time average of the cross sectional mean is equal to the stock average of the single stock mean, the result we show in Fig.~\ref{fig:Fig3E}(a) is exactly the same as the one shown in Fig.~\ref{fig:Fig2E}(a). The time average of the cross sectional volatility $\langle\sigma _{d}(k,t)\rangle$ (Fig.~\ref{fig:Fig3E}(b)) reveals a U-shaped pattern very similar to the stock average volatility, but less noisy (less pronounced peaks). The dispersion of stocks is stronger at the beginning of the day and decreases as the day proceeds. The average skewness $\langle\zeta _{d}(k,t)\rangle$ is noisy around zero without any particular pattern (Fig.~\ref{fig:Fig3E}(c)). The cross sectional kurtosis $\langle\kappa _{d}(k)\rangle$ (Fig.~\ref{fig:Fig3E}(d)) also exhibits an inverted U-pattern as in the case of the single stock kurtosis. It increases from around $2.5$ at the beginning of the day to around $4.5$ at mid day, and decreases again during the rest of the day. This means that at the beginning of the day the cross-sectional distribution of returns is on average closer to Gaussian.%
\bigskip 

\subsection[\quad U-Pattern Volatilities]{\quad U-Pattern Volatilities}
\label{sec:U}
\begin{figure}[t!]
        \begin{center}
        \subfigure {\includegraphics [width=0.48\textwidth] {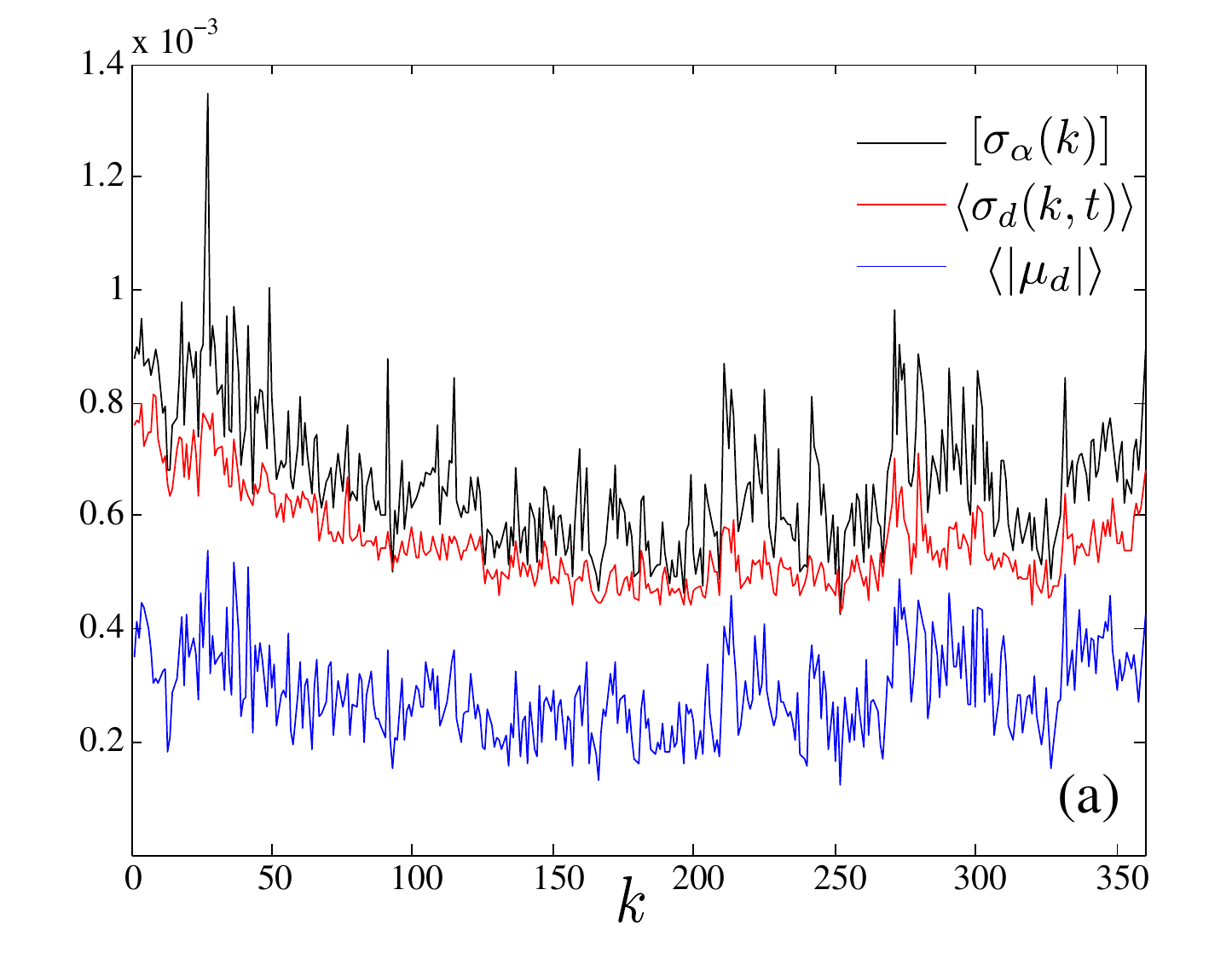}}
		\subfigure {\includegraphics [width=0.48\textwidth] {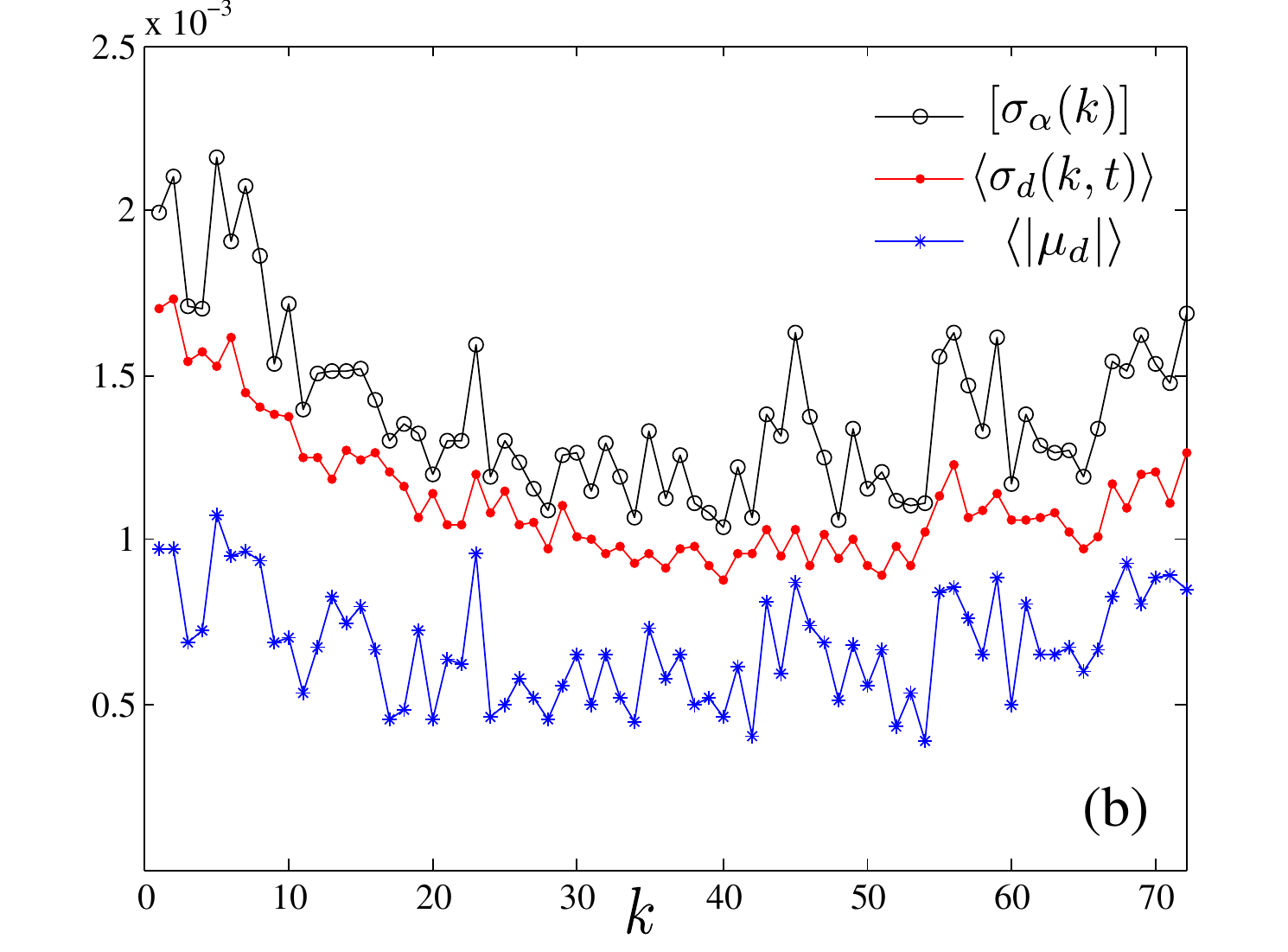}}           
    	\end{center}
    \caption[\quad U-Pattern Volatilities]{U-Pattern Volatilities: Stock average of single stock volatility $\left[\sigma _{\alpha }(k)\right]$ (black), time average of the cross-sectional volatility $\langle\sigma _{d}(k,t)\rangle$ (red) and the average absolute value of the equi-weighted index return $\langle|\mu _{d}|\rangle$ (blue) for the CAC\ $40$, for \textbf{(a)} $T=1$ minute bin and \textbf{(b)} $T=5$ minute bin. Similar results were obtained for the S\&P $500$.}
\label{fig:Fig4E}
\end{figure}

In Fig.~\ref{fig:Fig4E}, we compare the stock average of single stock volatility $\left[\sigma _{\alpha }(k)\right]$ (black), the time average of the cross-sectional volatility $ \langle \sigma _{d}(k,t)\rangle$ (red) and the average absolute value of the equi-weighted index return $\langle|\mu _{d}|\rangle$ (blue) for the CAC\ $40$, and for $T=1$ (left) and $T=5$ minute bin (right). Similar results were obtained for the S\&P $500$. As can be seen, the average absolute value of the equi-weighted index return also exhibits a U-shaped pattern and it is a proxy for the index volatility. One thing that results interesting to observe is that the values of these volatilities actually depends of the size of the bin that we consider. For $T=5$ minute bin, the volatilities double the values found for $T=1$ minute bin (we will discuss this result in the next sections).


\subsection[\quad Intra-day Seasonalities in the Stock Correlation]{\quad Intra-day Seasonalities in the Stock Correlation}
\begin{figure}[t!]
        \begin{center}
        \subfigure
        {\includegraphics [width=0.48\textwidth] {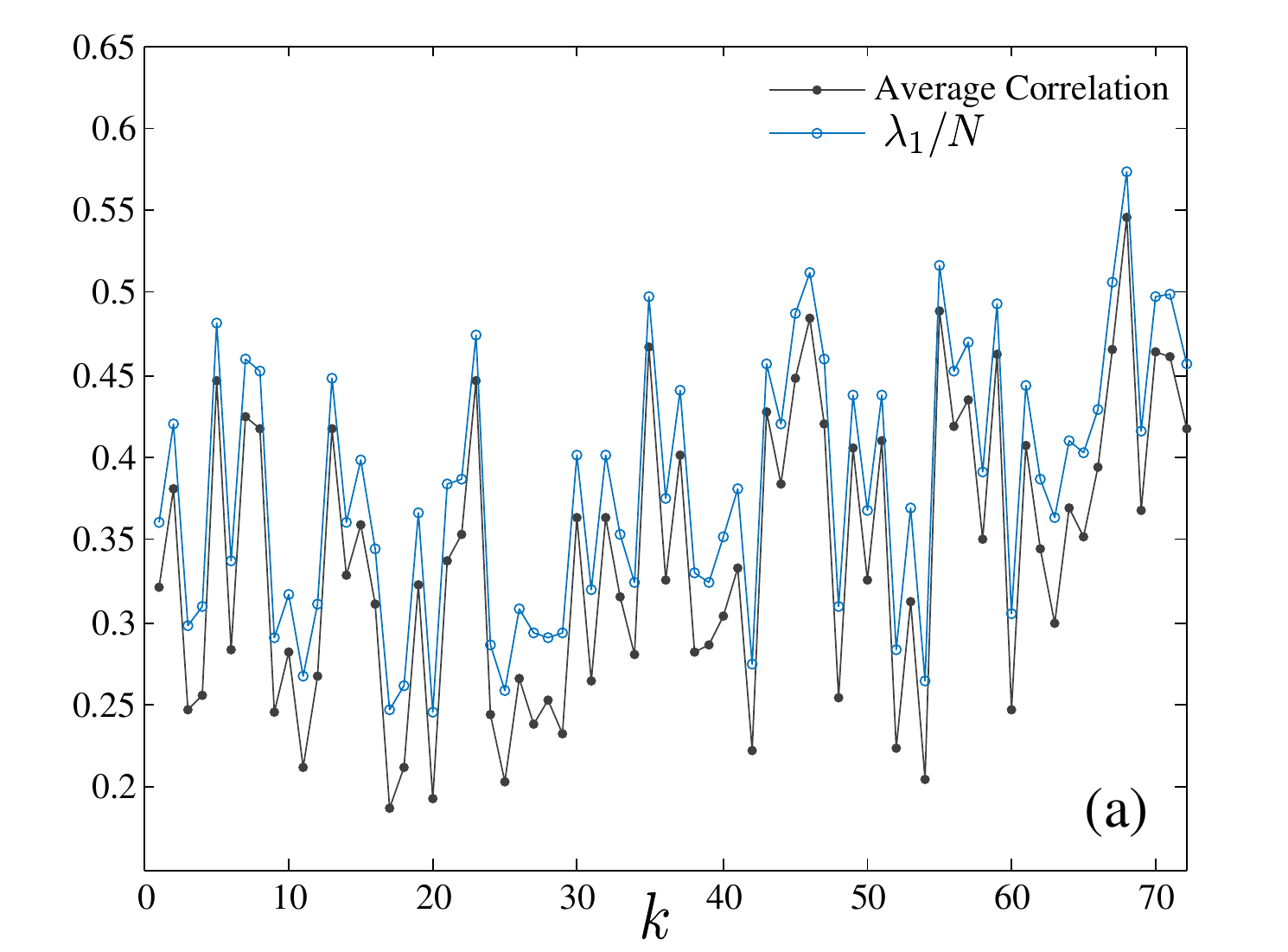}}
		\subfigure{\includegraphics [width=0.48\textwidth] {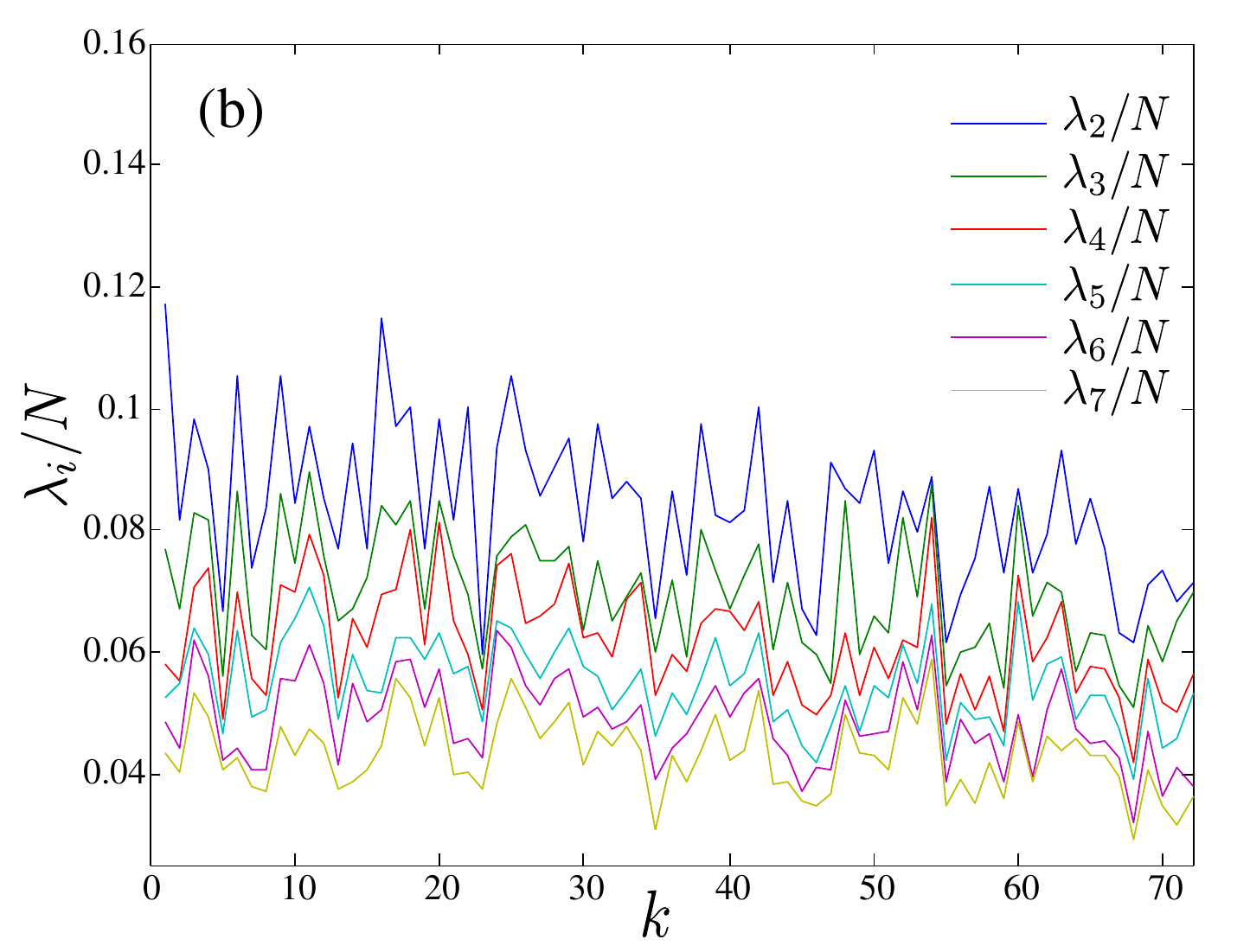}}           
    	\end{center}
    \caption[\quad Largest Eigenvalues Structure]{Largest eigenvalues structure for the CAC $40$, $T=5$ minute bin. \textbf{(a)} Average correlation between stocks (black) and top eigenvalue $\lambda _{1}/N$ (blue) of the correlation matrix $C_{\alpha \beta }(k)$. \textbf{(b)} Smaller eigenvalues.}
\label{fig:Fig5E}
\end{figure}

In order to compute the correlation between stocks, we first normalize the returns by the dispersion of the corresponding bin~\cite{7} i.e.,
\begin{equation}
\widehat{x}_{\alpha}(k,t) = x_{\alpha }^{(1)}(k,t)/\sigma _{d}(k,t)
\label{11}
\end{equation}%
The $N\times N$ correlation matrix for a given bin $k$ would be given by%
\begin{equation}
C_{\alpha \beta }(k)=\frac{\left\langle \widehat{x}_{\alpha }(k,t)%
\widehat{x}_{\beta}(k,t)\right\rangle -\left\langle \widehat{x}_{\alpha}(k,t)\right\rangle \left\langle \widehat{x}_{\beta}(k,t)\right\rangle }{\sigma _{\alpha}(k)\sigma _{\beta}(k)}. \label{12}
\end{equation}
In Fig.~\ref{fig:Fig5E}(a) we show the average correlation between stocks (blue) and top eigenvalue $\lambda_{1}/N$ (green) for the CAC $40$. As can be seen the largest eigenvalue is a measure of the average correlation between stocks~\cite{7, 13, 14, 15, 16}. This average correlation increases during the day from a value around $0.35$ to a value around $0.45$ when the market closes. For the case of smaller eigenvalues, what we can see is that the amplitude of risk factors decreases during the day (Fig.~\ref{fig:Fig5E}(b)), as more and more risk is carried by the market factor (Fig.~\ref{fig:Fig5E}(a))~\cite{7}.%

\begin{figure}[t!]
        \begin{center}
        \subfigure
        {\includegraphics [width=0.48\textwidth] {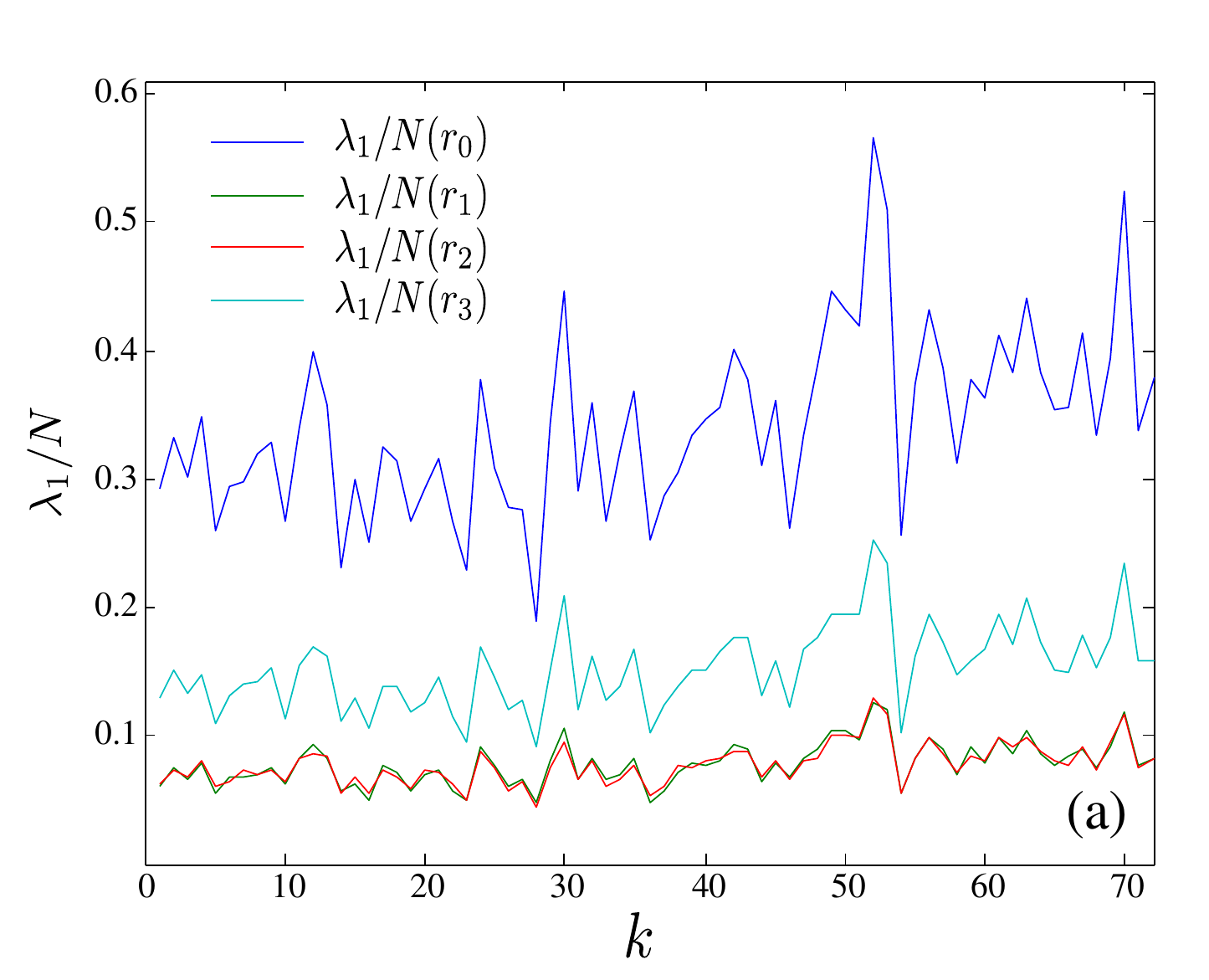}}
		\subfigure{\includegraphics [width=0.48\textwidth] {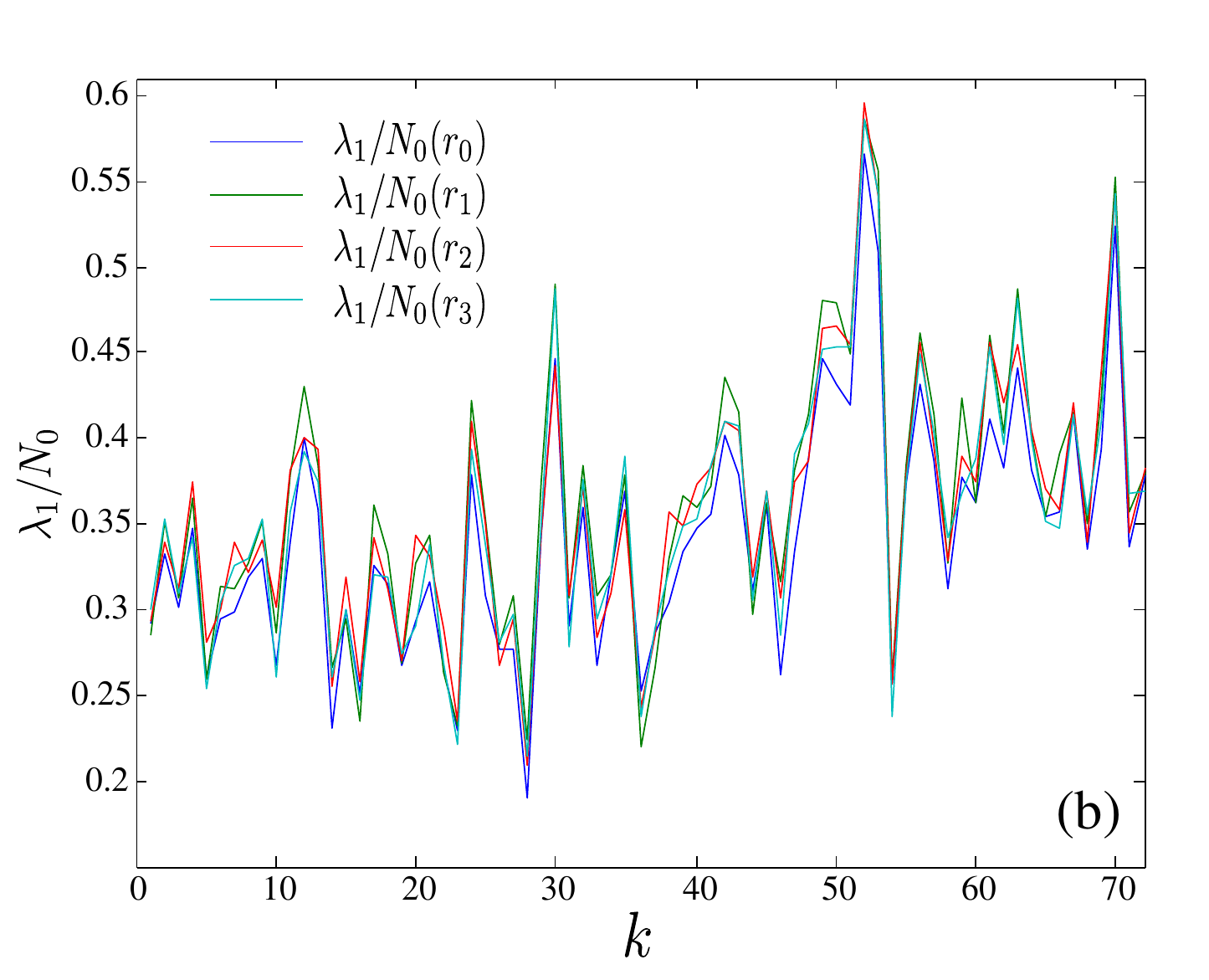}}           
    	\end{center}
    \caption[\quad Top Eigenvalues]{\textbf{(a)} Top eigenvalue $\lambda_{1}/N$ and \textbf{(b)} $\lambda_{1}/N_{0}$ for the S\&P $500$ for $4$ different sets of stocks: $r_{0}$ (blue), $r_{1}$ (green), $r_{2}$ (red) and $r_{3}$(clear blue). $T=5$ minute bin.}
\label{fig:Fig6E}
\end{figure}

In order to simplify the computation of the $N^{2}$ correlation matrices for each bin $k$ in the case of the S\&P $500$, we computed the correlation matrix $C_{\alpha \beta }$ for $4$ different sets of stocks: $r_{0}$: composed by the $100$ first stocks of the S\&P $500$; $r_{1,2}$: composed by  $100$ stocks randomly picked; and $r_{3}$: composed by $200$ stocks randomly picked. Figure~\ref{fig:Fig6E}(a) shows $\frac{\lambda _{1}}{N}$ as function of the bins. Although the values of the eigenvalues seem to be out of scale, it can be seen clearly that the average correlation increases during the day. This scale conflict is solved by normalizing the value of the top eigenvalue not by $N$ but by the sample size $N_{0}$ (i.e., $100$ or $200$) (Fig.~\ref{fig:Fig6E}(b)). As can be seen the average correlation of the index can be computed by taking a subset of it which means that actually just the more capitalized stocks in the index drive the rest of stocks.%

\newpage

\section[\quad Intra-day Seasonalities for Relative Prices]{\quad Intra-day Seasonalities for Relative Prices}
In this section, we will report the results we found for the S\&P $500$. Similar results were found also for the CAC $40$. We will see how in the case of the relative prices these intra-day seasonalities are independent of the size of the bin, also independent of the index we consider (but characteristic for each index) however this is not the case for the returns.

\subsection[\quad Single Stock Intra-day Seasonalities]{\quad Single Stock Intra-day Seasonalities}
\label{SSIDS}
\begin{figure}[th!]
        \begin{center}
        \subfigure[MEAN]{\includegraphics [width=0.48\textwidth] {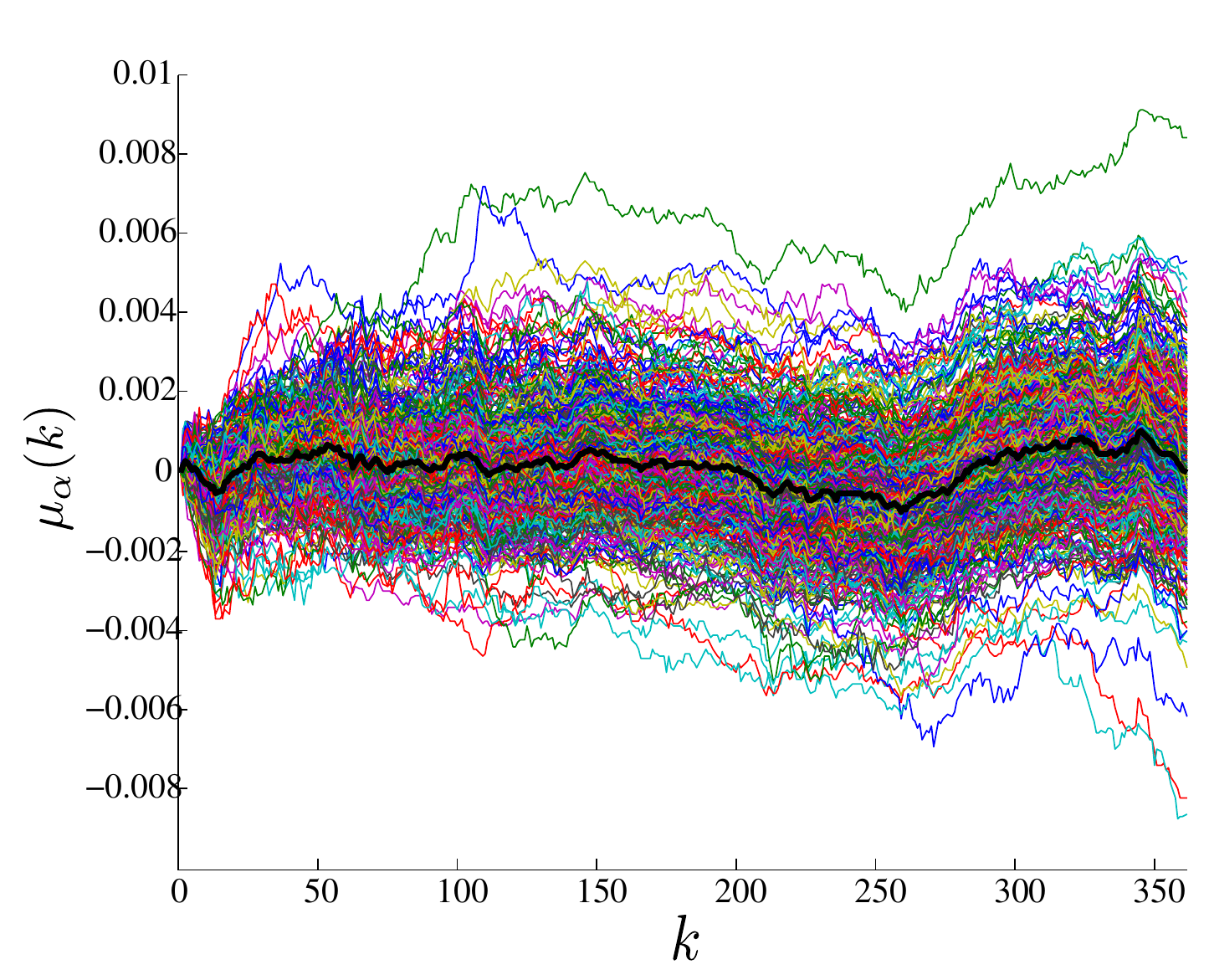}}
		\subfigure[VOLATILITY]{\includegraphics [width=0.48\textwidth] {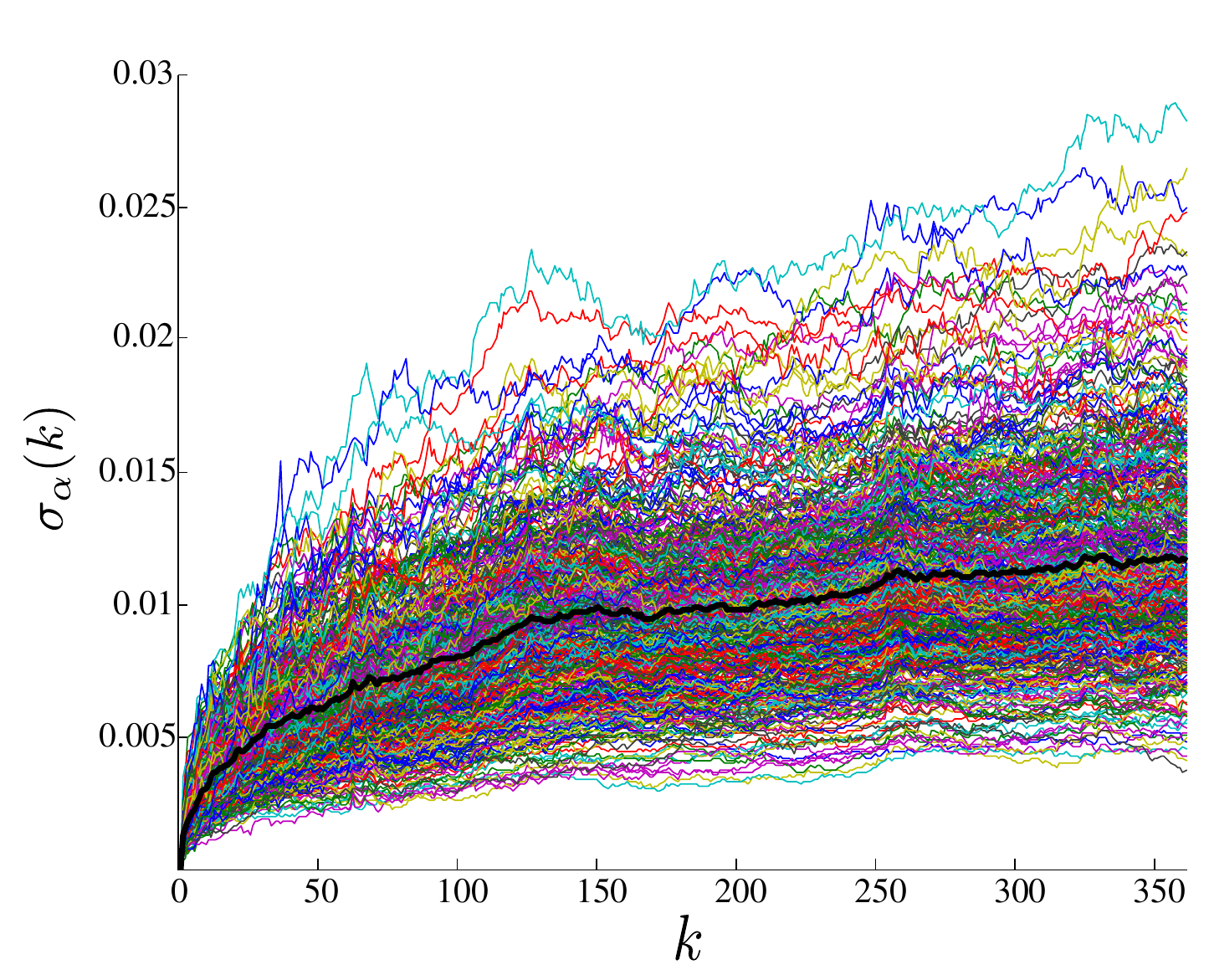}}\\ 
        \subfigure[SKEWNESS]{\includegraphics [width=0.48\textwidth] {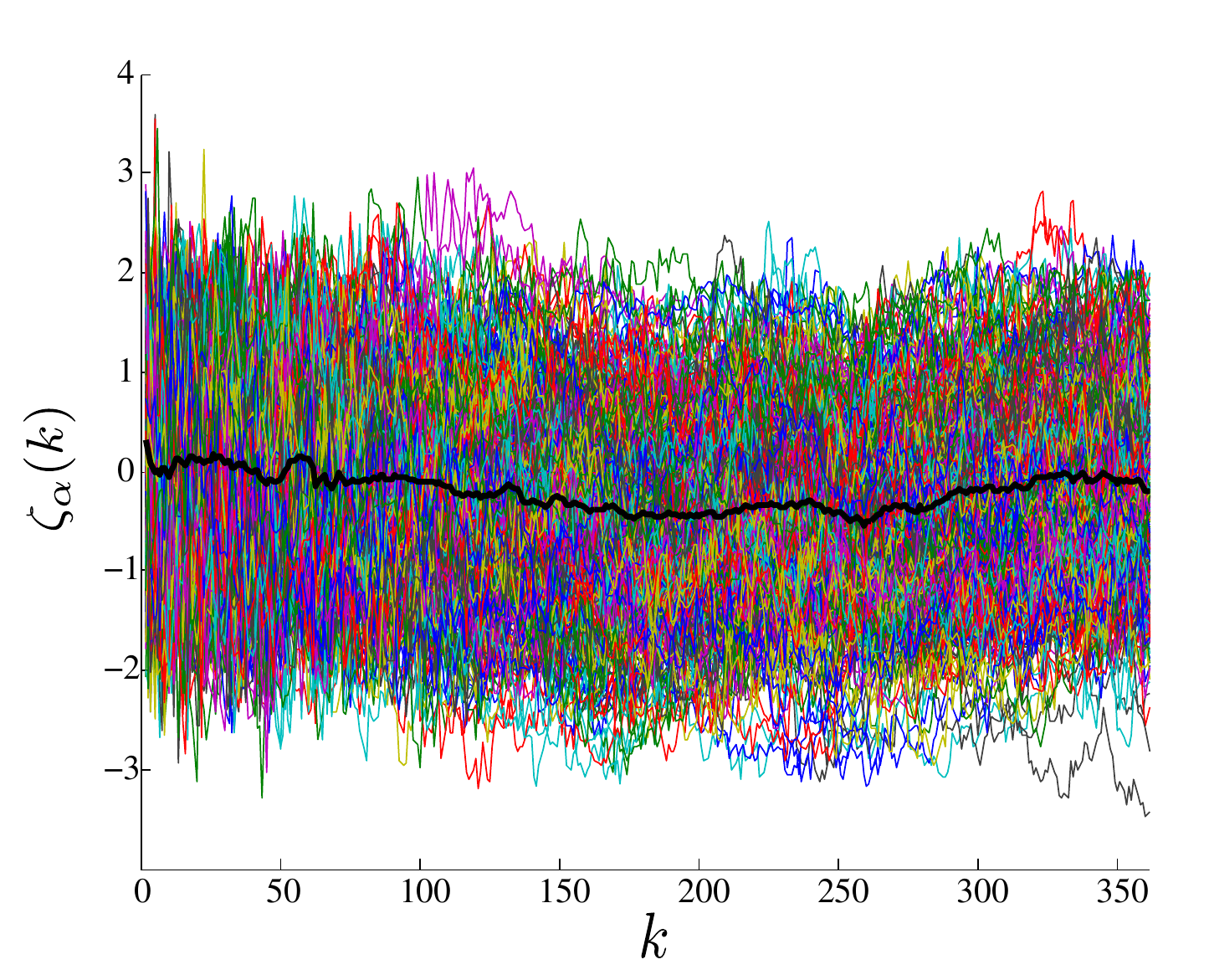}}
        \subfigure[KURTOSIS]{\includegraphics [width=0.48\textwidth]{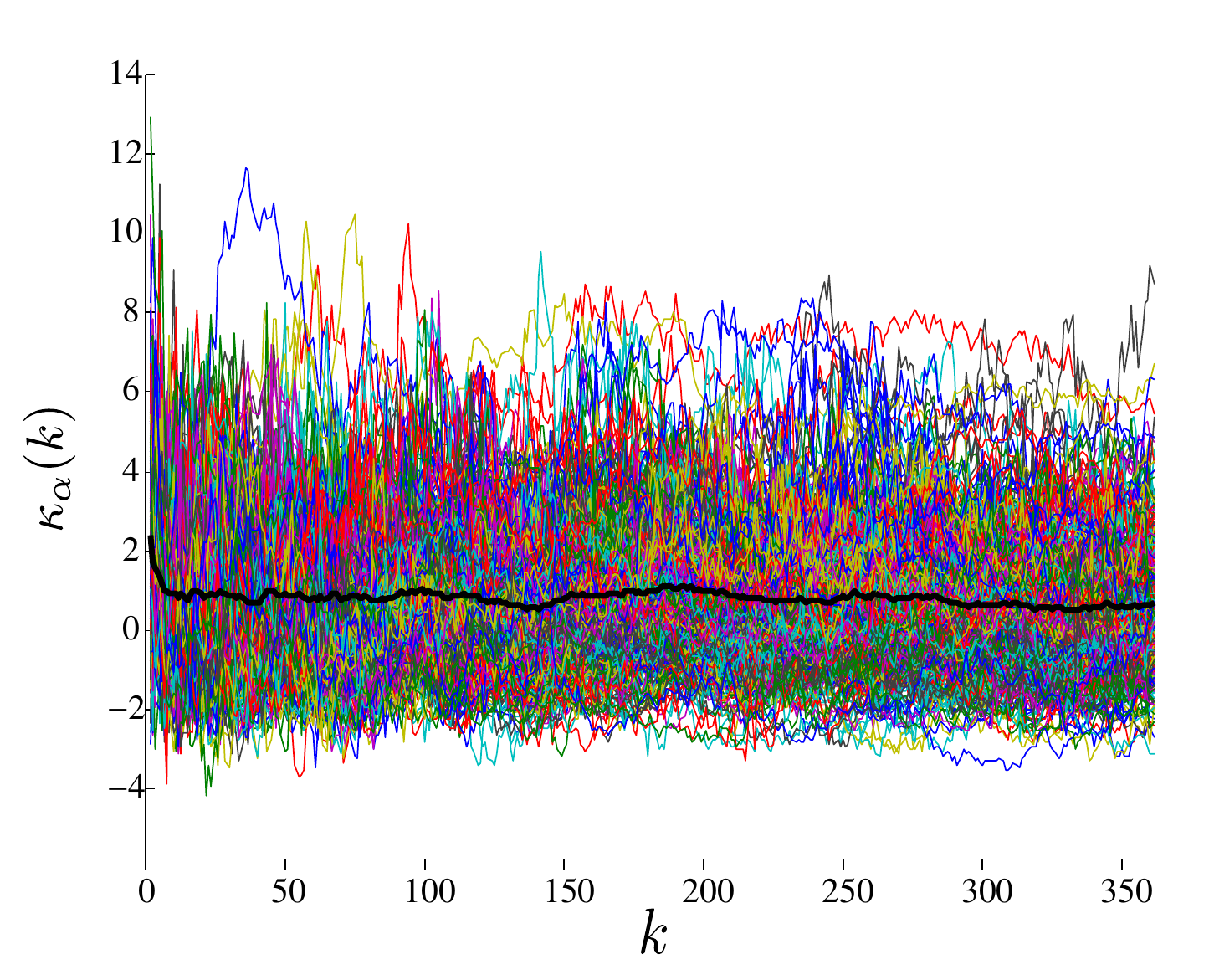}}                  
    	\end{center}
    \caption[\quad Single Stock Intra-day Seasonalities: Relative Prices]{Single Stock Intra-day Seasonalities: Stock average of the single stock mean, volatility, skewness and kurtosis for the S\&P $500$ (black). $T=1$ minute bin.}
\label{fig:Fig7E}
\end{figure}

Each path in Fig.~\ref{fig:Fig7E} represents the evolution of a particular moment of one of the stocks that compose the S\&P $500$ (i.e., one path, one stock moment). The stock average of the single stock mean $\left[\mu _{\alpha }(k)\right]$, volatility $\left[\sigma_{\alpha }(k)\right]$, skewness $\left[\zeta_{\alpha }(k)\right]$ and kurtosis $\left[\kappa_{\alpha }(k)\right]$ of the S\&P $500$ are shown in black. The stock average of the single stock mean varies around zero. The average volatility increases logarithmically with time. The skewness varies between $[-3,3]$ with an average value of zero. The single stock kurtosis takes values between $[-2,6]$ with an average value of one and its stock average starts from a value around $2$ in the very beginning of the day and decreases quickly to the mean value $1$ in the first minutes of the day.%

\subsection[\quad Cross-Sectional Intra-day Seasonalities]{\quad Cross-Sectional Intra-day Seasonalities}
\begin{figure}[th!]
        \begin{center}
        \subfigure[MEAN]{\includegraphics [width=0.48\textwidth] {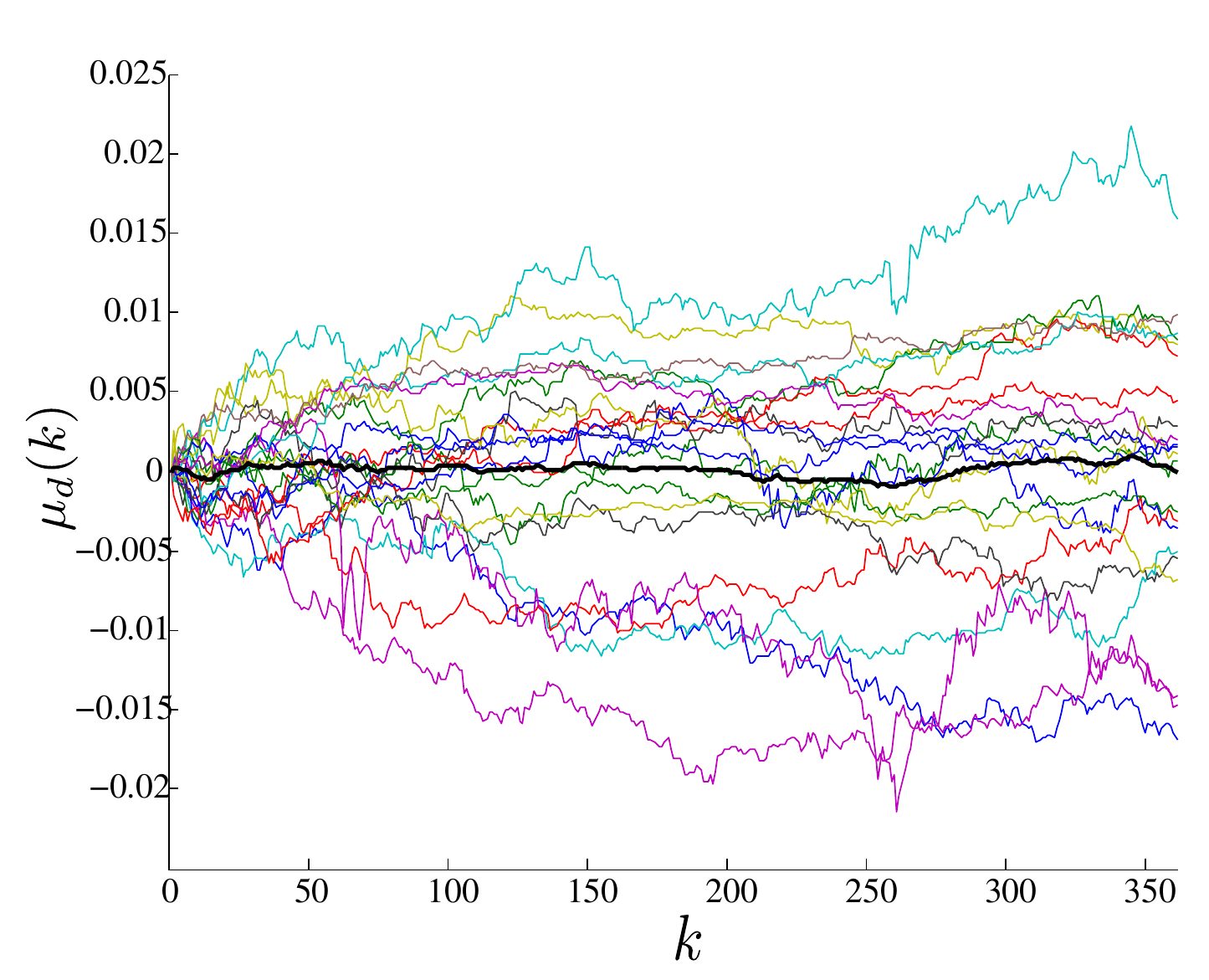}}
		\subfigure[VOLATILITY]{\includegraphics [width=0.48\textwidth] {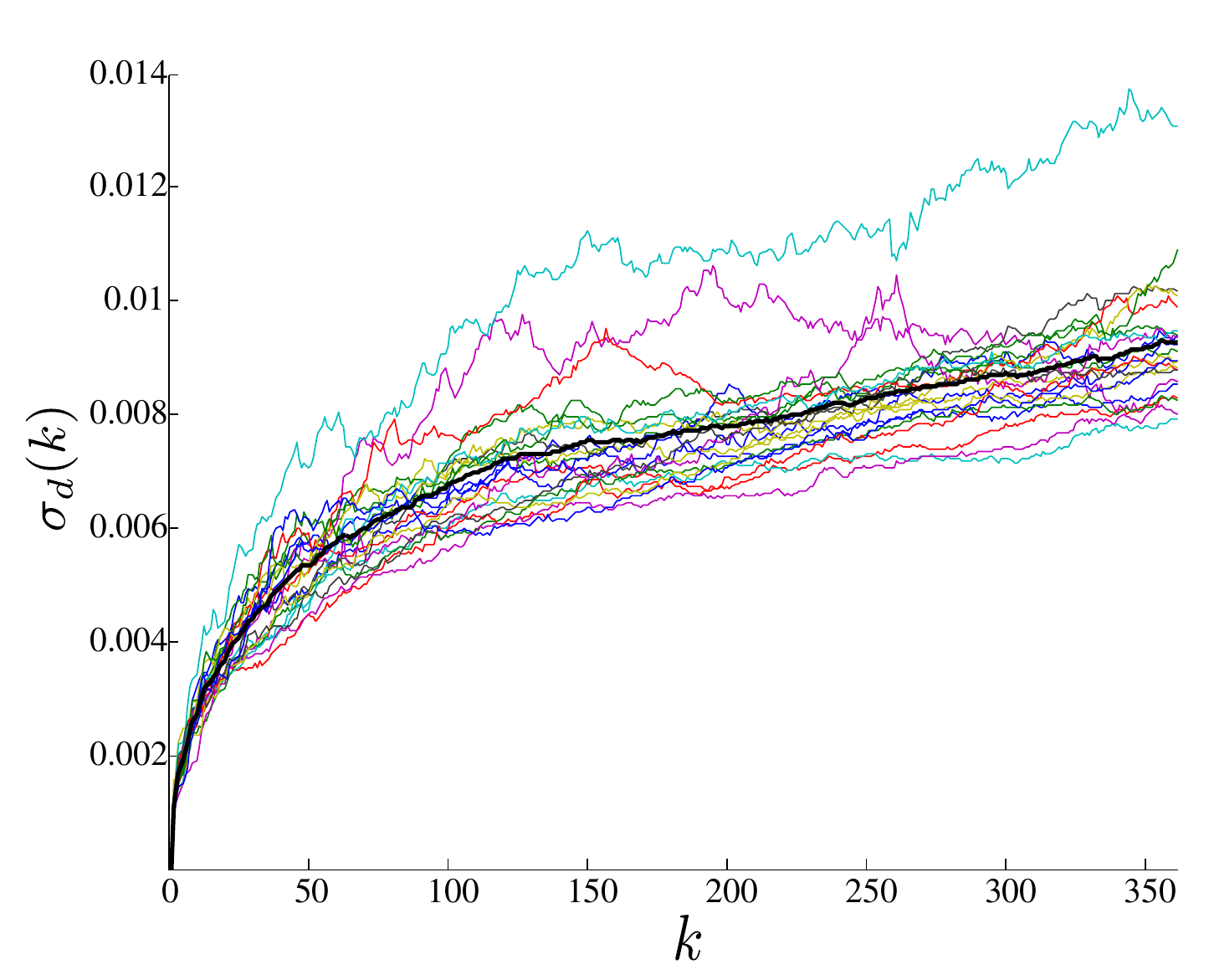}}\\ 
        \subfigure[SKEWNESS]{\includegraphics [width=0.48\textwidth] {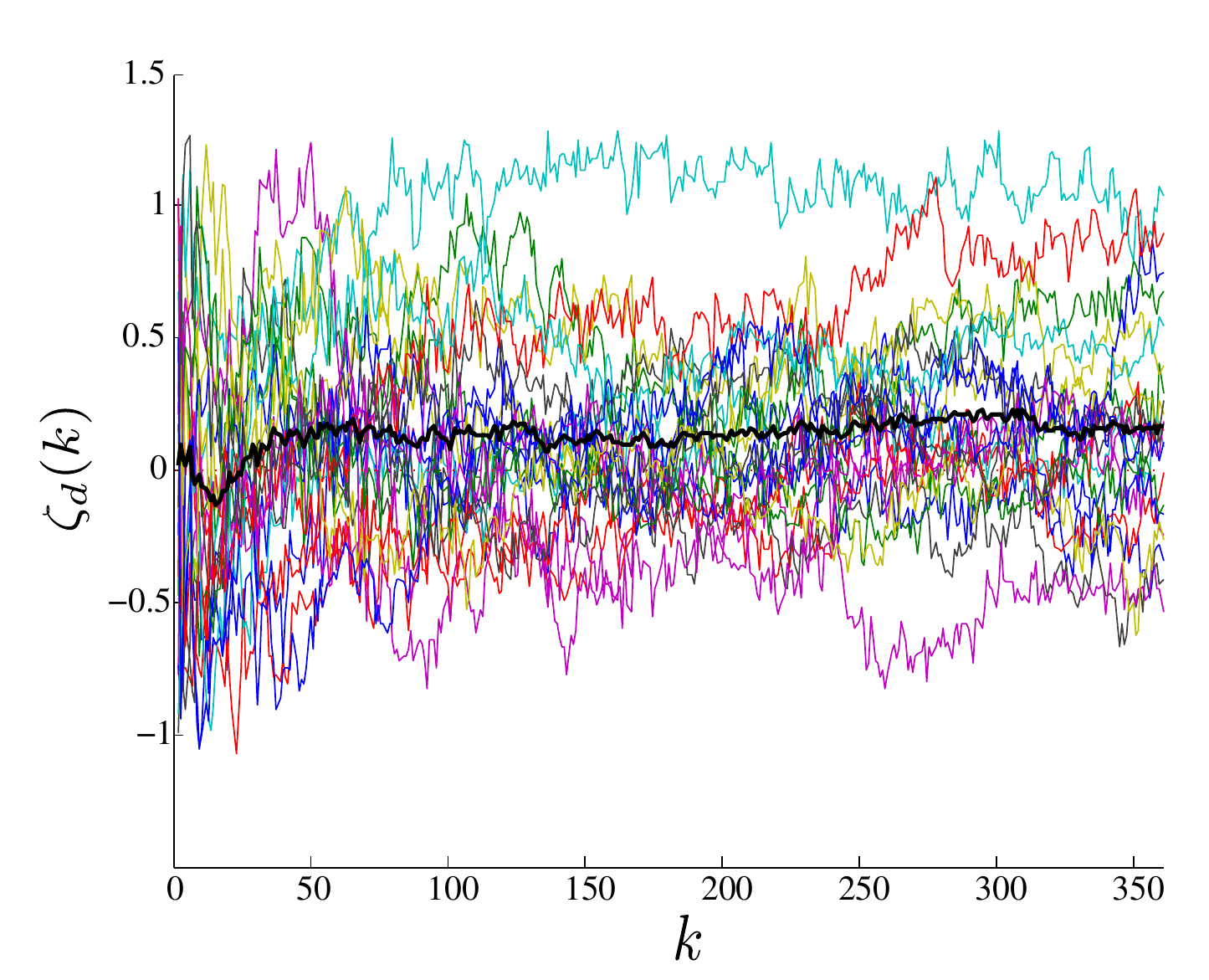}}
        \subfigure[KURTOSIS]{\includegraphics [width=0.48\textwidth] {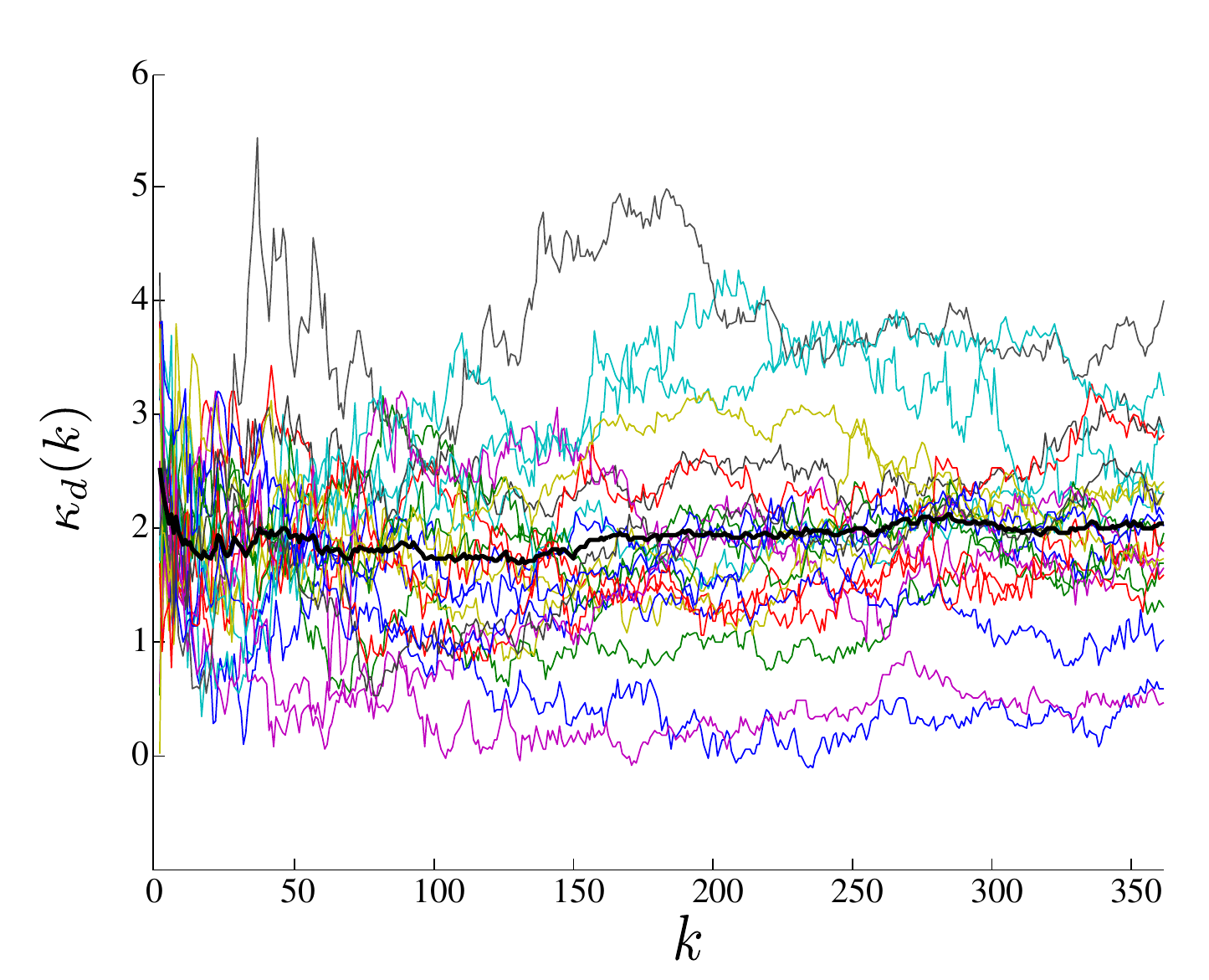}}                  
    	\end{center}
    \caption[\quad Cross-Sectional Intra-day Seasonalities: Relative Prices]{Cross-Sectional Intra-day Seasonalities: Time average of the cross-sectional mean, volatility, skewness and kurtosis for the S\&P $500$ (black). $T=1$ minute bin.}
\label{fig:Fig8E}
\end{figure}

Each path in Fig.~\ref{fig:Fig8E} represents the evolution of a particular index moment during a particular day (i.e., one path, one day moment). As in the case of the single stock volatility, the cross-sectional dispersion $\langle\sigma_{d}(k)\rangle$ increases logarithmically with respect to the time (Fig.~\ref{fig:Fig8E}(b)). The cross-sectional skewness $\langle\zeta_{d}(k)\rangle$ takes values in the interval $[-1,1]$ with an average value of zero (Fig.~\ref{fig:Fig8E}(c)). The average kurtosis $\langle\kappa _{d}(k)\rangle$ starts from a value around $2.5$ in the very beginning of the day and decreases quickly to the mean value $2$ in the first minutes of the day (Fig.~\ref{fig:Fig8E}(d)).

\subsection[\quad C-Pattern Volatilities]{\quad C-Pattern Volatilities}
\begin{figure}[th!]
\centering
\includegraphics [width=0.65\textwidth] {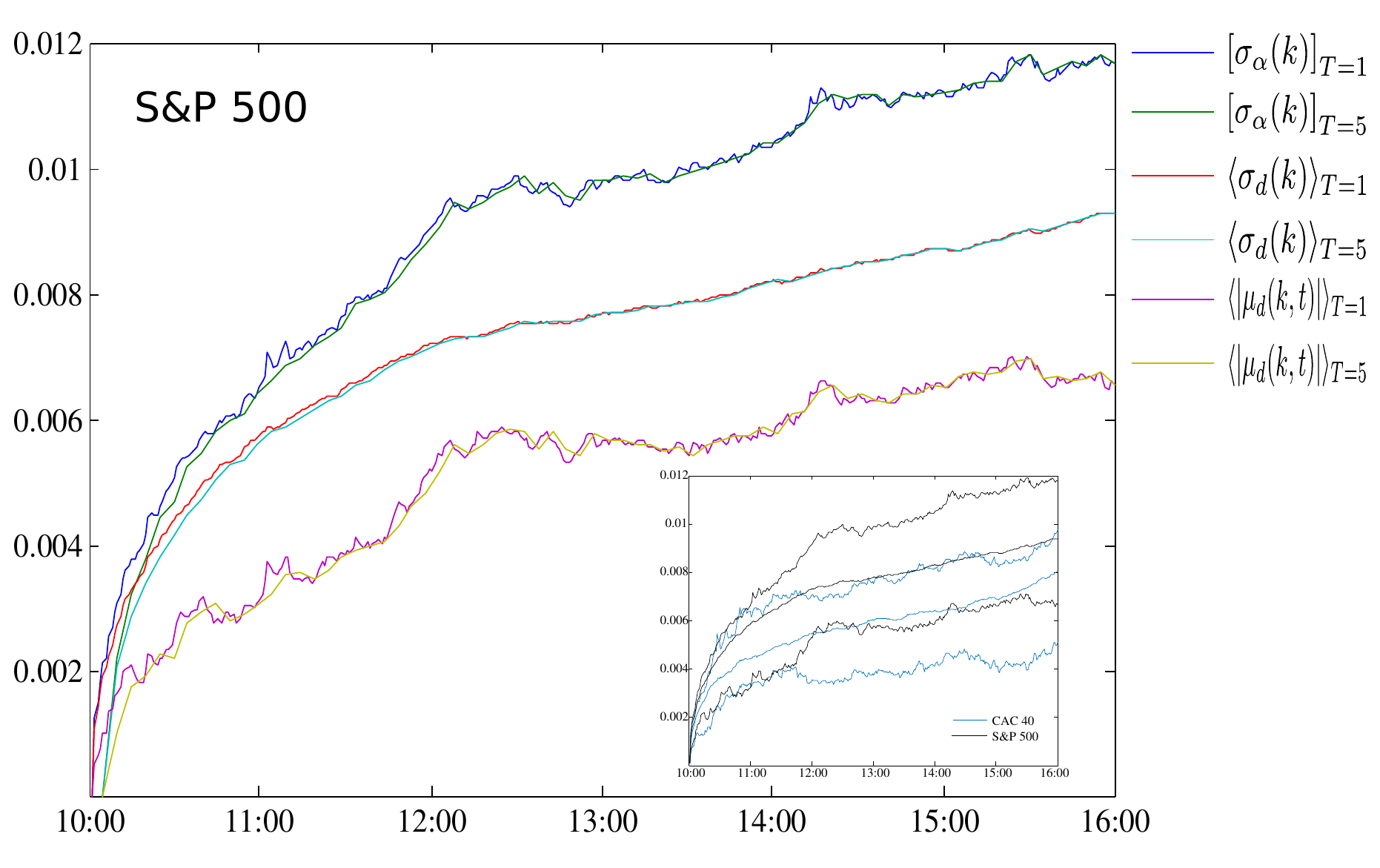}
\caption[\quad C-Pattern Volatilities]{C-Pattern Volatilities: Stock average of the single stock volatility $\left[\sigma _{\alpha }(k)\right]$, time average of the cross-sectional volatility $\langle\sigma_{d}(k,t)\rangle$ and the average absolute value of the cross-sectional mean $\langle|\mu_{d}|\rangle$ for the relative prices of the S\&P $500$. $T=1$ and $T=5$ minute bin. Inset: CAC $40$ (blue) and S\&P $500$ (black).}
\label{fig:Fig9E}
\end{figure}

Similarly as we did in Sec.~\ref{sec:U} for returns, in Fig.~\ref{fig:Fig9E} we show a comparative plot between the stock average of the single stock volatility $\left[\sigma _{\alpha }(k)\right]$, the time average of the cross-sectional volatility $\langle\sigma _{d}(k,t)\rangle$ and the average absolute value of the cross-sectional mean $\langle|\mu _{d}|\rangle$ for the relative prices
of the S\&P $500$, and for $T=1$ and $T=5$ minute bin. As can be seen, these three measures exhibit the same kind of intra-day pattern (as it did in the case of the returns). But the most important fact is to notice that this intra-day seasonality is independent of the size of the bin, also independent of the index we consider, but characteristic for each index (see inset Fig.~\ref{fig:Fig9E}). 

\section[\quad Intra-day Patterns and Bin Size]{\quad Intra-day Patterns and Bin Size}
\begin{figure}[b!]
        \begin{center}
        		\subfigure{\includegraphics [width=0.48\textwidth] {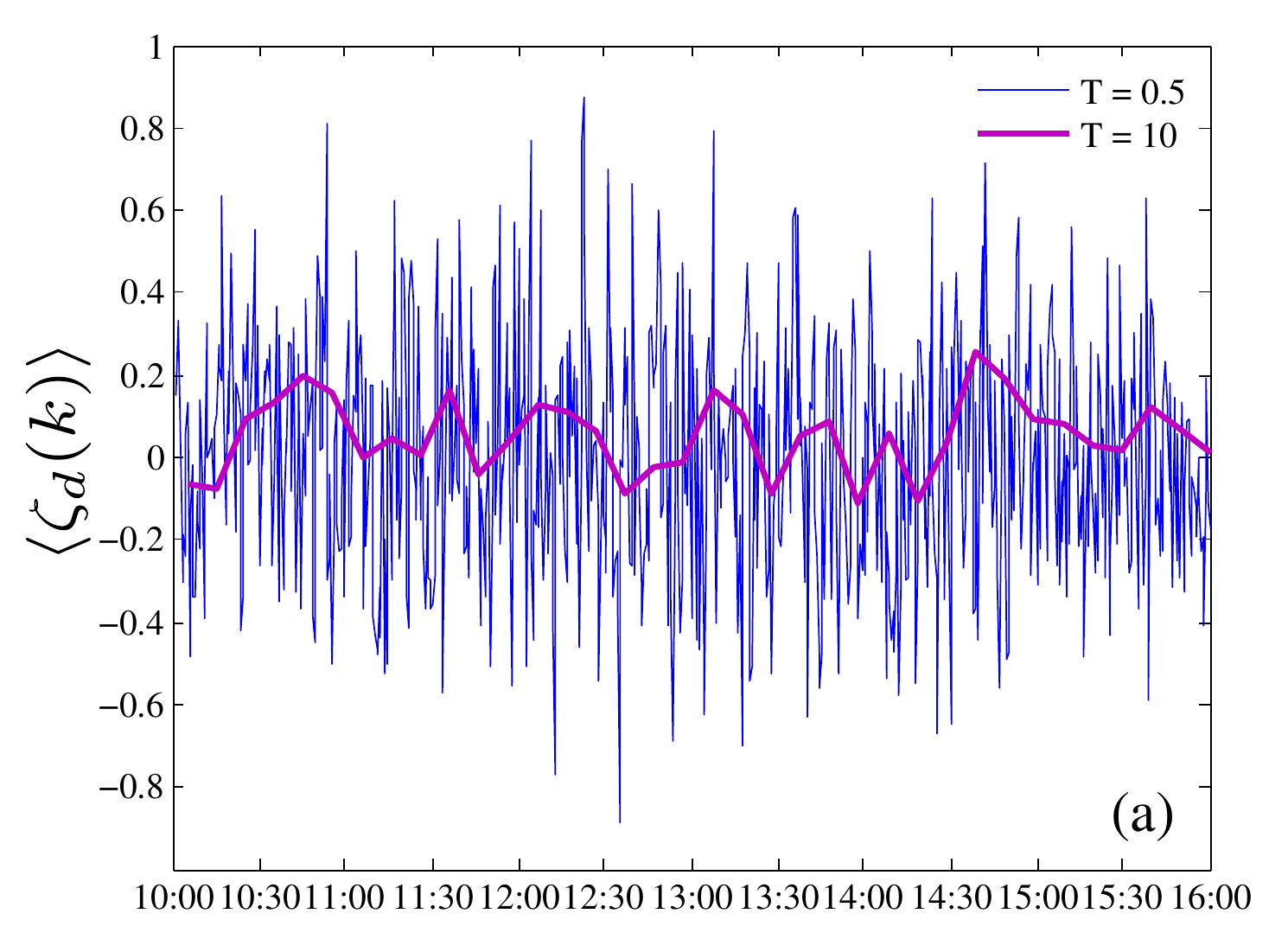}}
			\subfigure{\includegraphics [width=0.48\textwidth] {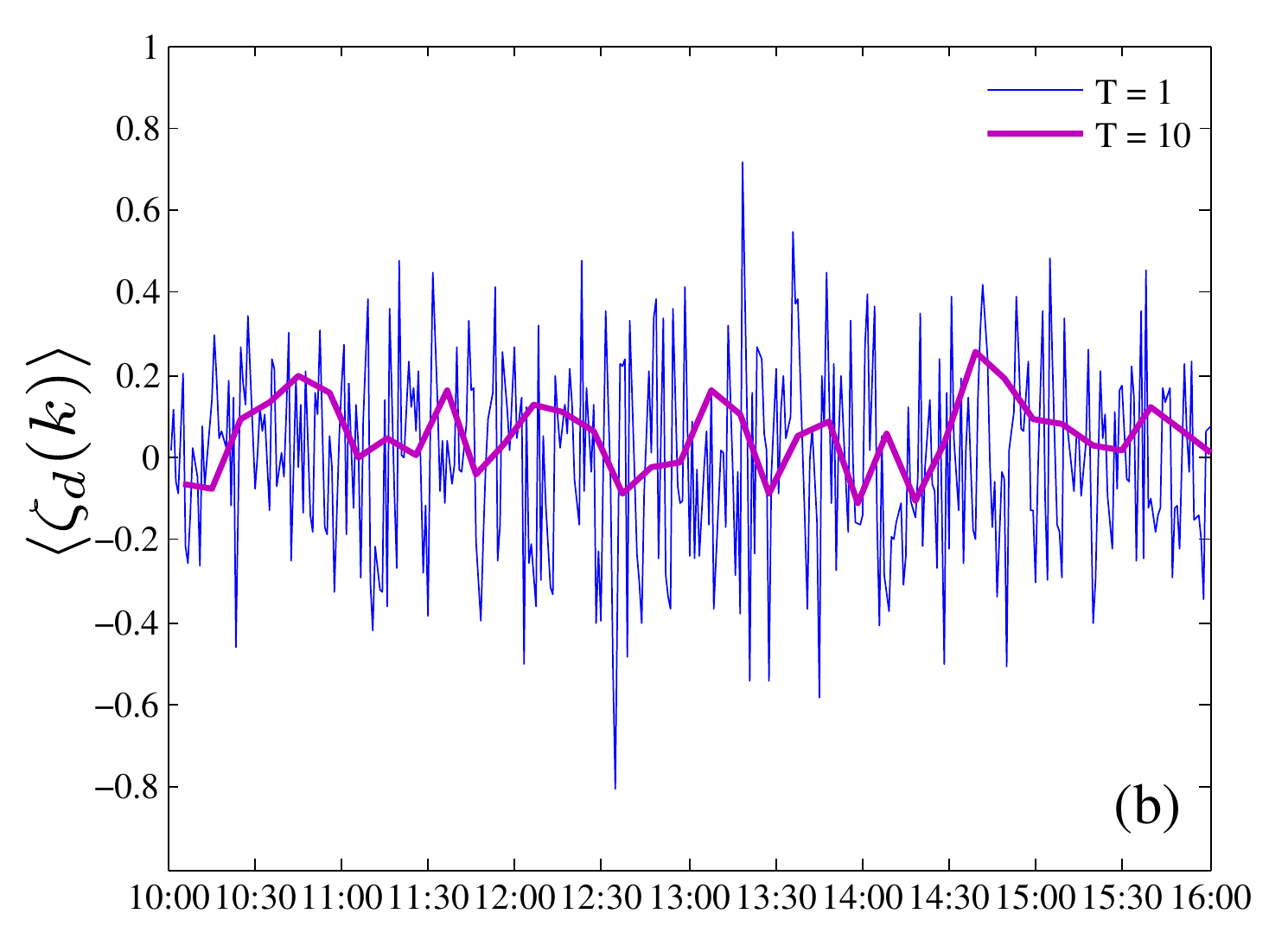}}                 
    		\end{center}
    		\caption[\quad Time Average of the Cross-Sectional Skewness]{Time average of the cross-sectional skewness: Comparison of the intra-day patterns for \textbf{(a)} $T=0.5$ and \textbf{(b)} $T=1$ against $T = 10$ minute bin for the S\&P $500$.}
\label{fig:Fig10E}
\end{figure}

\begin{figure}[h!]
\centering
\includegraphics [width=0.55\textwidth]{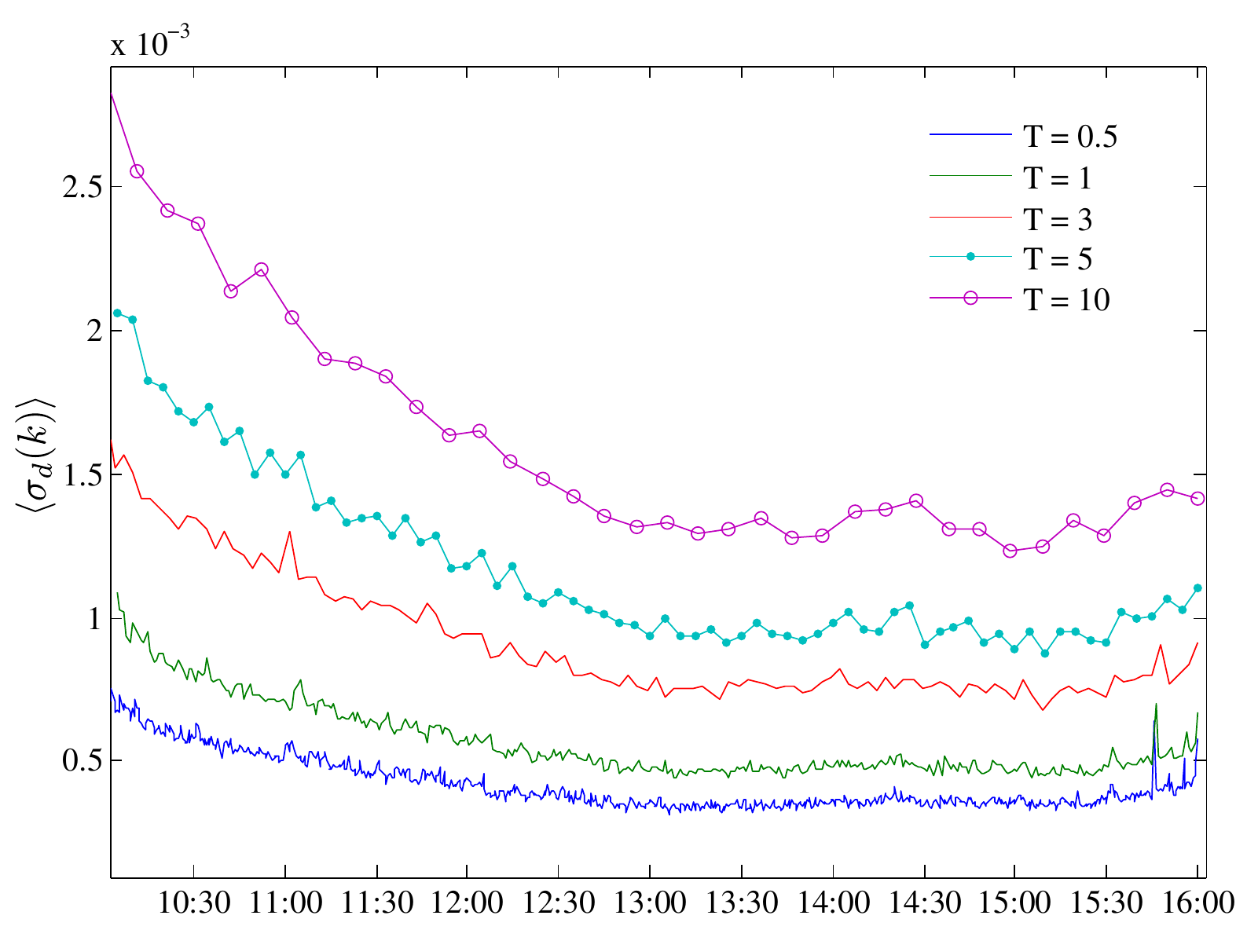}
\caption[\quad Bin Size Dependence in the U-Pattern Volatilities]{Bin size dependence in the U-pattern volatilities: Time average of the cross-sectional volatility for the S\&P $500$ for $5$ different values of bin size $T$.}
\label{fig:Fig11E}
\end{figure}

\begin{figure}[h!]
\centering
\includegraphics[width=0.55\textwidth] {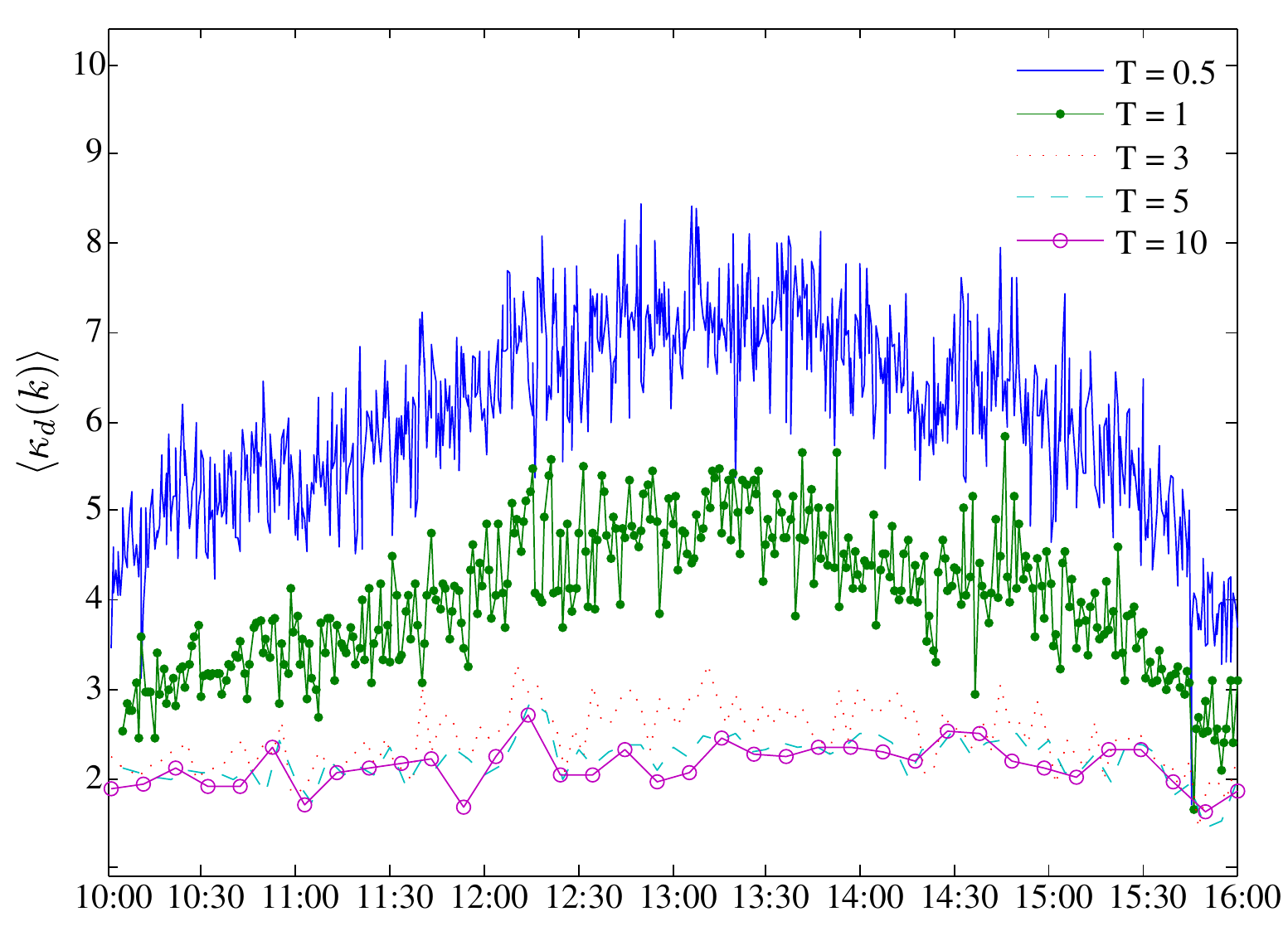}
\caption[\quad Bin Size Dependence in the Inverted U-Pattern Kurtosis]{Bin size dependence in the inverted U-pattern kurtosis: Time average of the cross-sectional kurtosis for the S\&P $500$ for $5$ different values of bin size $T$.}
\label{fig:Fig12E}
\end{figure}

As we saw in the last section, the volatilities for the relative prices
exhibit the same kind of intra-day pattern (Fig.~\ref{fig:Fig9E}). This intra-day seasonality is independent of the size of the bin, and the index we consider, but characteristic for each index. Actually, this is not true in the case of the returns as we already suggested in Sec.~\ref{sec:U} from Fig.~\ref{fig:Fig4E}. If we consider the odd moments (mean and skewness) of the returns, the behavior is basically the same (noisy around zero) and without any particular pattern, independently of the bin size (as can be seen in Figs.~\ref{fig:Fig2E},~\ref{fig:Fig3E} and~\ref{fig:Fig10E}). 
But for the case of the even moments of returns, although they exhibit the well known U and inverted U-patterns, these patterns depend on the bin size. This fact is well illustrated through Figs.~\ref{fig:Fig11E} and~\ref{fig:Fig12E} where we have chosen $5$ different values of bin size from $T=0.5$ to $T=10$ minutes. In these figures we show the time average of the cross-sectional volatility and kurtosis for the S\&P $500$  but a similar bin size dependence can be shown for the CAC $40$ or any other index and also for the time average of the single stock volatility and kurtosis.
By other hand the kurtosis is a decreasing function of the size of the bin and the inverted U-pattern is evident just when we consider ``small'' bin sizes, in our case this occurs for $T=1$ and $T=0.5$ minute bin (Fig.~\ref{fig:Fig12E}). This represents a confirmation that on small scales the returns have heavier tails, and on long time scales they are more Gaussian~\cite{4, 8, 9, 10}.

\section[\quad Intra-day Abnormal Patterns]{\quad Intra-day Abnormal Patterns}
\label{AP}
\begin{figure}[t!]
        \begin{center}
        \subfigure[MEAN]{\includegraphics [width=0.48\textwidth] {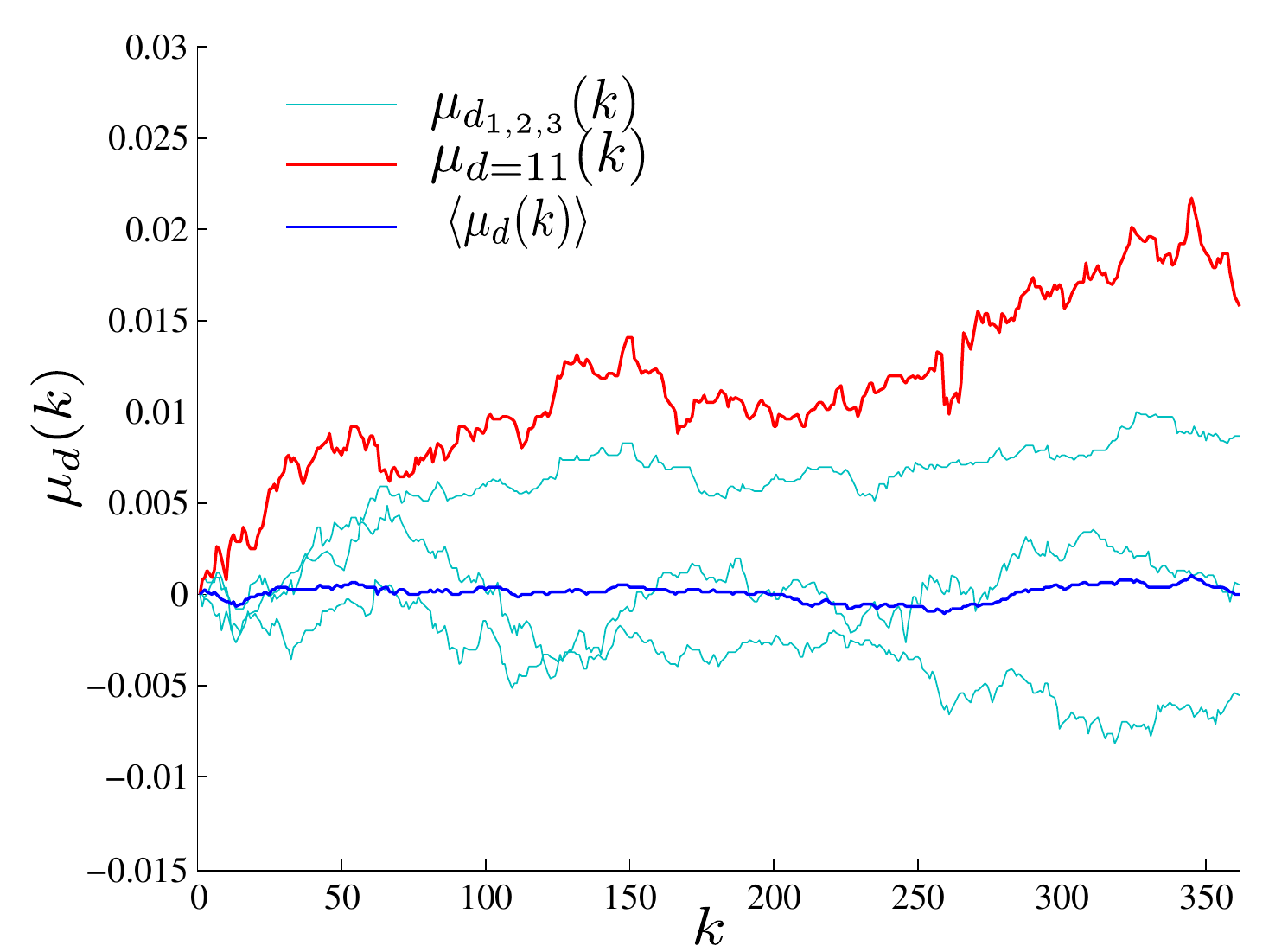}}
        \subfigure[VOLATILITY]{\includegraphics [width=0.48\textwidth]{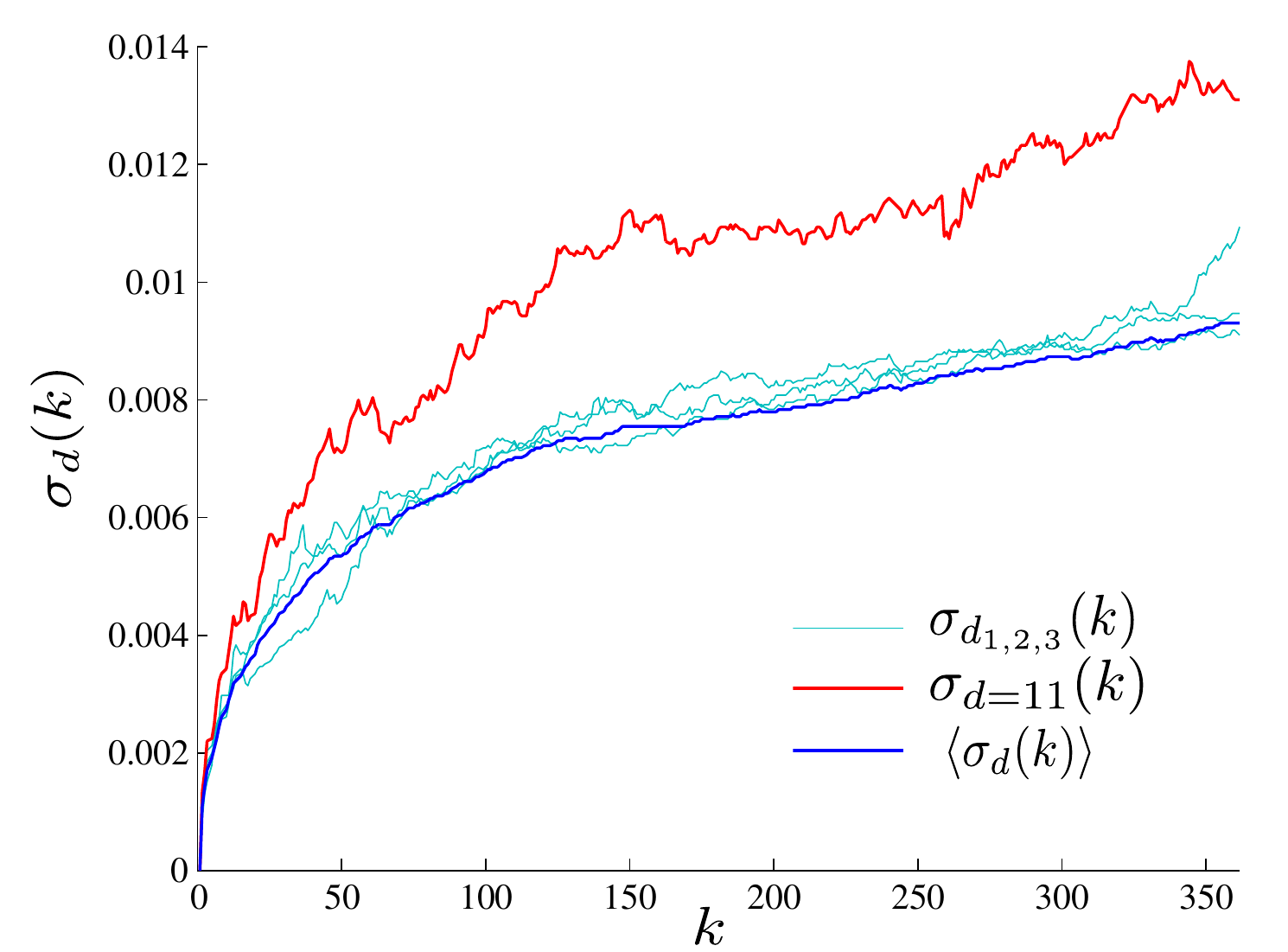}}                  
    	\end{center}
    \caption[\quad Atypical Day]{S\&P $500$ Atypical Day: Time average of the cross-sectional mean and volatility (blue), cross-sectional mean and volatility of the S\&P $500$ during day $11$ (red) and during three days chosen at random (clear blue).}
\label{fig:Fig13E}
\end{figure}

One of the motivations to explore into the intra-day seasonalities for relative prices was due to Kaisoji's previous work~\cite{12}. In his work he found that the upper tail of the complementary cumulative
distribution function of the ensemble of the relative prices in the high value of the price is well described by a power-law distribution which when its exponent approached two, the Japan's internet bubble burst.
Taking into consideration our recent findings we suggest the use of the bin size independence for intra-day patterns in relative prices in order to characterize ``atypical days'' for indexes and ``anomalous behaviors'' for stocks. 
\newpage
The time average of the cross-sectional moments represents the average behavior of a particular index moment during an average day. In Fig.~\ref{fig:Fig8E} each path represents the evolution of a particular index moment for one of the days of the period under analysis (i.e., one path, one day moment). If we look directly into the prices of the CAC $40$ and S\&P $500$, we can observe during day $11$ a fall of the prices of the stocks that compose both indexes. During the days before and following day $11$, the (index) moments move along our intra-day pattern. Moreover, if we pick randomly one day from our period of analysis, in most of the cases our index during that day will behave as our intra-day seasonality (as in Fig.~\ref{fig:Fig13E}), but the one for day $11$ will not. In Fig.~\ref{fig:Fig13E} we show the (cross-sectional) intra-day seasonalities  for the (a) mean and (b) volatility in blue and in clear blue the respective cross-sectional stock moments for 3 days randomly picked. The average behavior (of the moments) of our index during these days moves along with our intra-day pattern. This is not the case of the curve corresponding to the day $11$ shown in red which clearly diverges from the expected behavior. This is what could be called as an ``atypical day'' for the S\&P $500$.

\begin{figure}[t!]
        \begin{center}
        \subfigure[MEAN]{\includegraphics [width=0.48\textwidth] {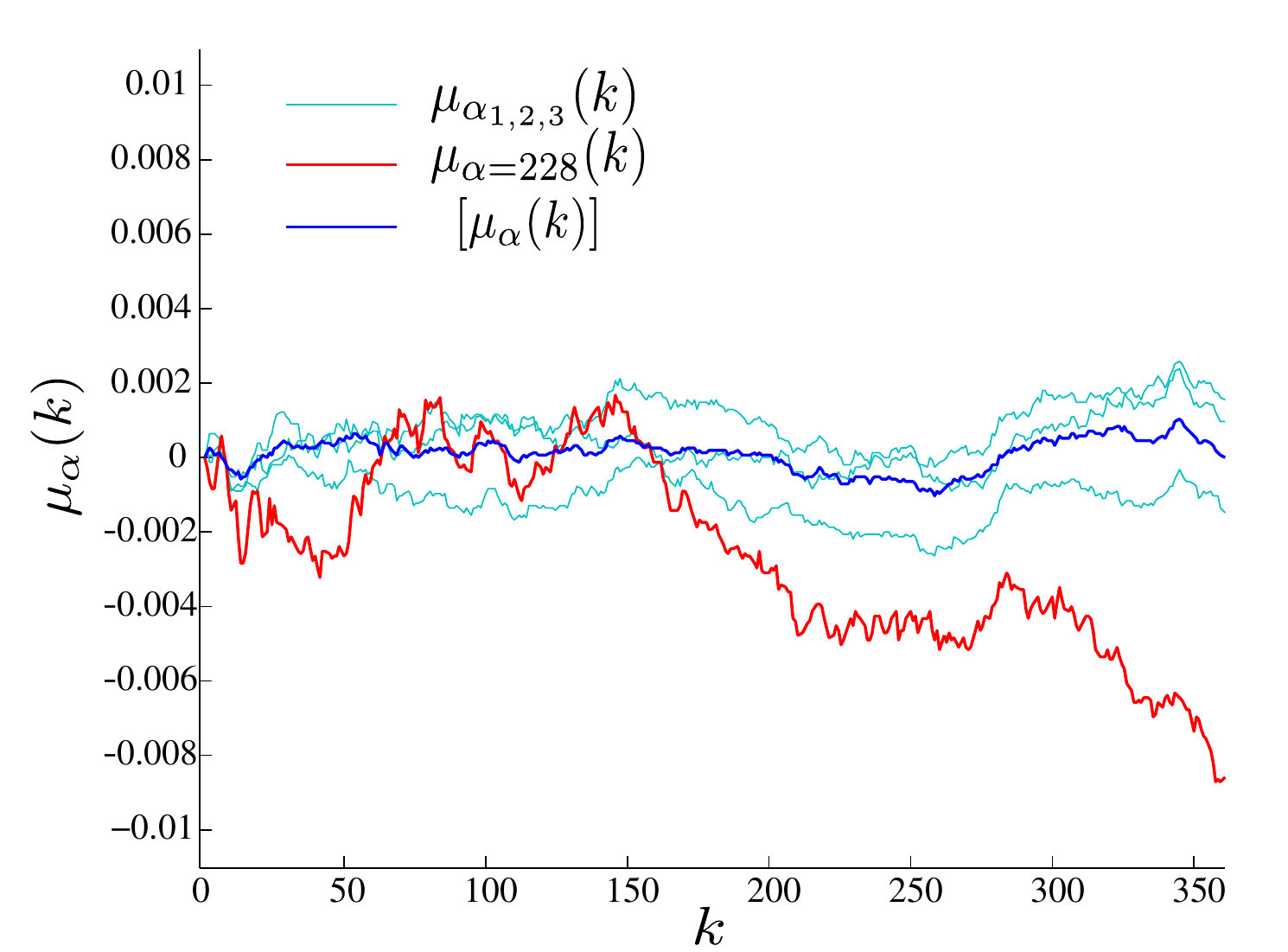}}
		\subfigure[VOLATILITY]{\includegraphics [width=0.48\textwidth] {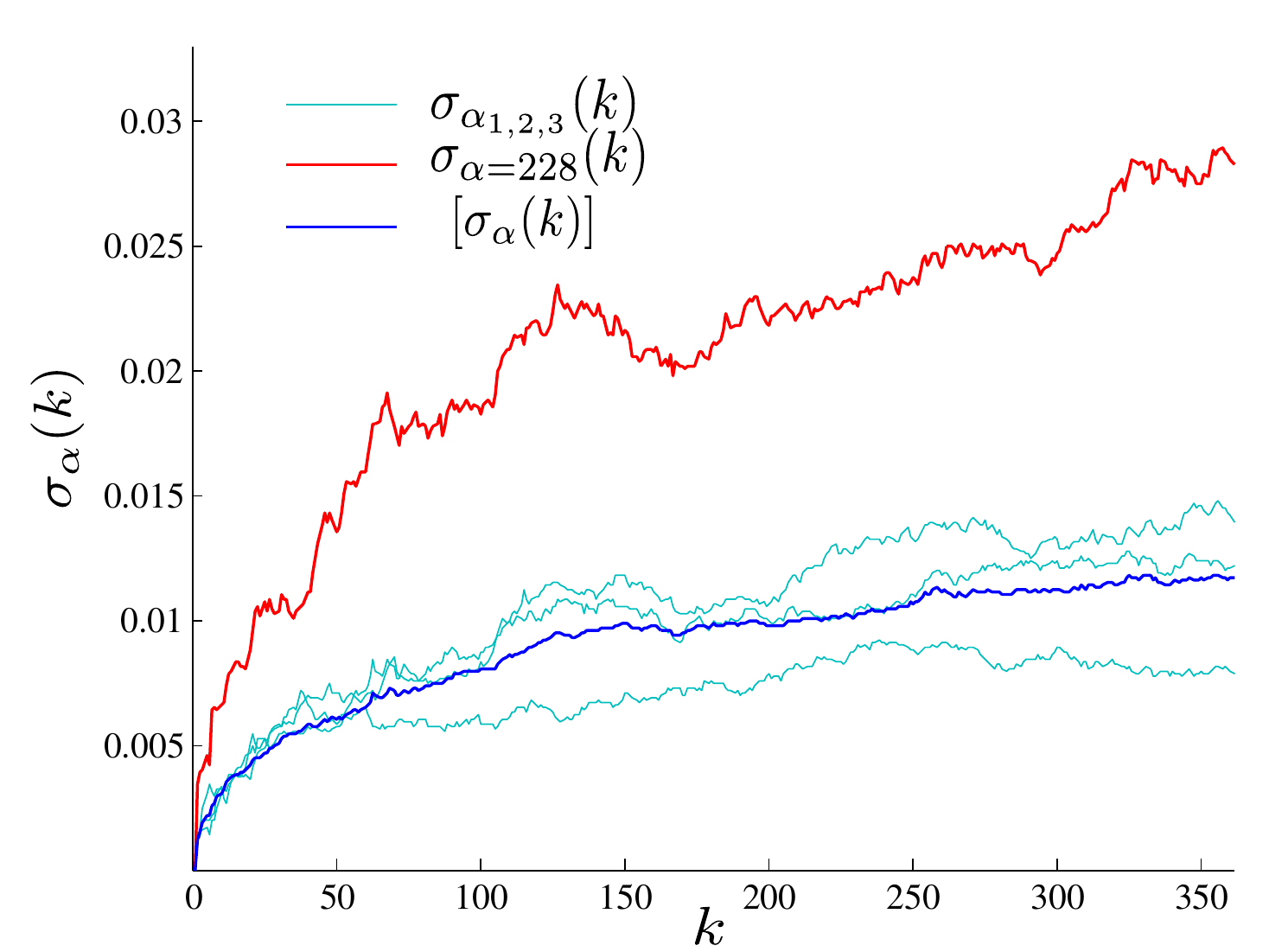}}\\ 
    	\end{center}
    \caption[\quad Anomalous Stock Behavior]{Anomalous Stock Behavior: Stock average of the single stock mean and volatility (blue), single stock mean and volatility of stock $228$ (red) and from three stocks chosen at random from the S\&P $500$ (clear blue).}
\label{fig:Fig14E}
\end{figure}

We could used the same reasoning as before in order to characterize ``anomalous behaviors'' in stocks. Each path in Fig.~\ref{fig:Fig7E} represents the average evolution of a particular moment of one of the stocks that compose the S\&P $500$. The stock average of those single stock moments represents the average behavior of that moment for an average stock during an average day of our period of analysis. 
Meaning that if we pick randomly one stock from our set of stocks, in most of the cases (its moments) will behave as the intra-day seasonality. This is clearly illustrated in Fig.~\ref{fig:Fig14E} where present the intra-day seasonalities for the (a) mean and (b) volatility in blue and the respective single stock moments for 3 stocks randomly picked in clear blue. As can be seen, the average behavior of the moments of these stocks move along with our intra-day patterns. However this is not the case for the curves shown in red which have been chosen on purpose to illustrate how in this case the stock $228$ behaves in an anomalous way with respect to what is expected.

\section[\quad Discussion]{\quad Discussion}
In this chapter, we have analyzed the intra-day seasonalities of the single and cross-sectional (or collective) stock dynamics by the evolution of the moments of
its returns (and relative prices) during a typical day. What we have
called ``single stock intra-day seasonalities'' is the average behavior of the
moments of the returns (and relative prices) of an average stock in an
average day. In the same way, the cross-sectional intra-day seasonality is
not more than the average day behavior of an index moment. We presented these intra-day seasonalities for returns (Figs.~\ref{fig:Fig2E} and~\ref{fig:Fig3E}) and relative prices (Figs.~\ref{fig:Fig7E} and~\ref{fig:Fig8E}) and compared the stock average of single stock volatility $\left[\sigma _{\alpha }(k)\right]$, the time average of the cross-sectional volatility $\langle\sigma _{d}(k,t)\rangle$ and the average absolute value of the equi-weighted index $\langle|\mu _{d}|\rangle$ (Figs.~\ref{fig:Fig4E} and~\ref{fig:Fig9E}). 

Notably, in the case of the returns, is that these ``patterns'' actually depend on the size of the bin. This fact was well illustrated with $5$ different values of bin size through Fig.~\ref{fig:Fig11E} for volatilities and Fig.~\ref{fig:Fig12E} for kurtosis in which its inverted U-pattern is evident just when we consider ``small'' bin sizes.

In the case of relative prices, the volatilities also  
exhibit the same kind of intra-day pattern (Fig.~\ref{fig:Fig9E}), but contrary with the returns, it is independent of the size of the bin, and the index we consider, but characteristic for each index. We suggested in Sec.~\ref{AP} how this bin size independence of intra-day patterns in relative prices could be used in order to characterize ``atypical days'' for indexes and ``anomalous behaviors'' in stocks. This was presented in Figs.~\ref{fig:Fig13E} and~\ref{fig:Fig14E} where we presented our intra-day seasonalities for the (a) mean and (b) volatility in blue and the respective the cross-sectional moments for 3 days (and the single stock moments for 3 stocks) randomly picked in clear blue and we saw how the average behavior of their moments move along with our intra-day patterns which was not the case for the day $11$ and the stock $228$.

\chapter[\quad Conclusion]{Conclusion}
\label{chap:Conclusion}

In this thesis, 
we studied issues that arise from the evaluation of the large deviation function (LDF) from population dynamics algorithms. 
Different versions of the cloning algorithm were used which differ among them in crucial aspects as
the way in which the selection mechanism is performed or on the restriction in the growth of the total population of copies of the system. 
The LDF behaviour 
for these different versions and its related features were analyzed. 
We gave particular attention to the dependence of the estimator with number of clones $N_c$ and the simulation time $t$ (the two parameters introduced by the method) by studying the finite-$t$ and finite-$N_c$ effects, its convergence in the infinite-$t$ and infinite-$N_c$ limit as well as its behavior in the large system size $L$ limit. Moreover, different ways and methods to improve the LDF estimation 
were proposed.

In chapter~\ref{chap:Discreteness}~\cite{hidalgo_discreteness_2016} using a non-constant population approach of the cloning algorithm, 
we analyzed the discreteness effects at initial times in population dynamics.
We show how these effects play an important role in the determination of the
large deviation function which may be obtained from the growth rate of an average log-population. Fluctuations at initial
times produce that some populations remain in their initial states for much longer than
others, producing a gap in their individual evolution. This induces a relative shift between populations that
lasts forever which supplemented with a short-time evolution affect strongly the average population and thus the LDF estimation. We argue in Sec.~\ref{Time Delay Correction} that these lags between populations could be compensated by performing a time translation (Eq.~\eqref{eq:13}) 
over populations in order to emphasize the effects of the exponential growth regime. This along with a discarding of initial regimes in the evolution of the population surpasses the influence of initial discreteness effects.

The finite-$t$ and finite-$N_c$ scalings in the evaluation of large deviation functions were studied in chapters~\ref{chap:DiscreteTime}~\cite{partI}, \ref{chap:ContinuousTime}~\cite{partII} and \ref{chap:CGF}~\cite{partII} following two different approaches:
an analytical one, in chapter~\ref{chap:DiscreteTime}, using a discrete-time version of the population dynamics algorithm~\cite{giardina_direct_2006}, and a numerical one, in chapters~\ref{chap:ContinuousTime} 
and~\ref{chap:CGF}, 
using a continuous-time version~\cite{lecomte_numerical_2007,tailleur_simulation_2009}. 
In both cases, we derived that the deviations of large deviation estimator from the desired value (which we called systematic errors) were small and behaved as $1/t$ and $1/N_c$ in the large-$t$ and large-$N_c$ asymptotics respectively. Importantly, in chapter~\ref{chap:ContinuousTime}~\cite{partII} we showed the validity of these results in more complex systems. 
Such scalings also provided a convergence criterion to the asymptotic regimes of the algorithm: In order to ensure a correct LDF evaluation, one has to confirm that the LDF estimator does present corrections (first) in $1/t$ and (second) in $1/N_c$ with respect to an asymptotic value.  
We discussed in Secs.~\ref{subsubsec:differenceContinuousTime} and~\ref{Discrete-time_algorithm} how these two versions 
differ on a crucial point which makes that an extension of the analysis developed in chapter~\ref{chap:DiscreteTime} cannot be done straightforwardly in order to comprehend the continuous-time case in chapter~\ref{chap:ContinuousTime} and thus the observation of these scalings themselves is also non-trivial.
This finite-$t$ and finite-$N_c$ scaling behavior was used in chapter~\ref{chap:ContinuousTime}~\cite{partII} in order interpolate the large-$t$ and large-$N_c$ asymptotic value of the LDF estimator from the measured values for finite and small $t$ and $N_c$. This allowed us to propose an improved version of the continuous-time cloning algorithm in Sec.~\ref{ssec:SM} providing more reliable results, 
%
less affected by finite-$t$ and -$N_c$ effects. 
We demonstrated numerically that the interpolation technique is very efficient, by a direct comparison of the resulting LDF estimation with the analytical value, which can be determined in the studied system.
However, the validity of the method and of these scalings were proved only for a simple one-site annihilation-creation dynamics and for a contact process with $L=6$ sites, leaving an analysis of the dependence 
with the system size (number of sites) $L$ pending.

In order to prove whether the finite-$t$ and -$N_c$ scalings 
observed in small (number of sites $L$) systems are also valid in the large-$L$ limit,
we redefined these scalings in a more general way in chapter~\ref{chap:LargeL}~\cite{largeLCP}.
We assumed a
$t^{-\gamma_{t}}$~\eqref{eq:tScal2} 
and a $N_{c}^{-\gamma_{N_{c}}}$~\eqref{eq:nScal2} -scaling behavior 
for the LDF estimator. 
This redefinition allowed us to verify in large-$L$ systems if effectively $\gamma_{t} \approx 1$ and $\gamma_{N_{c}} \approx 1$ and
whether the extracted quantities from the application of the scaling method 
represented the limits in $t \to \infty$ and $N_{c} \to \infty $. 
%
First, we considered a contact process with $L=100$ sites and two representative values of the parameter $s$.
Although the $t^{-1}$-scaling and $N_{c}^{-1}$-scaling were proved to hold for $s < 0$, this was not the case for $s > 0$, being this fact valid in general for large-$L$ systems. 
%
%
%
As the scaling method relied on the validity of the $t^{-1}$- and $N_{c}^{-1}$-scalings, 
in Sec.~\ref{sec: SML100} we showed how the determination of the infinite-$t$ and infinite-$N_c$ limit of the LDF estimator is affected.  
%
In order to have a clear picture of the change in the scalings of the LDF estimator, the analysis was extended to the plane $s-L$ 
where the exponents $\gamma_{t}$ and $\gamma_{N_c}$ were computed and characterized for a grid of values of the parameters $(s,L)$. 
Moreover, we discussed how this breakdown in the scalings in the large-$L$ limit could be related to the dynamical phase transition of the contact process. 

Although our study on the cloning algorithm is closed in chapter~\ref{chap:LargeL}~\cite{largeLCP}, the study of rare events is complemented with chapter~\ref{chap:Intraday}~\cite{binsize} using a completely different approach. 
This is the empirical 
study of the patterns that hide behind financial time series, known as stylized facts.
We analyzed the intra-day seasonalities of the single and cross-sectional (or collective) stock dynamics by characterizing the
dynamics of a stock (or a set of stocks) by the evolution of the moments of
its returns (and relative prices) during a typical day.
We showed how these patterns actually depend on the size of the bin in the case of the returns.
%
However, in the case of relative prices, 
these patterns are 
independent of the size of the bin, and the index we consider, but characteristic for each index. We suggested in Sec.~\ref{AP} how this bin size independence of intra-day patterns in relative prices could be used in order to characterize ``atypical days'' for indexes and ``anomalous behaviours'' in stocks.

\chapter[\quad Perspectives]{Perspectives}
\label{chap:Perspectives}
Below we mention some questions that arose from our study which remain open and may constitute possible directions for future research.

The analysis of the discreteness effects at initial times in population dynamics developed in chapter~\ref{chap:Discreteness}~\cite{hidalgo_discreteness_2016} (using a non-constant population approach of the cloning algorithm), was performed only on a simple system: a one-site annihilation-creation dynamics (Sec.~\ref{sec:bdp}). However we hope it can be extended to more complex phenomena so that our results can be verified or else, more interesting features can be found. Nevertheless even for that simple system there remain pending issues. Some of them related to the fact that the duration of the initial discrete-population regime could be understood from an analytical study of the population dynamics itself. On the other hand, the results presented support a power-law behaviour in time of the variance of the delays. Additionally, the distribution of the delays was found to take an universal form, after rescaling the variance. Both of which could be explored deeply.

From a constant population approach, as the one used in chapters~\ref{chap:DiscreteTime},~\ref{chap:ContinuousTime},~\ref{chap:CGF} and~\ref{chap:LargeL}, is still possible to reconstruct the evolution in time of the population of clones. Thus, it would result interesting to compare both approaches but importantly, the properties of the reconstructed populations in contrast with actual populations (obtained from a non-constant population as in chapter~\ref{chap:Discreteness}). 

From the analytical study of the finite-$t$ and -$N_c$ scalings of the LDF estimator developed in chapter~\ref{chap:DiscreteTime}~\cite{partI}, we mention two open questions. 
The first is related to the precise estimate of the error due to a non-infinitesimal time interval $\Delta t$ between cloning steps:
As explained in Sec.~\ref{subsubsec:dtDeltat} and Sec.~\ref{subsubsec:differenceContinuousTime}, 
taking the $\Delta t \rightarrow 0$ limit is important in our analysis, in order to make the estimator converge to the correct LDF. 
%
%
From a practical point of view, taking this limit can however be problematic, since it requires infinitely many cloning procedures per unit time (as $\Delta t\to 0$). 
%
 Interestingly, most of existing algorithms do not take such a limit (for instance the original version of the algorithm, Ref.~\cite{giardina_direct_2006}). Empirically, one thus expects that the error goes to zero as $N_c\to\infty$ while keeping $\Delta t$ finite. 
%
%
Within the method developed in chapter~\ref{chap:DiscreteTime}~\cite{partI} the analytical estimation of this error is challenging (see Sec.~\ref{subsubsec:differenceContinuousTime}) and remains an open problem. 

The second question is related with possible extensions of the formulation developed in chapter~\ref{chap:DiscreteTime}~\cite{partI}. As the cloning procedure is performed for a fixed time interval, the formulation cannot cover the case of algorithms where $\Delta t$ itself is statistically distributed, as in continuous-time cloning algorithms~\cite{lecomte_numerical_2007}. Moreover, the formulation is limited to Markov systems, although population dynamics algorithms are applied to chaotic deterministic dynamics~\cite{tailleur_probing_2007,1751-8121-46-25-254002_2013} or to non-Markovian evolutions~\cite{Non-MarkovPD}. Once one removes the Markov condition in the dynamics, developing analytical approaches becomes more challenging. However, as the physics of those systems are important scientifically and industrially~\cite{PhysRevLett.116.150002}, the understanding of such dynamics cannot be avoided for the further development of population algorithms.

Results evident to question whether  
the numerical study developed in (specially) chapter~\ref{chap:ContinuousTime}~\cite{partII} can be extended to systems presenting dynamical phase transition (DPT) in the form of a non-analyticity of the LDF. In particular, in this context, it would be useful to understand how the dynamical phase transition of the original system translates into anomalous features of the distribution of the LDF estimator in the cloning algorithm. 
Although the system used in chapter~\ref{chap:ContinuousTime}~\cite{partII} (the contact process, Sec.~\ref{sec:CP}) is know to exhibit a DPT in the $L \to \infty$ limit~\cite{lecomte_numerical_2007, Lecomte2007, ThermoCP, marro_dickman_1999}, the finite-$t$ and -$N_c$ scalings of the LDF estimator were studied on a small system with $L=6$ sites, for which the effects of the DPT cannot be observed. 
In chapter~\ref{chap:LargeL}~\cite{largeLCP} we extended our analysis to a large-$L$ contact process ($L=100$ sites, where DPT manifests~\cite{0295-5075-116-5-50009, lecomte_numerical_2007}) showing 
evidence of a changing in the LDF scalings with the size of the system $L$ which could be related to a DPT.
However, a study of the DPT effects
would require
a large-$N_c$ and -$t$ configuration, which under our approach
was a task not possible to fulfill (as the main objective in chapters~\ref{chap:ContinuousTime}~\cite{partII} and~\ref{chap:LargeL}~\cite{largeLCP} 
was the possibility of
extracting the infinite-$N_c$ infinite-$t$ limit of the LDF estimator 
from data for a small number of clones $N_c$ and time $t$). 

Taking in consideration that 
is well know that 
the existing methods
~\cite{giardina_direct_2006, Hedges1309, PitardDT, SpeckDPT, Speck}
perform poorly in the vicinity of a dynamical phase transition, 
or they are 
numerically expensive in order to obtain accurate estimations ~\cite{SpeckDPT, LimmerIce, DPTpath} developing if not important finite-size effects~\cite{hurtado_current_2009}, the analysis of this problem in the large-$t$ and -$N_c$ limits is not necessarily the best option. 
Recently has been proposed a promising method~\cite{nemoto_population-dynamics_2016, PhysRevLett.118.115702} which combines the existing cloning algorithm~\cite{giardina_direct_2006, giardina_simulating_2011, tailleur_simulation_2009, lecomte_numerical_2007, Hedges1309, PitardDT, SpeckDPT, Speck} with a modification of
the dynamics~\cite{jack_large_2010, 1742-5468-2010-10-P10007, PhysRevLett.111.120601, PhysRevLett.112.090602,Jack2015,RayImportance,RayExactFluct} resulting in a significant improvement of its computational efficiency. The method was successfully applied to the study of the dynamical phase transition of 1D FA model~\cite{FAmodel} using a relatively small $N_c$ and $L$. 
The implementation of this method will 
provide 
in a next stage
a clear contrast between the results obtained following the two different approaches and a better understanding of their limitations and advantages.

\clearpage
%

\backmatter

\renewcommand{\bibname}{Publications \label{pubs}}

\addcontentsline{toc}{chapter}{Publications}
\label{Publications}

\clearpage
\thispagestyle{empty}
\phantom{a}


\nocite{*} 
\cleardoublepage
\phantomsection

\renewcommand{\bibname}{Bibliography}
\addcontentsline{toc}{chapter}{Bibliography}
\bibliography{RefThesys,RefP0}
\end{document}